\documentclass[prl,aps,twocolumn,superscriptaddress,showpacs,showkeys]{revtex4-1}
\usepackage{amsmath,amsthm,amssymb,graphicx,subfigure,xcolor,bbm}
\usepackage{esint}
\usepackage[unicode=true]{hyperref}
\hypersetup{
     colorlinks=true,       		
     linkcolor=blue,          	
     citecolor=red,            
     urlcolor=magenta,           	
}
\newtheorem*{theorem*}{Theorem}

\newtheorem*{corollary*}{Corollary}

\newtheorem*{lemma*}{Lemma}

\newtheorem*{proposition*}{Proposition}

\newtheorem*{conjecture*}{Conjecture}
\theoremstyle{definition}

\newtheorem*{definition*}{Definition}
\theoremstyle{remark}

\newtheorem*{remark*}{Remark}

\newcommand{\Eq}[1]{Eq.~\eqref{#1}}

\begin{document}
\renewcommand{\figurename}{Fig.}
\renewcommand{\figureautorefname}{Fig.}
\renewcommand{\tablename}{Tab.}
\renewcommand{\tableautorefname}{Tab.}

\title{Predicting Angular-Momentum Waves Based on Yang-Mills Equations}

\author{Xing-Yan Fan}
\affiliation{Theoretical Physics Division, Chern Institute of Mathematics,
   Nankai University, Tianjin 300071, People's Republic of China}

\author{Xiang-Ru Xie}
\affiliation{School of Physics, Nankai University, Tianjin 300071, People's
   Republic of China}

\author{Jing-Ling Chen}
\email{chenjl@nankai.edu.cn}
\affiliation{Theoretical Physics Division, Chern Institute of Mathematics,
      Nankai University, Tianjin 300071, People's Republic of China}

\date{\today}
\begin{abstract}
   As one of the most elegant theories in physics, Yang-Mills (YM) theory not only incorporates Maxwell's equations unifying electromagnetism, but also underpins the standard model explaining the electroweak and strong interactions in a succinct way. Whereas the highly nonlinear terms in YM equations involving the interactions between potentials and fields retard the resolution for them. In the $U(1)$ case, the solutions of Maxwell's equations are the electromagnetic waves, which have been applied extensively in the modern communication networks all over the world. Likewise the operator solutions of the YM equations under the assumptions of weak-coupling and zero-coupling predict the $SU(2)$ angular-momentum waves, which is the staple of this work. Such angular-momentum waves are hopefully realized in the experiments through the oscillations of spin angular momentum, such as the ``spin Zitterbewegung'' of Dirac's electron.
\end{abstract}

\maketitle

\emph{Introduction.---}J. C. Maxwell finished the unification of the classical
      electromagnetic phenomena through a set of interconnected equations (i.e., \emph{Maxwell's equations}) after published his two volumes monumental selected papers \cite{Maxwell}, where the conception that energy consists in \emph{field} was first proposed \cite{2014PTodaYang}. Subsequently, Maxwell's equations were used as the bases for deducing electrodynamics, which can be found in every relevant textbook at the level of senior undergraduates in physics, such as \cite{GreinerCED,1999Jackson}. As a direct but significant prediction of Maxwell's equations, the \emph{electromagnetic waves} (EMW) excited by according electromagnetic fields are experimentally verified via H. R. Hertz \cite{Hertz}, and then lead the reformation in communication around the world.

      The investigation into Maxwell's equations draws much attention. For instance, the multipolar representation of Maxwell's equations was contributed to solve the problem of electromagnetic fields interacting with nonrelativistic electrons depicted by the Hamiltonian in multipolar form \cite{1982PRAHaller,1982ComtHealy,1982PRAPower}. The hydrodynamic Maxwell's equations were studied in nematic liquid crystals \cite{1994PRELiu,1995ComtBrand}. Moreover the essence of light wave is electromagnetic \cite{1888hertz}, thus Maxwell's equations are diffusely researched in optics, such as, cylindrical nonlinear Maxwell's equations in nonlinear optics \cite{2012PREXiong}, and the numerical simulation for the knotted solutions of Maxwell's equations \cite{2020PRE}.

      However all suchlike researches on Maxwell's equations are common resolution in certain situations, and no new physical insights are given for describing more interactions in nature. The cusp appears in 1954, inspired by the concept of \emph{gauge invariance} \cite{2001RMPGaugInvar} inherent in Maxwell's equations, Yang and Mills generalized it to the area of isotope \cite{1954YangMills}, which declared the birth of non-Abelian gauge theory. Further the standard model, as a type of $U(1)\times SU(2)\times SU(3)$ gauge field, accounted for the electroweak and strong interactions concisely \cite{SchwartzQFT}, claiming that three of four basic interactions in our universe so far have been consolidated through Yang-Mills (YM) theory.

      Henceforth the YM theory has been carried over a wide range of physics, one of the representatives is the lattice gauge theory. It can be applied in spin system \cite{1979RMPLatGauge}, modeling solids in the laboratory. With the advent of quantum simulation \cite{2014RMPQSimu}, the gauge fields become more realistic in neutral atoms \cite{2011RMPAGaugPoten}. In addition, the YM theory features in some interdisciplines, for examples, thermodynamics \cite{2022PRLThermYM} and cosmology \cite{2022PRDCosmYM}.

      The core problem in YM theory is to solve the \emph{Yang-Mills equations}. Although they are intractable to be solved for the existence of self-interaction terms between \emph{potentials} and fields, much work has been done in the classical solutions of YM equations \cite{1979RMPCYangMills,1979PLBOh,1985JMPOh,1986PRDColor1,1986PRDColor2,2023PRDShreecharan}. After that, in the light of Feynman, if one knows the information of all classical field configurations, then the knowledge pertaining to quantum theory can be attained via path integrals \cite{1995IntroQFT}.

       Since Maxwell's equations predict the well-known electromagnetic waves, and the YM equations are their natural generalization. It gives rise to a scientific question: What types of waves do the YM equations predict?
      The purpose of this work is to study operator solutions of YM equations and predict the angular-momentum waves (AMW). The work is organized as follows.
      First we review the YM equations in vacuum without sources and work out the conditions needed for the exact solutions. Then we explore the operator solutions of YM equations under weak-coupling approximation (WCA) and zero-coupling approximation (ZCA), respectively. Afterwards we discuss the $SU(2)$ angular-momentum waves and a possible way to detect them. Finally we conclude the whole work.


\emph{Yang-Mills Equations in Vacuum without Sources.---}
      Throughout the article, we adopt the convention of Jackson's \cite{1999Jackson}: the Greek indices take values $\{0,1,2,3\}$, and the Latin ones take $\{1,2,3\}$ (or $\{x,y,z\}$ without ambiguity). The contravariant coordinates in four-dimensional spacetime read $\bigl(x^\mu\bigr)=\bigl(c\,t,\:\vec{x}\bigr)$, while their covariant partner $\bigl(x_\mu\bigr)=\bigl(c\,t,\:-\vec{x}\bigr)$ with the rule of transformation $x_\mu =g_{\mu\nu} x^\nu$, and $x^\mu =g^{\mu\nu} x_\nu$, where the metric tensor $g_{\mu\nu} =g^{\mu\nu} ={\rm diag}(1,-1,-1,-1)$, and $c$ is the speed of light in vacuum. Similarly, one has the potentials as
      $\bigl(\mathbb{A}^\mu\bigr)=\bigl(\varphi,\:\vec{A}\bigr)$, $\bigl(\mathbb{A}_\mu\bigr)=\bigl(\varphi,\:-\vec{A}\bigr)$,
      and the partial differential operators as $\partial^\mu \equiv {\partial}/{\partial\,x_\mu} =({\partial}/{\partial\,(ct)},-\vec{\nabla})$, $\partial_\mu \equiv {\partial}/{\partial\,x^\mu} =({\partial}/{\partial\,(ct)},\vec{\nabla})$,
      with $\vec{\nabla}=(\partial/\partial\,x,\:\partial/\partial\,y,\:\partial/\partial\,z)$.

      The YM equations (in vacuum without sources) are given by \cite{2002CTGaugeT}
      \begin{subequations}
            \begin{align}
                  & D_\mu \mathcal{F}^{\mu\nu} \equiv\partial_\mu \mathcal{F}^{\mu\nu}
                        -{\rm i}\,g\Bigl[\mathbb{A}_\mu ,\ \mathcal{F}^{\mu\nu}\Bigr]=0,
                        \label{eq:YM1a} \\
                  & D_\mu \mathcal{F}_{\nu\gamma} +D_\nu \mathcal{F}_{\gamma\mu}
                        +D_\gamma \mathcal{F}_{\mu\nu} =0, \label{eq:YM1b}
            \end{align}
      \end{subequations}
      with the field-strength tensor
      \begin{equation}\label{eq:nonabelian}
            \mathcal{F}_{\mu\nu} =\partial_\mu \mathbb{A}_{\nu}
                  -\partial_\nu \mathbb{A}_{\mu}
                  +\mathrm{i}\,g[\mathbb{A}_{\mu},\ \mathbb{A}_{\nu}],
      \end{equation}
       and $g$ is the coupling parameter. Evidently, when the commutator $[\mathbb{A}_{\mu}, \mathbb{A}_{\nu}]=0$, the definition of $\mathcal{F}_{\mu\nu}$ reduces to the Abelian case (i.e., $F_{\mu\nu}=\partial_\mu A_\nu -\partial_\nu A_\mu$ for Maxwell's equations). According to Eq. (\ref{eq:nonabelian}), one may write down the ``magnetic" field as well as the ``electric'' field as
      \begin{subequations}
            \begin{eqnarray}
                  \vec{\mathcal{B}} &=& \vec{\nabla}\times\vec{\mathcal{A}}
                        -{\rm i}\,g\bigl(\vec{\mathcal{A}}\times\vec{\mathcal{A}}\bigr),
                        \\
                  \vec{\mathcal{E}} &=& -\dfrac{1}{c}
                              \dfrac{\partial\,\vec{\mathcal{A}}}{\partial\,t}
                        -\vec{\nabla}\varphi-{\rm i}\,g\left[\varphi,\ \vec{\mathcal{A}}
                        \right].
            \end{eqnarray}
      \end{subequations}
      Then in terms of $\big\{\vec{\mathcal{E}}, \vec{\mathcal{B}}, \vec{\mathcal{A}}, \varphi\big\}$, the YM equations can be recast to the following form
      \begin{subequations}
            \begin{eqnarray}
                  && \vec{\nabla}\cdot\vec{\mathcal{E}}\ +\boxed{{\rm i}\,g\Bigl(
                        \vec{\mathcal{A}}\cdot\vec{\mathcal{E}}
                        -\vec{\mathcal{E}}\cdot\vec{\mathcal{A}}\Bigr)}=0, \label{eq:DivEYM} \\
                  && -\dfrac{1}{c} \dfrac{\partial}{\partial\,t} \vec{\mathcal{B}}
                        -\vec{\nabla}\times\vec{\mathcal{E}} \notag \\
                        &&\quad\quad\;\; +\boxed{{\rm i}\,g\biggl(\Bigl[\varphi,\
                                    \vec{\mathcal{B}}\Bigr]
                              -\vec{\mathcal{A}}\times\vec{\mathcal{E}}
                              -\vec{\mathcal{E}}\times\vec{\mathcal{A}}\biggr)}=0, \label{eq:CurlEYM} \\
                  && \vec{\nabla}\cdot\vec{\mathcal{B}}\ +\boxed{{\rm i}\,g\Bigl(
                        \vec{\mathcal{A}}\cdot\vec{\mathcal{B}}
                        -\vec{\mathcal{B}}\cdot\vec{\mathcal{A}}\Bigr)}=0, \label{eq:DivBYM} \\
                  && -\dfrac{1}{c} \dfrac{\partial}{\partial\,t} \vec{\mathcal{E}}
                        +\vec{\nabla}\times\vec{\mathcal{B}} \notag \\
                        &&\quad\quad\;\; +\boxed{{\rm i}\,g\biggl(\Bigl[\varphi,\
                                    \vec{\mathcal{E}}\Bigr]
                              +\vec{\mathcal{A}}\times\vec{\mathcal{B}}
                              +\vec{\mathcal{B}}\times\vec{\mathcal{A}}\biggr)}=0, \label{eq:CurlBYM}
            \end{eqnarray}
      \end{subequations}
      where the ``boxed'' expressions imply the self-interaction terms of potentials and fields in YM equations, which come from the nonlinear terms in Eqs. (\ref{eq:YM1a}) and (\ref{eq:YM1b}), such as ${\rm i}\,g [\mathbb{A}_\mu ,\ \mathcal{F}^{\mu\nu}]$. Note that YM equations are both \emph{Lorentz covariant} and \emph{local gauge covariant} (see II.B and II.C in Supplemental Material (SM) \cite{SM} for detailed proof).


We now work out the conditions that needed to exactly solve YM equations. For simplicity, let us assume $\vec{\mathcal{A}}(\vec{r},t)= \vec{\mathcal{A}}(\vec{r})\,{\rm e}^{-{\rm i}\,\omega\,t}$, and $\varphi(\vec{r},t)=\varphi(\vec{r})\,{\rm e}^{-{\rm i}\,\omega\,t}$, which implies that spatial part $\vec{r}=(x,y,z)$ and time part $t$ can be separated in $\vec{\mathcal{A}}(\vec{r},t)$ and $\varphi(\vec{r},t)$. Next we define
      $\vec{\mathcal{A}}(\vec{r},t)\times\vec{\mathcal{A}}(\vec{r},t)=\vec{\mathcal{M}}\,{\rm e}^{-{\rm i}\,2\,\omega\,t}$, and $\vec{\mathcal{N}}\equiv\vec{\mathcal{N}}(\vec{r})=[\varphi(\vec{r}),\ \vec{\mathcal{A}}(\vec{r})]{\rm e}^{
      -{\rm i}\,2\,\omega\,t}$, then in this situation we have the ``magnetic'' field and the ``electric'' field as
      \begin{subequations}
            \begin{eqnarray}
                  \vec{\mathcal{B}} &=& \Big[\vec{\nabla}\times\vec{\mathcal{A}}(\vec{r})
                              \Big]{\rm e}^{-{\rm i}\,\omega\,t}
                        -{\rm i}\,g\,\vec{\mathcal{M}}\,{\rm e}^{
                              -{\rm i}\,2\,\omega\,t}, \\
                  \vec{\mathcal{E}} &=& \Big[{\rm i}\,k\,\vec{\mathcal{A}}(\vec{r})
                        -\vec{\nabla}\varphi(\vec{r})\Big]{\rm e}^{-{\rm i}\,\omega\,t}
                        -{\rm i}\,g\,\vec{\mathcal{N}}\,{\rm e}^{-{\rm i}\,2\,
                              \omega\,t},
            \end{eqnarray}
      \end{subequations}
      where $k=\omega/c$. After that, YM equations (\ref{eq:DivEYM})-(\ref{eq:CurlBYM}) yield the following conditions:
      \begin{subequations}
            \begin{eqnarray}
                  && {\rm i}\,k\Big[\vec{\nabla}\cdot\vec{\mathcal{A}}(
                        \vec{r})\Big]-\nabla^2 \varphi(\vec{r})=0, \label{eq:TICond1} \\
                  && g\,\Big[\varphi(\vec{r}),\
                                    \vec{\nabla}\cdot\vec{\mathcal{A}}(\vec{r})
                                    \Big]=0, \label{eq:TICond2} \\
                  && g^2 \biggr\{\vec{\mathcal{A}}(\vec{r})\cdot\vec{\mathcal{N}}
                                    -\vec{\mathcal{N}}\cdot\vec{\mathcal{A}}(
                                          \vec{r})\biggr\}=0, \label{eq:TICond3} \\
                  && g\biggl\{2\,k\,\vec{\mathcal{M}}
                                          +{\rm i}\,\vec{\nabla}\times\vec{\mathcal{N}}
                                          \biggr\}=0, \label{eq:TICond4} \\
                  && g \left(\vec{\nabla}\cdot\vec{\mathcal{M}}\right)=0, \label{eq:TICond5} \\
                  && \vec{\nabla}\Big[\vec{\nabla}\cdot
                                          \vec{\mathcal{A}}(\vec{r})\Big]
                                    -\nabla^2 \vec{\mathcal{A}}(\vec{r})
                                    -k^2 \vec{\mathcal{A}}(\vec{r})
                                    -{\rm i}\,k\Big[\vec{\nabla}\varphi(\vec{r})
                                          \Big]=0, \label{eq:TICond6} \\
                  && g\, \biggr\{ k\, \vec{\mathcal{N}}
                        +{\rm i}\,\vec{\mathcal{A}}(\vec{r})\times\Big[
                              \vec{\nabla}\times\vec{\mathcal{A}}(
                                    \vec{r})\Big]
                        +{\rm i}\,\Big[\vec{\nabla}\times\vec{\mathcal{A}}(
                              \vec{r})\Big]\times\vec{\mathcal{A}}(
                                    \vec{r}) \notag \\
                        &&\quad -{\rm i}\,\bigl(\vec{\nabla}\times\vec{\mathcal{M}}
                              \bigr) -{\rm i}\,\Big[\varphi(\vec{r}),\
                                    \vec{\nabla}\varphi(\vec{r})\Big]\biggr\}=0,
                                    \label{eq:TICond7} \\
                  && g^2 \, \biggr\{\Big[\varphi(\vec{r}),\
                              \vec{\mathcal{N}}\Big]
                        +\vec{\mathcal{A}}(\vec{r})\times
                              \vec{\mathcal{M}}
                        +\vec{\mathcal{M}}\times\vec{\mathcal{A}}(
                              \vec{r})\biggr\}=0. \label{eq:TICond8}
            \end{eqnarray}
      \end{subequations}
      These eight conditions contain four $g$-terms and two $g^2$-terms. If one wants to exactly solve YM equations, he needs to find out the explicit expressions of $\{\vec{\mathcal{A}}(\vec{r}), \varphi\}$ that satisfy Eqs. (\ref{eq:TICond1})-(\ref{eq:TICond8}) simultaneously. The two $g^2$-terms come from the self-interactions, whose existence has prevented people from achieving the exact and non-Abelian solutions for YM equations. In this work, we focus on YM equations under two types of approximations, i.e., weak-coupling approximation (WCA) and zero-coupling approximation (ZCA).


\emph{Solving Yang-Mills Equations under WCA.---}For small $g$, we omit the terms referring to $g^2$, then the eight conditions \eqref{eq:TICond1}-\eqref{eq:TICond8} reduce to the following six conditions:
      \begin{subequations}
            \begin{eqnarray}
                  && {\rm i}\,k\Big[\vec{\nabla}\cdot\vec{\mathcal{A}}(
                        \vec{r})\Big]-\nabla^2 \varphi(\vec{r})=0, \label{eq:TIGCond1} \\
                  && \Big[\varphi(\vec{r}),\ \vec{\nabla}\cdot\vec{\mathcal{A}}(
                        \vec{r})\Big]=0, \label{eq:TIGCond2} \\
                  && 2\,k\,\vec{\mathcal{M}}+{\rm i}\,\vec{\nabla}\times\Big[
                        \varphi(\vec{r}),\ \vec{\mathcal{A}}(\vec{r})\Big]=0,
                        \label{eq:TIGCond3} \\
                  && \vec{\nabla}\cdot\vec{\mathcal{M}}=0, \label{eq:TIGCond4} \\
                  && \vec{\nabla}\Big[\vec{\nabla}\cdot
                              \vec{\mathcal{A}}(\vec{r})\Big]
                        -\nabla^2 \vec{\mathcal{A}}(\vec{r})
                        -k^2 \vec{\mathcal{A}}(\vec{r})
                        -{\rm i}\,k\Big[\vec{\nabla}\varphi(\vec{r})
                              \Big]=0, \label{eq:TIGCond5} \\
                  && {\rm i}\,k\Big[\varphi(\vec{r}),\ \vec{\mathcal{A}}(\vec{r})
                              \Big]
                        -\vec{\mathcal{A}}(\vec{r})\times\Big[
                              \vec{\nabla}\times\vec{\mathcal{A}}(\vec{r})\Big]
                        -\Big[\vec{\nabla}\times\vec{\mathcal{A}}(
                              \vec{r})\Big]\times\vec{\mathcal{A}}(\vec{r}) \notag \\
                        &&\quad +\bigl(\vec{\nabla}\times\vec{\mathcal{M}}\bigr)
                        +\Big[\varphi(\vec{r}),\ \vec{\nabla}\varphi(\vec{r})\Big]
                        =0. \label{eq:TIGCond6}
            \end{eqnarray}
      \end{subequations}
      Fortunately, for Eqs. (\ref{eq:TIGCond1})-(\ref{eq:TIGCond6}), there are exact solutions for $\vec{\mathcal{A}}(\vec{r})$ and $\varphi(\vec{r})$. Precisely, the $SU(2)$ operator solutions of YM equations under WCA are given by \cite{SM}
      \begin{eqnarray}\label{eq:solu1a}
            \vec{\mathcal{A}}(\vec{r}, t) &=& \vec{\tau}\,{\rm e}^{
                  {\rm i}(\vec{k}\cdot\vec{r}-\omega\,t)}, \nonumber\\
            \varphi(\vec{r}, t) &=& \bigl(\vec{\tau}\cdot\hat{k}\bigr){\rm e}^{
                  {\rm i}(\vec{k}\cdot\vec{r}-\omega\,t)},\nonumber\\
            \vec{\mathcal{B}}(\vec{r}, t)
            &=& {\rm i}\bigl(\vec{k}\times\vec{\tau}\bigr){\rm e}^{
                        {\rm i}(\vec{k}\cdot\vec{r}-\omega\,t)}
                  +g\,\hbar\,\vec{\eta}\,{\rm e}^{
                        {\rm i}\,2(\vec{k}\cdot\vec{r}-\omega\,t)},
                        \\
            \vec{\mathcal{E}}(\vec{r}, t)
            &=& -{\rm i}\,k\Big[\hat{k}\times\bigl(\hat{k}
                              \times\vec{\tau}\bigr)\Big]{\rm e}^{
                                    {\rm i}(\vec{k}\cdot\vec{r}
                                          -\omega\,t)}
                        +g\,\hbar\,\vec{\xi}\,{\rm e}^{{\rm i}\,2(
                              \vec{k}\cdot\vec{r}-\omega t)},
                              \nonumber
      \end{eqnarray}
      with
      \begin{eqnarray}\label{eq:solu1b}
            &&\vec{\tau}=\vec{R}_0 \openone+\vec{R}_1 S_x +\vec{R}_2 S_y
                  +\vec{R}_3 S_z\nonumber\\
            && \vec{\eta}=\big(\vec{R}_2 \times\vec{R}_3\big)S_x
                  +\big(\vec{R}_3 \times\vec{R}_1\big)S_y
                  +\big(\vec{R}_1 \times\vec{R}_2\big)S_z,\nonumber\\
            &&  \vec{\xi}=-\hat{k}\times\vec{\eta}=\vec{\eta}\times \hat{k},
                  \\
            &&\vec{k}\cdot\bigl(\vec{R}_1 \times\vec{R}_2\bigr)
            =\vec{k}\cdot\bigl(\vec{R}_2 \times\vec{R}_3\bigr)
            =\vec{k}\cdot\bigl(\vec{R}_3 \times\vec{R}_1\bigr)
            =0. \notag
      \end{eqnarray}
      Here $\vec{k}$ is the propagation direction with $\hat{k}=\vec{k}/|\vec{k}|$, $\vec{S}=\{S_x, S_y, S_z\}$ is the angular momentum operator satisfying the $SU(2)$ commutative relations $[S_j, S_k]={\rm i}\hbar\,\epsilon_{jkl} S_l$, $j,k,l\in\{x,y,z\}$, $\vec{R}_\ell$'s $(\ell=0, 1, 2, 3)$ represent some constant vectors, with $\bigl\{\vec{k}, \vec{R}_1, \vec{R}_2, \vec{R}_3\bigr\}$ locating on the same plane, and $\vec{R}_0$ is arbitrary.

      Under WCA, $\vec{\mathcal{B}}(\vec{r}, t)$ and $\vec{\mathcal{E}}(\vec{r}, t)$ are the non-Abelian waves, because generally $\vec{\tau}\times\vec{\tau}\neq 0$. What will happen if we impose two additional conditions related to $g^2$ (i.e., Eq. (\ref{eq:TICond3}) and Eq. (\ref{eq:TICond8})) to the above solutions? We find that $\vec{\mathcal{B}}(\vec{r}, t)$ and $\vec{\mathcal{E}}(\vec{r}, t)$ reduce to some Abelian waves with  $\vec{\tau}=\vec{R}_0 \,\openone+\vec{R}\,\bigl(\vec{n}\cdot\vec{S}\bigr)$ (See Remark 5 in Sec. III. C of SM \cite{SM}, in this case one has $\vec{\tau}\times\vec{\tau}=0$ and $[\varphi, \vec{\tau}]=0$).

\emph{Solving Yang-Mills Equations under ZCA.---}Further, if we ignore the ``boxed'' parts in \Eq{eq:DivEYM}-\Eq{eq:CurlBYM}, then we attain YM equations under ZCA in the following form
      \begin{subequations}
            \begin{eqnarray}
                  && \vec{\nabla}\cdot\vec{\mathcal{E}} = 0, \label{eq:DivE-S} \\
                  && \vec{\nabla}\times\vec{\mathcal{E}} = -\dfrac{1}{c}
                        \dfrac{\partial\,\vec{\mathcal{B}}}{\partial\,t},
                        \label{eq:CurlE-S} \\
                  && \vec{\nabla}\cdot\vec{\mathcal{B}} = 0, \label{eq:DivB-S} \\
                  && \vec{\nabla}\times\vec{\mathcal{B}} = \dfrac{1}{c}
                        \dfrac{\partial\,\vec{\mathcal{E}}}{\partial\,t}.
                        \label{eq:CurlB-S}
            \end{eqnarray}
      \end{subequations}
      One may notice that Eqs. (\ref{eq:DivE-S})-(\ref{eq:CurlB-S}) and the famous Maxwell's equations share exactly the same forms, based on such a reason we would like to call them the \emph{Maxwell-type equations}. Analogously, in terms of potentials we may transform Eqs. (\ref{eq:DivE-S})-(\ref{eq:CurlB-S}) to the following conditions:
      \begin{subequations}
            \begin{eqnarray}
                  && \vec{\nabla}\cdot\left(\vec{\mathcal{A}}\times\vec{\mathcal{A}}\right)=0,
                        \label{eq:fourcond-1}\\
                  && \vec{\nabla}\times[\varphi,\ \vec{\mathcal{A}}] =-\dfrac{1}{c}
                        \dfrac{\partial}{\partial\,t} \bigl(
                              \vec{\mathcal{A}}\times\vec{\mathcal{A}}\bigr),\label{eq:fourcond-2}\\
                  &&  \dfrac{1}{c} \dfrac{\partial(\vec{\nabla}\cdot\vec{\mathcal{A}})}{\partial\,t}
                        +{\nabla}^2\varphi+{\rm i}\,g \;\vec{\nabla}\cdot \left([
                              \varphi,\ \vec{\mathcal{A}}]\right)=0, \label{eq:fourcond-3}\\
                  &&  -\vec{\nabla}\left(\vec{\nabla}\cdot\vec{\mathcal{A}}\right)
                        +\nabla^2 \vec{\mathcal{A}}
                              +{\rm i}\,g\,\vec{\nabla}
                                    \times\left(\vec{\mathcal{A}}\times\vec{\mathcal{A}}\right) \notag \\
                  &&\;\;\;\; =\dfrac{1}{c^2} \dfrac{\partial^2 \vec{\mathcal{A}}}{\partial\,t^2}
                        +\dfrac{1}{c} \dfrac{\partial(\vec{\nabla}\varphi)}{\partial\,t}
                        +\dfrac{\mathrm{i}\,g}{c} \dfrac{\partial}{\partial\,t} [\varphi,
                              \ \vec{\mathcal{A}}]. \label{eq:fourcond-4}
            \end{eqnarray}
      \end{subequations}

      In Sec. IV. A. 1 of SM \cite{SM}, we can prove that the $SU(2)$ operator solutions of Maxwell-type equations are just Eq. (\ref{eq:solu1a}) and Eq. (\ref{eq:solu1b}), which are the same with YM equations under WCA.  Broadly, the potentials $\vec{\mathcal{A}}(\vec{r})$ and $\varphi(\vec{r})$ can be expanded using the generators of $SU(N)$ group. For the general $SU(N)$ operator solutions of Yang-Mills under two types of approximations, one may refer to Sec. VI in SM \cite{SM}.

\emph{$SU(2)$ Angular-Momentum Waves.---}Without loss of generality, we concern ourselves mainly with the $SU(2)$ operator solutions of YM equations under WCA and ZCA. Actually, their physical meanings are just the $SU(2)$ angular-momentum waves. To see this point cleanly, let us consider the following wave equations
      \begin{eqnarray}
            &&\nabla^2 \vec{\mathcal{B}}-\dfrac{1}{c^2} \dfrac{\partial\,\vec{\mathcal{B}}}{\partial\,t^2} =0, \nonumber\\
            &&\nabla^2 \vec{\mathcal{E}}-\dfrac{1}{c^2} \dfrac{\partial\,\vec{\mathcal{E}}}{\partial\,t^2} =0,
      \end{eqnarray}
which holds for $\vec{\mathcal{B}}(\vec{r}, t)$ and $\vec{\mathcal{E}}(\vec{r}, t)$ as shown in Eq. (\ref{eq:solu1a}).  Therefore, the ``magnetic'' field $\vec{\mathcal{B}}(\vec{r}, t)$ and the ``electric'' field $\vec{\mathcal{E}}(\vec{r}, t)$ can be regarded as some types of waves. Because the amplitudes of $\vec{\mathcal{B}}(\vec{r}, t)$ and $\vec{\mathcal{E}}(\vec{r}, t)$ contain the components of the $SU(2)$ angular-momentum operator, we call them the $SU(2)$ angular-momentum waves.

One can check the following properties
            \begin{eqnarray}
            &&\vec{\mathcal{B}}(\vec{r}, t)=\hat{k}\times\vec{\mathcal{E}}(\vec{r}, t), \;\;\;\vec{\mathcal{E}}(\vec{r}, t)=\vec{\mathcal{B}}(\vec{r}, t)\times\hat{k},  \nonumber\\
            &&\hat{k}\cdot\vec{\mathcal{B}}(\vec{r}, t)= \hat{k}\cdot\vec{\mathcal{E}}(\vec{r}, t)= \vec{\mathcal{B}}(\vec{r}, t)\cdot \vec{\mathcal{E}}(\vec{r}, t)=0,
            \end{eqnarray}
that is, $\vec{\mathcal{B}}(\vec{r}, t)$, $\vec{\mathcal{E}}(\vec{r}, t)$ and the propagation direction $\vec{k}$ are mutually perpendicular. It is worth mentioning that the Maxwell-type equations are Lorentz covariant~\cite{SM}, which suggests that the propagation velocity of the AMW is precisely $c$. Here we provide a nontrivial example with $\vec{\mathcal{A}}\times \vec{\mathcal{A}}\neq 0$, $[\varphi, \vec{\mathcal{A}}]\neq 0$.

      \emph{Example 1.---}Let $\vec{k}=k\hat{z}$, $\vec{R_0}=\vec{R_2}=0$, $\vec{R}_1=\hat{x}$, and
            $\vec{R}_3=\hat{z}$ (i.e., the vectors in $\{\vec{R}_1, \vec{R}_3, \vec{k}\}$ locate on the $xz$-plane). In this case, we have
            $\vec{\tau}=\vec{R}_1 S_x +\vec{R}_3 S_z =\hat{x} S_x+ \hat{z} S_z$,
            $\vec{\eta}=(\vec{R}_3 \times\vec{R}_1)S_y=\hat{y}S_y$. Then the AMW are given by
            \begin{eqnarray}\label{eq:EgBE}
                  \vec{\mathcal{B}}(\vec{r}, t) &=& \hat{y} \biggr[{\rm i} \, k S_x {\rm e}^{{\rm i}(k z-\omega t)}+g\,\hbar\,S_y\;{\rm e}^{2{\rm i}(kz-\omega t)}\biggr],\nonumber\\
                  \vec{\mathcal{E}}(\vec{r}, t) &=& \hat{x} \biggr[{\rm i}\, k S_x {\rm e}^{{\rm i}(k z-\omega t)}+g\,\hbar\,S_y\;{\rm e}^{2{\rm i}(kz-\omega t)}\biggr].
            \end{eqnarray}
            Notice that the waves in \Eq{eq:EgBE} are in the superposition of ${\rm e}^{{\rm i}(k z-\omega t)}$ and ${\rm e}^{2{\rm i}(kz-\omega t)}$, and their corresponding amplitudes are non-commutative.

\emph{Thought Experiment to Detect AMW.---}It is well-known that the EMW
      are produced via electromagnetic oscillation in spacetime. In general, an oscillating charged particle can emit (or radiate) the EMW \cite{GreinerCED}. Likewise, the ``\emph{oscillation of angular momentum}'' can be regarded as a mechanism to emit the AMW.

\begin{figure}[t]
            \centering
            \includegraphics[width=80mm]{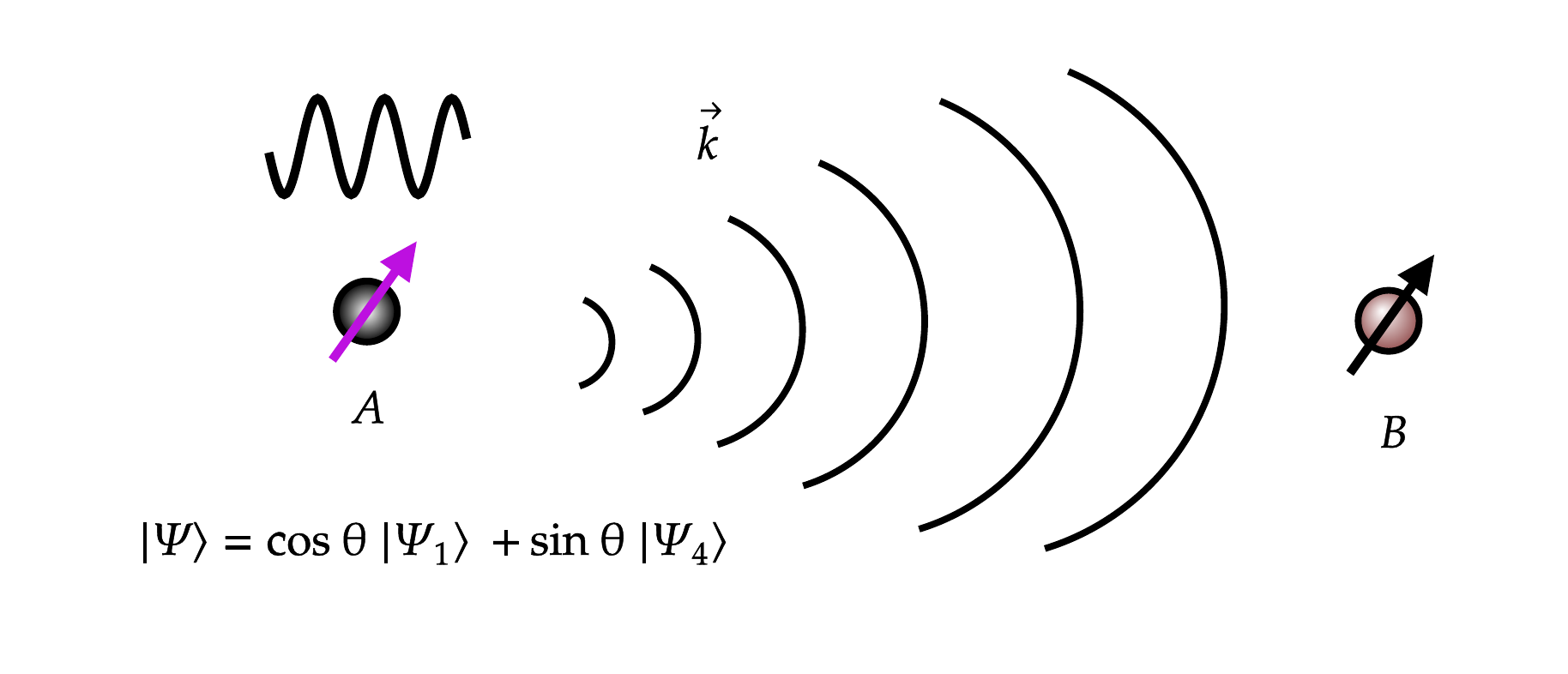}
            \caption{Illustration of a thought experiment on detecting the spin AMW. Particle $A$ is a free Dirac's electron, which is prepared in a superposition state of $|\Psi_1\rangle$ (positive energy and positive helicity) and $|\Psi_4\rangle$ (negative energy and negative helicity). In such a superposition state, the spin of the electron oscillates rapidly due to the relativistic phenomenon of ``spin Zitterbewegung''. Because of the spin oscillation, particle $A$ emits the spin AMW along the direction $\vec{k}$. Due to the Pauli-like interaction $\vec{S}_b \cdot\vec{\mathcal{B}}$, the distanced particle $B$ can experience the existence of the spin AMW emitted by particle $A$.}
            \label{fig:AMWave}
      \end{figure}

       When Schr\"odinger first examined the expectation value of the position operator $\vec{r}$ of a relativistic free electron (i.e., Dirac's electron), he discovered a surprising highly frequent oscillation if Dirac's electron was in a superposition state of positive-energy and negative-energy. Such a relativistic phenomenon was called the (position) ``Zitterbewegung'' \cite{1931Zitter}. Accordingly, in Sec. V. B of SM \cite{SM}, the ``spin Zitterbewegung'' of Dirac's electron is showed when the expectation value of the spin operator $\vec{S}$ is considered. The ``spin Zitterbewegung'' is a relativistic phenomenon of spin oscillation, which can be used to emit the spin AMW. In Fig. \ref{fig:AMWave}, we illustrate a possible way to detect the spin AMW. A Dirac's electron (particle $A$) is prepared in the superposition state of different energy and different helicity (marked as $|\Psi_1\rangle$ and
      $|\Psi_4\rangle$), i.e., $|\Psi\rangle=\cos\theta|\Psi_1\rangle +\sin\theta|\Psi_4\rangle$. For such a state, there exists a phenomenon of ``spin Zitterbewegung'', whose expectation value
      $\mathcal{Z}_s =\langle\Psi|\hat{\mathcal{Z}}_s|\Psi\rangle \neq 0$ in general, with $\hat{\mathcal{Z}}_s$ referring to the ``spin Zitterbewegung'' operator \cite{SM}. The spin oscillation of the Dirac's electron can be viewed as a source to emit AMW along the direction of the wave ket $\vec{k}$. Particle $B$ with spin $\vec{S}_b$ is used to detect the spin AMW. Due to the the Pauli-like interaction
      $\vec{S}_b\cdot\vec{\mathcal{B}}(\vec{r}, t)$, the spin of particle $B$ can experience the existence of the spin AMW emitted by particle $A$.

\emph{Conclusion.---}
      In conclusion, we have studied the $SU(2)$ operator solutions (as well as the
      $SU(N)$ operator solutions) of YM equations under tow types of approximations (WCA and ZCA) in vacuum without sources. Similar to that Maxwell's equations predict the EMW, the $SU(2)$ operator solutions of YM equations under approximations naturally predict the AMW. The comparisons between Maxwell's equations and Yang-Mills equations under approximations are listed in Tab.~\ref{tab:Maxwell}. We also presented a thought experiment to detect the AMW due to the spin oscillation. As a natural generalization of its classical partner (i.e., EMW), the AMW may be devoted to probing spin interactions in microscopic world.

      Eventually, as is well-known that, for a solid physical theory, its predictions are often reliable. For examples, Maxwell's equations predicted the EMW \cite{Maxwell}, special relativity predicted the mass-energy relation \cite{EinsteinSR}, general relativity predicted the gravitational waves
      \cite{EinsteinGR1916,EinsteinGR1918,1990GWave,1996GWave,2012GWave,2017GWave}, and quantum mechanics predicted Heisenberg uncertainty relations \cite{Heisenberg}, Aharonov-Bohm effect \cite{AB}, Berry phase \cite{Berry}, etc. YM theory is undoubtedly a fundamental theory in modern physics. It is reasonable to believe that the angular-momentum waves predicted by YM equation really exist in the physical world. We expect that such AMW will be found in laboratory in the near future.

\vspace{3mm}

\noindent\textbf{Acknowledgments}\\
      J.L.C. was supported by the National Natural Science Foundation of China (Grant Nos. 12275136 and 12075001) and the 111 Project of B23045. X.Y.F. was supported by the Nankai Zhide Foundations.

\vspace{3mm}

\noindent{\bf Author contributions}\\
J.L.C supervised the project. All authors read the paper and discussed the results.
\vspace{3mm}

\noindent{\bf Competing interests}\\
The authors declare no competing interests.
\vspace{3mm}

\noindent{\bf Data Availability}\\
All data that support the findings of this study are included within the article (and any supplementary files).

\vspace{3mm}

\noindent\textbf{Keywords}: Maxwell's equations, Yang-Mills equations, Angular-momentum waves, Spin, Spin oscillation

\newpage

\onecolumngrid
\begin{widetext}
      \begin{table}[htbp]
                  \caption{
                        The comparison between Maxwell's equations and Yang-Mills equations under approximations (sourceless and in vacuum). The former predict the well-known electromagnetic waves $\{\vec{E}(\vec{r}, t), \vec{B}(\vec{r}, t)\}$, while the latter predict the $SU(2)$ angular-momentum waves
                        $\{\vec{\mathcal{E}}(\vec{r}, t), \vec{\mathcal{B}}(\vec{r}, t)\}$.}\label{tab:Maxwell}
                  \centering\begin{tabular}{|c|l|l|}
                  \hline
                  \cline{2-3}
                  & \quad \quad \quad Maxwell's Equations & \qquad
                        Yang-Mills Equations under Approximations \\
                  \hline
                   &  &  WCA: \;\;  Omitting the $g^2$ terms in YM equations  \\
                  Expressions & $\partial_\mu F^{\mu\nu} = 0$\ \
                        & ZCA: \;\; $   \partial_\mu \mathcal{F}^{\mu\nu}=0$ \\
                  &$\partial_\mu F_{\nu\gamma}+\partial_\nu F_{\gamma\mu}
                              +\partial_\gamma F_{\mu\nu} = 0$ &
                        $\quad \quad\;\;\;\;\;\; \partial_\mu \mathcal{F}_{\nu\gamma}
                                    +\partial_\nu \mathcal{F}_{\gamma\mu}
                                    +\partial_\gamma \mathcal{F}_{\mu\nu} =0$ \\
                  \hline
                  Tensors & $F_{\mu\nu} =\partial_\mu A_{\nu} -\partial_\nu A_{\mu}$\ \
                        & $\mathcal{F}_{\mu\nu}=\partial_\mu \mathbb{A}_{\nu}
                              -\partial_\nu \mathbb{A}_{\mu}
                              +{\rm i}\,g[\mathbb{A}_{\mu},\
                                    \mathbb{A}_{\nu}]$ \\
                  \hline
                  &  & WCA: see Eq. (4a)-Eq. (4d)\\
                  &  & ZCA: \\
                  & $\vec{\nabla}\cdot\vec{E}=0$\qquad\qquad(1) & $\vec{\nabla}\cdot\vec{\mathcal{E}}=0$
                        \qquad\qquad(1) \\
                  Expressions & $\vec{\nabla}\times\vec{E}=-\dfrac{1}{c}
                        \dfrac{\partial\,\vec{B}}{\partial\,t}$\ \ (2)
                        & $\vec{\nabla}\times\vec{\mathcal{E}}=-\dfrac{1}{c}
                        \dfrac{\partial\,\vec{\mathcal{B}}}{\partial\,t}$\ \ \, (2) \\
                  & $\vec{\nabla}\cdot\vec{B}=0$\qquad\qquad(3) &
                        $\vec{\nabla}\cdot\vec{\mathcal{B}}=0$\qquad\qquad\ (3) \\
                  & $\vec{\nabla}\times\vec{B}=\dfrac{1}{c} \dfrac{\partial\,\vec{E}}{\partial\,t}$
                        \quad\,(4) & $\vec{\nabla}\times\vec{\mathcal{B}}=\dfrac{1}{c}
                        \dfrac{\partial\,\vec{\mathcal{E}}}{\partial\,t}$\quad\ \ \ (4) \\
                  \hline
                  Magnetic Field
                  & $\vec{B}=\vec{\nabla}\times\vec{A}$ & $\vec{\mathcal{B}}
                        =\vec{\nabla}\times\vec{\mathcal{A}} - {\rm i}\,g\,
                        \vec{\mathcal{A}}\times\vec{\mathcal{A}}$ \\
                  Electric Field & $\vec{E}=-\dfrac{1}{c} \dfrac{\partial\,\vec{A}}{\partial\,t} -\vec{\nabla}\varphi$
                        & $\vec{\mathcal{E}}=-\dfrac{1}{c} \dfrac{\partial\,\vec{\mathcal{A}}}{\partial\,t}
                              -\vec{\nabla}\varphi-{\rm i}\,g\, \left[
                                    \varphi,\ \vec{\mathcal{A}}\right]$ \\
                  \hline
                  Vector Potential & $\vec{A}(\vec{r}, t)=\vec{A}_{01}\;{\rm e}^{{\rm i}(\vec{k}\cdot\vec{r}
                        -\omega t)}$
                        & $\vec{\mathcal{A}}(\vec{r}, t)=\vec{\tau}\;{\rm e}^{
                              {\rm i}(\vec{k}\cdot\vec{r}-\omega t)}$\\
                  Scalar Potential & $\varphi(\vec{r}, t)=0$\;\;\; (Coulomb gauge)
                  & $\varphi(\vec{r}, t)
                        =\left(\hat{k}\cdot\vec{\tau}\right)\;{\rm e}^{
                              {\rm i}(\vec{k}\cdot\vec{r}-\omega t)}$ \\
                  & & $\vec{\tau}=\vec{R}_0 {\bf 1}
                        +\vec{R}_1 S_x+\vec{R}_2 S_y
                        +\vec{R}_3 S_z$ \\
                  \hline
                  Solutions of Fields& $\vec{B}(\vec{r}, t)={\rm i}(\vec{k}\times\vec{A}_{01})\,{\rm e}^{
                        {\rm i}(\vec{k}\cdot\vec{r}-\omega t)}$
                  & $\vec{\mathcal{B}}(\vec{r}, t) = {\rm i}\left(\vec{k}\times\vec{\tau}\right){\rm e}^{{\rm i}(\vec{k}\cdot\vec{r}-\omega t)}+g \hbar \; \vec{\eta}\;{\rm e}^{2{\rm i}(\vec{k}\cdot\vec{r}-\omega t)}$ \\
                  & $\vec{E}(\vec{r}, t)={\rm i}\,k\,\vec{A}_{01}\,{\rm e}^{
                        {\rm i}(\vec{k}\cdot\vec{r}-\omega t)}$ & $\vec{\mathcal{E}}(\vec{r}, t) = -\hat{k}\times \vec{\mathcal{B}}(\vec{r}, t)$\\
                  &  &  $\vec{\eta}=
                              (\vec{R}_2 \times\vec{R}_3)S_x
                              +(\vec{R}_3 \times\vec{R}_1)S_y
                              +(\vec{R}_1 \times\vec{R}_2)S_z$\\
                  \hline
                  Constraints & $\vec{k}\cdot \vec{A}_{01}=0$ &  $\vec{k}\cdot\bigl(\vec{R}_1 \times\vec{R}_2\bigr)
                  =\vec{k}\cdot\bigl(\vec{R}_2 \times\vec{R}_3\bigr)
                  =\vec{k}\cdot\bigl(\vec{R}_3 \times\vec{R}_1\bigr)=0$\\
                  \hline
                  Properties & $\vec{k}\perp\vec{B}, \; \vec{k}\perp\vec{E},\; \vec{B}\perp\vec{E},$
                        & $\vec{k}\perp\vec{\mathcal{B}}, \; \vec{k}\perp\vec{\mathcal{E}},\; \vec{\mathcal{B}}\perp\vec{\mathcal{E}},$ \\
                  & $\vec{B}=\hat{k}\times \vec{E}, \;\vec{E}=-\hat{k}\times \vec{B},\;\hat{k}\propto \vec{E}\times \vec{B}$
                        & $ \vec{\mathcal{B}}=\hat{k}\times \vec{\mathcal{E}},\; \vec{\mathcal{E}}=-\hat{k}\times \vec{\mathcal{B}}$, \; $\hat{k}\propto \vec{\mathcal{E}}\times \vec{\mathcal{B}}$, \\
                  &    & $ \vec{S}\times\vec{S}={\rm i} \hbar\vec{S}$, \; $\vec{\tau}\times \vec{\tau}={\rm i} \hbar \vec{\eta}$ \\
                  \hline
                  Predictions & \ Electromagnetic waves
                        & The $SU(2)$ angular-momentum waves  \\
                  & $\nabla^2 \vec{E}-\dfrac{1}{c^2} \dfrac{\partial\,\vec{E}}{\partial\,t^2} =0$
                        & $\nabla^2 \vec{\mathcal{E}}-\dfrac{1}{c^2} \dfrac{\partial\,\vec{\mathcal{E}}}{\partial\,t^2} =0$  \\
                  & $\nabla^2 \vec{B}-\dfrac{1}{c^2} \dfrac{\partial\,\vec{B}}{\partial\,t^2} =0$
                        & $\nabla^2 \vec{\mathcal{B}}-\dfrac{1}{c^2} \dfrac{\partial\,\vec{\mathcal{B}}}{\partial\,t^2} =0$  \\
                  \hline
                  Experimental Verification & Verified by German physicist Hertz (1886)
                        & Possibly be observed in spin-oscillation experiments. \\
                  &
                        & Waiting for experimental discovery   \\
                  \hline
                  \end{tabular}
                 \end{table}
\end{widetext}
\twocolumngrid



\begin{thebibliography}{10}
      \bibitem{Maxwell}
            J. C. Maxwell, \emph{The Scientific Papers of James Clerk Maxwell},
            Vol. 1-2 (Cambridge University Press, Cambridge, 2011).
      \bibitem{2014PTodaYang}
            C. N. Yang,
            \emph{The Conceptual Origins of Maxwell's Equations and Gauge Theory},
            \href{https://physicstoday.scitation.org/doi/full/10.1063/PT.3.2585}{
                  Physics Today {\bf 67}, 45 (2014)}.
      \bibitem{GreinerCED}
            W. Greiner, \emph{Classical Electrodynamics}, edited by W. Greiner
            (Springer New York, New York, NY, 1998).
      \bibitem{1999Jackson}
            J. D. Jackson, \emph{Classical Electrodynamics: Third Edition},
            (Wiley, New York, 1999).
      \bibitem{Hertz}
            J. F. Mulligan, editor, \emph{Heinrich Rudolf Hertz (1857-1894):
                  A Collection of Articles and Addresses} (Routledge, London, 2019).
      \bibitem{1982PRAHaller}
            K. Haller,
            \emph{Maxwell's Equations in the Multipolar Representation},
            \href{https://link.aps.org/doi/10.1103/PhysRevA.26.1796}{
                  Phys. Rev. A {\bf 26}, 1796 (1982)}.
      \bibitem{1982ComtHealy}
            W. P. Healy,
            \emph{Comment on ``Maxwell's Equations in the Multipolar Representation''},
            \href{https://link.aps.org/doi/10.1103/PhysRevA.26.1798}{
                  Phys. Rev. A {\bf 26}, 1798 (1982)}.
      \bibitem{1982PRAPower}
            E. A. Power and T. Thirunamachandran,
            \emph{Maxwell's Equations and the Multipolar Hamiltonian},
            \href{https://link.aps.org/doi/10.1103/PhysRevA.26.1800}{
                  Phys. Rev. A {\bf 26}, 1800 (1982)}.
      \bibitem{1994PRELiu}
            M. Liu, \emph{Maxwell Equations in Nematic Liquid Crystals},
            \href{https://link.aps.org/doi/10.1103/PhysRevE.50.2925}{
                  Phys. Rev. E {\bf 50}, 2925 (1994)}.
      \bibitem{1995ComtBrand}
            H. R. Brand and H. Pleiner,
            \emph{Macroscopic Dynamics and Hydrodynamic Maxwell Equations},
            \href{https://link.aps.org/doi/10.1103/PhysRevLett.74.1883}{
                  Phys. Rev. Lett. {\bf 74}, 1883 (1995)}.
      \bibitem{1888hertz}
            H. Hertz,
            \emph{Ueber die Ausbreitungsgeschwindigkeit der electrodynamischen Wirkungen}
            \href{https://doi.org/10.1002/andp.18882700708}{
                  Annalen der Physik und Chemie. Band 270, Joh. Ambr. Barth, Leipzig 1888, S. 551-569}.
      \bibitem{2012PREXiong}
            H. Xiong, L.-G. Si, C. Ding, X.-Y. L{\" u}, X. Yang, and Y. Wu,
            \emph{Solutions of the Cylindrical Nonlinear Maxwell Equations},
            \href{https://link.aps.org/doi/10.1103/PhysRevE.85.016602}{
                  Phys. Rev. E {\bf 85}, 016602 (2012)}.
      \bibitem{2020PRE}
            A. M. Valverde, L. D. Angulo, M. R. Cabello, S. G. Garc{\' i}a, J. J. Omiste,
                  and J. Luo,
            \emph{Numerical Simulation of Knotted Solutions for Maxwell Equations},
            \href{https://link.aps.org/doi/10.1103/PhysRevE.101.063305}{
                  Phys. Rev. E {\bf 101}, 063305 (2020)}.
      \bibitem{2001RMPGaugInvar}
            J. D. Jackson and L. B. Okun, \emph{Historical Roots of Gauge Invariance}, \href{https://link.aps.org/doi/10.1103/RevModPhys.73.663}{
                  Rev. Mod. Phys. {\bf 73}, 663 (2001)}.
      \bibitem{1954YangMills}
            C. N. Yang and R. L. Mills,
            \emph{Conservation of Isotopic Spin and Isotopic Gauge Invariance},
            \href{https://link.aps.org/doi/10.1103/PhysRev.96.191}{
                  Phys. Rev. {\bf 96}, 191 (1954)}.
      \bibitem{SchwartzQFT}
            M. D. Schwartz, \emph{Quantum Field Theory and the Standard Model}
            (Cambridge University Press, Cambridge, 2013).
      \bibitem{1979RMPLatGauge}
            J. B. Kogut,
            \emph{An Introduction to Lattice Gauge Theory and Spin Systems}, \href{https://link.aps.org/doi/10.1103/RevModPhys.51.659}{
                        Rev. Mod. Phys. {\bf 51}, 659 (1979)}.
      \bibitem{2014RMPQSimu}
            I. M. Georgescu, S. Ashhab, and F. Nori, \emph{Quantum Simulation}, \href{https://link.aps.org/doi/10.1103/RevModPhys.86.153}{
                        Rev. Mod. Phys. {\bf 86}, 153 (2014)}.
      \bibitem{2011RMPAGaugPoten}
            J. Dalibard, F. Gerbier, G. Juzeliunas, and P. {\" O}hberg,
            \emph{Colloquium: Artificial Gauge Potentials for Neutral Atoms}, \href{https://link.aps.org/doi/10.1103/RevModPhys.83.1523}{
                        Rev. Mod. Phys. {\bf 83}, 1523 (2011)}.
      \bibitem{2022PRLThermYM}
            S. Chen, K. Fukushima, and Y. Shimada, \emph{Perturbative Confinement in
                  Thermal Yang-Mills Theories Induced by Imaginary Angular Velocity},
            \href{https://link.aps.org/doi/10.1103/PhysRevLett.129.242002}{
                  Phys. Rev. Lett. {\bf 129}, 242002 (2022)}.
      \bibitem{2022PRDCosmYM}
            M. Yamada and K. Yonekura,
            \emph{Cosmic Strings from Pure Yang--Mills Theory}, \href{https://link.aps.org/doi/10.1103/PhysRevD.106.123515}{
                        Phys. Rev. D {\bf 106}, 123515 (2022)}.
      \bibitem{1979RMPCYangMills}
            A. Actor,
            \emph{Classical Solutions of $\mathrm{SU}(2)$ Yang-Mills Theories}, \href{https://link.aps.org/doi/10.1103/RevModPhys.51.461}{
                        Rev. Mod. Phys. {\bf 51}, 461 (1979)}.
      \bibitem{1979PLBOh}
            C. H. Oh and R. Teh, \emph{Periodic Solutions of the Yang-Mills Field Equations}, \href{https://doi.org/10.1016/0370-2693(79)90025-X}{Physics Letters B {\bf 87}, 83 (1979)}.
      \bibitem{1985JMPOh}
            C. H. Oh and R. Teh, \emph{Nonabelian Progressive Waves},
            \href{https://doi.org/10.1063/1.526576}{Journal of Mathematical Physics {\bf 26}, 841 (1985)}.
      \bibitem{1986PRDColor1}
            C. H. Lai and C. H. Oh, \emph{Color Screening and Topological Index in the Classical Yang-Mills Theory with Sources},
            \href{https://link.aps.org/doi/10.1103/PhysRevD.33.1825}{Phys. Rev. D
                  {\bf 33}, 1825 (1986)}.
      \bibitem{1986PRDColor2}
            C. H. Oh, C. H. Lai, and R. Teh, \emph{Color Radiation in the Classical Yang-Mills Theory}, \href{https://link.aps.org/doi/10.1103/PhysRevD.33.1133}{Phys. Rev. D {\bf 33}, 1133 (1986)}.
      \bibitem{2023PRDShreecharan}
            T. Shreecharan and T. S. Raju, \emph{Time-Dependent Non-Abelian Waves and Their Stochastic Regimes for Gauge Fields Coupled to External Sources},
            \href{https://link.aps.org/doi/10.1103/PhysRevD.107.056011}{Phys. Rev. D {\bf 107}, 056011 (2023)}.
      \bibitem{1995IntroQFT}
            M. E. Peskin, \emph{An Introduction To Quantum Field Theory: First Edition}(
                  CRC Press, 1995).
      \bibitem{2002CTGaugeT}
            V. Rubakov, \emph{Classical Theory of Gauge Fields}(
                  Princeton University Press, 2002).
      \bibitem{SM}
            See Supplemental Material for detailed proofs.
      \bibitem{1931Zitter}
            E. Schr{\" o}dinger, Sitzungsb. Preuss. Akad. Wiss. Phys. Math. Kl.
                  {\bf 24}, 418 (1930); {\bf 3}, 1 (1931).

      \bibitem{EinsteinSR}
            A. Einstein, \emph{The Principle of Relativity}, (Dover Publications, 1923).
      \bibitem{EinsteinGR1916}
            A. Einstein,
            \emph{Approximative integration of the field equations of gravitation},
            \href{https://onlinelibrary.wiley.com/doi/10.1002/3527608958.ch7}{
                  Sitzungsber. K. Preuss. Akad. Wiss. {\bf 1} 688 (1916)}.
      \bibitem{EinsteinGR1918}
            A. Einstein,
            \emph{On gravitational waves},
            \href{https://onlinelibrary.wiley.com/doi/10.1002/3527608958.ch12}{
                  Sitzungsber. K. Preuss. Akad. Wiss. 154-167 (1918)}.
      \bibitem{1990GWave}
            V. B. Braginsky, N. S. Kardashev, A. G. Polnarev, and I. D. Novikov, \emph{
                  Propagation of Electromagnetic Radiation in a Random Field of
                  Gravitational Waves and Space Radio Interferometry},
            \href{https://link.springer.com/article/10.1007/BF02827323}{Nuovo Cimento B
                  {\bf 105}, 1141 (1990)}.
      \bibitem{1996GWave}
            T. Pyne,
            \emph{Gravitational Radiation and Very Long Baseline Interferometry}, \href{https://ui.adsabs.harvard.edu/abs/1996ApJ...465..566P/abstract}{
                  Astrophys. J. {\bf 465}, 566 (1996)}.
      \bibitem{2012GWave}
            J. A. Ellis, F. A. Jenet, and M. A. McLaughlin, \emph{Practical Methods for
                  Continuous Gravitational Wave Detection Using Pulsar Timing Data},
            \href{https://iopscience.iop.org/article/10.1088/0004-637X/753/2/96}{
                  Astrophys. J. {\bf 753}, 96 (2012)}.
        \bibitem{2017GWave}
            B. P. Abbott \emph{et al.}  (LIGO Scientific Collaboration and Virgo Collaboration), \emph{GW170817: Observation of Gravitational Waves from a Binary Neutron Star Inspiral},
            \href{https://journals.aps.org/prl/abstract/10.1103/PhysRevLett.119.161101}{
                  Phys. Rev. Lett. {\bf 119}, 161101 (2017)}.

      \bibitem{Heisenberg}
            W. Heisenberg, \emph{{\" U}ber Den Anschaulichen Inhalt Der Quantentheoretischen
                  Kinematik Und Mechanik}, \href{https://doi.org/10.1007/BF01397280}{
                        Zeitschrift F{\" u}r Physik {\bf 43}, 172 (1927)}.
      \bibitem{AB}
            Y. Aharonov and D. Bohm,
            \emph{Significance of Electromagnetic Potentials in the Quantum Theory}, \href{https://link.aps.org/doi/10.1103/PhysRev.115.485}{
                  Phys. Rev. {\bf 115}, 485 (1959)}.
      \bibitem{Berry}
            M. V. Berry, \emph{Quantal Phase Factors Accompanying Adiabatic Changes}, \href{https://doi.org/10.1098/rspa.1984.0023}{Proceedings of the Royal Society of London. A. Mathematical and Physical Sciences {\bf 392}, 45 (1984)}.
\end{thebibliography}
\end{document}


\renewcommand{\figureautorefname}{Fig.}
\renewcommand{\figurename}{Fig.}

\title{Supplemental Material for \\
      ``Predicting Angular-Momentum Waves Based on Yang-Mills Equations''}

\author{Xing-Yan Fan}
\affiliation{Theoretical Physics Division, Chern Institute of Mathematics, Nankai
      University, Tianjin 300071, People's Republic of China}

\author{Xiang-Ru Xie}
\affiliation{School of Physics, Nankai University, Tianjin 300071, People's Republic of
      China}

\author{Jing-Ling Chen}
\email{chenjl@nankai.edu.cn}
\affiliation{Theoretical Physics Division, Chern Institute of Mathematics, Nankai
      University, Tianjin 300071, People's Republic of China}

\date{\today}
\maketitle

\tableofcontents

\newpage

Throughout the supplemental material, we adopt the convention identical to Jackson \cite{1999Jackson}: the Greek indices take values $\{0,1,2,3\}$, and the Latin ones take $\{1,2,3\}$ (or $\{x,y,z\}$ without ambiguity). The contravariant coordinates in four-dimensional spacetime read $\bigl(x^\mu\bigr)=\bigl(c\,t,\:\vec{x}\bigr)$, while their covariant partner $\bigl(x_\mu\bigr)=\bigl(c\,t,\:-\vec{x}\bigr)$ with the rule of transformation $x_\mu =g_{\mu\nu} x^\nu$, and $x^\mu =g^{\mu\nu} x_\nu$, where the metric tensor
\begin{equation}
      g_{\mu\nu} =g^{\mu\nu} =\begin{bmatrix}
            1 & 0 & 0 & 0 \\
            0 & -1 & 0 & 0 \\
            0 & 0 & -1 & 0 \\
            0 & 0 & 0 & -1
      \end{bmatrix},
\end{equation}
and $c$ is the speed of light in vacuum. Similarly, one has the potentials as
\begin{equation}
     \bigl(\mathbb{A}^\mu\bigr)=\bigl(\varphi,\:\vec{A}\bigr), \;\;\;\;\; \bigl(\mathbb{A}_\mu\bigr)=\bigl(\varphi,\:-\vec{A}\bigr),
\end{equation}
and the partial differential operators as
\begin{equation}
      \partial^\mu \equiv\dfrac{\partial}{\partial\,x_\mu} =\left(
            \dfrac{1}{c} \dfrac{\partial}{\partial\,t},-\vec{\nabla}\right),\qquad
      \partial_\mu \equiv\dfrac{\partial}{\partial\,x^\mu} =\left(
            \dfrac{1}{c} \dfrac{\partial}{\partial\,t},\vec{\nabla}\right),
\end{equation}
with
\begin{equation}
     \vec{\nabla}=(\dfrac{\partial}{\partial x}, \dfrac{\partial}{\partial y}, \dfrac{\partial}{\partial z}).
\end{equation}

\section{Maxwell's Equations and Electromagnetic Waves}

      \subsection{Three Forms of Maxwell's Equations in Vacuum}
            \emph{First Form.---}In terms of the field-strength tensors, Maxwell's equations (sourceless and in vacuum) can be written in a covariant (Lorentz-covariant) form as follows~\cite{1999Jackson},
            \begin{subequations}\label{eq:MaxwFMuNu}
                  \begin{eqnarray}
                        && \partial_\mu F^{\mu\nu} = 0, \label{eq:PartAlphaF} \\
                        && \partial_\mu F_{\nu\gamma}+\partial_\nu F_{\gamma\mu}
                              +\partial_\gamma F_{\mu\nu} = 0, \label{eq:3PartAlphaF}
                  \end{eqnarray}
            \end{subequations}
            where the field-strength tensors are given by
            \begin{equation}
                  F^{\mu\nu} =\begin{bmatrix}
                        0 & -E_x & -E_y & -E_z \\
                        E_x & 0 & -B_z & B_y \\
                        E_y & B_z & 0 & -B_x \\
                        E_z & -B_y & B_x & 0
                  \end{bmatrix},
            \end{equation}
            and
            \begin{equation}
                  F_{\mu\nu} =\begin{bmatrix}
                        0 & E_x & E_y & E_z \\
                        -E_x & 0 & -B_z & B_y \\
                        -E_y & B_z & 0 & -B_x \\
                        -E_z & -B_y & B_x & 0
                  \end{bmatrix}.
            \end{equation}
            Here the electric field $\vec{E}=(F^{10}, F^{20}, F^{30})$ and the magnetic field $\vec{B}=(F^{32}, F^{13}, F^{21})$. The definition of $F_{\mu\nu}$ through the vector potential $\vec{A}$ and the scalar potential $\varphi$ is given by
            \begin{equation}\label{eq:FMuNu}
                  F_{\mu\nu} =\partial_\mu A_{\nu} -\partial_\nu A_{\mu}.
            \end{equation}

            \emph{Second Form.---}After that, in terms of electromagnetic fields we can recast Maxwell's equations in the original form (in Gaussian units, see e.g. \cite{EMWave}) as
            \begin{subequations}
                  \begin{eqnarray}
                  && \vec{\nabla}\cdot\vec{E} = 0, \label{eq:DivE} \\
                  && \vec{\nabla}\times\vec{E} = -\dfrac{1}{c}
                              \dfrac{\partial\,\vec{B}}{\partial\,t}, \label{eq:CurlE} \\
                  && \vec{\nabla}\cdot\vec{B} = 0, \label{eq:DivB} \\
                  && \vec{\nabla}\times\vec{B} = \dfrac{1}{c}
                              \dfrac{\partial\,\vec{E}}{\partial\,t}, \label{eq:CurlB}
                  \end{eqnarray}
            \end{subequations}
            Note that \Eq{eq:PartAlphaF} leads to Eqs. (\ref{eq:DivE}) and (\ref{eq:CurlB}), while \Eq{eq:3PartAlphaF} brings about Eqs. (\ref{eq:CurlE}) and (\ref{eq:DivB}).

             \emph{Third Form.---}Maxwell's equations written in terms of vector potential $\vec{A}$ and scalar potential $\varphi$. Since
            \begin{equation}
                  \vec{\nabla}\cdot (\vec{\nabla}\times\vec{A})=0
            \end{equation}
            holds for any vector $\vec{A}$, if one expresses the magnetic filed $\vec{B}$ via the vector potential $\vec{A}$ in the following form
            \begin{equation}\label{eq:BA}
                  \vec{B}=\vec{\nabla}\times\vec{A},
            \end{equation}
            he immediately has Eq. (\ref{eq:DivB}). After substituting Eq. (\ref{eq:BA}) into Eq. (\ref{eq:CurlE}), one obtains
            \begin{eqnarray}
            && \vec{\nabla}\times\left(\vec{E}+\dfrac{1}{c} \dfrac{\partial\,\vec{A}}{\partial\,t}
                  \right)=0,
            \end{eqnarray}
            which yields the electric filed as
            \begin{equation}\label{eq:EVarphiA}
                  \vec{E}=-\vec{\nabla}\varphi-\dfrac{1}{c} \dfrac{\partial\,\vec{A}}{\partial\,t},
            \end{equation}
            where $\varphi$ is the scalar potential and
            \begin{equation}
                  \vec{\nabla}\times (\vec{\nabla}\varphi)=0
            \end{equation}
            holds for any function $\varphi$. In this way, \Eq{eq:DivE} can be rewritten as
            \begin{equation}\label{eq:LaplaPhi-1}
                  -\nabla^2 \varphi-\dfrac{1}{c}
                        \dfrac{\partial\bigl(\vec{\nabla}\cdot\vec{A}\bigr)}{\partial\,t}=0.
            \end{equation}
            Substitute \Eq{eq:BA} and \Eq{eq:EVarphiA} into \Eq{eq:CurlB}, we have
            \begin{equation}\label{eq:C-1}
                  \vec{\nabla}\times(\vec{\nabla}\times\vec{A})+\dfrac{1}{c} \dfrac{\partial}{\partial\,t}
                        \left(\vec{\nabla}\varphi+\dfrac{1}{c} \dfrac{\partial\,\vec{A}}{\partial\,t}
                        \right)=0.
            \end{equation}
            Due to
            \begin{equation}
                  \vec{\nabla}\times(\vec{\nabla}\times\vec{A})
                  =\vec{\nabla}(\vec{\nabla}\cdot\vec{A})-\nabla^2 \vec{A},
            \end{equation}
            Eq. (\ref{eq:C-1}) becomes
            \begin{equation}\label{eq:DalembA-1}
                  \nabla^2 \vec{A}- \dfrac{1}{c^2} \dfrac{\partial^2 \vec{A}}{\partial\,t^2}
                  = \vec{\nabla}(\vec{\nabla}\cdot\vec{A})+\dfrac{1}{c} \vec{\nabla}\left(
                        \dfrac{\partial\,\varphi}{\partial\,t}\right).
            \end{equation}
            Recapitulate the results above: Maxwell's equations can be rephrased as the following potential form.
            \begin{subequations}
                  \begin{eqnarray}
                        && \nabla^2 \varphi+\dfrac{1}{c} \dfrac{\partial\bigl(
                              \vec{\nabla}\cdot\vec{A}\bigr)}{\partial\,t}=0, \label{eq:ME1a}\\
                        && \nabla^2 \vec{A}-\vec{\nabla}(\vec{\nabla}\cdot\vec{A})
                              -\dfrac{1}{c^2} \dfrac{\partial^2 \vec{A}}{\partial\,t^2}
                              -\dfrac{1}{c} \vec{\nabla}\left(\dfrac{\partial\,\varphi}{
                                    \partial\,t}\right)=0.\label{eq:ME1b}
                  \end{eqnarray}
            \end{subequations}

            In the following, we shall review how to use the potential method to solve Maxwell's equations as shown in Eq. (\ref{eq:ME1a}) and Eq. (\ref{eq:ME1b}).

      \subsection{Gauge Invariance and Solutions of Maxwell's Equations}
            Notice that if we set
            \begin{eqnarray}\label{eq:GaugeAPhi}
                  \vec{A}'=\vec{A}+\vec{\nabla}\,f,\qquad
                  \varphi'=\varphi-\dfrac{1}{c} \dfrac{\partial\,f}{\partial\,t},
            \end{eqnarray}
            with $f\equiv f(\vec{r},t)$ being an arbitrary function of $\vec{r}=(x,y,z)$ and $t$, we may find that $\vec{B}$ and
            $\vec{E}$ remain invariant, i.e.,
            \begin{eqnarray}
                  &&\vec{B}'=\vec{\nabla}\times\vec{A}'=\vec{\nabla}\times\vec{A}
                        =\vec{B},\nonumber\\
                  &&\vec{E}'=-\vec{\nabla}\varphi'-\dfrac{1}{c} \dfrac{\partial\,\vec{A}'}
                              {\partial\,t}
                        = -\vec{\nabla}\varphi-\dfrac{1}{c} \dfrac{\partial\,\vec{A}}
                              {\partial\,t}
                        =\vec{E}.
            \end{eqnarray}
            Therefore, Maxwell's equations in terms of $\vec{B}$ and $\vec{E}$ remain unchanged under the transformations as shown in \eqref{eq:GaugeAPhi} (which actually serves as a $U(1)$ gauge transformation).

            To determine the magnetic field $\vec{B}$ and the electric field $\vec{E}$, for simplicity we choose
            \begin{equation}\label{eq:con}
                  f=0, \quad \varphi=0,\quad \vec{\nabla}\cdot\vec{A}=0,
            \end{equation}
            then \Eq{eq:ME1a} holds automatically, and \Eq{eq:ME1b} becomes
            \begin{eqnarray}\label{eq:DPrEqA-1}
            && \nabla^2 \vec{A}-\dfrac{1}{c^2}\dfrac{\partial^2 \vec{A}}{\partial\,t^2} =0,
            \end{eqnarray}
            with $\vec{A}\equiv \vec{A}(\vec{r}, t)$. We now set
            \begin{equation}
                  \vec{A}(\vec{r}, t)=\vec{A}(\vec{r})\,{\rm e}^{-{\rm i}\,\omega\,t},
            \end{equation}
            where $\omega$ is the frequency, then \Eq{eq:DPrEqA-1} reduces to
            \begin{equation}
                  \nabla^2 \vec{A}(\vec{r})
                  +\dfrac{\omega^2}{c^2} \vec{A}(\vec{r})=0,
            \end{equation}
            whose solution reads
            \begin{equation}
                  \vec{A}(\vec{r})=\vec{A}_{01}\,{\rm e}^{
                              {\rm i}\vec{k}\cdot\vec{r}}
                        +\vec{A}_{02}\,{\rm e}^{-{\rm i}\vec{k}\cdot\vec{r}},
            \end{equation}
            where $k=|\vec{k}|=\omega/c$, $\vec{A}_{01}$ and $\vec{A}_{02}$ are two constant vectors. As a result, one obtains
            \begin{equation}
                  \vec{A}=\vec{A}(\vec{r})\,{\rm e}^{-{\rm i}\,\omega\,t}
                  =\left(\vec{A}_{01}\,{\rm e}^{{\rm i}\vec{k}\cdot\vec{r}}
                        +\vec{A}_{02}\,{\rm e}^{-{\rm i}\vec{k}\cdot\vec{r}}\right){\rm e}^{
                              -{\rm i}\,\omega\,t}.
            \end{equation}
            Notice that the condition $\vec{\nabla}\cdot\vec{A}=0$ in Eq. (\ref{eq:con}) leads to the following constraint
            \begin{equation}
                  (\vec{A}_{01} \cdot\vec{k}) {\rm e}^{
            {\rm i}\vec{k}\cdot\vec{r}}-\bigl(\vec{A}_{02} \cdot\vec{k}\bigr) {\rm e}^{
            -{\rm i}\vec{k}\cdot\vec{r}}=0,
            \end{equation}
            thus
            \begin{eqnarray}\label{eq:A-3}
                  \vec{A}_{01}\cdot \vec{k} =0, \;\;\;\;\;\vec{A}_{02}\cdot \vec{k}=0,
            \end{eqnarray}
            i.e., $\vec{A}_{01} \perp\vec{k} $ and $\vec{A}_{02} \perp\vec{k} $.

            In this case, from Eq. (\ref{eq:BA}) and Eq. (\ref{eq:EVarphiA}) we have the magnetic field and the electric field  as
            \begin{eqnarray}\label{eq:BB-1}
                  \vec{B} &=& {\rm i} \left[(\vec{k} \times \vec{A}_{01})\,{\rm e}^{{\rm i}\vec{k}\cdot\vec{r}}
                              - (\vec{k} \times \vec{A}_{02})\,{\rm e}^{-{\rm i}\vec{k}\cdot\vec{r}}\right] {\rm e}^{
                                    -{\rm i} \omega t},\nonumber\\
                  \vec{E} &=& {\rm i}\,k\left[\vec{A}_{01}\,{\rm e}^{{\rm i}\vec{k}\cdot\vec{r}}
                        +\vec{A}_{02}\,{\rm e}^{-{\rm i}\vec{k}\cdot\vec{r}}\right] {\rm e}^{
                              -{\rm i} \omega t}.
            \end{eqnarray}
            By using Eq. (\ref{eq:A-3}), we easily verify that
            \begin{eqnarray}\label{eq:A-4}
                  \vec{B} \cdot \vec{k} =0, \;\;\;\;\; \vec{E} \cdot \vec{k}=0,
                  \end{eqnarray}
            which means that the vectors $\vec{B}$ and $\vec{E}$ locate on the plane perpendicular to the vector $\vec{k}$.
            Moreover, we may obtain
            \begin{eqnarray}\label{eq:A-5}
                  \vec{B}\cdot\vec{E} &=& -k\,{\rm e}^{-{\rm i}\,2 \omega t} \left[
                        (\vec{k}\times\vec{A}_{01})\,{\rm e}^{{\rm i}\vec{k}\cdot\vec{r}}
                        -(\vec{k}\times\vec{A}_{02})\,{\rm e}^{-{\rm i}\vec{k}\cdot\vec{r}}\right]\cdot\left[\vec{A}_{01}\,{\rm e}^{{\rm i}\vec{k}\cdot\vec{r}}
                              +\vec{A}_{02}\,{\rm e}^{-{\rm i}\vec{k}\cdot\vec{r}}\right]\nonumber\\
                  &=&  -k\,{\rm e}^{-{\rm i}\,2 \omega t} \left[(\vec{k}\times\vec{A}_{01})
                              \cdot\vec{A}_{02}
                        -(\vec{k}\times\vec{A}_{02})\cdot\vec{A}_{01}\right] \nonumber\\
                  &=& -2\,k\,{\rm e}^{-{\rm i}\,2 \omega t}
                        (\vec{k} \times \vec{A}_{01})\cdot\vec{A}_{02}.
            \end{eqnarray}
            When $\vec{A}_{01}$ and $\vec{A}_{02}$ are not parallel, we have
            $\vec{B}\cdot\vec{E}\neq 0$. When $\vec{A}_{01}$ and $\vec{A}_{02}$ are
            parallel, $\vec{B}\perp\vec{E}$. Alternatively, let $\vec{A}_{02}$ be the zero vector, we also have $\vec{B}\perp\vec{E}$. In this case we simply have
            \begin{eqnarray}\label{eq:A-6}
                  \vec{B} &=& {\rm i} (\vec{k} \times \vec{A}_{01})\,{\rm e}^{
                        {\rm i}(\vec{k}\cdot\vec{r}-\omega t)}, \nonumber\\
                  \vec{E} &=& {\rm i}\, k\; \vec{A}_{01}\, {\rm e}^{
                        {\rm i}(\vec{k}\cdot\vec{r}-\omega t)},
            \end{eqnarray}
            accordingly the three vectors in the set $\{\vec{k}, \vec{E}, \vec{B}\}$ are mutually orthogonal, with $\hat{k}\times \vec{E}=\vec{B}$ and $\hat{k}=\vec{k}/k$.
      \subsection{Electromagnetic Waves}
            The physical meanings of $\vec{E}$ and $\vec{B}$ are just the electromagnetic waves, which are perpendicular to the propagation direction $\vec{k}$, and $c=\omega/k$ is the speed of propagation. To see this point clearly, let us derive the wave equations of $\vec{E}$ and $\vec{B}$ from Maxwell's equations as follows.

            From Eq. (\ref{eq:CurlE}), we have
            \begin{equation}
                  \vec{\nabla}\times\bigl(\vec{\nabla}\times\vec{E}\bigr)
                  =\vec{\nabla}\bigl(\vec{\nabla}\cdot\vec{E}\bigr)-\nabla^2 \vec{E}
                  =-\dfrac{1}{c} \dfrac{\partial\bigl(
                        \vec{\nabla}\times\vec{B}\bigr)}{\partial\,t},
            \end{equation}
            due to Eq. (\ref{eq:DivE}) and Eq. (\ref{eq:CurlB}), we arrive at
            \begin{equation}\label{eq:WavE-1}
                  \nabla^2 \vec{E}-\dfrac{1}{c^2} \dfrac{\partial\,\vec{E}}{\partial\,t^2} =0.
            \end{equation}
            Likewise, based on Eqs. (\ref{eq:CurlB}), (\ref{eq:DivB}) and (\ref{eq:CurlE}) we obtain
            \begin{equation}\label{eq:WavB-1}
                  \nabla^2 \vec{B}-\dfrac{1}{c^2} \dfrac{\partial\,\vec{B}}{\partial\,t^2} =0.
            \end{equation}
            Here, \Eq{eq:WavE-1} and \Eq{eq:WavB-1} are just the wave equations for the electromagnetic fields $\vec{E}$ and $\vec{B}$ respectively, whose solutions are given in Eq. (\ref{eq:BB-1}) or Eq. (\ref{eq:A-6}).
            \newpage

\section{Yang-Mills Equations and Two Kinds of Approximations}

      In 1954, Yang and Mills \cite{1954YangMills} extended Maxwell's equations to the non-Abelian cases, which are both local gauge covariant (see \ref{subsec:GagInvYM}) and Lorentz covariant (see \ref{subsec:LorenzYM}).

      \subsection{Yang-Mills Equations}

            The Yang-Mills equations are given by
            \begin{subequations}
                  \begin{eqnarray}
                  && D_\mu \mathcal{F}^{\mu\nu} =\partial_\mu \mathcal{F}^{\mu\nu}
                        -{\rm i}\,g\Bigl[\mathbb{A}_\mu ,\ \mathcal{F}^{\mu\nu}\Bigr]=0,
                        \label{eq:YM1a}\\
                  && D_\mu \mathcal{F}_{\nu\gamma} +D_\nu \mathcal{F}_{\gamma\mu}
                        +D_\gamma \mathcal{F}_{\mu\nu} =0, \label{eq:YM1b}
           \end{eqnarray}
           \end{subequations}
           with the gauge field-strength tensors
                  \begin{equation}
                        \mathcal{F}^{\mu\nu} =\begin{bmatrix}
                              0 & -\mathcal{E}_x & -\mathcal{E}_y & -\mathcal{E}_z \\
                              \mathcal{E}_x & 0 & -\mathcal{B}_z & \mathcal{B}_y \\
                              \mathcal{E}_y & \mathcal{B}_z & 0 & -\mathcal{B}_x \\
                              \mathcal{E}_z & -\mathcal{B}_y & \mathcal{B}_x & 0
                        \end{bmatrix},
                  \end{equation}
                  and
                  \begin{equation}
                        \mathcal{F}_{\mu\nu} =\begin{bmatrix}
                              0 & \mathcal{E}_x & \mathcal{E}_y & \mathcal{E}_z \\
                              -\mathcal{E}_x & 0 & -\mathcal{B}_z & \mathcal{B}_y \\
                              -\mathcal{E}_y & \mathcal{B}_z & 0 & -\mathcal{B}_x \\
                              -\mathcal{E}_z & -\mathcal{B}_y & \mathcal{B}_x & 0
                        \end{bmatrix}.
                  \end{equation}
                  Here the ``electric'' field $\vec{\mathcal{E}}=(\mathcal{F}^{10}, \mathcal{F}^{20}, \mathcal{F}^{30})$ and  the ``magnetic'' field $\vec{\mathcal{B}}=(\mathcal{F}^{32}, \mathcal{F}^{13}, \mathcal{F}^{21})$, respectively.
                  Note that the forms of $\mathcal{F}^{\mu\nu}$ and $\mathcal{F}_{\mu\nu}$ are exactly the same as those of  $F^{\mu\nu}$ and  $F_{\mu\nu}$, but the fields $\vec{\mathcal{E}}$ and $\vec{\mathcal{B}}$ are different from $\vec{E}$ and $\vec{B}$. The gauge field-strength tensors are defined by
                  \begin{eqnarray}\label{eq:nonabelian-1}
                        \mathcal{F}_{\mu\nu}=\partial_\mu \mathbb{A}_{\nu}
                              -\partial_\nu \mathbb{A}_{\mu}
                              +{\rm i}\,g[\mathbb{A}_{\mu},\
                                    \mathbb{A}_{\nu}],
                  \end{eqnarray}
                  where $g$ is the coupling parameter. Evidently, when $g=0$ or
                  $[\mathbb{A}_{\mu}, \mathbb{A}_{\nu}]=0$, the definition of $\mathcal{F}_{\mu\nu}$ reduces to the Abelian case, i.e. \Eq{eq:FMuNu}.

            Based on the definition (\ref{eq:nonabelian-1}),  one can have the ``magnetic'' field
            \begin{eqnarray}
                  \vec{\mathcal{B}}=\vec{\nabla}\times\vec{\mathcal{A}}
                        -{\rm i}\,g\bigl(\vec{\mathcal{A}}\times\vec{\mathcal{A}}\bigr),
            \end{eqnarray}
            together with the ``electric'' field
            \begin{equation}
                  \vec{\mathcal{E}}
                  =-\dfrac{1}{c} \dfrac{\partial\,\vec{\mathcal{A}}}{\partial\,t}
                        -\vec{\nabla}\varphi-{\rm i}\,g\left[\varphi,\ \vec{\mathcal{A}}
                        \right].
            \end{equation}
            Then in terms of $\{\vec{\mathcal{E}}, \vec{\mathcal{B}}, \vec{\mathcal{A}}, \varphi\}$ the Yang-Mills equations can be rewritten as the follow form
            \begin{subequations}
                  \begin{eqnarray}
                  & \vec{\nabla}\cdot\vec{\mathcal{E}}{\color{orange}+{\rm i}\,g\Bigl(
                        \vec{\mathcal{A}}\cdot\vec{\mathcal{E}}
                        -\vec{\mathcal{E}}\cdot\vec{\mathcal{A}}\Bigr)}=0, \label{eq:DivEYM} \\
                  & -\dfrac{1}{c} \dfrac{\partial}{\partial\,t} \vec{\mathcal{B}}
                        -\vec{\nabla}\times\vec{\mathcal{E}}
                        {\color{orange}+{\rm i}\,g\biggl(\Bigl[\varphi ,\ \vec{\mathcal{B}}\Bigr]
                              -\vec{\mathcal{A}}\times\vec{\mathcal{E}}
                              -\vec{\mathcal{E}}\times\vec{\mathcal{A}}\biggr)}=0, \label{eq:CurlEYM} \\
                  & \vec{\nabla}\cdot\vec{\mathcal{B}}{\color{orange}+{\rm i}\,g\Bigl(
                        \vec{\mathcal{A}}\cdot\vec{\mathcal{B}}
                        -\vec{\mathcal{B}}\cdot\vec{\mathcal{A}}\Bigr)}=0, \label{eq:DivBYM} \\
                  & -\dfrac{1}{c} \dfrac{\partial}{\partial\,t} \vec{\mathcal{E}}
                        +\vec{\nabla}\times\vec{\mathcal{B}}
                        {\color{orange}+{\rm i}\,g\biggl(\Bigl[\varphi ,\ \vec{\mathcal{E}}\Bigr]
                              +\vec{\mathcal{A}}\times\vec{\mathcal{B}}
                              +\vec{\mathcal{B}}\times\vec{\mathcal{A}}\biggr)}=0, \label{eq:CurlBYM}
           \end{eqnarray}
           \end{subequations}
            where the expressions in orange imply the self-interaction terms of potentials and fields in Yang-Mills equations, which come from the nonlinear terms in Eqs. (\ref{eq:YM1a}) and (\ref{eq:YM1b}), such as ${\rm i}\,g [\mathbb{A}_\mu ,\ \mathcal{F}^{\mu\nu}]$.

            The terms involving self-interaction above can be further colored via the order of $g$, i.e.
            \begin{subequations}
                  \begin{eqnarray}
                  && \vec{\nabla}\cdot\vec{\mathcal{E}}\textcolor{red}{
                        -{\rm i}\,g\Biggl\{\bigg[\vec{\mathcal{A}}\cdot\Big(
                                          \dfrac{1}{c} \dfrac{\partial\,
                                                \vec{\mathcal{A}}}{\partial\,t}\Big)
                                    -\Big(\dfrac{1}{c} \dfrac{\partial\,
                                          \vec{\mathcal{A}}}{\partial\,t}\Big)\cdot
                                                \vec{\mathcal{A}}\bigg]
                              +\vec{\mathcal{A}}\cdot\bigl(\vec{\nabla}\varphi\bigr)
                              -\bigl(\vec{\nabla}\varphi\bigr)\cdot\vec{\mathcal{A}}
                              \Biggr\}}
                        {\color{blue}+g^2 \biggl\{\vec{\mathcal{A}}\cdot\left[
                                    \varphi,\ \vec{\mathcal{A}}\right]
                              -\left[\varphi,\ \vec{\mathcal{A}}\right]
                                    \cdot\vec{\mathcal{A}}\biggr\}}=0, \label{eq:DivEYM3} \\
                  && -\dfrac{1}{c} \dfrac{\partial}{\partial\,t} \vec{\mathcal{B}}
                        -\vec{\nabla}\times\vec{\mathcal{E}}
                        \textcolor{red}{+{\rm i}\,g\Biggl\{\bigg[\vec{\mathcal{A}}
                                          \times\Big(\dfrac{1}{c}
                                                \dfrac{\partial\,\vec{\mathcal{A}}}{
                                                      \partial\,t}\Big)
                                    +\Big(\dfrac{1}{c} \dfrac{\partial\,
                                          \vec{\mathcal{A}}}{\partial\,t}\Big)\times
                                                \vec{\mathcal{A}}\bigg]
                              +\Bigl[\varphi,\ \big(\vec{\nabla}\times\vec{\mathcal{A}}
                                    \big)\Bigr]
                              +\bigl(\vec{\nabla}\varphi\bigr)\times\vec{\mathcal{A}}
                              +\vec{\mathcal{A}}\times\bigl(\vec{\nabla}\varphi\bigr)
                              \Biggr\}} \notag \\
                        &&\qquad\qquad\qquad\qquad\ {\color{blue}+g^2 \biggl\{\Bigl[\varphi,\ \bigl(
                                    \vec{\mathcal{A}}\times\vec{\mathcal{A}}\bigr)\Bigr]
                              -\vec{\mathcal{A}}\times\left[\varphi,\
                                    \vec{\mathcal{A}}\right]
                              -\left[\varphi,\ \vec{\mathcal{A}}\right]
                                    \times\vec{\mathcal{A}}\biggr\}}=0, \label{eq:CurlEYM3} \\
                  && \vec{\nabla}\cdot\vec{\mathcal{B}}
                        \textcolor{red}{-{\rm i}\,g\,\vec{\nabla}\cdot\bigl(\vec{\mathcal{A}}
                              \times\vec{\mathcal{A}}\bigr)}=0, \label{eq:DivBYM3} \\
                  && -\dfrac{1}{c} \dfrac{\partial}{\partial\,t} \vec{\mathcal{E}}
                        +\vec{\nabla}\times\vec{\mathcal{B}}
                        \textcolor{red}{-{\rm i}\,g\Biggl\{\bigg[\varphi,\
                                    \dfrac{1}{c} \dfrac{\partial\,\vec{\mathcal{A}}}{
                                          \partial\,t}\bigg]
                              +\Big[\varphi,\ \bigl(\vec{\nabla}\varphi\bigr)\Big]
                              -\vec{\mathcal{A}}\times\bigl(\vec{\nabla}\times
                                    \vec{\mathcal{A}}\bigr)
                              -\bigl(\vec{\nabla}\times\vec{\mathcal{A}}\bigr)\times
                                    \vec{\mathcal{A}}\Biggr\}} \notag \\
                        &&\qquad\qquad\qquad\qquad\ {\color{blue}+g^2 \biggl\{\Big[\varphi,\ \big[\varphi,\
                                    \vec{\mathcal{A}}\big]\Big]
                              +\vec{\mathcal{A}}\times\bigl(\vec{\mathcal{A}}
                                    \times\vec{\mathcal{A}}\bigr)
                              +\bigl(\vec{\mathcal{A}}\times\vec{\mathcal{A}}\bigr)
                                    \times\vec{\mathcal{A}}\biggr\}}=0, \label{eq:CurlBYM3}
            \end{eqnarray}
            \end{subequations}
            with the terms of self-interaction involving $g$ and $g^2$ in red and blue respectively.

      \subsection{Local Gauge Covariance of Yang-Mills Equations}\label{subsec:GagInvYM}

            The Yang-Mills equations are local gauge covariant \cite{1954YangMills}. Here we provide a brief review. The gauge transformation for the (non-Abelian) potentials is as follows \cite{2005QParadox}
            \begin{equation}\label{eq:GaugeTransfA}
                  \begin{cases}
                        & \vec{\mathcal{A}}'=U\,\vec{\mathcal{A}}\,U^\dagger
                              +\dfrac{{\rm i}\hbar\,c}{q} U\,\vec{\nabla}\left(
                                    U^\dagger\right), \\
                        & \\
                        & \varphi'=U\,\varphi\,U^\dagger -\dfrac{{\rm i}\hbar}{q}\,U\left(
                              \dfrac{\partial\,U^\dagger}{\partial\,t}\right)
                        =U\,\varphi\,U^\dagger -\dfrac{{\rm i}\hbar\,c}{q} U\left(
                              \dfrac{1}{c} \dfrac{\partial\,U^\dagger}{\partial\,t}
                              \right),
                  \end{cases}
            \end{equation}
            then \Eq{eq:GaugeTransfA} implies
            \begin{equation}
                  \mathbb{A}'=U\,\mathbb{A}\,U^\dagger
                        -\dfrac{{\rm i}\hbar\,c}{q} U\,(\partial_\mu U^\dagger).
            \end{equation}

            Let $g=-q/(\hbar\,c)$, then the (non-Abelian) potential $\mathbb{A}$ transforms as follows:
            \begin{equation}
                  \mathbb{A}'=U\,\mathbb{A}\,U^\dagger
                        +\dfrac{\rm i}{g} U\,(\partial_\mu U^\dagger)
                  =U\,\mathbb{A}\,U^\dagger
                        -\dfrac{\rm i}{g} (\partial_\mu U)U^\dagger,
            \end{equation}
            here we have used $U\,(\partial_\mu U^\dagger)=-(\partial_\mu U)U^\dagger$ (because $\partial_\mu (UU^\dagger)=0$).
            If we let
            \begin{equation}
                  U={\rm e}^{{\rm i}\,g\sum_a \theta_a \hat{G}_a}
                        ={\bf 1}+{\rm i}\,g\Big(\sum_a \theta_a \hat{G}_a\Big)+O(\theta^2),
                  \qquad\mathbb{A}_\mu =\sum_a \mathbb{A}_\mu^a \hat{G}_a,
            \end{equation}
            where $\theta_a \equiv\theta_a (x)$,
            $\mathbb{A}_\mu^a \equiv\mathbb{A}_\mu^a (x)$, ${\bf 1}$ is the identity operator, and $\hat{G}_a$'s $\big(a\in\{1,2,...,N^2 -1\}\big)$ represent the generators of the $SU(N)$ group, with $[\hat{G}_a,\ \hat{G}_b]={\rm i}\sum_c f^{abc} \hat{G}_c$, then
            \begin{eqnarray}
                    {\mathbb{A}'}_\mu &&=U\,\mathbb{A}_\mu U^\dagger
                              -\dfrac{\rm i}{g} (\partial_\mu U)U^\dagger \nonumber\\
                       && = \bigg[{\bf 1}+{\rm i}\,g\Big(\sum_a \theta_a \hat{G}_a\Big)
                                          +O(\theta^2)\bigg]
                                    \Bigl(\sum_b \mathbb{A}_\mu^b \hat{G}_b\Bigr)
                                    \bigg[{\bf 1}-{\rm i}\,g\Big(
                                          \sum_a \theta_a \hat{G}_a\Big)+O(\theta^2)\bigg]
                              -\dfrac{\rm i}{g} \partial_\mu \Big[{\bf 1}
                                    +{\rm i}\,g\sum_a (\theta_a \hat{G}_a)+O(\theta^2)
                                    \Big] U^\dagger \nonumber\\
                        &&= \Bigl(\sum_b \mathbb{A}_\mu^b \hat{G}_b\Bigr)
                              +{\rm i}\,g\Big(\sum_{a,b,c} \theta_a \mathbb{A}_\mu^b
                                    {\rm i}\,f^{abc} \hat{G}_c\Big)+O(\theta^2)
                              +\Bigl[\sum_a \partial_\mu (\theta_a \hat{G}_a)\Bigr]
                              +O(\theta^2) \nonumber\\
                        &&= \mathbb{A}_\mu -g\Big(\sum_{a,b,c} f^{abc} \theta_a
                                    \mathbb{A}_\mu^b \hat{G}_c\Big)
                              +\Bigl[\sum_a \partial_\mu (\theta_a)\hat{G}_a\Bigr]
                              +O(\theta^2)
                        =U\,\mathbb{A}_\mu U^\dagger +\Bigl[
                              \sum_a \partial_\mu (\theta_a)\hat{G}_a\Bigr]+O(\theta^2).
            \end{eqnarray}
            Since
            \begin{eqnarray}
                  &&  \partial_\mu {\mathbb{A}'}_{\nu} =\partial_\mu \left[
                                    U\,\mathbb{A}_\nu\,U^\dagger
                              -\dfrac{\rm i}{g} (\partial_\nu U)U^\dagger\right]
                        =(\partial_\mu U)\mathbb{A}_\nu\,U^\dagger
                              +U\,\partial_\mu (\mathbb{A}_\nu\,U^\dagger)
                              -\dfrac{\rm i}{g} \textcolor{black}{[}\partial_\mu (\partial_\nu U)\textcolor{black}{]}U^\dagger
                              -\dfrac{\rm i}{g} (\partial_\nu U)\textcolor{black}{(}\partial_\mu U^\dagger\textcolor{black}{)},\nonumber\\
                  &&  \partial_\nu {\mathbb{A}'}_{\mu} =\partial_\nu \left[
                                    U\,\mathbb{A}_\mu\,U^\dagger
                              -\dfrac{\rm i}{g} (\partial_\mu U)U^\dagger\right]
                        =(\partial_\nu U)\mathbb{A}_\mu\,U^\dagger
                              +U\,\partial_\nu (\mathbb{A}_\mu\,U^\dagger)
                              -\dfrac{\rm i}{g} \textcolor{black}{[}\partial_\nu (\partial_\mu U)\textcolor{black}{]}U^\dagger
                              -\dfrac{\rm i}{g} (\partial_\mu U)\textcolor{black}{(}\partial_\nu U^\dagger\textcolor{black}{)},
            \end{eqnarray}
            and
            \begin{eqnarray}
                       [{\mathbb{A}'}_{\mu},\ {\mathbb{A}'}_{\nu}]
                        &&=\left[U\,\mathbb{A}_\mu\,U^\dagger
                                    -\dfrac{\rm i}{g} U\,
                                          \partial_\mu U^\dagger,\
                              U\,\mathbb{A}_\nu\,U^\dagger
                                    -\dfrac{\rm i}{g} U\,
                                          \partial_\nu U^\dagger\right] \nonumber\\
                       &&= \left(U\,\mathbb{A}_\mu\,U^\dagger
                                    -\dfrac{\rm i}{g} U\,
                                          \partial_\mu U^\dagger\right)
                              \left(U\,\mathbb{A}_\nu\,U^\dagger
                                    -\dfrac{\rm i}{g} U\,
                                          \partial_\nu U^\dagger\right)
                              -\left(U\,\mathbb{A}_\nu\,U^\dagger
                                    -\dfrac{\rm i}{g} U\,
                                          \partial_\nu U^\dagger\right)
                              \left(U\,\mathbb{A}_\mu\,U^\dagger
                                    -\dfrac{\rm i}{g} U\,
                                          \partial_\mu U^\dagger\right) \nonumber\\
                        &&= U\,\mathbb{A}_\mu \mathbb{A}_\nu U^\dagger
                              -\dfrac{\rm i}{g} \big[
                                    U(\partial_\mu U^\dagger)
                                          U\,\mathbb{A}_\nu\,U^\dagger
                                    +U\,\mathbb{A}_\mu (\partial_\nu U^\dagger)
                                    \big]
                              -\dfrac{1}{g^2} U(\partial_\mu U^\dagger)U(
                                    \partial_\nu U^\dagger) \nonumber\\
                              &&\ -\left\{U\,\mathbb{A}_\nu \mathbb{A}_\mu
                                          U^\dagger
                                    -\dfrac{\rm i}{g} \big[U(
                                                \partial_\nu U^\dagger)U\,
                                                      \mathbb{A}_\mu\,U^\dagger
                                          +U\,\mathbb{A}_\nu (
                                                \partial_\mu U^\dagger)\big]
                                    -\dfrac{1}{g^2} U(
                                          \partial_\nu U^\dagger)U(
                                          \partial_\mu U^\dagger)\right\} \nonumber\\
                        &&= U\,\mathbb{A}_\mu \mathbb{A}_\nu U^\dagger
                              -\dfrac{\rm i}{g} \big[
                                    -(\partial_\mu U)U^\dagger
                                          U\,\mathbb{A}_\nu\,U^\dagger
                                    +U\,\mathbb{A}_\mu (\partial_\nu U^\dagger)
                                    \big]
                              +\dfrac{1}{g^2} (\partial_\mu U)
                                    U^\dagger U(\partial_\nu U^\dagger) \nonumber\\
                              &&\ -\left\{U\,\mathbb{A}_\nu \mathbb{A}_\mu
                                          U^\dagger
                                    -\dfrac{\rm i}{g} \big[
                                                -(\partial_\nu U)U^\dagger U\,
                                                      \mathbb{A}_\mu\,U^\dagger
                                          +U\,\mathbb{A}_\nu (
                                                \partial_\mu U^\dagger)\big]
                                    +\dfrac{1}{g^2} (\partial_\nu U)
                                          U^\dagger U(\partial_\mu U^\dagger)
                                          \right\} \nonumber\\
                        &&= U\,\mathbb{A}_\mu \mathbb{A}_\nu U^\dagger
                              -\dfrac{\rm i}{g} \big[
                                    -(\partial_\mu U)\mathbb{A}_\nu\,U^\dagger
                                    +U\,\mathbb{A}_\mu (\partial_\nu U^\dagger)
                                    \big]
                              +\dfrac{1}{g^2} (\partial_\mu U)      (
                                    \partial_\nu U^\dagger) \nonumber\\
                              &&\ -\left\{U\,\mathbb{A}_\nu \mathbb{A}_\mu
                                          U^\dagger
                                    -\dfrac{\rm i}{g} \big[-
                                                (\partial_\nu U)\mathbb{A}_\mu\,
                                                      U^\dagger
                                          +U\,\mathbb{A}_\nu (
                                                \partial_\mu U^\dagger)\big]
                                    +\dfrac{1}{g^2} (\partial_\nu U)(
                                          \partial_\mu U^\dagger)\right\},
            \end{eqnarray}
            we then have
            \begin{eqnarray}
                         \mathcal{F}_{\mu\nu}' &&=\partial_\mu {\mathbb{A}'}_{\nu}
                              -\partial_\nu {\mathbb{A}'}_{\mu}
                              +{\rm i}\,g[{\mathbb{A}'}_{\mu},\
                                    {\mathbb{A}'}_{\nu}] \nonumber\\
                        &&= (\partial_\mu U)\mathbb{A}_\nu\,U^\dagger
                                    +U\,\partial_\mu (\mathbb{A}_\nu\,U^\dagger)
                                    -\dfrac{\rm i}{g} (\partial_\mu U)
                                          \partial_\nu U^\dagger
                                    -\dfrac{\rm i}{g} U\,
                                          \partial_\mu (\partial_\nu U^\dagger) \nonumber\\
                              && -\Big[(\partial_\nu U)\mathbb{A}_\mu\,U^\dagger
                                    +U\,\partial_\nu (\mathbb{A}_\mu\,U^\dagger)
                                    -\dfrac{\rm i}{g} (\partial_\nu U)
                                          \partial_\mu U^\dagger
                                    -\dfrac{\rm i}{g} U\,
                                          \partial_\nu (\partial_\mu U^\dagger)
                                    \Big]
                              +{\rm i}\,g U\big[
                                    \mathbb{A}_\mu,\ \mathbb{A}_\nu\big] U^\dagger
                                    \nonumber\\
                              && +\big[-(\partial_\mu U)\mathbb{A}_\nu\,U^\dagger
                                    +U\,\mathbb{A}_\mu (\partial_\nu U^\dagger)
                                    +(\partial_\nu U)\mathbb{A}_\mu\,U^\dagger
                                    -U\,\mathbb{A}_\nu (\partial_\mu U^\dagger)
                                    \big]
                              +\dfrac{\rm i}{g} \Big[
                                    (\partial_\mu U)(\partial_\nu U^\dagger)
                                    -(\partial_\nu U)(\partial_\mu U^\dagger)\Big]
                                    \nonumber\\
                        &&= U\,\partial_\mu (\mathbb{A}_\nu\,U^\dagger)
                              -U\,\partial_\nu (\mathbb{A}_\mu\,U^\dagger)
                              +{\rm i}\,g U\big[
                                    \mathbb{A}_\mu,\ \mathbb{A}_\nu\big] U^\dagger
                              +U\,\mathbb{A}_\mu (\partial_\nu U^\dagger)
                                    -U\,\mathbb{A}_\nu (\partial_\mu U^\dagger) \nonumber\\
                        &&= U\Big[\partial_\mu (\mathbb{A}_\nu\,U^\dagger)
                                    -\mathbb{A}_\nu (\partial_\mu U^\dagger)\Big]
                              -U\Big[\partial_\nu (\mathbb{A}_\mu\,U^\dagger)
                                    -\mathbb{A}_\mu (\partial_\nu U^\dagger)\Big]
                              +{\rm i}\,g U\big[
                                    \mathbb{A}_\mu,\ \mathbb{A}_\nu\big]U^\dagger
                                    \nonumber\\
                        &&= U\big(\partial_\mu \mathbb{A}_\nu
                                    -\partial_\nu \mathbb{A}_\mu\big)U^\dagger
                              +{\rm i}\,g U\big[
                                    \mathbb{A}_\mu,\ \mathbb{A}_\nu\big]U^\dagger
                        =U\,\mathcal{F}_{\mu\nu} U^\dagger.
            \end{eqnarray}
            Because
            \begin{eqnarray}
                         \partial_\mu {\mathcal{F}'}^{\mu\nu}
                        &&=\partial_\mu \bigl(U\,\mathcal{F}^{\mu\nu} U^\dagger\bigr)
                        =(\partial_\mu U)\mathcal{F}^{\mu\nu} U^\dagger
                              +U\bigl(\partial_\mu \mathcal{F}^{\mu\nu}\bigr)U^\dagger
                              +U\,\mathcal{F}^{\mu\nu} (\partial_\mu U^\dagger) \nonumber\\
                        &&= {\rm i}\,g\Big[\sum_a (\partial_\mu\theta_a) \hat{G}_a\Big]U\,
                                    \mathcal{F}^{\mu\nu} U^\dagger
                              +U\bigl(\partial_\mu \mathcal{F}^{\mu\nu}\bigr)U^\dagger
                              -{\rm i}\,g\,U\,\mathcal{F}^{\mu\nu} \Big[
                                    \sum_a (\partial_\mu\theta_a) \hat{G}_a\Big]U^\dagger
                              +O(\theta^2) \nonumber\\
                        &&= {\rm i}\,g\Big[\sum_a (\partial_\mu\theta_a) \hat{G}_a\Big]
                                    \mathcal{F}^{\mu\nu} U^\dagger
                              +U\bigl(\partial_\mu \mathcal{F}^{\mu\nu}\bigr)U^\dagger
                              -{\rm i}\,g\,U\,\mathcal{F}^{\mu\nu} \Big[
                                    \sum_a (\partial_\mu\theta_a) \hat{G}_a\Big]
                              +O(\theta^2) \nonumber\\
                        &&= {\rm i}\,g\Big[\sum_a (\partial_\mu\theta_a) \hat{G}_a\Big]
                                    \mathcal{F}^{\mu\nu}
                              +U\bigl(\partial_\mu \mathcal{F}^{\mu\nu}\bigr)U^\dagger
                              -{\rm i}\,g\,\mathcal{F}^{\mu\nu} \Big[
                                    \sum_a (\partial_\mu\theta_a) \hat{G}_a\Big]
                              +O(\theta^2) \nonumber\\
                        &&= U\bigl(\partial_\mu \mathcal{F}^{\mu\nu}\bigr)U^\dagger
                              +{\rm i}\,g\Big[\sum_a (\partial_\mu\theta_a) \hat{G}_a ,\
                                    \sum_b \mathcal{F}^{\mu\nu}_b \hat{G}_b\Big]
                              +O(\theta^2).
            \end{eqnarray}
            and
            \begin{eqnarray}
                         D'_\mu {\mathcal{F}'}^{\mu\nu}
                        &&=\partial_\mu {\mathcal{F}'}^{\mu\nu}
                              -{\rm i}\,g\Bigl[{\mathbb{A}'}_\mu ,\
                                    {\mathcal{F}'}^{\mu\nu}\Bigr]
                              \nonumber\\
                        &&= U\bigl(\partial_\mu \mathcal{F}^{\mu\nu}\bigr)U^\dagger
                              +{\rm i}\,g\Big[\sum_a (\partial_\mu\theta_a) \hat{G}_a ,\
                                    \sum_b \mathcal{F}^{\mu\nu}_b \hat{G}_b\Big]
                              -{\rm i}\,g\bigg[U\,\mathbb{A}_\mu U^\dagger
                                          +\sum_a \partial_\mu (\theta_a)\hat{G}_a ,\
                                    U\,\mathcal{F}^{\mu\nu} U^\dagger\bigg] \nonumber\\
                        &&= U\bigl(\partial_\mu \mathcal{F}^{\mu\nu}\bigr)U^\dagger
                              -{\rm i}\,g\bigg[U\,\mathbb{A}_\mu U^\dagger ,\
                                    U\,\mathcal{F}^{\mu\nu} U^\dagger\bigg]
                              +{\rm i}\,g\Big[\sum_a (\partial_\mu\theta_a) \hat{G}_a ,\
                                    \sum_b \mathcal{F}^{\mu\nu}_b \hat{G}_b\Big]
                              -{\rm i}\,g\bigg[\sum_a \partial_\mu (\theta_a)\hat{G}_a ,\
                                    U\,\mathcal{F}^{\mu\nu} U^\dagger\bigg] \nonumber\\
                        &&= U\biggl\{\partial_\mu \mathcal{F}^{\mu\nu}
                                    -{\rm i}\,g\Bigl[\mathbb{A}_\mu ,\
                                          \mathcal{F}^{\mu\nu}\Bigr]\biggr\}U^\dagger
                              +{\rm i}\,g\Big[\sum_a (\partial_\mu\theta_a) \hat{G}_a ,\
                                    \sum_b \mathcal{F}^{\mu\nu}_b \hat{G}_b\Big]
                              -{\rm i}\,g\bigg[\sum_a \partial_\mu (\theta_a)\hat{G}_a ,\
                                    \sum_b \mathcal{F}^{\mu\nu}_b \hat{G}_b\bigg] \nonumber\\
                        &&= U(D_\mu \mathcal{F}^{\mu\nu})U^\dagger.
            \end{eqnarray}
            Thus one can obtain the following relations
            \begin{equation}
                  \begin{cases}
                        & D_\mu \mathcal{F}^{\mu\nu} =0\xrightarrow{
                                    \rm local\ gauge\ transformation}
                              D'_\mu {\mathcal{F}'}^{\mu\nu} =0, \\
                        & D_\mu \mathcal{F}_{\nu\gamma} +D_\nu \mathcal{F}_{\gamma\mu}
                              +D_\gamma \mathcal{F}_{\mu\nu} =0\xrightarrow{
                                    \rm local\ gauge\ transformation}
                              D'_\mu \mathcal{F}'_{\nu\gamma}
                              +D'_\nu \mathcal{F}'_{\gamma\mu}
                              +D'_\gamma \mathcal{F}'_{\mu\nu} =0,
                  \end{cases}
            \end{equation}
            which mean the Yang-Mills equations are local gauge covariant.

      \subsection{Lorentz Covariant of Yang-Mills Equations}\label{subsec:LorenzYM}

             The Yang-Mills equations are Lorentz covariant. Here we provide a brief review. Assume there is another inertial reference system $(t', x',y',z')$ moves uniformly with the velocity $v$ along the $\hat{z}$-axis with respect to the former one $(t,x,y,z)$. Then we have the Lorentz transformation:
            \begin{equation}\label{eq:XPrimX1}
                  x'^\mu=c_{\mu\nu} x^\nu,
            \end{equation}
            with $c_{\mu\nu}$ signifying the elements of the Lorentz transformation matrix
            \begin{equation}
                  (c_{\mu\nu})=C\equiv\begin{bmatrix}
                        \gamma & 0 & 0 & -\gamma\dfrac{v}{c} \\
                        0 & 1 & 0 & 0 \\
                        0 & 0 & 1 & 0 \\
                        -\gamma\dfrac{v}{c} & 0 & 0 & \gamma
                  \end{bmatrix},
            \end{equation}
            and the Lorentz factor
            \begin{equation}
                  \gamma\equiv\dfrac{1}{\sqrt{1-v^2 /c^2}}.
            \end{equation}
            \begin{remark}
                  Since $\gamma^2 -\gamma^2 (v^2 /c^2)=1$, then
                  \begin{equation}
                        \begin{split}
                              & (c_{\mu\nu})(c^{\alpha\beta}) \equiv
                                    \begin{bmatrix}
                                          \gamma & 0 & 0 & -\gamma\dfrac{v}{c} \\
                                          0 & 1 & 0 & 0 \\
                                          0 & 0 & 1 & 0 \\
                                          -\gamma\dfrac{v}{c} & 0 & 0 & \gamma
                                    \end{bmatrix}\begin{bmatrix}
                                          \gamma & 0 & 0 & \gamma\dfrac{v}{c} \\
                                          0 & 1 & 0 & 0 \\
                                          0 & 0 & 1 & 0 \\
                                          \gamma\dfrac{v}{c} & 0 & 0 & \gamma
                                    \end{bmatrix}
                              ={\bf 1},
                        \end{split}
                  \end{equation}
                  namely
                  \begin{equation}
                        C^{-1} \equiv\begin{bmatrix}
                              \gamma & 0 & 0 & \gamma\dfrac{v}{c} \\
                              0 & 1 & 0 & 0 \\
                              0 & 0 & 1 & 0 \\
                              \gamma\dfrac{v}{c} & 0 & 0 & \gamma
                        \end{bmatrix}.
                  \end{equation}
                  According to the invariance of relativistic interval
                  $x'^\mu x'_\mu =c_{\mu\nu} x^\nu c^{\mu\alpha} x_\alpha
                  =c_{\mu\nu} c^{\mu\alpha} x^\nu x_\alpha =x^\nu x_\nu$, which leads to $c_{\mu\nu} c^{\mu\alpha} =\delta^\alpha_\nu$.
            \end{remark}
            Note \Eq{eq:XPrimX1} can be generalized to differential operations under the Lorentz transformation $C$. Because
            \begin{equation}
                  x'^\mu=c_{\mu\nu} x^\nu \Rightarrow
                  \dfrac{\partial\,x'^\mu}{\partial\,x^\nu} =c_{\mu\nu}.
            \end{equation}
            After that,
            \begin{equation}
                  \partial'_\mu=\dfrac{\partial}{\partial\,x'^\mu}
                  =\dfrac{\partial\,x^\nu}{\partial\,x'^\mu}
                        \dfrac{\partial}{\partial\,x^\nu}
                  =\dfrac{1}{\frac{\partial\,x'^\mu}{\partial\,x^\nu}}
                        \dfrac{\partial}{\partial\,x^\nu}
                  =\dfrac{1}{c_{\mu\nu}} \dfrac{\partial}{\partial\,x^\nu}
                  =\dfrac{\delta^\mu_\mu}{c_{\mu\nu}} \partial_\nu
                  =c^{\mu\nu} \partial_\nu.
            \end{equation}
            Similarly, we obtain
            \begin{equation}
                  \mathbb{A}_\mu'=c^{\mu\nu} \mathbb{A}_\nu.
            \end{equation}

            After that, we arrive at
            \begin{eqnarray}
                 \mathcal{F}'_{\mu\nu} & =& \partial'_\mu \mathbb{A}'_\nu
                                    -\partial'_\nu \mathbb{A}'_\mu
                              +{\rm i}\,g[\mathbb{A}'_\mu,\
                                    \mathbb{A}'_\nu]
                        =c^{\mu\alpha} c^{\nu\beta} F_{\alpha\beta}
                              +{\rm i}\,g[
                                    c^{\mu\alpha} \mathbb{A}_\alpha,\
                                    c^{\nu\beta} \mathbb{A}_\beta] \nonumber\\
                       & =& c^{\mu\alpha} c^{\nu\beta} \biggl\{F_{\alpha\beta}
                              +{\rm i}\,g[\mathbb{A}_\alpha,\
                                    \mathbb{A}_\beta]\biggr\}
                        =c^{\mu\alpha} c^{\nu\beta} \mathcal{F}_{\alpha\beta}.
            \end{eqnarray}
            Likewise, ${\mathcal{F}'}^{\mu\nu} =c_{\mu\alpha}\,c_{\nu\beta} \mathcal{F}^{\alpha\beta}$. Further, we attain
            \begin{equation}
                  \partial'_\mu {\mathcal{F}'}^{\mu\nu}
                  =c_{\nu\xi} \partial_\alpha \mathcal{F}^{\alpha\xi}.
            \end{equation}
            \begin{equation}
                  \partial'_\mu \mathcal{F}'_{\nu\alpha}
                        +\partial'_\nu \mathcal{F}'_{\alpha\mu}
                        +\partial'_\alpha \mathcal{F}'_{\mu\nu}
                  =c^{\mu\beta} c^{\nu\zeta} c^{\alpha\xi} \bigl(
                        \partial_\beta \mathcal{F}_{\zeta\xi}
                        +\partial_\zeta \mathcal{F}_{\xi\beta}
                        +\partial_\xi \mathcal{F}_{\beta\zeta}\bigr).
            \end{equation}
            Hence for \Eq{eq:YM1a}, one has
            \begin{eqnarray}
                         D'_\mu {\mathcal{F}'}^{\mu\nu}
                        &=&\partial'_\mu {\mathcal{F}'}^{\mu\nu} -{\rm i}\,g\Bigl[
                              \mathbb{A}'_\mu ,\ {\mathcal{F}'}^{\mu\nu}\Bigr]
                        =c_{\nu\xi} \partial_\alpha \mathcal{F}^{\alpha\xi}
                              -{\rm i}\,g\Bigl[c^{\mu\beta} \mathbb{A}_\beta ,\
                                    c_{\mu\kappa}\,c_{\nu\lambda}
                                          \mathcal{F}^{\kappa\lambda}\Bigr] \nonumber\\
                        &=& c_{\nu\xi} \partial_\alpha \mathcal{F}^{\alpha\xi}
                              -c^{\mu\beta} c_{\mu\kappa}\,c_{\nu\lambda} \Bigl\{
                                    {\rm i}\,g\bigl[\mathbb{A}_\beta ,\
                                          \mathcal{F}^{\kappa\lambda}\bigr]\Bigr\}
                        =c_{\nu\xi} \partial_\alpha \mathcal{F}^{\alpha\xi}
                              -\delta^\beta_\kappa\,c_{\nu\lambda} \Bigl\{
                                    {\rm i}\,g\bigl[\mathbb{A}_\beta ,\
                                          \mathcal{F}^{\kappa\lambda}\bigr]\Bigr\} \nonumber\\
                        &=& c_{\nu\xi} \partial_\alpha \mathcal{F}^{\alpha\xi}
                              -c_{\nu\lambda} \Bigl\{{\rm i}\,g\bigl[\mathbb{A}_\beta,\
                                    \mathcal{F}^{\beta\lambda}\bigr]\Bigr\}
                        =c_{\nu\xi} \Bigl\{\partial_\alpha \mathcal{F}^{\alpha\xi}
                              -{\rm i}\,g\bigl[\mathbb{A}_\alpha,\
                                    \mathcal{F}^{\alpha\xi}\bigr]\Bigr\}
                        =c_{\nu\xi} D_\alpha \mathcal{F}^{\alpha\xi}.
           \end{eqnarray}
            With the same argument, for \Eq{eq:YM1b}, one has
           \begin{eqnarray}
                        && D'_\mu \mathcal{F}'_{\nu\alpha}
                              +D'_\nu \mathcal{F}'_{\alpha\mu}
                              +D'_\alpha \mathcal{F}'_{\mu\nu}\nonumber\\
                       && =\partial'_\mu \mathcal{F}'_{\nu\alpha} -{\rm i}\,g\Bigl[
                                    \mathbb{A}'_\mu ,\ \mathcal{F}'_{\nu\alpha}\Bigr]
                              +\partial'_\nu \mathcal{F}'_{\alpha\mu} -{\rm i}\,g\Bigl[
                                    \mathbb{A}'_\nu ,\ \mathcal{F}'_{\alpha\mu}\Bigr]
                              +\partial'_\alpha \mathcal{F}'_{\mu\nu} -{\rm i}\,g\Bigl[
                                    \mathbb{A}'_\alpha ,\ \mathcal{F}'_{\mu\nu}\Bigr] \nonumber\\
                       &&= c^{\mu\beta} c^{\nu\zeta} c^{\alpha\xi} \bigl(
                                    \partial_\beta \mathcal{F}_{\zeta\xi}
                                    +\partial_\zeta \mathcal{F}_{\xi\beta}
                                    +\partial_\xi \mathcal{F}_{\beta\zeta}\bigr)
                                    -{\rm i}\,g\biggl\{\Bigl[c^{\mu\beta} \mathbb{A}_\beta ,
                                    \ c^{\nu\kappa}\,c^{\alpha\lambda}
                                                \mathcal{F}_{\kappa\lambda}\Bigr]
                                    +\Bigl[c^{\nu\rho} \mathbb{A}_\rho ,\
                                          c^{\alpha\phi}\,c^{\mu\psi}
                                                \mathcal{F}_{\phi\psi}\Bigr]
                                    +\Bigl[c^{\alpha\lambda} \mathbb{A}_\lambda ,\
                                          c^{\mu\zeta}\,c^{\nu\eta}
                                                \mathcal{F}_{\zeta\eta}\Bigr]\biggr\} \nonumber\\
                        &&= c^{\mu\beta} c^{\nu\zeta} c^{\alpha\xi} \bigl(
                                    \partial_\beta \mathcal{F}_{\zeta\xi}
                                    +\partial_\zeta \mathcal{F}_{\xi\beta}
                                    +\partial_\xi \mathcal{F}_{\beta\zeta}\bigr)
                                    -{\rm i}\,g\biggl\{c^{\mu\beta} c^{\nu\kappa}\,
                                          c^{\alpha\lambda}\Bigl[\mathbb{A}_\beta ,\
                                                \mathcal{F}_{\kappa\lambda}\Bigr]
                                    +c^{\nu\rho} c^{\alpha\phi}\,c^{\mu\psi} \Bigl[
                                          \mathbb{A}_\rho ,\ \mathcal{F}_{\phi\psi}
                                          \Bigr]
                                    +c^{\alpha\lambda} c^{\mu\zeta}\,c^{\nu\eta} \Bigl[
                                          \mathbb{A}_\lambda ,\ \mathcal{F}_{\zeta\eta}
                                          \Bigr]\biggr\} \nonumber\\
                        &&= c^{\mu\beta} c^{\nu\zeta} c^{\alpha\xi} \bigl(
                                    \partial_\beta \mathcal{F}_{\zeta\xi}
                                    +\partial_\zeta \mathcal{F}_{\xi\beta}
                                    +\partial_\xi \mathcal{F}_{\beta\zeta}\bigr)
                                    -{\rm i}\,g\biggl\{c^{\mu\beta} c^{\nu\zeta}\,
                                          c^{\alpha\xi}\Bigl[\mathbb{A}_\beta ,\
                                                \mathcal{F}_{\zeta\xi}\Bigr]
                                    +c^{\nu\zeta} c^{\alpha\xi}\,c^{\mu\beta} \Bigl[
                                          \mathbb{A}_\zeta ,\ \mathcal{F}_{\xi\beta}
                                          \Bigr]
                                    +c^{\alpha\xi} c^{\mu\beta}\,c^{\nu\zeta} \Bigl[
                                          \mathbb{A}_\xi ,\ \mathcal{F}_{\beta\zeta}
                                          \Bigr]\biggr\} \nonumber\\
                        &&= c^{\mu\beta} c^{\nu\zeta} c^{\alpha\xi} \Biggl\{\bigl(
                                    \partial_\beta \mathcal{F}_{\zeta\xi}
                                    +\partial_\zeta \mathcal{F}_{\xi\beta}
                                    +\partial_\xi \mathcal{F}_{\beta\zeta}\bigr)
                              -{\rm i}\,g\biggl(\Bigl[\mathbb{A}_\beta ,\
                                                \mathcal{F}_{\zeta\xi}\Bigr]
                                    +\Bigl[\mathbb{A}_\zeta ,\ \mathcal{F}_{\xi\beta}
                                          \Bigr]
                                    +\Bigl[\mathbb{A}_\xi ,\ \mathcal{F}_{\beta\zeta}
                                          \Bigr]\biggr)\Biggr\} \nonumber\\
                        &&= c^{\mu\beta} c^{\nu\zeta} c^{\alpha\xi} \bigl(
                              D_\beta \mathcal{F}_{\zeta\xi}
                              +D_\zeta \mathcal{F}_{\xi\beta}
                              +D_\xi \mathcal{F}_{\beta\zeta}\bigr).
           \end{eqnarray}
            In summary, we have
            \begin{subequations}
                  \begin{eqnarray}
                  && D'_\mu {\mathcal{F}'}^{\mu\nu} =0, \\
                  && D'_\mu \mathcal{F}'_{\nu\gamma} +D'_\nu \mathcal{F}'_{\gamma\mu}
                        +D'_\gamma \mathcal{F}'_{\mu\nu} =0,
            \end{eqnarray}
            \end{subequations}
            i.e., the Yang-Mills equations are Lorentz covariant.

            The existence of highly nonlinear terms retards the resolution for the Yang-Mills equations, namely, it is hard to obtain the exact solutions for the case of non-Abelian potentials. Hence, in this work we shall consider two kinds of significant approximations.

      \subsection{The First Kind of Approximation: Weak-Coupling Approximation}

            When the coupling parameter $g$ is sufficient small, one can set $g^2\approx 0$. In this case, we can omit the terms referring to $g^2$ in \Eq{eq:DivEYM3}-\Eq{eq:CurlBYM3}, then they reduce to
            \begin{subequations}
                  \begin{eqnarray}
                  && \vec{\nabla}\cdot\vec{\mathcal{E}}\textcolor{red}{
                        -{\rm i}\,g\Biggl\{\bigg[\vec{\mathcal{A}}\cdot\Big(
                                          \dfrac{1}{c} \dfrac{\partial\,
                                                \vec{\mathcal{A}}}{\partial\,t}\Big)
                                    -\Big(\dfrac{1}{c} \dfrac{\partial\,
                                          \vec{\mathcal{A}}}{\partial\,t}\Big)\cdot
                                                \vec{\mathcal{A}}\bigg]
                              +\vec{\mathcal{A}}\cdot\bigl(\vec{\nabla}\varphi\bigr)
                              -\bigl(\vec{\nabla}\varphi\bigr)\cdot\vec{\mathcal{A}}
                              \Biggr\}}=0, \label{eq:DivEYM4} \\
                  && -\dfrac{1}{c} \dfrac{\partial}{\partial\,t} \vec{\mathcal{B}}
                        -\vec{\nabla}\times\vec{\mathcal{E}}
                        \textcolor{red}{+{\rm i}\,g\Biggl\{\bigg[\vec{\mathcal{A}}
                                          \times\Big(\dfrac{1}{c}
                                                \dfrac{\partial\,\vec{\mathcal{A}}}{
                                                      \partial\,t}\Big)
                                    +\Big(\dfrac{1}{c} \dfrac{\partial\,
                                          \vec{\mathcal{A}}}{\partial\,t}\Big)\times
                                                \vec{\mathcal{A}}\bigg]
                              +\Bigl[\varphi,\ \big(\vec{\nabla}\times\vec{\mathcal{A}}
                                    \big)\Bigr]
                              +\bigl(\vec{\nabla}\varphi\bigr)\times\vec{\mathcal{A}}
                              +\vec{\mathcal{A}}\times\bigl(\vec{\nabla}\varphi\bigr)
                              \Biggr\}}=0, \label{eq:CurlEYM4} \\
                  && \vec{\nabla}\cdot\vec{\mathcal{B}}
                        \textcolor{red}{-{\rm i}\,g\,\vec{\nabla}\cdot\bigl(\vec{\mathcal{A}}
                              \times\vec{\mathcal{A}}\bigr)}=0, \label{eq:DivBYM4} \\
                  && -\dfrac{1}{c} \dfrac{\partial}{\partial\,t} \vec{\mathcal{E}}
                        +\vec{\nabla}\times\vec{\mathcal{B}}
                        \textcolor{red}{-{\rm i}\,g\Biggl\{\bigg[\varphi,\
                                    \dfrac{1}{c} \dfrac{\partial\,\vec{\mathcal{A}}}{
                                          \partial\,t}\bigg]
                              +\Big[\varphi,\ \bigl(\vec{\nabla}\varphi\bigr)\Big]
                              -\vec{\mathcal{A}}\times\bigl(\vec{\nabla}\times
                                    \vec{\mathcal{A}}\bigr)
                              -\bigl(\vec{\nabla}\times\vec{\mathcal{A}}\bigr)\times
                                    \vec{\mathcal{A}}\Biggr\}}=0. \label{eq:CurlBYM4}
            \end{eqnarray}
            \end{subequations}
            For convenient, we would like to call Eq. (\ref{eq:DivEYM4})-Eq. (\ref{eq:CurlBYM4}) as the Yang-Mills equations under weak-coupling approximation (The WCA-YM equations).

      \subsection{The Second  Kind of Approximation: Zero-Coupling Approximation}

            Further, if we omit the red terms in \Eq{eq:DivEYM4}-\Eq{eq:CurlBYM4}, then they degenerate into
            \begin{subequations}
                  \begin{eqnarray}
                  && \vec{\nabla}\cdot\vec{\mathcal{E}}=0 \label{eq:DivEYM5} \\
                  && -\dfrac{1}{c} \dfrac{\partial}{\partial\,t} \vec{\mathcal{B}}
                        -\vec{\nabla}\times\vec{\mathcal{E}}=0, \label{eq:CurlEYM5} \\
                  && \vec{\nabla}\cdot\vec{\mathcal{B}}=0, \label{eq:DivBYM5} \\
                  && -\dfrac{1}{c} \dfrac{\partial}{\partial\,t} \vec{\mathcal{E}}
                        +\vec{\nabla}\times\vec{\mathcal{B}}=0. \label{eq:CurlBYM5}
            \end{eqnarray}
            \end{subequations}
            Eq. (\ref{eq:DivEYM5})-Eq. (\ref{eq:CurlBYM5}) are equivalent to
            \begin{subequations}\label{eq:MaxwFMuNu-1}
                  \begin{eqnarray}
                        && \partial_\mu \mathcal{F}^{\mu\nu} = 0, \label{eq:PartAlphaF-1} \\
                        && \partial_\mu \mathcal{F}_{\nu\gamma}+\partial_\nu \mathcal{F}_{\gamma\mu}
                              +\partial_\gamma \mathcal{F}_{\mu\nu} = 0, \label{eq:3PartAlphaF-1}
                  \end{eqnarray}
            \end{subequations}
            which can be obtained from Yang-Mills equations Eqs. (\ref{eq:YM1a})-(\ref{eq:YM1b}) by omitting the self-interaction terms. In this sense, we would like to call Eq. (\ref{eq:PartAlphaF-1})-Eq. (\ref{eq:3PartAlphaF-1}) as the Yang-Mills equations under zero-coupling approximation, which means the Yang-Mills equations without self-interactions.

            Furthermore, by comparing Eqs. (\ref{eq:DivEYM5})-(\ref{eq:CurlBYM5}) with the original Maxwell's equations (i.e., Eqs. (\ref{eq:DivE})-(\ref{eq:CurlB})), one may find that they share exactly the same forms. Thus, for convenient, we would like to call Eqs. (\ref{eq:DivEYM5})-(\ref{eq:CurlBYM5}) as the Maxwell-type equations.
\newpage

\section{Solutions of Yang-Mills Equations under Weak-Coupling Approximation}

\subsection{Eight Conditions that Need to Exactly Solve Yang-Mills Equations}
      Let us focus on the Yang-Mills equations as shown in \Eq{eq:DivEYM}-\Eq{eq:CurlBYM}. In this subsection, we derive the eight conditions that in general need to exactly solve the Yang-Mills equations. They are

      \begin{subequations}
      \begin{eqnarray}
            && {\rm i}\,k\Big[\vec{\nabla}\cdot\vec{\mathcal{A}}(
                  \vec{r})\Big]-\nabla^2 \varphi(\vec{r})=0, \label{eq:TICond1} \\
            && g\,\Big[\varphi(\vec{r}),\
                              \vec{\nabla}\cdot\vec{\mathcal{A}}(\vec{r})
                              \Big]=0, \label{eq:TICond2} \\
            && g^2 \biggr\{\vec{\mathcal{A}}(\vec{r})\cdot\vec{\mathcal{N}}
                              -\vec{\mathcal{N}}\cdot\vec{\mathcal{A}}(
                                    \vec{r})\biggr\}=0, \label{eq:TICond3} \\
            && g\biggl\{2\,k\,\vec{\mathcal{M}}
                                    +{\rm i}\,\vec{\nabla}\times\vec{\mathcal{N}}
                                    \biggr\}=0, \label{eq:TICond4} \\
            && g \left(\vec{\nabla}\cdot\vec{\mathcal{M}}\right)=0, \label{eq:TICond5} \\
            && \vec{\nabla}\Big[\vec{\nabla}\cdot
                                    \vec{\mathcal{A}}(\vec{r})\Big]
                              -\nabla^2 \vec{\mathcal{A}}(\vec{r})
                              -k^2 \vec{\mathcal{A}}(\vec{r})
                              -{\rm i}\,k\Big[\vec{\nabla}\varphi(\vec{r})
                                    \Big]=0, \label{eq:TICond6} \\
            && g\, \biggr\{ k\, \vec{\mathcal{N}}
                              +{\rm i}\,\vec{\mathcal{A}}(\vec{r})\times\Big[
                                    \vec{\nabla}\times\vec{\mathcal{A}}(
                                          \vec{r})\Big]
                              +{\rm i}\,\Big[\vec{\nabla}\times\vec{\mathcal{A}}(
                                    \vec{r})\Big]\times\vec{\mathcal{A}}(
                                          \vec{r})
                              -{\rm i}\,\bigl(\vec{\nabla}\times\vec{\mathcal{M}}
                                    \bigr)
                              -{\rm i}\,\Big[\varphi(\vec{r}),\ \vec{\nabla}
                                    \varphi(\vec{r})\Big] \biggr\}=0,  \label{eq:TICond7} \\
            && g^2 \, \biggr\{\Big[\varphi(\vec{r}),\
                                    \vec{\mathcal{N}}\Big]
                              +\vec{\mathcal{A}}(\vec{r})\times
                                    \vec{\mathcal{M}}
                              +\vec{\mathcal{M}}\times\vec{\mathcal{A}}(
                                    \vec{r})\biggr\}=0, \label{eq:TICond8}
      \end{eqnarray}
      \end{subequations}
      where $k=\omega/c$, and
      \begin{eqnarray}
            &&  \vec{\mathcal{M}}\equiv\vec{\mathcal{M}}(\vec{r})=\vec{\mathcal{A}}(\vec{r})
                  \times\vec{\mathcal{A}}(\vec{r}), \\
           && \vec{\mathcal{N}}\equiv\vec{\mathcal{N}}(\vec{r})= \left[ \varphi(\vec{r}),\ \vec{\mathcal{A}}(\vec{r})
                        \right].
      \end{eqnarray}

     \emph{The Details of Derivation.---}Let us study the Yang-Mills equations as shown in \Eq{eq:DivEYM}-\Eq{eq:CurlBYM}. For the ``exact solutions'', it means that there are explicit expressions for $\vec{\mathcal{A}}(\vec{r},t)$ and $\varphi(\vec{r},t)$, which satisfy \Eq{eq:DivEYM}-\Eq{eq:CurlBYM}.

     For simplicity, we consider the following  simple case,
      \begin{eqnarray}
            \vec{\mathcal{A}}(\vec{r},t) &=& \vec{\mathcal{A}}(\vec{r})\,{\rm e}^{
                  -{\rm i}\,\omega\,t}, \\
            \varphi (\vec{r},t)&=& \varphi(\vec{r})\,{\rm e}^{-{\rm i}\,\omega\,t},
      \end{eqnarray}
      which implies that spatial part and time part can be separated in $\vec{\mathcal{A}}(\vec{r},t)$ and $\varphi(\vec{r},t)$. We then have
      \begin{equation}
            \vec{\mathcal{A}}(\vec{r},t)\times\vec{\mathcal{A}}(\vec{r},t)=\vec{\mathcal{M}}\,
                  {\rm e}^{-{\rm i}\,2\,\omega\,t},
      \end{equation}
      After that, we have the ``magnetic'' field
      \begin{eqnarray}
            && \vec{\mathcal{B}}=\vec{\nabla}\times\vec{\mathcal{A}}(\vec{r},t)
                  -{\rm i}\,g\; \vec{\mathcal{A}}(\vec{r},t)\times\vec{\mathcal{A}}(\vec{r},t)
            =\Big[\vec{\nabla}\times\vec{\mathcal{A}}(\vec{r})\Big]
                        {\rm e}^{-{\rm i}\,\omega\,t}
                  -{\rm i}\,g\,\vec{\mathcal{M}}\,{\rm e}^{
                        -{\rm i}\,2\,\omega\,t}
      \end{eqnarray}
      and the ``electric'' field
      \begin{eqnarray}
                   \vec{\mathcal{E}}&=&-\dfrac{1}{c}
                        \dfrac{\partial\,\vec{\mathcal{A}}}{\partial\,t}
                              -\vec{\nabla}\varphi-{\rm i}\,g\left[
                                    \varphi,\ \vec{\mathcal{A}}\right]
                  ={\rm i}\,k\,\vec{\mathcal{A}}(\vec{r})\,{\rm e}^{
                              -{\rm i}\,\omega\,t}
                        -\Big[\vec{\nabla}\varphi(\vec{r})\Big]{\rm e}^{
                              -{\rm i}\,\omega\,t}
                        - {\rm i}g\,\vec{\mathcal{N}}\,{\rm e}^{
                              -{\rm i}\,2\,\omega\,t}.
      \end{eqnarray}
      In the next step, we come to derive the eight conditions based on \Eq{eq:DivEYM}-\Eq{eq:CurlBYM}.

      \begin{itemize}
            \item [(i).] For \Eq{eq:DivEYM}, we attain
                  \begin{eqnarray}
                         &&\vec{\nabla}\cdot\vec{\mathcal{E}}+{\rm i}\,
                              g\Bigl(\vec{\mathcal{A}}\cdot
                                          \vec{\mathcal{E}}
                                    -\vec{\mathcal{E}}\cdot
                                          \vec{\mathcal{A}}\Bigr) \nonumber\\
                        &=& \vec{\nabla}\cdot\bigg\{{\rm i}\,k\,
                                          \vec{\mathcal{A}}(\vec{r})\,
                                          {\rm e}^{-{\rm i}\,\omega\,t}
                                    -\Big[\vec{\nabla}\varphi(\vec{r})\Big]
                                                {\rm e}^{-{\rm i}\,
                                                      \omega\,t}
                                          - {\rm i}g\,\vec{\mathcal{N}}\,
                                                {\rm e}^{-{\rm i}\,2\,
                                                      \omega\,t}\bigg\} \nonumber\\
                              && +{\rm i}\,g\,\vec{\mathcal{A}}(\vec{r})\,
                                          {\rm e}^{-{\rm i}\,\omega\,t}
                                    \cdot\bigg\{{\rm i}\,k\,
                                          \vec{\mathcal{A}}(\vec{r})\,
                                          {\rm e}^{-{\rm i}\,\omega\,t}
                                    -\Big[\vec{\nabla}\varphi(\vec{r})\Big]
                                                {\rm e}^{-{\rm i}\,
                                                      \omega\,t}
                                          - {\rm i}g\,\vec{\mathcal{N}}\,
                                                {\rm e}^{-{\rm i}\,2\,
                                                      \omega\,t}\bigg\} \nonumber\\
                              && -{\rm i}\,g\bigg\{{\rm i}\,k\,
                                                \vec{\mathcal{A}}(\vec{r})\,
                                                {\rm e}^{-{\rm i}\,
                                                      \omega\,t}
                                          -\Big[\vec{\nabla}\varphi(\vec{r})
                                                \Big]{\rm e}^{-{\rm i}\,
                                                            \omega\,t}
                                                - {\rm i}g\,\vec{\mathcal{N}}
                                                \,{\rm e}^{-{\rm i}\,2\,
                                                      \omega\,t}\bigg\}
                                    \cdot\vec{\mathcal{A}}(\vec{r})\,
                                          {\rm e}^{-{\rm i}\,\omega\,t} \nonumber\\
                        &=& {\rm i}\,k\Big[\vec{\nabla}\cdot
                                    \vec{\mathcal{A}}(\vec{r})\Big]{\rm e}^{
                                          -{\rm i}\,\omega\,t}
                              -\Big[\nabla^2 \varphi(\vec{r})\Big]{\rm e}^{
                                    -{\rm i}\,\omega\,t}
                              - {\rm i}g\,\Bigl(\vec{\nabla}\cdot
                                    \vec{\mathcal{N}}\Bigr){\rm e}^{
                                          -{\rm i}\,2\,\omega\,t} \nonumber\\
                              && -g\,k\Big[\vec{\mathcal{A}}(\vec{r})\cdot
                                    \vec{\mathcal{A}}(\vec{r})\Big]{\rm e}^{
                                          -{\rm i}\,2\,\omega\,t}
                                    -{\rm i}\,g\bigg\{\vec{\mathcal{A}}(
                                          \vec{r})\cdot\Big[\vec{\nabla}
                                                \varphi(\vec{r})\Big]
                                          \bigg\}{\rm e}^{-{\rm i}\,2\,
                                                \omega\,t}
                                    +g^2 \Big[
                                          \vec{\mathcal{A}}(\vec{r})\cdot
                                          \vec{\mathcal{N}}\Big]{\rm e}^{
                                                -{\rm i}\,3\,\omega\,t} \nonumber\\
                              && +g\,k\Big[\vec{\mathcal{A}}(\vec{r})\cdot
                                    \vec{\mathcal{A}}(\vec{r})\Big]{\rm e}^{
                                          -{\rm i}\,2\,\omega\,t}
                                    +{\rm i}\,g\bigg\{\Big[\vec{\nabla}
                                                \varphi(\vec{r})\Big]\cdot
                                                \vec{\mathcal{A}}(\vec{r})
                                          \bigg\}{\rm e}^{-{\rm i}\,2\,
                                                \omega\,t}
                                    -g^2 \Big[
                                          \vec{\mathcal{N}}\cdot
                                          \vec{\mathcal{A}}(\vec{r})
                                          \Big]{\rm e}^{
                                                -{\rm i}\,3\,\omega\,t} \nonumber\\
                        &=& \biggl\{{\rm i}\,k\Big[\vec{\nabla}\cdot
                                          \vec{\mathcal{A}}(\vec{r})\Big]
                                    -\nabla^2 \varphi(\vec{r})\biggr\}
                                          {\rm e}^{-{\rm i}\,\omega\,t}\nonumber\\
                             && +\Biggl(-{\rm i}\; g\,\Bigl(\vec{\nabla}\cdot
                                          \vec{\mathcal{N}}\Bigr)
                                    +{\rm i}\,g\bigg\{\Big[\vec{\nabla}
                                                \varphi(\vec{r})\Big]\cdot
                                                \vec{\mathcal{A}}(\vec{r})
                                          -\vec{\mathcal{A}}(\vec{r})\cdot
                                                \Big[\vec{\nabla}\varphi(
                                                      \vec{r})\Big]\bigg\}
                                    \Biggr){\rm e}^{-{\rm i}\,2\,\omega\,t}
                                    \nonumber\\
                              && +g^2 \Big[\vec{\mathcal{A}}(
                                          \vec{r})\cdot\vec{\mathcal{N}}
                                    -\vec{\mathcal{N}}\cdot
                                          \vec{\mathcal{A}}(\vec{r})
                                    \Big]{\rm e}^{-{\rm i}\,3\,\omega\,t}
                                    \nonumber\\
                        &=& \biggl\{{\rm i}\,k\Big[\vec{\nabla}\cdot
                                          \vec{\mathcal{A}}(\vec{r})\Big]
                                    -\nabla^2 \varphi(\vec{r})\biggr\}
                                          {\rm e}^{-{\rm i}\,\omega\,t} \nonumber\\
                              && +\Biggl(-{\rm i}\,g\biggl\{
                                    \Big[\vec{\nabla}\varphi(\vec{r})\Big]
                                          \cdot\vec{\mathcal{A}}(\vec{r})
                                    +\varphi(\vec{r})\Big[\vec{\nabla}
                                          \cdot\vec{\mathcal{A}}(\vec{r})
                                          \Big]
                                    -\Big[\vec{\nabla}\cdot
                                          \vec{\mathcal{A}}(\vec{r})\Big]
                                                \varphi(\vec{r})
                                    -\vec{\mathcal{A}}(\vec{r})\cdot\Big[
                                          \vec{\nabla}\varphi(\vec{r})\Big]
                                    \biggr\} \nonumber\\
                                    &&+{\rm i}\,g\bigg\{
                                          \Big[\vec{\nabla}\varphi(\vec{r})
                                                \Big]\cdot\vec{\mathcal{A}}(
                                                      \vec{r})
                                          -\vec{\mathcal{A}}(\vec{r})\cdot
                                                \Big[\vec{\nabla}\varphi(
                                                      \vec{r})\Big]\bigg\}
                                    \Biggr){\rm e}^{-{\rm i}\,2\,\omega\,t}
                              +g^2 \Big[\vec{\mathcal{A}}(
                                          \vec{r})\cdot\vec{\mathcal{N}}
                                    -\vec{\mathcal{N}}\cdot
                                          \vec{\mathcal{A}}(\vec{r})
                                    \Big]{\rm e}^{-{\rm i}\,3\,\omega\,t} \nonumber\\
                        &=& \biggl\{{\rm i}\,k\Big[\vec{\nabla}\cdot
                                          \vec{\mathcal{A}}(\vec{r})\Big]
                                    -\nabla^2 \varphi(\vec{r})\biggr\}
                                          {\rm e}^{-{\rm i}\,\omega\,t}
                              -{\rm i}\,g\biggl\{\varphi(\vec{r})\Big[
                                          \vec{\nabla}\cdot
                                                \vec{\mathcal{A}}(\vec{r})
                                          \Big]
                                    -\Big[\vec{\nabla}\cdot
                                          \vec{\mathcal{A}}(\vec{r})\Big]
                                                \varphi(\vec{r})
                                    \biggr\}{\rm e}^{-{\rm i}\,2\,
                                          \omega\,t} \nonumber\\
                              && +g^2 \Big[\vec{\mathcal{A}}(
                                          \vec{r})\cdot\vec{\mathcal{N}}
                                    -\vec{\mathcal{N}}\cdot
                                          \vec{\mathcal{A}}(\vec{r})
                                    \Big]{\rm e}^{-{\rm i}\,3\,\omega\,t}=0,
                  \end{eqnarray}
                  which holds for arbitrary $\omega$, then we arrive at
                  \begin{eqnarray}
                        && {\rm i}\,k\Big[\vec{\nabla}\cdot\vec{\mathcal{A}}(
                              \vec{r})\Big]-\nabla^2 \varphi(\vec{r})=0, \\
                        && g\,\Big[\varphi(\vec{r}),\
                              \vec{\nabla}\cdot\vec{\mathcal{A}}(\vec{r})
                              \Big]=0, \\
                        && g^2 \biggr\{\vec{\mathcal{A}}(\vec{r})\cdot\vec{\mathcal{N}}
                              -\vec{\mathcal{N}}\cdot\vec{\mathcal{A}}(
                                    \vec{r})\biggr\}=0,
                 \end{eqnarray}
      which correspond to conditions (\ref{eq:TICond1}), (\ref{eq:TICond2}) and (\ref{eq:TICond3}), respectively.
            \item [(ii).] For \Eq{eq:CurlEYM}, since
                  \begin{eqnarray}
                              && -\dfrac{1}{c} \dfrac{\partial}{\partial\,t}
                                          \vec{\mathcal{B}}
                              =-\dfrac{1}{c} \dfrac{\partial}{\partial\,t}
                                    \biggl\{\Big[\vec{\nabla}\times
                                          \vec{\mathcal{A}}(\vec{r})\Big]
                                          {\rm e}^{-{\rm i}\,\omega\,t}
                                    -{\rm i}\,g\,\vec{\mathcal{M}}\,
                                          {\rm e}^{-{\rm i}\,2\,\omega\,t}
                                    \biggr\}
                              =k\biggl\{{\rm i}\Big[\vec{\nabla}\times
                                          \vec{\mathcal{A}}(\vec{r})\Big]
                                          {\rm e}^{-{\rm i}\,\omega\,t}
                                    +2\,g\,\vec{\mathcal{M}}\,{\rm e}^{
                                          -{\rm i}\,2\,\omega\,t}\biggr\},
                  \end{eqnarray}
                  \begin{eqnarray}
                               \vec{\nabla}\times\vec{\mathcal{E}}
                              &=&\vec{\nabla}\times\biggl\{{\rm i}\,k\,
                                          \vec{\mathcal{A}}(\vec{r})\,
                                          {\rm e}^{-{\rm i}\,\omega\,t}
                                    -\Big[\vec{\nabla}\varphi(\vec{r})\Big]
                                          {\rm e}^{-{\rm i}\,\omega\,t}
                                    -{\rm i}g\,\vec{\mathcal{N}}\,{\rm e}^{
                                          -{\rm i}\,2\,\omega\,t}\biggr\} \nonumber\\
                              &=& {\rm i}\,k\Big[\vec{\nabla}\times
                                          \vec{\mathcal{A}}(\vec{r})\Big]
                                          {\rm e}^{-{\rm i}\,\omega\,t}
                                    -\vec{\nabla}\times\Big[\vec{\nabla}
                                          \varphi(\vec{r})\Big]{\rm e}^{
                                                -{\rm i}\,\omega\,t}
                                    -{\rm i}g\,\Big[\vec{\nabla}\times
                                          \vec{\mathcal{N}}\Big]{\rm e}^{
                                                -{\rm i}\,2\,\omega\,t}\nonumber\\
                              &=& {\rm i}\,k\Big[\vec{\nabla}\times
                                          \vec{\mathcal{A}}(\vec{r})\Big]
                                          {\rm e}^{-{\rm i}\,\omega\,t}
                                    -{\rm i}g\,\Big[\vec{\nabla}\times
                                          \vec{\mathcal{N}}\Big]{\rm e}^{
                                                -{\rm i}\,2\,\omega\,t},
                  \end{eqnarray}
                  \begin{eqnarray}
                               \Bigl[\varphi,\ \vec{\mathcal{B}}\Bigr]
                              &=& \bigg[\varphi(\vec{r})\,{\rm e}^{
                                                -{\rm i}\,\omega\,t},\
                                    \Big\{\vec{\nabla}\times
                                          \vec{\mathcal{A}}(\vec{r})\Big\}
                                          {\rm e}^{-{\rm i}\,\omega\,t}
                                    -{\rm i}\,g\,\vec{\mathcal{M}}\,
                                          {\rm e}^{-{\rm i}\,2\,\omega\,t}
                                    \bigg]\nonumber\\
                             & =&\Big[\varphi(\vec{r}),\ \vec{\nabla}\times
                                          \vec{\mathcal{A}}(\vec{r})\Big]
                                          {\rm e}^{-{\rm i}\,2\,\omega\,t}
                                    -{\rm i}\,g\Big[\varphi(\vec{r}),\
                                          \vec{\mathcal{M}}\Big]{\rm e}^{
                                                -{\rm i}\,3\,\omega\,t},
                  \end{eqnarray}
                  \begin{eqnarray}
                              && \vec{\mathcal{A}}\times\vec{\mathcal{E}}
                                    +\vec{\mathcal{E}}
                                          \times\vec{\mathcal{A}} \nonumber\\
                              &=& \vec{\mathcal{A}}(\vec{r})\,{\rm e}^{
                                          -{\rm i}\,\omega\,t}\times
                                    \biggl\{{\rm i}\,k\,
                                          \vec{\mathcal{A}}(\vec{r})\,
                                          {\rm e}^{-{\rm i}\,\omega\,t}
                                    -\Big[\vec{\nabla}\varphi(\vec{r})\Big]
                                          {\rm e}^{-{\rm i}\,\omega\,t}
                                    -{\rm i}\,g\,\vec{\mathcal{N}}\,{\rm e}^{
                                          -{\rm i}\,2\,\omega\,t}\biggr\} \nonumber\\
                                    && +\biggl\{{\rm i}\,k\,
                                          \vec{\mathcal{A}}(\vec{r})\,
                                          {\rm e}^{-{\rm i}\,\omega\,t}
                                    -\Big[\vec{\nabla}\varphi(\vec{r})\Big]
                                          {\rm e}^{-{\rm i}\,\omega\,t}
                                    -{\rm i}\,g\,\vec{\mathcal{N}}\,{\rm e}^{
                                          -{\rm i}\,2\,\omega\,t}\biggr\}\times\vec{\mathcal{A}}(\vec{r})\,
                                                {\rm e}^{-{\rm i}\,
                                                      \omega\,t} \nonumber\\
                              &=& {\rm i}\,k\,\vec{\mathcal{M}}\,{\rm e}^{
                                          -{\rm i}\,2\,\omega\,t}
                                    -\vec{\mathcal{A}}(\vec{r})\times\Big[
                                          \vec{\nabla}\varphi(\vec{r})\Big]
                                          {\rm e}^{-{\rm i}\,2\,\omega\,t}
                                    -{\rm i}\,g\,\Big[\vec{\mathcal{A}}(
                                          \vec{r})\times\vec{\mathcal{N}}
                                          \Big]{\rm e}^{-{\rm i}\,3\,
                                                \omega\,t} \\
                                    && +{\rm i}\,k\,\vec{\mathcal{M}}\,
                                          {\rm e}^{-{\rm i}\,2\,\omega\,t}
                                    -\Big[\vec{\nabla}\varphi(\vec{r})\Big]
                                          \times\vec{\mathcal{A}}(\vec{r})
                                          {\rm e}^{-{\rm i}\,2\,\omega\,t}
                                    -{\rm i}\,g\, \Big[\vec{\mathcal{N}}\times
                                          \vec{\mathcal{A}}(\vec{r})\Big]
                                                {\rm e}^{-{\rm i}\,3\,
                                                      \omega\,t} \nonumber\\
                              &=& {\rm i}\,2\,k\,\vec{\mathcal{M}}\,{\rm e}^{
                                          -{\rm i}\,2\,\omega\,t}
                                    -\biggl\{\vec{\mathcal{A}}(\vec{r})
                                          \times\Big[\vec{\nabla}\varphi(
                                                \vec{r})\Big]
                                          +\Big[\vec{\nabla}\varphi(\vec{r})
                                          \Big]\times\vec{\mathcal{A}}(
                                                \vec{r})\biggr\}
                                          {\rm e}^{-{\rm i}\,2\,\omega\,t}
                                    -{\rm i}\,g\,\Big[\vec{\mathcal{A}}(
                                                \vec{r})\times
                                                      \vec{\mathcal{N}}
                                          +\vec{\mathcal{N}}\times
                                                \vec{\mathcal{A}}(\vec{r})
                                          \Big]{\rm e}^{-{\rm i}\,3\,
                                                \omega\,t},
                  \end{eqnarray}
                  thus
                  \begin{eqnarray}
                              && -\dfrac{1}{c} \dfrac{\partial}{\partial\,t}
                                          \vec{\mathcal{B}}
                                    -\vec{\nabla}\times\vec{\mathcal{E}}
                                    +{\rm i}\,g\biggl(\Bigl[\varphi,\
                                                \vec{\mathcal{B}}\Bigr]
                                          -\vec{\mathcal{A}}\times
                                                \vec{\mathcal{E}}
                                          -\vec{\mathcal{E}}\times
                                                \vec{\mathcal{A}}\biggr) \nonumber\\
                              &=& k\biggl\{{\rm i}\Big[\vec{\nabla}\times
                                                \vec{\mathcal{A}}(\vec{r})
                                                \Big]{\rm e}^{-{\rm i}\,
                                                      \omega\,t}
                                          +2\,g\,\vec{\mathcal{M}}\,
                                                {\rm e}^{-{\rm i}\,2\,
                                                      \omega\,t}\biggr\}
                                    -{\rm i}\,k\Big[\vec{\nabla}\times
                                          \vec{\mathcal{A}}(\vec{r})\Big]
                                          {\rm e}^{-{\rm i}\,\omega\,t}
                                    +{\rm i}\, g\,\Big[\vec{\nabla}\times
                                          \vec{\mathcal{N}}\Big]{\rm e}^{
                                                -{\rm i}\,2\,\omega\,t} \nonumber\\
                                   & & +{\rm i}\,g\Big[\varphi(\vec{r}),\
                                          \vec{\nabla}\times
                                                \vec{\mathcal{A}}(\vec{r})
                                          \Big]{\rm e}^{-{\rm i}\,2\,
                                                \omega\,t}
                                    +g^2 \Big[\varphi(\vec{r}),\
                                          \vec{\mathcal{M}}\Big]{\rm e}^{
                                                -{\rm i}\,3\,\omega\,t}
                                    +2\,g\,k\,\vec{\mathcal{M}}\,{\rm e}^{
                                          -{\rm i}\,2\,\omega\,t} \nonumber\\
                                    && +{\rm i}\,g\biggl\{\vec{\mathcal{A}}(
                                          \vec{r})\times\Big[\vec{\nabla}
                                                \varphi(\vec{r})\Big]
                                          +\Big[\vec{\nabla}\varphi(\vec{r})
                                          \Big]\times\vec{\mathcal{A}}(
                                                \vec{r})\biggr\}
                                          {\rm e}^{-{\rm i}\,2\,\omega\,t}
                                    -g^2 \,\Big[
                                          \vec{\mathcal{A}}(\vec{r})\times
                                                \vec{\mathcal{N}}
                                          +\vec{\mathcal{N}}\times
                                                \vec{\mathcal{A}}(\vec{r})
                                          \Big]{\rm e}^{-{\rm i}\,3\,
                                                \omega\,t} \nonumber\\
                              &=& 4\,g\,k\,\vec{\mathcal{M}}\,{\rm e}^{
                                          -{\rm i}\,2\,\omega\,t}
                                    +{\rm i}\,g\,\Big[\vec{\nabla}\times
                                          \vec{\mathcal{N}}\Big]{\rm e}^{
                                                -{\rm i}\,2\,\omega\,t}
                                    +{\rm i}\,g\Big[\varphi(\vec{r}),\
                                          \vec{\nabla}\times
                                                \vec{\mathcal{A}}(\vec{r})
                                          \Big]{\rm e}^{-{\rm i}\,2\,
                                                \omega\,t} \nonumber\\
                                  &&  +{\rm i}\,g\biggl\{\vec{\mathcal{A}}(
                                          \vec{r})\times\Big[\vec{\nabla}
                                                \varphi(\vec{r})\Big]
                                          +\Big[\vec{\nabla}\varphi(\vec{r})
                                          \Big]\times\vec{\mathcal{A}}(
                                                \vec{r})\biggr\}
                                          {\rm e}^{-{\rm i}\,2\,\omega\,t}
                                          \nonumber\\
                                    && +g^2 \Big[\varphi(\vec{r}),\
                                          \vec{\mathcal{M}}\Big]{\rm e}^{
                                                -{\rm i}\,3\,\omega\,t}
                                    -g^2 \,\Big[
                                          \vec{\mathcal{A}}(\vec{r})\times
                                                \vec{\mathcal{N}}
                                          +\vec{\mathcal{N}}\times
                                                \vec{\mathcal{A}}(\vec{r})
                                          \Big]{\rm e}^{-{\rm i}\,3\,
                                                \omega\,t} \nonumber\\
                              &=& g\Biggl(4\,k\,\vec{\mathcal{M}}
                                    +{\rm i}\Big[\vec{\nabla}\times
                                          \vec{\mathcal{N}}\Big]
                                    +{\rm i}\biggl\{\Big[\varphi(\vec{r}),\
                                          \vec{\nabla}\times
                                                \vec{\mathcal{A}}(\vec{r})
                                          \Big]
                                    +\vec{\mathcal{A}}(
                                          \vec{r})\times\Big[\vec{\nabla}
                                                \varphi(\vec{r})\Big]
                                          +\Big[\vec{\nabla}\varphi(\vec{r})
                                          \Big]\times\vec{\mathcal{A}}(
                                                \vec{r})\biggr\}\Biggr)
                                                {\rm e}^{-{\rm i}\,2\,
                                                      \omega\,t} \nonumber\\
                                    && +g^2 \biggl\{\Big[\varphi(\vec{r}),\
                                                \vec{\mathcal{M}}\Big]
                                          -\Big[
                                                \vec{\mathcal{A}}(\vec{r})
                                                \times\vec{\mathcal{N}}
                                                +\vec{\mathcal{N}}\times
                                                      \vec{\mathcal{A}}(
                                                            \vec{r})\Big]
                                          \biggr\}{\rm e}^{-{\rm i}\,3\,
                                                \omega\,t} \nonumber\\
                             & =& g\biggl\{4\,k\,\vec{\mathcal{M}}
                                    +{\rm i}\Big(\vec{\nabla}\times
                                          \vec{\mathcal{N}}\Big)
                                    +\textcolor{black}{{\rm i}\,\vec{\nabla}\times\Big[
                                          \varphi(\vec{r})\,
                                                \vec{\mathcal{A}}(\vec{r})
                                          -\vec{\mathcal{A}}(\vec{r})\,
                                                \varphi(\vec{r})\Big]
                                                }
                                    \biggr\}{\rm e}^{-{\rm i}\,2\,
                                          \omega\,t} \nonumber\\
                                    && +g^2 \biggl\{\Big[\varphi(\vec{r}),\
                                                \vec{\mathcal{M}}\Big]
                                          -\Big[
                                                \vec{\mathcal{A}}(\vec{r})
                                                \times\vec{\mathcal{N}}
                                                +\vec{\mathcal{N}}\times
                                                      \vec{\mathcal{A}}(
                                                            \vec{r})\Big]
                                          \biggr\}{\rm e}^{-{\rm i}\,3\,
                                                \omega\,t} \nonumber\\
                              &=& 2g\biggl\{2\,k\,\vec{\mathcal{M}}
                                    +{\rm i}\,\vec{\nabla}\times\vec{\mathcal{N}}
                                    \biggr\}{\rm e}^{-{\rm i}\,2\,
                                          \omega\,t}
                                    +g^2 \textcolor{black}{\biggl\{\Big[\varphi(\vec{r}),\
                                                \vec{\mathcal{M}}\Big]
                                          -\Big[
                                                \vec{\mathcal{A}}(\vec{r})
                                                \times\vec{\mathcal{N}}
                                                +\vec{\mathcal{N}}\times
                                                      \vec{\mathcal{A}}(
                                                            \vec{r})\Big]
                                          \biggr\}
                                          }{\rm e}^{-{\rm i}\,3\,
                                                \omega\,t}
                                                 \nonumber\\
                              &=& 2g\biggl\{2\,k\,\vec{\mathcal{M}}
                                    +{\rm i}\,\vec{\nabla}\times\vec{\mathcal{N}}
                                    \biggr\}{\rm e}^{-{\rm i}\,2\,
                                          \omega\,t}
                                    -g^2 \biggl\{\vec{\mathcal{A}}(
                                                \vec{r})\times\Big[
                                                      \varphi(\vec{r})
                                                      \,\vec{\mathcal{A}}(
                                                            \vec{r})\Big]
                                          -\Big[\vec{\mathcal{A}}(\vec{r})\,
                                                \varphi(\vec{r})\Big]\times
                                                      \vec{\mathcal{A}}(
                                                            \vec{r})
                                          \biggr\}
                                          {\rm e}^{-{\rm i}\,3\,
                                                \omega\,t} \nonumber\\
                                    &=& 2g\biggl\{2\,k\,\vec{\mathcal{M}}
                                    +{\rm i}\,\vec{\nabla}\times\vec{\mathcal{N}}
                                    \biggr\}{\rm e}^{-{\rm i}\,2\,
                                          \omega\,t}=0.
                  \end{eqnarray}
                  Here we have used the relation $\vec{\mathcal{A}}(
                                                \vec{r})\times\Big[
                                                      \varphi(\vec{r})
                                                      \,\vec{\mathcal{A}}(
                                                            \vec{r})\Big]
                                          -\Big[\vec{\mathcal{A}}(\vec{r})\,
                                                \varphi(\vec{r})\Big]\times
                                                      \vec{\mathcal{A}}(
                                                            \vec{r})=0$.
                  Then we obtain
                  \begin{eqnarray}
                        && g\biggl\{2\,k\,\vec{\mathcal{M}}
                                    +{\rm i}\,\vec{\nabla}\times\vec{\mathcal{N}}
                                    \biggr\}=0,
                  \end{eqnarray}
                  which corresponds to condition (\ref{eq:TICond4}).
            \item [(iii).] For \Eq{eq:DivBYM}, we obtain
                  \begin{eqnarray}
                              && \vec{\nabla}\cdot\vec{\mathcal{B}}
                                    +{\rm i}\,g\Bigl(\vec{\mathcal{A}}
                                          \cdot\vec{\mathcal{B}}
                                    -\vec{\mathcal{B}}\cdot\vec{\mathcal{A}}
                                          \Bigr) \nonumber\\
                              &=& \vec{\nabla}\cdot\biggl\{\Big[\vec{\nabla}
                                                \times\vec{\mathcal{A}}(
                                                      \vec{r})\Big]{\rm e}^{
                                                            -{\rm i}\,
                                                                  \omega\,t}
                                          -{\rm i}\,g\,\vec{\mathcal{M}}\,
                                                {\rm e}^{-{\rm i}\,2\,
                                                      \omega\,t}\biggr\}  +{\rm i}\,g\Bigg(\vec{\mathcal{A}}(
                                                \vec{r})\,{\rm e}^{
                                                      -{\rm i}\,\omega\,t}
                                          \cdot\biggl\{\Big[\vec{\nabla}
                                                \times\vec{\mathcal{A}}(
                                                      \vec{r})\Big]{\rm e}^{
                                                            -{\rm i}\,
                                                                  \omega\,t}
                                          -{\rm i}\,g\,\vec{\mathcal{M}}\,
                                                {\rm e}^{-{\rm i}\,2\,
                                                      \omega\,t}\biggr\} \nonumber\\
                                    && -\biggl\{\Big[\vec{\nabla}
                                                \times\vec{\mathcal{A}}(
                                                      \vec{r})\Big]{\rm e}^{
                                                            -{\rm i}\,
                                                                  \omega\,t}
                                          -{\rm i}\,g\,\vec{\mathcal{M}}\,
                                                {\rm e}^{-{\rm i}\,2\,
                                                      \omega\,t}\biggr\}
                                          \cdot\Big[\vec{\mathcal{A}}(
                                                \vec{r})\,{\rm e}^{
                                                      -{\rm i}\,\omega\,t}
                                                \Big]\Bigg) \nonumber\\
                              &=& {\rm i}\,g\biggl\{\vec{\mathcal{A}}(
                                                \vec{r})\cdot\Big[
                                                      \vec{\nabla}\times
                                                      \vec{\mathcal{A}}(
                                                            \vec{r})\Big]
                                          -\Big[\vec{\nabla}\times
                                                \vec{\mathcal{A}}(\vec{r})
                                                \Big]\cdot\vec{\mathcal{A}}(
                                                      \vec{r})
                                          -\vec{\nabla}\cdot\vec{
                                                \mathcal{M}}
                                          \biggr\}{\rm e}^{
                                                -{\rm i}\,2\,\omega\,t}
                                    +g^2 \Big[\vec{\mathcal{A}}(\vec{r})
                                                \cdot\vec{\mathcal{M}}
                                          -\vec{\mathcal{M}}\cdot
                                                \vec{\mathcal{A}}(\vec{r})
                                          \Big]{\rm e}^{
                                                -{\rm i}\,3\,\omega\,t} \nonumber\\
                              &=& {\rm i}\,g\biggl\{\vec{\mathcal{A}}(
                                                \vec{r})\cdot\Big[
                                                      \vec{\nabla}\times
                                                      \vec{\mathcal{A}}(
                                                            \vec{r})\Big]
                                          -\Big[\vec{\nabla}\times
                                                \vec{\mathcal{A}}(\vec{r})
                                                \Big]\cdot\vec{\mathcal{A}}(
                                                      \vec{r})
                                          -\Big[\vec{\nabla}\times
                                                \vec{\mathcal{A}}(\vec{r})
                                                \Big]\cdot\vec{\mathcal{A}}(
                                                      \vec{r})
                                                +\vec{\mathcal{A}}(\vec{r})
                                                      \cdot\Big[\vec{\nabla}
                                                      \times\vec{
                                                            \mathcal{A}}(
                                                                  \vec{r})
                                                      \Big]
                                          \biggr\}{\rm e}^{
                                                -{\rm i}\,2\,\omega\,t} \nonumber\\
                                   & & +g^2 \biggl\{\vec{\mathcal{A}}(
                                                \vec{r})\cdot\Big[
                                                      \vec{\mathcal{A}}(
                                                      \vec{r})
                                                      \times\vec{
                                                            \mathcal{A}}(
                                                                  \vec{r})\Big]
                                          -\Big[\vec{\mathcal{A}}(\vec{r})
                                                \times\vec{
                                                \mathcal{A}}(\vec{r})\Big]\cdot\vec{\mathcal{A}}(
                                                      \vec{r})
                                          \biggr\}{\rm e}^{
                                                -{\rm i}\,3\,\omega\,t} \nonumber\\
                              &=& {\rm i} \,g \left\{-2\; \vec{\nabla}\cdot\vec{\mathcal{M}}\right\}{\rm e}^{ -{\rm i}\,2\,\omega\,t}
                              =0,
                  \end{eqnarray}
                  then we arrive at
                  \begin{align}
                       g \left( \vec{\nabla}\cdot\vec{\mathcal{M}}\right)=0,
                  \end{align}
                   which corresponds to condition (\ref{eq:TICond5}).

            \item [(iv).] For \Eq{eq:CurlBYM}, because
                  \begin{eqnarray}
                               \dfrac{1}{c} \dfrac{\partial}{\partial\,t}
                                    \vec{\mathcal{E}}
                              &=&\dfrac{1}{c} \dfrac{\partial}{\partial\,t}
                                    \biggl\{{\rm i}\,k\,
                                          \vec{\mathcal{A}}(\vec{r})\,
                                          {\rm e}^{-{\rm i}\,\omega\,t}
                                    -\Big[\vec{\nabla}\varphi(\vec{r})\Big]
                                          {\rm e}^{-{\rm i}\,\omega\,t}
                                    -{\rm i}g\,\vec{\mathcal{N}}\,{\rm e}^{
                                          -{\rm i}\,2\,\omega\,t}\biggr\} \nonumber\\
                              &=& k^2 \vec{\mathcal{A}}(\vec{r})\,{\rm e}^{
                                          -{\rm i}\,\omega\,t}
                                    +{\rm i}\,k\Big[\vec{\nabla}\varphi(
                                          \vec{r})\Big]{\rm e}^{-{\rm i}\,
                                                \omega\,t}
                                    -2\,k\,g\,
                                          \vec{\mathcal{N}}\,{\rm e}^{
                                                -{\rm i}\,2\,\omega\,t},
                  \end{eqnarray}
                  \begin{eqnarray}
                               \vec{\nabla}\times\vec{\mathcal{B}}
                              &=&\vec{\nabla}\times\biggl\{
                                    \Big[\vec{\nabla}\times
                                                \vec{\mathcal{A}}(\vec{r})
                                                \Big]{\rm e}^{-{\rm i}\,
                                                      \omega\,t}
                                          -{\rm i}\,g\,\vec{\mathcal{M}}\,
                                                {\rm e}^{-{\rm i}\,2\,
                                                      \omega\,t}\biggr\}
                              =\vec{\nabla}\times\Big[\vec{\nabla}
                                          \times\vec{\mathcal{A}}(\vec{r})
                                          \Big]{\rm e}^{-{\rm i}\,\omega\,t}
                                    -{\rm i}\,g\bigl(\vec{\nabla}\times
                                          \vec{\mathcal{M}}\bigr){\rm e}^{
                                                -{\rm i}\,2\,\omega\,t} \nonumber\\
                              &=& \biggl\{\vec{\nabla}\Big[\vec{\nabla}\cdot
                                                \vec{\mathcal{A}}(\vec{r})
                                                \Big]
                                          -\nabla^2 \vec{\mathcal{A}}(
                                                \vec{r})\biggr\}{\rm e}^{
                                                      -{\rm i}\,\omega\,t}
                                    -{\rm i}\,g\bigl(\vec{\nabla}\times
                                          \vec{\mathcal{M}}\bigr){\rm e}^{
                                                -{\rm i}\,2\,\omega\,t}.
                  \end{eqnarray}
                  \begin{eqnarray}
                               \Bigl[\varphi,\ \vec{\mathcal{E}}\Bigr]
                              &=&\bigg[\varphi(\vec{r})\,{\rm e}^{
                                          -{\rm i}\,\omega\,t},\
                                    {\rm i}\,k\,\vec{\mathcal{A}}(\vec{r})\,
                                          {\rm e}^{-{\rm i}\,\omega\,t}
                                    -\Big\{\vec{\nabla}\varphi(\vec{r})
                                          \Big\}{\rm e}^{-{\rm i}\,
                                                \omega\,t}
                                    -{\rm i}\, g\,\vec{\mathcal{N}}\,{\rm e}^{
                                          -{\rm i}\,2\,\omega\,t}\bigg] \nonumber\\
                              &=& {\rm i}\,k\Big[\varphi(\vec{r}),\
                                          \vec{\mathcal{A}}(\vec{r})\Big]
                                          {\rm e}^{-{\rm i}\,2\,\omega\,t}
                                    -\Big[\varphi(\vec{r}),\ \vec{\nabla}
                                          \varphi(\vec{r})\Big]{\rm e}^{
                                          -{\rm i}\,2\,\omega\,t}
                                    -{\rm i}\,g\,\Big[\varphi(\vec{r}),\
                                          \vec{\mathcal{N}}\Big]{\rm e}^{
                                                -{\rm i}\,3\,\omega\,t},
                  \end{eqnarray}
                  \begin{eqnarray}
                              && \vec{\mathcal{A}}\times\vec{\mathcal{B}}
                                    +\vec{\mathcal{B}}\times
                                          \vec{\mathcal{A}} \nonumber\\
                              &=& \vec{\mathcal{A}}(\vec{r})\,{\rm e}^{
                                          -{\rm i}\,\omega\,t} \times
                                    \biggl\{\Big[\vec{\nabla}\times
                                                \vec{\mathcal{A}}(\vec{r})
                                                \Big]{\rm e}^{-{\rm i}\,
                                                      \omega\,t}
                                          -{\rm i}\,g\,\vec{\mathcal{M}}\,
                                                {\rm e}^{-{\rm i}\,2\,
                                                      \omega\,t}\biggr\}
                                    +\biggl\{\Big[\vec{\nabla}\times
                                                      \vec{\mathcal{A}}(
                                                            \vec{r})
                                                      \Big]{\rm e}^{-{\rm i}
                                                            \,\omega\,t}
                                                -{\rm i}\,g\,
                                                      \vec{\mathcal{M}}\,
                                                      {\rm e}^{-{\rm i}\,2\,
                                                            \omega\,t}
                                          \biggr\}\times\vec{\mathcal{A}}(
                                                \vec{r})\,{\rm e}^{-{\rm i}
                                                      \,\omega\,t} \nonumber\\
                              &=& \biggl\{\vec{\mathcal{A}}(\vec{r})\times
                                          \Big[\vec{\nabla}\times
                                                \vec{\mathcal{A}}(\vec{r})
                                                \Big]
                                          +\Big[\vec{\nabla}\times
                                                \vec{\mathcal{A}}(\vec{r})
                                                \Big]\times
                                                      \vec{\mathcal{A}}(
                                                            \vec{r})
                                          \biggr\}{\rm e}^{-{\rm i}\,2\,
                                                \omega\,t}
                                    -{\rm i}\,g\Big[\vec{\mathcal{A}}(
                                          \vec{r})\times\vec{\mathcal{M}}
                                          +\vec{\mathcal{M}}\times
                                                \vec{\mathcal{A}}(\vec{r})
                                          \Big]{\rm e}^{-{\rm i}\,3\,
                                                \omega\,t},
                  \end{eqnarray}
                  then we have
                  \begin{eqnarray}
                              && -\dfrac{1}{c} \dfrac{\partial}{\partial\,t}
                                          \vec{\mathcal{E}}
                                    +\vec{\nabla}\times\vec{\mathcal{B}}
                                    +{\rm i}\,g\biggl(\Bigl[\varphi,\
                                                \vec{\mathcal{E}}\Bigr]
                                          +\vec{\mathcal{A}}\times
                                                \vec{\mathcal{B}}
                                          +\vec{\mathcal{B}}\times
                                                \vec{\mathcal{A}}\biggr) \nonumber\\
                              &=& -k^2 \vec{\mathcal{A}}(\vec{r})\,{\rm e}^{
                                          -{\rm i}\,\omega\,t}
                                    -{\rm i}\,k\Big[\vec{\nabla}\varphi(
                                          \vec{r})\Big]{\rm e}^{-{\rm i}\,
                                                \omega\,t}
                                    +2\,kg\,
                                          \vec{\mathcal{N}}\,{\rm e}^{
                                                -{\rm i}\,2\,\omega\,t}\nonumber\\
                                  &&  +\biggl\{\vec{\nabla}\Big[
                                          \vec{\nabla}\cdot
                                                \vec{\mathcal{A}}(\vec{r})
                                          \Big]
                                          -\nabla^2 \vec{\mathcal{A}}(
                                                \vec{r})\biggr\}{\rm e}^{
                                                      -{\rm i}\,\omega\,t}
                                    -{\rm i}\,g\bigl(\vec{\nabla}\times
                                          \vec{\mathcal{M}}\bigr){\rm e}^{
                                                -{\rm i}\,2\,\omega\,t} \nonumber\\
                                   & & -g\,k\Big[\varphi(\vec{r}),\
                                          \vec{\mathcal{A}}(\vec{r})\Big]
                                          {\rm e}^{-{\rm i}\,2\,\omega\,t}
                                    -{\rm i}\,g\Big[\varphi(\vec{r}),\
                                          \vec{\nabla}\varphi(\vec{r})\Big]
                                          {\rm e}^{-{\rm i}\,2\,\omega\,t}
                                    +g^2 \,\Big[\varphi(
                                          \vec{r}),\ \vec{\mathcal{N}}\Big]
                                          {\rm e}^{-{\rm i}\,3\,\omega\,t}
                                          \nonumber\\
                                    && +{\rm i}\,g\biggl\{\vec{\mathcal{A}}(
                                          \vec{r})\times\Big[\vec{\nabla}
                                                \times\vec{\mathcal{A}}(
                                                      \vec{r})\Big]
                                          +\Big[\vec{\nabla}\times
                                                \vec{\mathcal{A}}(\vec{r})
                                                \Big]\times
                                                      \vec{\mathcal{A}}(
                                                            \vec{r})
                                          \biggr\}{\rm e}^{-{\rm i}\,2\,
                                                \omega\,t}
                                    +g^2 \Big[\vec{\mathcal{A}}(
                                          \vec{r})\times\vec{\mathcal{M}}
                                          +\vec{\mathcal{M}}\times
                                                \vec{\mathcal{A}}(\vec{r})
                                          \Big]{\rm e}^{-{\rm i}\,3\,
                                                \omega\,t} \nonumber\\
                              &=& -k^2 \vec{\mathcal{A}}(\vec{r})\,{\rm e}^{
                                          -{\rm i}\,\omega\,t}
                                    -{\rm i}\,k\Big[\vec{\nabla}\varphi(
                                          \vec{r})\Big]{\rm e}^{-{\rm i}\,
                                                \omega\,t}
                                    +\biggl\{\vec{\nabla}\Big[
                                          \vec{\nabla}\cdot
                                                \vec{\mathcal{A}}(\vec{r})
                                          \Big]-\nabla^2 \vec{\mathcal{A}}(
                                                \vec{r})\biggr\}{\rm e}^{
                                                      -{\rm i}\,\omega\,t} \nonumber\\
                               & & +2\,k\,g\,
                                          \vec{\mathcal{N}}\,{\rm e}^{
                                                -{\rm i}\,2\,\omega\,t}
                                    +{\rm i}\,g\biggl\{\vec{\mathcal{A}}(
                                          \vec{r})\times\Big[\vec{\nabla}
                                                \times\vec{\mathcal{A}}(
                                                      \vec{r})\Big]
                                          +\Big[\vec{\nabla}\times
                                                \vec{\mathcal{A}}(\vec{r})
                                                \Big]\times
                                                      \vec{\mathcal{A}}(
                                                            \vec{r})
                                          \biggr\}{\rm e}^{-{\rm i}\,2\,
                                                \omega\,t}
                                    -{\rm i}\,g\bigl(\vec{\nabla}\times
                                          \vec{\mathcal{M}}\bigr){\rm e}^{
                                                -{\rm i}\,2\,\omega\,t} \nonumber\\
                                &    & - k\,g\,\Big[\varphi(\vec{r}),\
                                          \vec{\mathcal{A}}(\vec{r})\Big]
                                          {\rm e}^{-{\rm i}\,2\,\omega\,t}
                                    -{\rm i}\,g\Big[\varphi(\vec{r}),\
                                          \vec{\nabla}\varphi(\vec{r})\Big]
                                          {\rm e}^{-{\rm i}\,2\,\omega\,t}
                                          \nonumber\\
                                 &   & +g^2\, \Big[
                                          \varphi(\vec{r}),\
                                          \vec{\mathcal{N}}\Big]{\rm e}^{
                                                -{\rm i}\,3\,\omega\,t}
                                    +g^2 \Big[\vec{\mathcal{A}}(
                                          \vec{r})\times\vec{\mathcal{M}}
                                          +\vec{\mathcal{M}}\times
                                                \vec{\mathcal{A}}(\vec{r})
                                          \Big]{\rm e}^{-{\rm i}\,3\,
                                                \omega\,t} \nonumber\\
                              &=& -k^2 \vec{\mathcal{A}}(\vec{r})\,{\rm e}^{
                                          -{\rm i}\,\omega\,t}
                                    -{\rm i}\,k\Big[\vec{\nabla}\varphi(
                                          \vec{r})\Big]{\rm e}^{-{\rm i}\,
                                                \omega\,t}
                                    +\biggl\{\vec{\nabla}\Big[
                                          \vec{\nabla}\cdot
                                                \vec{\mathcal{A}}(\vec{r})
                                          \Big]-\nabla^2 \vec{\mathcal{A}}(
                                                \vec{r})\biggr\}{\rm e}^{
                                                      -{\rm i}\,\omega\,t} \nonumber\\
                               &     & +g\,k \,
                                          \vec{\mathcal{N}}
                                          {\rm e}^{-{\rm i}\,2\,\omega\,t}
                                    +{\rm i}\,g\biggl\{\vec{\mathcal{A}}(
                                          \vec{r})\times\Big[\vec{\nabla}
                                                \times\vec{\mathcal{A}}(
                                                      \vec{r})\Big]
                                          +\Big[\vec{\nabla}\times
                                                \vec{\mathcal{A}}(\vec{r})
                                                \Big]\times
                                                      \vec{\mathcal{A}}(
                                                            \vec{r})
                                          \biggr\}{\rm e}^{-{\rm i}\,2\,
                                                \omega\,t} \nonumber\\
                                &    & -{\rm i}\,g\bigl(\vec{\nabla}\times
                                          \vec{\mathcal{M}}\bigr){\rm e}^{
                                                -{\rm i}\,2\,\omega\,t}
                                    -{\rm i}\,g\Big[\varphi(\vec{r}),\
                                          \vec{\nabla}\varphi(\vec{r})\Big]
                                          {\rm e}^{-{\rm i}\,2\,\omega\,t}
                                          \nonumber\\
                                 &   & +g^2 \,\Big[
                                          \varphi(\vec{r}),\
                                          \vec{\mathcal{N}}\Big]{\rm e}^{
                                                -{\rm i}\,3\,\omega\,t}
                                    +g^2 \Big[\vec{\mathcal{A}}(
                                          \vec{r})\times\vec{\mathcal{M}}
                                          +\vec{\mathcal{M}}\times
                                                \vec{\mathcal{A}}(\vec{r})
                                          \Big]{\rm e}^{-{\rm i}\,3\,
                                                \omega\,t}=0.
                  \end{eqnarray}
                  For the arbitrariness of $\omega$, we have
                  \begin{eqnarray}
                        && \vec{\nabla}\Big[\vec{\nabla}\cdot
                                    \vec{\mathcal{A}}(\vec{r})\Big]
                              -\nabla^2 \vec{\mathcal{A}}(\vec{r})
                              -k^2 \vec{\mathcal{A}}(\vec{r})
                              -{\rm i}\,k\Big[\vec{\nabla}\varphi(\vec{r})
                                    \Big]=0,
                              \label{eq:CurlBCond1} \\
                        && g\, \biggr\{ k\, \vec{\mathcal{N}}
                              +{\rm i}\,\vec{\mathcal{A}}(\vec{r})\times\Big[
                                    \vec{\nabla}\times\vec{\mathcal{A}}(
                                          \vec{r})\Big]
                              +{\rm i}\,\Big[\vec{\nabla}\times\vec{\mathcal{A}}(
                                    \vec{r})\Big]\times\vec{\mathcal{A}}(
                                          \vec{r})
                              -{\rm i}\,\bigl(\vec{\nabla}\times\vec{\mathcal{M}}
                                    \bigr)
                              -{\rm i}\,\Big[\varphi(\vec{r}),\ \vec{\nabla}
                                    \varphi(\vec{r})\Big] \biggr\}=0, \label{eq:CurlBCond2} \\
                        && g^2 \, \biggr\{\Big[\varphi(\vec{r}),\
                                    \vec{\mathcal{N}}\Big]
                              +\vec{\mathcal{A}}(\vec{r})\times
                                    \vec{\mathcal{M}}
                              +\vec{\mathcal{M}}\times\vec{\mathcal{A}}(
                                    \vec{r})\biggr\}=0,
                  \end{eqnarray}
      \end{itemize}
       which correspond to conditions (\ref{eq:TICond6}), (\ref{eq:TICond7}) and (\ref{eq:TICond8}), respectively. This ends the derivation.

      \subsection{The Solutions under Weak-Coupling Approximation}
      For small $g$, we omit the terms referring to $g^2$, then the eight conditions \eqref{eq:TICond1}-\eqref{eq:TICond8} reduce to the following six conditions:
      \begin{subequations}
      \begin{eqnarray}
                  && {\rm i}\,k\Big[\vec{\nabla}\cdot\vec{\mathcal{A}}(
                        \vec{r})\Big]-\nabla^2 \varphi(\vec{r})=0, \label{eq:TIGCond1} \\
                  && \Big[\varphi(\vec{r}),\ \vec{\nabla}\cdot\vec{\mathcal{A}}(
                        \vec{r})\Big]=0, \label{eq:TIGCond2} \\
                  && 2\,k\,\vec{\mathcal{M}}+{\rm i}\,\vec{\nabla}\times\Big[
                        \varphi(\vec{r}),\ \vec{\mathcal{A}}(\vec{r})\Big]=0,
                        \label{eq:TIGCond3} \\
                  && \vec{\nabla}\cdot\vec{\mathcal{M}}=0, \label{eq:TIGCond4} \\
                  && \vec{\nabla}\Big[\vec{\nabla}\cdot
                              \vec{\mathcal{A}}(\vec{r})\Big]
                        -\nabla^2 \vec{\mathcal{A}}(\vec{r})
                        -k^2 \vec{\mathcal{A}}(\vec{r})
                        -{\rm i}\,k\Big[\vec{\nabla}\varphi(\vec{r})
                              \Big]=0, \label{eq:TIGCond5} \\
                  && {\rm i}\,k\Big[\varphi(\vec{r}),\ \vec{\mathcal{A}}(\vec{r})
                              \Big]
                        -\vec{\mathcal{A}}(\vec{r})\times\Big[
                              \vec{\nabla}\times\vec{\mathcal{A}}(\vec{r})\Big]
                        -\Big[\vec{\nabla}\times\vec{\mathcal{A}}(
                              \vec{r})\Big]\times\vec{\mathcal{A}}(\vec{r})
                        +\bigl(\vec{\nabla}\times\vec{\mathcal{M}}\bigr)
                        +\Big[\varphi(\vec{r}),\ \vec{\nabla}\varphi(\vec{r})\Big]
                        =0. \label{eq:TIGCond6}
            \end{eqnarray}
            \end{subequations}
            Fortunately, for Eqs. (\ref{eq:TIGCond1})-(\ref{eq:TIGCond6}), there are exact solutions for $\vec{\mathcal{A}}(\vec{r})$ and $\varphi(\vec{r})$.

            \emph{Proof.---}For simplicity, we assume that
             \begin{eqnarray}
                  &&\vec{\mathcal{A}}(\vec{r},t)=\vec{\tau}\,{\rm e}^{
                        {\rm i}(\vec{k}\cdot\vec{r}-\omega\,t)}, \nonumber\\
                 && \varphi(\vec{r},t)=\tilde{\varphi}\,{\rm e}^{
                        {\rm i}(\vec{k}\cdot\vec{r}-\omega\,t)},
             \end{eqnarray}
             namely
             \begin{eqnarray}
                  &&\vec{\mathcal{A}}(\vec{r})=\vec{\tau}\,{\rm e}^{
                        {\rm i}\,\vec{k}\cdot\vec{r}}, \nonumber\\
                 && \varphi(\vec{r})=\tilde{\varphi}\,{\rm e}^{
                        {\rm i}\,\vec{k}\cdot\vec{r}},
             \end{eqnarray}
             with
            \begin{eqnarray}
                  \vec{\tau} &=& \vec{R}_0 \openone+\vec{R}_1 S_x
                              +\vec{R}_2 S_y +\vec{R}_3 S_z,\nonumber \\
                  \tilde{\varphi} &=& \tilde{\varphi}_0 \openone
                        +\tilde{\varphi}_1 S_x +\tilde{\varphi}_2 S_y
                        +\tilde{\varphi}_3 S_z,
            \end{eqnarray}
            and $\vec{R}_j$'s ($j=0,1,2,3$) are some constant vectors that do not depend on $\vec{r}$.
            Here $\vec{\tau}$ and $\tilde{\varphi}$ have been expanded by the identity operator $\openone$ and three $SU(2)$ generators, which satisfy the following commutative relations
            \begin{eqnarray}
                        [S_x, S_y]={\rm i}\hbar S_z, \;[S_y, S_z]={\rm i}\hbar S_x, \; [S_z, S_y]={\rm i}\hbar S_x,
            \end{eqnarray}
            or
            \begin{eqnarray}
                        \vec{S} \times \vec{S} ={\rm i} \hbar \, \vec{S},
            \end{eqnarray}
            with $\vec{S}=(S_x, S_y, S_z)$, $\hbar=h/2\pi$, and $h$ being Planck's constant.

            Let us define
            \begin{equation}
                  \vec{\tau}\times\vec{\tau}={\rm i}\hbar\,\vec{\eta}
                  \equiv{\rm i}\hbar\left[\big(\vec{R}_2 \times\vec{R}_3\big)S_x
                        +\big(\vec{R}_3 \times\vec{R}_1\big)S_y
                        +\big(\vec{R}_1 \times\vec{R}_2\big)S_z\right],
            \end{equation}
            or
            \begin{equation}
                  \vec{\eta}=\frac{1}{{\rm i}\hbar}\vec{\tau}\times\vec{\tau}=\big(\vec{R}_2 \times\vec{R}_3\big)S_x
                        +\big(\vec{R}_3 \times\vec{R}_1\big)S_y
                        +\big(\vec{R}_1 \times\vec{R}_2\big)S_z,
            \end{equation}
            then we have the ``magnetic'' field as
            \begin{eqnarray}
                   \vec{\mathcal{B}}&=&\vec{\nabla}\times\vec{\mathcal{A}}-{\rm i}
                        \,g\bigl(\vec{\mathcal{A}}\times\vec{\mathcal{A}}\bigr)
                  =\vec{\nabla}\times\Big[\vec{\tau}\,{\rm e}^{
                              {\rm i}(\vec{k}\cdot\vec{r}-\omega\,t)}\Big]
                        -{\rm i}\,g\biggl\{\Big[\vec{\tau}\,{\rm e}^{
                                    {\rm i}(\vec{k}\cdot\vec{r}-\omega\,t)}\Big]
                              \times\Big[\vec{\tau}\,{\rm e}^{
                                    {\rm i}(\vec{k}\cdot\vec{r}-\omega\,t)}\Big]
                              \biggr\} \nonumber \\
                  &=& {\rm i}\bigl(\vec{k}\times\vec{\tau}\bigr){\rm e}^{
                              {\rm i}(\vec{k}\cdot\vec{r}-\omega\,t)}
                        -{\rm i}\,g\bigl(\vec{\tau}\times\vec{\tau}\bigr){\rm e}^{
                              {\rm i}\,2(\vec{k}\cdot\vec{r}-\omega\,t)}
                  ={\rm i}\bigl(\vec{k}\times\vec{\tau}\bigr){\rm e}^{
                              {\rm i}(\vec{k}\cdot\vec{r}-\omega\,t)}
                        +g\,\hbar\,\vec{\eta}\,{\rm e}^{
                              {\rm i}\,2(\vec{k}\cdot\vec{r}-\omega\,t)}.
            \end{eqnarray}
            Moreover
            \begin{eqnarray}
                         \left[\varphi(\vec{r},t),\ \vec{\mathcal{A}}(\vec{r},t)\right]
                        &=& \left[ \varphi(\vec{r}),\ \vec{\mathcal{A}}(\vec{r})
                        \right]{\rm e}^{
                              -{\rm i}\,2\omega t}=\left[\tilde{\varphi},\ \vec{\tau}\right]{\rm e}^{
                              {\rm i}\,2(\vec{k}\cdot\vec{r}-\omega t)}\nonumber\\
                        &=&{\rm i}\hbar\biggl[\left(\tilde{\varphi}_2 \vec{R}_3
                                    -\tilde{\varphi}_3 \vec{R}_2\right)S_x
                              +\left(\tilde{\varphi}_3 \vec{R}_1
                                    -\tilde{\varphi}_1 \vec{R}_3\right)S_y
                              +\left(\tilde{\varphi}_1 \vec{R}_2
                                    -\tilde{\varphi}_2 \vec{R}_1\right)S_z
                              \biggr]{\rm e}^{{\rm i}\,2(\vec{k}\cdot\vec{r}
                                    -\omega t)} \nonumber\\
                        &=& {\rm i}\hbar\,\vec{\xi}\,{\rm e}^{
                              {\rm i}\,2(\vec{k}\cdot\vec{r}-\omega t)},
            \end{eqnarray}
            or
            \begin{eqnarray}
            \vec{\xi} &=& \left(\tilde{\varphi}_2 \vec{R}_3
                                    -\tilde{\varphi}_3 \vec{R}_2\right)S_x
                              +\left(\tilde{\varphi}_3 \vec{R}_1
                                    -\tilde{\varphi}_1 \vec{R}_3\right)S_y
                              +\left(\tilde{\varphi}_1 \vec{R}_2
                                    -\tilde{\varphi}_2 \vec{R}_1\right)S_z,
            \end{eqnarray}
            or
            \begin{eqnarray}\label{eq:xi-1a}
            \vec{\mathcal{N}}(\vec{r})= \left[ \varphi(\vec{r}),\ \vec{\mathcal{A}}(\vec{r})
                        \right]={\rm i}\hbar\,\vec{\xi}\,{\rm e}^{{\rm i}\,2\vec{k}\cdot\vec{r}}.
            \end{eqnarray}
            Hence the ``electric'' field reads
            \begin{eqnarray}
                         \vec{\mathcal{E}}&=&-\dfrac{1}{c}
                              \dfrac{\partial\,\vec{\mathcal{A}}}{\partial\,t}
                                    -\vec{\nabla}\varphi-{\rm i}\,g\left[
                                          \varphi,\ \vec{\mathcal{A}}\right] \nonumber\\
                        &=& -\dfrac{1}{c} \dfrac{\partial}{\partial\,t} \Bigl[
                              \vec{\tau}\,{\rm e}^{
                                    {\rm i}(\vec{k}\cdot\vec{r}-\omega\,t)}\Bigr]
                              -\vec{\nabla}\Big[\bigl(\tilde{\varphi}_0 \openone
                                    +\tilde{\varphi}_1 S_x +\tilde{\varphi}_2 S_y
                                    +\tilde{\varphi}_3 S_z\bigr){\rm e}^{
                                          {\rm i}(\vec{k}\cdot\vec{r}-\omega\,t)}
                                    \Big] \nonumber\\
                         &     & +g\,\hbar\biggl[\left(\tilde{\varphi}_2 \vec{R}_3
                                          -\tilde{\varphi}_3 \vec{R}_2\right)S_x
                                    +\left(\tilde{\varphi}_3 \vec{R}_1
                                          -\tilde{\varphi}_1 \vec{R}_3\right)S_y
                                    +\left(\tilde{\varphi}_1 \vec{R}_2
                                          -\tilde{\varphi}_2 \vec{R}_1\right)S_z
                                    \biggr]{\rm e}^{{\rm i}\,2(\vec{k}\cdot\vec{r}
                                          -\omega t)} \nonumber\\
                        &=& {\rm i}\,k\,\vec{\tau}\,{\rm e}^{
                                    {\rm i}(\vec{k}\cdot\vec{r}-\omega\,t)}
                              -{\rm i}\,\vec{k}\bigl(\tilde{\varphi}_0 \openone
                                    +\tilde{\varphi}_1 S_x +\tilde{\varphi}_2 S_y
                                    +\tilde{\varphi}_3 S_z\bigr){\rm e}^{
                                          {\rm i}(\vec{k}\cdot\vec{r}-\omega\,t)}
                                    \nonumber\\
                         &     & +g\,\hbar\biggl[\left(\tilde{\varphi}_2 \vec{R}_3
                                          -\tilde{\varphi}_3 \vec{R}_2\right)S_x
                                    +\left(\tilde{\varphi}_3 \vec{R}_1
                                          -\tilde{\varphi}_1 \vec{R}_3\right)S_y
                                    +\left(\tilde{\varphi}_1 \vec{R}_2
                                          -\tilde{\varphi}_2 \vec{R}_1\right)S_z
                                    \biggr]{\rm e}^{{\rm i}\,2(\vec{k}\cdot\vec{r}
                                          -\omega t)} \nonumber\\
                        &=& {\rm i}\,k\,\vec{\tau}\,{\rm e}^{
                                    {\rm i}(\vec{k}\cdot\vec{r}-\omega\,t)}
                              -{\rm i}\,\vec{k}\,\tilde{\varphi}\,{\rm e}^{
                                    {\rm i}(\vec{k}\cdot\vec{r}-\omega\,t)}
                              +g\,\hbar\,\vec{\xi}\,{\rm e}^{{\rm i}\,2(
                                    \vec{k}\cdot\vec{r}-\omega t)} \nonumber\\
                        &=&-{\rm i}\,k\Big[\hat{k}\times\bigl(\hat{k}
                                    \times\vec{\tau}\bigr)\Big]{\rm e}^{
                                          {\rm i}(\vec{k}\cdot\vec{r}
                                                -\omega\,t)}
                              +g\,\hbar\,\vec{\xi}\,{\rm e}^{{\rm i}\,2(
                                    \vec{k}\cdot\vec{r}-\omega t)}.
            \end{eqnarray}
            After that, \Eq{eq:TIGCond1}-\Eq{eq:TIGCond3} can be recast to
            \begin{align}
                  & k^2 \tilde{\varphi}-k\bigl(\vec{\tau}\cdot\vec{k}
                        \bigr)=0, \label{eq:PhiTuaK} \\
                  & \Big[\tilde{\varphi}, \ \bigl(\vec{\tau}\cdot\vec{k}\bigr)\Big]
                        =0, \label{eq:TauKPhi} \\
                  & k\,\vec{\eta}-\vec{k}\times\vec{\xi}=0.\label{eq:EtaXi1}
            \end{align}
            \Eq{eq:PhiTuaK} tells us
            \begin{equation}\label{eq:phi1a}
                  \tilde{\varphi}=\vec{\tau}\cdot\hat{k},
            \end{equation}
            with $\hat{k}=\vec{k}/k$ and $k=|\vec{k}|$, hence
            \begin{equation}
                  \varphi=\bigl(\vec{\tau}\cdot\hat{k}\bigr){\rm e}^{
                        {\rm i}(\vec{k}\cdot\vec{r}-\omega\,t)}.
            \end{equation}
            Based on Eq. (\ref{eq:phi1a}), one finds that Eq. (\ref{eq:TauKPhi}) is automatically satisfied. Eq. (\ref{eq:EtaXi1}) leads to
            \begin{equation}\label{eq:phi1b}
                  \vec{\eta}=\hat{k}\times\vec{\xi}.
            \end{equation}


            \Eq{eq:TIGCond4} leads to
            \begin{equation}\label{eq:phi1c}
                 \vec{k}\cdot \vec{\eta}=0,
            \end{equation}
            namely
            \begin{equation}
                  \vec{k}\cdot\bigl(\vec{R}_1 \times\vec{R}_2\bigr)
                  =\vec{k}\cdot\bigl(\vec{R}_2 \times\vec{R}_3\bigr)
                  =\vec{k}\cdot\bigl(\vec{R}_3 \times\vec{R}_1\bigr)
                  =0,
            \end{equation}
            which indicates that four vectors $\{\vec{R}_1, \vec{R}_2, \vec{R}_3, \vec{k}\}$ locate on the same plane, and $\vec{R}_0$ is arbitrary.

            Note \Eq{eq:TIGCond5} holds naturally, and \Eq{eq:TIGCond6} reduces to
            \begin{align}
                  \vec{\xi}=-\hat{k}\times\vec{\eta}=\vec{\eta}\times \hat{k},
            \end{align}
            which is consistent with Eq. (\ref{eq:phi1b}) and Eq. (\ref{eq:phi1c}).

            In summary,  the solutions of the Yang-Mills equations under the weak-coupling approximation are as follows:
            \begin{eqnarray}\label{eq:WCA-1a}
                 && \vec{\mathcal{A}}(\vec{r}, t)=\vec{\tau}\,{\rm e}^{
                        {\rm i}(\vec{k}\cdot\vec{r}-\omega\,t)}, \nonumber\\
                 && \varphi(\vec{r}, t) =\bigl(\vec{\tau}\cdot\hat{k}\bigr){\rm e}^{
                        {\rm i}(\vec{k}\cdot\vec{r}-\omega\,t)},\nonumber\\
                 && \vec{\mathcal{B}}(\vec{r}, t)
                  ={\rm i}\bigl(\vec{k}\times\vec{\tau}\bigr){\rm e}^{
                              {\rm i}(\vec{k}\cdot\vec{r}-\omega\,t)}
                        +g\,\hbar\,\vec{\eta}\,{\rm e}^{
                              {\rm i}\,2(\vec{k}\cdot\vec{r}-\omega\,t)}, \nonumber\\
                 && \vec{\mathcal{E}}(\vec{r}, t)
                  =-{\rm i}\,k\Big[\hat{k}\times\bigl(\hat{k}
                                    \times\vec{\tau}\bigr)\Big]{\rm e}^{
                                          {\rm i}(\vec{k}\cdot\vec{r}
                                                -\omega\,t)}
                              +g\,\hbar\,\vec{\xi}\,{\rm e}^{{\rm i}\,2(
                                    \vec{k}\cdot\vec{r}-\omega t)},
            \end{eqnarray}
            with
            \begin{eqnarray}\label{eq:WCA-1b}
                  &&\vec{\tau}=\vec{R}_0 \openone+\vec{R}_1 S_x +\vec{R}_2 S_y
                        +\vec{R}_3 S_z\nonumber\\
                  && \vec{\eta}=\big(\vec{R}_2 \times\vec{R}_3\big)S_x
                        +\big(\vec{R}_3 \times\vec{R}_1\big)S_y
                        +\big(\vec{R}_1 \times\vec{R}_2\big)S_z,\nonumber\\
                  &&  \vec{\xi}=-\hat{k}\times\vec{\eta}=\vec{\eta}\times \hat{k}, \nonumber\\
                  &&\vec{k}\cdot\bigl(\vec{R}_1 \times\vec{R}_2\bigr)
                  =\vec{k}\cdot\bigl(\vec{R}_2 \times\vec{R}_3\bigr)
                  =\vec{k}\cdot\bigl(\vec{R}_3 \times\vec{R}_1\bigr)
                  =0.
            \end{eqnarray}

            \begin{remark}
          For the spin-1/2 particle, the angular momentum operator is given by
 \begin{eqnarray}
 \label{S-1}
  && \vec{S}=\frac{\hbar}{2} \vec{\sigma},
  \end{eqnarray}
where $\vec{\sigma}$ is the vector of Pauli matrices, whose three components read
 \begin{eqnarray}
 \label{S-3}
  && \sigma_x=
  \left(
    \begin{array}{cc}
      0 & 1 \\
      1 & 0 \\
    \end{array}
  \right),\;\;\;\sigma_y=
  \left(
    \begin{array}{cc}
      0 & -{\rm i} \\
      {\rm i} & 0 \\
    \end{array}
  \right),\;\;\;  \sigma_z=
  \left(
    \begin{array}{cc}
      1 & 0 \\
      0 & -1 \\
    \end{array}
  \right).
   \end{eqnarray}
   For the spin-1 particle, the angular momentum operator is given by
      \begin{eqnarray}
 \label{Spin1}
  && S_x=\frac{\hbar}{\sqrt{2}}\left(
           \begin{array}{ccc}
             0 & 1 & 0 \\
             1 & 0 & 1 \\
             0 & 1 & 0 \\
           \end{array}
         \right),\;\;\;\;\;\;\;\;
  S_y=\frac{{\rm i}\hbar}{\sqrt{2}}\left(
           \begin{array}{ccc}
             0 & -1 & 0 \\
             1 & 0 & -1 \\
             0 & 1 & 0 \\
           \end{array}
         \right),\;\;\;\;\;\;\;\;
   S_z=\hbar \left(
           \begin{array}{ccc}
             1 & 0 & 0 \\
             0 & 0 & 0 \\
             0 & 0 & -1 \\
           \end{array}
         \right).
  \end{eqnarray}
\end{remark}

      \subsection{Physical Meaning of the Solutions: The Angular-Momentum Waves}

      Let us focus on the ``magnetic'' field $\vec{\mathcal{B}}(\vec{r}, t)$ and the ``electric'' field $\vec{\mathcal{E}}(\vec{r}, t)$, whose expressions are given as follows:
      \begin{eqnarray}
      && \vec{\mathcal{B}}(\vec{r}, t)
                  ={\rm i}\bigl(\vec{k}\times\vec{\tau}\bigr){\rm e}^{
                              {\rm i}(\vec{k}\cdot\vec{r}-\omega\,t)}
                        +g\,\hbar\,\vec{\eta}\,{\rm e}^{
                              {\rm i}\,2(\vec{k}\cdot\vec{r}-\omega\,t)}, \label{eq:BB}\\
      && \vec{\mathcal{E}}(\vec{r}, t)
                  =-{\rm i}\,k\Big[\hat{k}\times\bigl(\hat{k}
                                    \times\vec{\tau}\bigr)\Big]{\rm e}^{
                                          {\rm i}(\vec{k}\cdot\vec{r}
                                                -\omega\,t)}
                              +g\,\hbar\,\vec{\xi}\,{\rm e}^{{\rm i}\,2(
                                    \vec{k}\cdot\vec{r}-\omega t)}. \label{eq:EE}
            \end{eqnarray}
      It is easy to verify that the ``magnetic'' field $\vec{\mathcal{B}}(\vec{r}, t)$ and  the ``electric'' field $\vec{\mathcal{E}}(\vec{r}, t)$ satisfy the following wave equations:
      \begin{eqnarray}
           && \nabla^2 \vec{\mathcal{E}}-\dfrac{1}{c^2} \dfrac{\partial\,\vec{\mathcal{E}}}{\partial\,t^2} =0,\label{eq:WavE-aa}\\
           && \nabla^2 \vec{\mathcal{B}}-\dfrac{1}{c^2} \dfrac{\partial\,\vec{\mathcal{B}}}{\partial\,t^2} =0, \label{eq:WavB-bb}
      \end{eqnarray}
     therefore, $\vec{\mathcal{B}}(\vec{r}, t)$ and $\vec{\mathcal{E}}(\vec{r}, t)$ can be regarded as some types of waves. Because the amplitudes of $\vec{\mathcal{B}}(\vec{r}, t)$ and $\vec{\mathcal{E}}(\vec{r}, t)$ contain the components of the $SU(2)$ angular-momentum operator $\vec{S}$ (i.e., $S_x, S_y, S_z$), thus we would like to call $\vec{\mathcal{B}}(\vec{r}, t)$ and $\vec{\mathcal{E}}(\vec{r}, t)$ as the $SU(2)$ angular-momentum waves.

     \begin{remark}
     One can check that
            \begin{eqnarray}
            &&\vec{\mathcal{B}}(\vec{r}, t)=\hat{k}\times\vec{\mathcal{E}}(\vec{r}, t), \nonumber\\
            &&\vec{\mathcal{E}}(\vec{r}, t)=\vec{\mathcal{B}}(\vec{r}, t)\times\hat{k},  \nonumber\\
            &&\hat{k}\cdot\vec{\mathcal{B}}(\vec{r}, t)=0, \;\; \hat{k}\cdot\vec{\mathcal{E}}(\vec{r}, t)=0. \;\;
            \vec{\mathcal{B}}(\vec{r}, t)\cdot \vec{\mathcal{E}}(\vec{r}, t)=0,
            \end{eqnarray}
    that is the ``magnetic'' field $\vec{\mathcal{B}}(\vec{r}, t)$, the ``electric'' field $\vec{\mathcal{E}}(\vec{r}, t)$ and the propagation direction $\vec{k}$ are mutually perpendicular.
            Precisely we have
            \begin{equation}
                  \begin{split}
                        & \vec{\mathcal{B}}(\vec{r},t)\cdot\vec{\mathcal{E}}(\vec{r},t) \\
                        =& \Big[{\rm i}\bigl(\vec{k}\times\vec{\tau}\bigr){\rm e}^{{\rm i}(
                                    \vec{k}\cdot\vec{r}-\omega\,t)}
                              -{\rm i}\,g\bigl(\vec{\tau}\times\vec{\tau}\bigr){\rm e}^{
                                    {\rm i}\,2(\vec{k}\cdot\vec{r}-\omega\,t)}
                              \Big]\cdot\biggl\{-{\rm i}\,k\Big[\hat{k}\times\bigl(\hat{k}
                                          \times\vec{\tau}\bigr)\Big]{\rm e}^{{\rm i}(
                                                \vec{k}\cdot\vec{r}-\omega\,t)}
                                    -{\rm i}\,g\,\Big[\bigl(\hat{k}\cdot\vec{\tau}\bigr),\
                                          \vec{\tau}\Big]{\rm e}^{{\rm i}\,2(\vec{k}\cdot
                                                \vec{r}-\omega t)}\biggr\} \\
                        =& \Big[k\bigl(\hat{k}\times\vec{\tau}\bigr){\rm e}^{{\rm i}(
                                    \vec{k}\cdot\vec{r}-\omega\,t)}
                              -g\bigl(\vec{\tau}\times\vec{\tau}\bigr){\rm e}^{
                                    {\rm i}\,2(\vec{k}\cdot\vec{r}-\omega\,t)}
                              \Big]\cdot\biggl\{k\Big[\hat{k}\times\bigl(\hat{k}
                                          \times\vec{\tau}\bigr)\Big]{\rm e}^{{\rm i}(
                                                \vec{k}\cdot\vec{r}-\omega\,t)}
                                    +g\,\Big[\bigl(\hat{k}\cdot\vec{\tau}\bigr),\ \vec{\tau}
                                    \Big]{\rm e}^{{\rm i}\,2(\vec{k}\cdot\vec{r}-\omega t)}
                                    \biggr\} \\
                        =& \Big[k\bigl(\hat{k}\times\vec{\tau}\bigr){\rm e}^{{\rm i}(
                                    \vec{k}\cdot\vec{r}-\omega\,t)}
                              -g\bigl(\vec{\tau}\times\vec{\tau}\bigr){\rm e}^{
                                    {\rm i}\,2(\vec{k}\cdot\vec{r}-\omega\,t)}
                              \Big]\cdot\biggl\{k\Big[\bigl(\hat{k}\cdot\vec{\tau}\bigr)\hat{
                                          k}-\vec{\tau}\Big]{\rm e}^{{\rm i}(\vec{k}\cdot
                                                \vec{r}-\omega\,t)}
                                    +g\,\Big[\bigl(\hat{k}\cdot\vec{\tau}\bigr),\ \vec{\tau}
                                    \Big]{\rm e}^{{\rm i}\,2(\vec{k}\cdot\vec{r}-\omega t)}
                                    \biggr\} \\
                        =& k^2 \bigl(\hat{k}\times\vec{\tau}\bigr)\cdot\Big[\bigl(\hat{k}
                                    \cdot\vec{\tau}\bigr)\hat{k}-\vec{\tau}\Big]{\rm e}^{
                                          {\rm i}\,2(\vec{k}\cdot\vec{r}-\omega\,t)}
                              +g\,k\biggl\{\bigl(\hat{k}\times\vec{\tau}\bigr)\cdot\Big[\bigl(
                                          \hat{k}\cdot\vec{\tau}\bigr),\ \vec{\tau}\Big]
                                    -\bigl(\vec{\tau}\times\vec{\tau}\bigr)\cdot\Big[\bigl(
                                          \hat{k}\cdot\vec{\tau}\bigr)\hat{k}-\vec{\tau}\Big]
                                    \biggr\}{\rm e}^{{\rm i}\,3(\vec{k}\cdot\vec{r}
                                          -\omega\,t)} \\
                              & -g^2 \bigl(\vec{\tau}\times\vec{\tau}\bigr)\cdot\Big[\bigl(
                                    \hat{k}\cdot\vec{\tau}\bigr),\ \vec{\tau}\Big]{\rm e}^{
                                          {\rm i}\,4(\vec{k}\cdot\vec{r}-\omega t)} \\
                        =& 0.
                  \end{split}
            \end{equation}
            For
            \begin{itemize}
                  \item [(i).] the first term,
                        \begin{equation}
                              \begin{split}
                                    & \bigl(\hat{k}\times\vec{\tau}\bigr)\cdot\Big[\bigl(
                                          \hat{k}\cdot\vec{\tau}\bigr)\hat{k}-\vec{\tau}\Big]
                                    =\Big[\bigl(\hat{k}\times\vec{\tau}\bigr)\cdot\hat{k}\Big]
                                                \bigl(\hat{k}\cdot\vec{\tau}\bigr)
                                          -\bigl(\hat{k}\times\vec{\tau}\bigr)\cdot\vec{\tau}
                                    =-\hat{k}\cdot\bigl(\vec{\tau}\times\vec{\tau}\bigr)
                                    =-{\rm i}\hbar\,\hat{k}\cdot\vec{\eta}=0;
                              \end{split}
                        \end{equation}
                  \item [(ii).] the second term,
                        \begin{equation}
                              \begin{split}
                                    & \bigl(\hat{k}\times\vec{\tau}\bigr)\cdot\Big[\bigl(
                                                \hat{k}\cdot\vec{\tau}\bigr),\ \vec{\tau}\Big]
                                          -\bigl(\vec{\tau}\times\vec{\tau}\bigr)\cdot\Big[
                                                \bigl(\hat{k}\cdot\vec{\tau}\bigr)\hat{k}
                                          -\vec{\tau}\Big]
                                    ={\rm i}\hbar\bigl(\hat{k}\times\vec{\tau}\bigr)\cdot
                                                \vec{\xi}
                                          -\Big[\bigl(\vec{\tau}\times\vec{\tau}\bigr)\cdot
                                                \hat{k}\Big]\bigl(\hat{k}\cdot\vec{\tau}\bigr)
                                          +\bigl(\vec{\tau}\times\vec{\tau}\bigr)\cdot\vec{
                                                \tau} \\
                                    =& -{\rm i}\hbar\bigl(\vec{\tau}\times\hat{k}\bigr)\cdot
                                                \vec{\xi}
                                          +\bigl(\vec{\tau}\times\vec{\tau}\bigr)\cdot\vec{
                                                \tau}
                                    =-{\rm i}\hbar\,\vec{\tau}\cdot\bigl(\hat{k}\times\vec{
                                                \xi}\bigr)
                                          +\vec{\tau}\cdot\bigl(\vec{\tau}\times\vec{\tau}
                                                \bigr)
                                    ={\rm i}\hbar\,\vec{\tau}\cdot\Big[\vec{\eta}-\bigl(
                                          \hat{k}\times\vec{\xi}\bigr)\Big]=0;
                              \end{split}
                        \end{equation}
                  \item [(iii).] the third term, notice
                        \begin{equation}
                              \begin{split}
                                    & \dfrac{1}{{\rm i}\hbar} \vec{\eta}\times\vec{\eta}
                                    =\Big[\big(\vec{R}_2 \times\vec{R}_3\big)S_x
                                                +\big(\vec{R}_3 \times\vec{R}_1\big)S_y
                                                +\big(\vec{R}_1 \times\vec{R}_2\big)S_z\Big]
                                          \times\Big[\big(\vec{R}_2 \times\vec{R}_3\big)S_x
                                                +\big(\vec{R}_3 \times\vec{R}_1\big)S_y
                                                +\big(\vec{R}_1 \times\vec{R}_2\big)S_z\Big]
                                                \\
                                    =& \Big[\big(\vec{R}_3 \times\vec{R}_1\big)\times\big(
                                                \vec{R}_1 \times\vec{R}_2\big)\Big]S_x
                                          +\Big[\big(\vec{R}_1 \times\vec{R}_2\big)\times\big(
                                                \vec{R}_2 \times\vec{R}_3\big)\Big]S_y
                                          +\Big[\big(\vec{R}_2 \times\vec{R}_3\big)\times\big(
                                                \vec{R}_3 \times\vec{R}_1\big)\Big]S_z \\
                                    =& 0.
                              \end{split}
                        \end{equation}
                        Hence
                        \begin{equation}
                              \bigl(\vec{\tau}\times\vec{\tau}\bigr)\cdot\Big[\bigl(
                                    \hat{k}\cdot\vec{\tau}\bigr),\ \vec{\tau}\Big]
                              =-\hbar^2 \vec{\eta}\cdot\vec{\xi}
                              =-\hbar^2 \vec{\eta}\cdot\bigl(\vec{\eta}\times\hat{k}
                                    \bigr)
                              =-\hbar^2 \bigl(\vec{\eta}\times\vec{\eta}\bigr)\cdot
                                    \hat{k}=0.
                        \end{equation}
            \end{itemize}
     \end{remark}
     \begin{remark}
            \begin{equation}
                  \begin{split}
                        & \bigl(\vec{k}\times\vec{\tau}\bigr)\times\bigl(\vec{k}
                              \times\vec{\tau}\bigr)
                        =\Big[\bigl(\vec{k}\times\vec{\tau}\bigr)\cdot\vec{\tau}\Big]
                                    \vec{k}
                              -\Big[\bigl(\vec{k}\times\vec{\tau}\bigr)\cdot\vec{k}\Big]
                                    \vec{\tau}
                        =\Big[\vec{k}\cdot\bigl(\vec{\tau}\times\vec{\tau}\bigr)\Big]
                                    \vec{k}
                        =0.
                  \end{split}
            \end{equation}
            \begin{equation}
                  \mathcal{B}_x ={\rm i}\bigl(\vec{k}\times\vec{\tau}\bigr)_x {\rm e}^{
                              {\rm i}(\vec{k}\cdot\vec{r}-\omega\,t)}
                        +g\,\hbar\,\eta_x {\rm e}^{{\rm i}\,2(\vec{k}\cdot\vec{r}
                              -\omega\,t)} ,\quad
                  \mathcal{B}_y ={\rm i}\bigl(\vec{k}\times\vec{\tau}\bigr)_y {\rm e}^{
                              {\rm i}(\vec{k}\cdot\vec{r}-\omega\,t)}
                        +g\,\hbar\,\eta_y {\rm e}^{{\rm i}\,2(\vec{k}\cdot\vec{r}
                              -\omega\,t)}.
            \end{equation}
            Then
            \begin{equation}
                  \begin{split}
                        & \mathcal{B}_x \mathcal{B}_y
                        =\Big[{\rm i}\bigl(\vec{k}\times\vec{\tau}\bigr)_x {\rm e}^{
                                          {\rm i}(\vec{k}\cdot\vec{r}-\omega\,t)}
                                    +g\,\hbar\,\eta_x {\rm e}^{{\rm i}\,2(\vec{k}\cdot
                                          \vec{r}-\omega\,t)}\Big]
                              \Big[{\rm i}\bigl(\vec{k}\times\vec{\tau}\bigr)_y
                                          {\rm e}^{{\rm i}(\vec{k}\cdot\vec{r}
                                                -\omega\,t)}
                                    +g\,\hbar\,\eta_y {\rm e}^{{\rm i}\,2(\vec{k}\cdot
                                          \vec{r}-\omega\,t)}\Big] \\
                        =& -\bigl(\vec{k}\times\vec{\tau}\bigr)_x \bigl(\vec{k}\times
                              \vec{\tau}\bigr)_y {\rm e}^{{\rm i}\,2(\vec{k}\cdot\vec{r}
                                    -\omega\,t)}
                              +{\rm i}\,g\,\hbar\Big[\bigl(\vec{k}\times\vec{\tau}
                                          \bigr)_x \eta_y
                                    +\bigl(\vec{k}\times\vec{\tau}\bigr)_y \eta_x\Big]
                                          {\rm e}^{{\rm i}\,3(\vec{k}\cdot\vec{r}
                                                -\omega\,t)}
                              +g^2 \hbar^2 \eta_x \eta_y {\rm e}^{{\rm i}\,4(
                                    \vec{k}\cdot\vec{r}-\omega\,t)},
                  \end{split}
            \end{equation}
            likewise we obtain
            \begin{equation}
                  \mathcal{B}_y \mathcal{B}_x
                  =-\bigl(\vec{k}\times\vec{\tau}\bigr)_y \bigl(\vec{k}\times\vec{\tau}
                              \bigr)_x {\rm e}^{{\rm i}\,2(\vec{k}\cdot\vec{r}
                                          -\omega\,t)}
                        +{\rm i}\,g\,\hbar\Big[\bigl(\vec{k}\times\vec{\tau}
                                    \bigr)_y \eta_x
                              +\bigl(\vec{k}\times\vec{\tau}\bigr)_x \eta_y\Big]
                                    {\rm e}^{{\rm i}\,3(\vec{k}\cdot\vec{r}
                                          -\omega\,t)}
                        +g^2 \hbar^2 \eta_y \eta_x {\rm e}^{{\rm i}\,4(
                              \vec{k}\cdot\vec{r}-\omega\,t)} ,
            \end{equation}
            thus
            \begin{equation}
                  \begin{split}
                        & \bigl[\mathcal{B}_x ,\ \mathcal{B}_y\bigr]
                        =\mathcal{B}_x \mathcal{B}_y -\mathcal{B}_y \mathcal{B}_x
                        =-\Big[\bigl(\vec{k}\times\vec{\tau}\bigr)_x \bigl(
                                          \vec{k}\times\vec{\tau}\bigr)_y
                                    -\bigl(\vec{k}\times\vec{\tau}\bigr)_y \bigl(
                                          \vec{k}\times\vec{\tau}\bigr)_x\Big] {\rm e}^{
                                                {\rm i}\,2(\vec{k}\cdot\vec{r}
                                                      -\omega\,t)}
                              +g^2 \hbar^2 \bigl(\eta_x \eta_y -\eta_y \eta_x\bigr)
                                    {\rm e}^{{\rm i}\,4(\vec{k}\cdot\vec{r}-\omega\,t)}
                                    \\
                        =& -\Big[\bigl(\vec{k}\times\vec{\tau}\bigr)\times\bigl(\vec{k}
                                    \times\vec{\tau}\bigr)\Big]_z {\rm e}^{{\rm i}\,2(
                                          \vec{k}\cdot\vec{r}-\omega\,t)}
                              +g^2 \hbar^2 \bigl(\vec{\eta}\times\vec{\eta}\bigr)_z
                                    {\rm e}^{{\rm i}\,4(\vec{k}\cdot\vec{r}-\omega\,t)},
                  \end{split}
            \end{equation}
            i.e.
            \begin{equation}
                  \vec{\mathcal{B}}\times\vec{\mathcal{B}}=0.
            \end{equation}
            Similarly, we have
            \begin{equation}
                  \begin{split}
                        & \vec{\mathcal{E}}\times\vec{\mathcal{E}}
                        =\bigl(\vec{\mathcal{B}}\times\hat{k}\bigr)\times\bigl(
                              \vec{\mathcal{B}}\times\hat{k}\bigr)
                        =\Big[\bigl(\vec{\mathcal{B}}\times\hat{k}\bigr)\cdot\hat{k}
                                    \Big]\vec{\mathcal{B}}
                              -\Big[\bigl(\vec{\mathcal{B}}\times\hat{k}\bigr)\cdot
                                    \vec{\mathcal{B}}\Big]\hat{k}
                        =\Big[\bigl(\hat{k}\times\vec{\mathcal{B}}\bigr)\cdot
                              \vec{\mathcal{B}}\Big]\hat{k}
                        =\Big[\hat{k}\cdot\bigl(\vec{\mathcal{B}}\times\vec{\mathcal{B}}
                              \bigr)\Big]\hat{k}=0.
                  \end{split}
            \end{equation}
            \begin{equation}
                  \begin{split}
                        & \vec{\mathcal{E}}\times\vec{\mathcal{B}}
                        =\bigl(\vec{\mathcal{B}}\times\hat{k}\bigr)\times
                              \vec{\mathcal{B}}
                        =\bigl(\vec{\mathcal{B}}\cdot\vec{\mathcal{B}}\bigr)\hat{k}
                              -\vec{\mathcal{B}}\bigl(\hat{k}\cdot\vec{\mathcal{B}}
                                    \bigr)
                        =\bigl(\vec{\mathcal{B}}\cdot\vec{\mathcal{B}}\bigr)\hat{k}
                        \propto\hat{k}.
                  \end{split}
            \end{equation}
      \end{remark}
      In the following we give an example.
      \begin{proposition}
            Under the assumption of
            $\vec{\mathcal{A}}\times\vec{\mathcal{A}}\neq 0$, and
            $[\varphi, \vec{\mathcal{A}}]\neq 0$, let $\vec{k}=k\hat{z}$ (i.e., the propagation direction is along the $z$-axis), $\vec{R_0}=\vec{R_2}=0$,
            $\vec{R}_1=\hat{x}$, and $\vec{R}_3=\hat{z}$ (i.e., the vectors
            $\vec{R}_1,\,\vec{R}_3,\,\vec{k}$ locate on the $xz$-plane). In this case, we have
            $\vec{\tau}=\vec{R}_1 S_x +\vec{R}_3 S_z =\hat{x} S_x+ \hat{z} S_z$,
            $\vec{\eta}=(\vec{R}_3 \times\vec{R}_1)S_y=\hat{y}S_y$. Then $SU(2)$ angular-momentum waves are given by
            \begin{eqnarray}\label{eq:EGAMW}
                  \vec{\mathcal{B}}(\vec{r}, t) &=& {\rm i}\left(\vec{k}\times\vec{\tau}
                              \right){\rm e}^{{\rm i}(k\,z-\omega t)}
                        +g\hbar \vec{\eta}\;{\rm e}^{2{\rm i}(k\,z-\omega t)}\nonumber\\
            &=&{\rm i} k \left[\hat{z}\times \left( \hat{x} S_x+ \hat{z} S_z\right)
                              \right]{\rm e}^{{\rm i}(k z-\omega t)}
                        +g\hbar \hat{y}S_y\;{\rm e}^{2{\rm i}(k\,z-\omega t)}\nonumber\\
                  &=& \hat{y} \biggr[{\rm i} k S_x {\rm e}^{{\rm i}(k z-\omega t)}
                        +g \hbar S_y\;{\rm e}^{2{\rm i}(k\,z-\omega t)}\biggr],\nonumber\\
                  \vec{\mathcal{E}}(\vec{r}, t) &=& -\hat{k}\times
                        \vec{\mathcal{B}}(\vec{r}, t)\nonumber\\
                  & =&\hat{x}\biggr[{\rm i} k S_x {\rm e}^{{\rm i}(k z-\omega t)}
                        +g\hbar S_y\;{\rm e}^{2{\rm i}(k\,z-\omega t)}\biggr].
            \end{eqnarray}
      \end{proposition}
   \begin{remark}
     Under the weak-coupling approximation, $\vec{\mathcal{B}}(\vec{r}, t)$ and $\vec{\mathcal{E}}(\vec{r}, t)$ are the non-Abelian waves, because generally $\vec{\tau}\times \vec{\tau} \neq 0$. What will happen if we impose two additional conditions related to $g^2$ to the above solutions, i.e.,
      \begin{eqnarray}
            && g^2 \biggr\{\vec{\mathcal{A}}(\vec{r})\cdot\vec{\mathcal{N}}
                              -\vec{\mathcal{N}}\cdot\vec{\mathcal{A}}(
                                    \vec{r})\biggr\}=0, \label{eq:TICond3-a} \\
                        && g^2 \, \biggr\{\Big[\varphi(\vec{r}),\
                                    \vec{\mathcal{N}}\Big]
                              +\vec{\mathcal{A}}(\vec{r})\times
                                    \vec{\mathcal{M}}
                              +\vec{\mathcal{M}}\times\vec{\mathcal{A}}(
                                    \vec{r})\biggr\}=0, \label{eq:TICond8-a}
      \end{eqnarray}
      which correspond to Eq. (\ref{eq:TICond3}) and Eq. (\ref{eq:TICond8}). We find that $\vec{\mathcal{B}}(\vec{r}, t)$ and $\vec{\mathcal{E}}(\vec{r}, t)$ reduce to some Abelian waves.

      \emph{Proof.---}From Eq. (\ref{eq:TICond3-a}), we have
      \begin{eqnarray}
                   \vec{\mathcal{A}}(\vec{r})\cdot\vec{\mathcal{N}}
                        -\vec{\mathcal{N}}\cdot\vec{\mathcal{A}}(\vec{r})
                  &=&\biggl\{\vec{\tau}\cdot\left[\varphi(\vec{r}),\
                              \vec{\mathcal{A}}(\vec{r})\right]
                        -\left[\varphi(\vec{r}),\ \vec{\mathcal{A}}(\vec{r})
                        \right]\cdot\vec{\tau}\biggr\}{\rm e}^{{\rm i}\,3(
                              \vec{k}\cdot\vec{r}-\omega\,t)} \nonumber\\
                  &=& \biggl\{\vec{\tau}\cdot\left[\bigl(\vec{\tau}\cdot\hat{k}\bigr),\
                              \vec{\tau}\right]
                        -\left[\bigl(\vec{\tau}\cdot\hat{k}\bigr),\ \vec{\tau}\right]
                              \cdot\vec{\tau}\biggr\}{\rm e}^{{\rm i}\,3(
                                    \vec{k}\cdot\vec{r}-\omega\,t)} \nonumber\\
                  &= &\biggl\{\vec{\tau}\cdot\left[\bigl(\vec{\tau}\cdot\hat{k}\bigr)
                              \vec{\tau}\right]
                        -\bigl(\vec{\tau}\cdot\vec{\tau}\bigr)\bigl(
                              \vec{\tau}\cdot\hat{k}\bigr)
                        -\bigl(\vec{\tau}\cdot\hat{k}\bigr)\bigl(
                              \vec{\tau}\cdot\vec{\tau}\bigr)
                        +\left[\vec{\tau}\bigl(\vec{\tau}\cdot\hat{k}\bigr)\right]
                              \cdot\vec{\tau}\biggr\}{\rm e}^{{\rm i}\,3(
                                    \vec{k}\cdot\vec{r}-\omega\,t)} \nonumber\\
                  &=& \biggl(\vec{\tau}\cdot\left[\bigl(\vec{\tau}\cdot\hat{k}\bigr)
                              \vec{\tau}\right]
                        +\left[\vec{\tau}\bigl(\vec{\tau}\cdot\hat{k}\bigr)\right]
                              \cdot\vec{\tau}
                        -\Bigl\{\bigl(\vec{\tau}\cdot\vec{\tau}\bigr),\ \bigl(
                              \vec{\tau}\cdot\hat{k}\bigr)\Bigr\}\biggr){\rm e}^{
                                    {\rm i}\,3(\vec{k}\cdot\vec{r}-\omega\,t)}=0,
      \end{eqnarray}
      which means
      \begin{equation}
            \left[\vec{\tau}\cdot\hat{k}, \vec{\tau}\right]=0,
      \end{equation}
      i.e.,
      \begin{equation}
            \left[\vec{\tau}\cdot\hat{k}, \vec{\tau}\right]=\left[\varphi(\vec{r}),\
                              \vec{\mathcal{A}}(\vec{r})\right]=\vec{\mathcal{N}}=0,
      \end{equation}
      then based on Eq. (\ref{eq:xi-1a}) one has $\vec{\xi}=0$, thus $\vec{\eta}=\hat{k}\times\vec{\xi}=0$. Since
      \begin{equation}
           \vec{\eta}= \big(\vec{R}_2 \times\vec{R}_3\big)S_x
            +\big(\vec{R}_3 \times\vec{R}_1\big)S_y
            +\big(\vec{R}_1 \times\vec{R}_2\big)S_z =0,
      \end{equation}
      then we attain
      \begin{equation}
            \vec{R}_2 \times\vec{R}_3 =\vec{R}_3 \times\vec{R}_1
            =\vec{R}_1 \times\vec{R}_2 =0,
      \end{equation}
      which leads to
      \begin{equation}
            \vec{R}_1 = n_x\,\vec{R}, \;\; \vec{R}_2 = n_y\,\vec{R}, \;\;\vec{R}_3 = n_z\,\vec{R},
      \end{equation}
      and
      \begin{equation}
            \vec{\tau}=\vec{R}_0 \,\openone+\vec{R}\,\left(n_x S_x+n_y S_y+n_z S_z\right)=\vec{R}_0 \,\openone+\vec{R}\,\bigl(\vec{n}\cdot\vec{S}\bigr),
      \end{equation}
      with $\vec{n}=(n_x, n_y, n_z)$. In this way, \Eq{eq:TICond8-a} holds naturally, for $\vec{\mathcal{N}}=0$ and $\vec{\mathcal{M}}\propto \vec{\tau}\times \vec{\tau}=0$.
      In summary, although it is hard to obtain the exact non-Abelian solutions of the Yang-Mills equations without any approximation, one does have an exact Abelian solution, which is
       \begin{eqnarray}
       && \vec{\mathcal{A}}(\vec{r}, t)=\vec{\tau}\,{\rm e}^{
                        {\rm i}(\vec{k}\cdot\vec{r}-\omega\,t)}, \nonumber\\
                 && \varphi(\vec{r}, t) =\bigl(\vec{\tau}\cdot\hat{k}\bigr){\rm e}^{
                        {\rm i}(\vec{k}\cdot\vec{r}-\omega\,t)},\nonumber\\
                 && \vec{\mathcal{B}}(\vec{r}, t)
                  ={\rm i}\bigl(\vec{k}\times\vec{\tau}\bigr){\rm e}^{
                              {\rm i}(\vec{k}\cdot\vec{r}-\omega\,t)}, \nonumber\\
                 && \vec{\mathcal{E}}(\vec{r}, t)
                  =-{\rm i}\,k\Big[\hat{k}\times\bigl(\hat{k}
                                    \times\vec{\tau}\bigr)\Big]{\rm e}^{
                                          {\rm i}(\vec{k}\cdot\vec{r}
                                                -\omega\,t)},\nonumber\\
          &&  \vec{\tau}=\vec{R}_0 \,\openone+\vec{R}\,\bigl(\vec{n}\cdot\vec{S}\bigr).
      \end{eqnarray}
      In this case, $\vec{\mathcal{B}}(\vec{r}, t)$ and $\vec{\mathcal{E}}(\vec{r}, t)$ become the Abelian waves.
      \end{remark}
\newpage
      \section{Solutions of Yang-Mills Equations under Zero-Coupling Approximation}

      \subsection{Zero-Coupling Approximation: The Maxwell-Type Equations}

      As mentioned above, we have called Yang-Mills equations under zero-coupling approximation as the Maxwell-type equations. For the zero-coupling approximation, we preserve the (Lorentz-covariant) structure of Maxwell's equations in \Eq{eq:MaxwFMuNu} (see {\bf Subsec.} \ref{subsec:LorenzMaxwTyp} for details), but jettison the constraint of (non-Abelian) local gauge covariance, then it leads to the Maxwell-type equations:
            \begin{subequations}
                  \begin{eqnarray}
                        && \vec{\nabla}\cdot\vec{\mathcal{E}} = 0, \label{eq:DivE-S} \\
                        && \vec{\nabla}\times\vec{\mathcal{E}} = -\dfrac{1}{c} \dfrac{\partial\,\vec{\mathcal{B}}}{\partial\,t},
                              \label{eq:CurlE-S} \\
                              && \vec{\nabla}\cdot\vec{\mathcal{B}} = 0, \label{eq:DivB-S} \\
                        && \vec{\nabla}\times\vec{\mathcal{B}} = \dfrac{1}{c} \dfrac{\partial\,\vec{\mathcal{E}}}{\partial\,t},
                              \label{eq:CurlB-S}
                  \end{eqnarray}
            \end{subequations}
            which look exactly the same as Maxwell's equations in Eqs. (\ref{eq:DivE})-(\ref{eq:CurlB}). For Maxwell's equations, the vectors $\{\vec{E}, \vec{B}\}$ are the ``usual'' vectors, while for the Maxwell-type equations, the vectors $\{\vec{\mathcal{E}},\vec{\mathcal{B}}\}$ can be some operator-type vectors, whose components (i.e., $\mathcal{E}_i,\ \mathcal{B}_j$, $i, j=x, y, z$) are generally matrices. Based on the definition (\ref{eq:nonabelian-1}), one has the ``magnetic'' field and the ``electric'' field as
            \begin{eqnarray}
                  && \vec{\mathcal{B}} = \vec{\nabla}\times\vec{\mathcal{A}} -{\rm i}\,g
                        \bigl(\vec{\mathcal{A}}\times\vec{\mathcal{A}}\bigr), \label{eq:GenerSpinB} \\
                  && \vec{\mathcal{E}}=-\dfrac{1}{c} \dfrac{\partial\,\vec{\mathcal{A}}}{\partial\,t}
                        -\vec{\nabla}\varphi-{\rm i}\,g\left[\varphi,\ \vec{\mathcal{A}}\right].\label{eq:GenerSpinE}
            \end{eqnarray}
            Commonly, $\vec{\mathcal{A}}\times\vec{\mathcal{A}}\neq 0$ and $\bigl[\varphi,\ \vec{\mathcal{A}}\bigr]\neq 0$. When $g=0$, \Eq{eq:GenerSpinB} and \Eq{eq:GenerSpinE} reduce to the forms of \Eq{eq:BA} and \Eq{eq:EVarphiA} respectively.

            After substituting Eq. (\ref{eq:GenerSpinB}) into Eq. (\ref{eq:DivB-S}), we have
            \begin{eqnarray}
                  \vec{\nabla}\cdot \left(\vec{\nabla}\times\vec{\mathcal{A}}
                        {-}{\rm i}\,g
                              \vec{\mathcal{A}}\times\vec{\mathcal{A}}\right)=0.
            \end{eqnarray}
            Due to $\vec{\nabla}\cdot\bigl(\vec{\nabla}\times\vec{\mathcal{A}}\bigr)=0$, we obtain \emph{the first condition} for $\{\vec{\mathcal{A}}(\vec{r},t),\varphi(\vec{r},t)\}$ as
            \begin{eqnarray}\label{eq:B-1d}
                  \vec{\nabla}\cdot\left(\vec{\mathcal{A}}\times\vec{\mathcal{A}}\right)=0.
            \end{eqnarray}
            Substitute Eq. (\ref{eq:GenerSpinB}) and \Eq{eq:GenerSpinE} into
            Eq. (\ref{eq:CurlE-S}), then we have
            \begin{eqnarray}
                  \vec{\nabla}\times\Biggl\{-\dfrac{1}{c}
                              \dfrac{\partial\,\vec{\mathcal{A}}}{\partial\,t}
                        -\vec{\nabla}\varphi-{\rm i}\,g [
                              \varphi,\ \vec{\mathcal{A}}]\Biggr\}
                  =-\dfrac{1}{c} \dfrac{\partial}{\partial\,t}\biggl[
                              \vec{\nabla}\times\vec{\mathcal{A}}
                        -{\rm i}\,g \vec{\mathcal{A}}\times
                              \vec{\mathcal{A}}\biggr].
            \end{eqnarray}
            By using $\vec{\nabla}\times(\vec{\nabla}\varphi)=0$, we obtain \emph{the second condition} for
            $\{\vec{\mathcal{A}}(\vec{r},t),\varphi(\vec{r},t)\}$ as
            \begin{equation}
                  \vec{\nabla}\times[\varphi,\ \vec{\mathcal{A}}]
                  =-\dfrac{1}{c} \dfrac{\partial}{\partial\,t} \bigl(
                        \vec{\mathcal{A}}\times\vec{\mathcal{A}}\bigr).
            \end{equation}
            Substitute Eq. (\ref{eq:GenerSpinE}) in to Eq. (\ref{eq:DivE-S}), we have
            \begin{equation}\label{eq:SpinE-a}
                  \vec{\nabla}\cdot \left(
                  -\dfrac{1}{c} \dfrac{\partial\,\vec{\mathcal{A}}}{\partial\,t}
                        -\vec{\nabla}\varphi-{\rm i}\,g [\varphi,\ \vec{\mathcal{A}}]\right)=0,
            \end{equation}
            then we have \emph{the third condition} for $\{\vec{\mathcal{A}}(\vec{r},t),\varphi(\vec{r},t)\}$ as
            \begin{equation}\label{eq:SpinE-b}
                  \dfrac{1}{c} \dfrac{\partial\,(\vec{\nabla}\cdot \vec{\mathcal{A}})}
                        {\partial\,t}
                  +{\nabla}^2\varphi+{\rm i}\,g \;\vec{\nabla}\cdot \left([
                              \varphi,\ \vec{\mathcal{A}}]\right)=0.
            \end{equation}
            Substitute \Eq{eq:GenerSpinB} and \Eq{eq:GenerSpinE} into
            \Eq{eq:CurlB-S}, then we have
            \begin{eqnarray}
                  \vec{\nabla}\times\left(\vec{\nabla}\times\vec{\mathcal{A}}
                        -{\rm i}\,g
                              \vec{\mathcal{A}}\times\vec{\mathcal{A}}\right)
                  =\dfrac{1}{c} \dfrac{\partial}{\partial\,t} \Biggl\{
                        -\dfrac{1}{c} \dfrac{\partial\,\vec{\mathcal{A}}}{
                              \partial\,t}
                        -\vec{\nabla}\varphi-{\rm i}\,g [
                              \varphi,\ \vec{\mathcal{A}}]\Biggr\}.
            \end{eqnarray}
            Due to $\vec{\nabla}\times\left(\vec{\nabla}\times\vec{\mathcal{A}}\right)
                  =\vec{\nabla}\left(\vec{\nabla}\cdot\vec{\mathcal{A}}\right)
                        -\nabla^2 \vec{\mathcal{A}}$, we attain \emph{the fourth condition} for
            $\{\vec{\mathcal{A}}(\vec{r},t),\varphi(\vec{r},t)\}$ as
            \begin{eqnarray}\label{eq:cond-4}
                  -\vec{\nabla}\left(\vec{\nabla}\cdot\vec{\mathcal{A}}\right)
                        +\nabla^2 \vec{\mathcal{A}}
                              +{\rm i}\,g\,\vec{\nabla}
                                    \times\left(\vec{\mathcal{A}}\times\vec{\mathcal{A}}\right)
                  =\dfrac{1}{c^2} \dfrac{\partial^2 \vec{\mathcal{A}}}{
                              \partial\,t^2}
                        +\dfrac{1}{c} \dfrac{\partial(\vec{\nabla}\varphi)}{
                              \partial\,t}
                        +\frac{\mathrm{i}\,g}{c} \dfrac{\partial}
                                    {\partial\,t} [\varphi,\ \vec{\mathcal{A}}].
            \end{eqnarray}

            In summary, we express the Maxwell-type equations (\ref{eq:DivE-S})-(\ref{eq:CurlB-S}) in terms of the vector potential $\vec{\mathcal{A}}\equiv\vec{\mathcal{A}}(\vec{r},t)$ and the scalar potential $\varphi\equiv\varphi(\vec{r},t)$ as follows.
            \begin{subequations}
                  \begin{eqnarray}
                        && \vec{\nabla}\cdot\left(\vec{\mathcal{A}}\times\vec{\mathcal{A}}\right)=0,
                              \label{eq:fourcond-1}\\
                        && \vec{\nabla}\times[\varphi,\ \vec{\mathcal{A}}] =-\dfrac{1}{c}
                              \dfrac{\partial}{\partial\,t} \bigl(
                                    \vec{\mathcal{A}}\times\vec{\mathcal{A}}\bigr),\label{eq:fourcond-2}\\
                        &&  \dfrac{1}{c} \dfrac{\partial(\vec{\nabla}\cdot\vec{\mathcal{A}})}{\partial\,t}
                              +{\nabla}^2\varphi+{\rm i}\,g \;\vec{\nabla}\cdot \left([
                                    \varphi,\ \vec{\mathcal{A}}]\right)=0, \label{eq:fourcond-3}\\
                        &&  -\vec{\nabla}\left(\vec{\nabla}\cdot\vec{\mathcal{A}}\right)
                              +\nabla^2 \vec{\mathcal{A}}
                                    +{\rm i}\,g\,\vec{\nabla}
                                          \times\left(\vec{\mathcal{A}}\times\vec{\mathcal{A}}\right)
                        = \dfrac{1}{c^2} \dfrac{\partial^2 \vec{\mathcal{A}}}{\partial\,t^2}
                              +\dfrac{1}{c} \dfrac{\partial(\vec{\nabla}\varphi)}{\partial\,t}
                              +\frac{\mathrm{i}\,g}{c} \dfrac{\partial}
                                    {\partial\,t} [\varphi,\ \vec{\mathcal{A}}].
                                    \label{eq:fourcond-4}
                  \end{eqnarray}
            \end{subequations}
            Luckily, Eq. (\ref{eq:fourcond-1})-Eq. (\ref{eq:fourcond-4}) can be exactly solved.  For the $SU(2)$ operator solutions of the Maxwell-type equations (see {\bf Subsec.} \ref{secMTE} for details), we find that they are the same as those shown in Eq. (\ref{eq:WCA-1a}) and Eq. (\ref{eq:WCA-1b}). Therefore, for the Maxwell-type equations, or the Yang-Mills equations under zero-coupling approximation, their corresponding $SU(2)$ operator solutions (i.e.,  $\vec{\mathcal{B}}(\vec{r}, t)$ and $\vec{\mathcal{E}}(\vec{r}, t)$) are also the $SU(2)$ angular-momentum waves (see {\bf Subsec.} \ref{secAMW} for details).

            \subsubsection{$SU(2)$ Operator Solutions of the Maxwell-Type Equations}\label{secMTE}

                  For $SU(2)$ operator solutions of the Maxwell-type equations, we have the following result:

                  \emph{Result 1.---}Let $\vec{S}=(S_x , S_y , S_z)$ be the angular-momentum operator, then the operator solutions of the Maxwell-type equations are given by
                  \begin{eqnarray}\label{eq:BE}
                  \vec{\mathcal{B}}(\vec{r}, t) &=& {\rm i}\left(\vec{k}\times\vec{\tau}\right){\rm e}^{{\rm i}(\vec{k}\cdot\vec{r}-\omega t)}+g \hbar \vec{\eta}\;{\rm e}^{2{\rm i}(\vec{k}\cdot\vec{r}-\omega t)}, \nonumber\\
                  \vec{\mathcal{E}}(\vec{r}, t) &=& -\hat{k}\times \vec{\mathcal{B}}(\vec{r}, t),
                  \end{eqnarray}
                  where
                  \begin{eqnarray}\label{eq:taueta}
                  \vec{\tau} &=& \vec{R}_0 \openone+\vec{R}_1 S_x
                              +\vec{R}_2 S_y +\vec{R}_3 S_z,\nonumber\\
                        \vec{\eta}&=& (\vec{R}_2 \times\vec{R}_3)S_x
                              +(\vec{R}_3 \times\vec{R}_1)S_y +(\vec{R}_1 \times\vec{R}_2)S_z,
                  \end{eqnarray}
                  and $\vec{r}$ is the position operator, $\vec{k}$ is the propagation direction, $\hat{k}=\vec{k}/|\vec{k}|$, $\omega$ is the frequency, and $t$ is time. Here, $\vec{S}=\{S_x, S_y, S_z\}$ satisfies the SU(2) commutative relations:
                  $[S_x, S_y]={\rm i}\hbar S_z, \;[S_y, S_z]={\rm i}\hbar S_x, \; [S_z, S_y]={\rm i}\hbar S_x,$
                  $\vec{S}$ commutes with the position operator $\vec{r}$, i.e., $\bigl[S_j ,\ \vec{r}\bigr]=0$, $(j=x, y, z)$, $\openone$ indicates the identity operator, $\vec{R}_\ell$'s $(\ell=0,1,2,3)$ represent some constant vectors, with $\bigl\{\vec{k}, \vec{R}_1, \vec{R}_2, \vec{R}_3\bigr\}$ locating on the same plane, and $\vec{R}_0$ is arbitrary.

                  \begin{proof}
                        \emph{Result 1} will be proved in 5 steps.

                        \emph{Step 1.---}By setting the vector potential and the scalar potential be the following forms
                        \begin{eqnarray}\label{eq:WaveAPhi}
                              &&\vec{\mathcal{A}}(\vec{r}, t)=\vec{\mathcal{A}}(\vec{r}) \;{\rm e}^{
                                    -{\rm i} \omega t}, \nonumber\\
                              && \varphi(\vec{r}, t)=\varphi(\vec{r}) \;{\rm e}^{-{\rm i} \omega t},
                        \end{eqnarray}
                        based on Eqs. (\ref{eq:fourcond-1})-(\ref{eq:fourcond-4}), we establish the conditions for
                        $\{\vec{\mathcal{A}}(\vec{r}), \varphi(\vec{r})\}$. After substituting
                        Eq. (\ref{eq:WaveAPhi}) into Eq. (\ref{eq:fourcond-1}), we have
                        \begin{eqnarray}\label{eq:D-1a}
                              &&  \vec{\nabla}\cdot\left\{\vec{\mathcal{A}}(\vec{r})\times\vec{\mathcal{A}}(\vec{r})\right\}=0.
                        \end{eqnarray}
                        Substitute Eq. (\ref{eq:WaveAPhi}) into Eq. (\ref{eq:fourcond-2}), we obtain
                        \begin{eqnarray}\label{eq:D-1b}
                              &&  \vec{\nabla}\times[\varphi(\vec{r}),\ \vec{\mathcal{A}}(\vec{r})]
                              =2 {\rm i} \dfrac{\omega}{c}\left\{
                                    \vec{\mathcal{A}}(\vec{r})\times\vec{\mathcal{A}}(\vec{r})\right\}.
                        \end{eqnarray}
                        Substitute Eq. (\ref{eq:WaveAPhi}) into Eq. (\ref{eq:fourcond-3}), we have
                        \begin{eqnarray}
                              & -{\rm i}\dfrac{\omega}{c} {\rm e}^{-{\rm i} \omega t} \left[
                                          \vec{\nabla}\cdot \vec{\mathcal{A}}(\vec{r})\right]
                                    +{\rm e}^{-{\rm i} \omega t} {\nabla}^2\varphi(\vec{r})
                                    +{\rm i}\,g {
                                          \rm e}^{-{\rm i}\,2\,\omega t} \vec{\nabla}\cdot\left(
                                                [\varphi(\vec{r}),\ \vec{\mathcal{A}}(\vec{r})]\right)=0,
                        \end{eqnarray}
                        Because $\omega\neq 0$, and the above condition is valid for any time $t$, then we must have
                        \begin{eqnarray}
                              &&  {\rm i} k\left[\vec{\nabla}\cdot\vec{\mathcal{A}}(\vec{r})\right]
                                    -{\nabla}^2\varphi(\vec{r})=0, \nonumber\\
                              && \vec{\nabla}\cdot\left([\varphi(\vec{r}),\ \vec{\mathcal{A}}(\vec{r})]
                                    \right)=0,
                        \end{eqnarray}
                        where we have denoted $k=\omega/c$. Substitute Eq. (\ref{eq:WaveAPhi}) into Eq. (\ref{eq:fourcond-4}), then one has
                        \begin{eqnarray}\label{eq:D-1d}
                              &&  {\rm e}^{-{\rm i} \omega t}\;\left\{-\vec{\nabla}\left(\vec{\nabla}
                                                \cdot\vec{\mathcal{A}}(\vec{r})\right)
                                          +\nabla^2 \vec{\mathcal{A}}(\vec{r})\right\}
                                    +{\rm e}^{-2 {\rm i} \omega t}\;\left\{{\rm i}\,g
                                          \,\vec{\nabla}\times\left(\vec{\mathcal{A}}(\vec{r})\times
                                                \vec{\mathcal{A}}(\vec{r})\right)\right\}\nonumber\\
                              &&= -\dfrac{\omega^2}{c^2} {\rm e}^{-{\rm i}\omega t}\;
                                                \vec{\mathcal{A}}(\vec{r})
                                          -{\rm i}\dfrac{\omega}{c} {\rm e}^{-{\rm i} \omega t} \vec{
                                                \nabla} \varphi(\vec{r})
                                    -2\,{\rm i}\dfrac{\omega}{c} {\rm e}^{-2{\rm i} \omega t}\;\left\{{\rm i}\,g[\varphi(\vec{r}),\
                                          \vec{\mathcal{A}}(\vec{r})]\right\},
                        \end{eqnarray}
                        i.e.,
                        \begin{eqnarray}\label{eq:D-1g}
                              -\vec{\nabla}\left(\vec{\nabla}\cdot\vec{\mathcal{A}}(\vec{r})\right)
                                    +\nabla^2 \vec{\mathcal{A}}(\vec{r})+k^2\; \vec{\mathcal{A}}(\vec{r})
                                    +{\rm i}\,k\bigl[\vec{\nabla}\varphi(\vec{r})\bigr]
                              =-{\rm e}^{-{\rm i}\omega t} {\rm i}\,g \biggl\{
                                    2\,{\rm i}\,k\bigl[\varphi(\vec{r}),\ \vec{\mathcal{A}}(\vec{r})
                                          \bigr]
                                    +\vec{\nabla}\times\left[\vec{\mathcal{A}}(\vec{r})\times
                                          \vec{\mathcal{A}}(\vec{r})\right]\biggr\},
                        \end{eqnarray}
                        Because $\omega\neq 0$, and the above condition is valid for any time $t$, thus we must have
                        \begin{eqnarray}\label{eq:D-1h}
                              &&  -\vec{\nabla}\left(\vec{\nabla}\cdot\vec{\mathcal{A}}(\vec{r})\right)
                                    +\nabla^2 \vec{\mathcal{A}}(\vec{r})+k^2\; \vec{\mathcal{A}}(\vec{r})
                                    +{\rm i}\,k\;\vec{\nabla}\varphi(\vec{r})=0, \nonumber\\
                              &&  2\,{\rm i}\,k\bigl[\varphi(\vec{r}),\ \vec{\mathcal{A}}(\vec{r})\bigr]
                                    +\vec{\nabla}\times\left[\vec{\mathcal{A}}(\vec{r})\times
                                          \vec{\mathcal{A}}(\vec{r})\right]=0.
                        \end{eqnarray}
                        In summary, we have the following six conditions for
                        $\{\vec{\mathcal{A}}(\vec{r}), \varphi(\vec{r})\}$:
                        \begin{subequations}
                              \begin{eqnarray}
                                    && \vec{\nabla}\cdot\left[\vec{\mathcal{A}}(\vec{r})\times
                                          \vec{\mathcal{A}}(\vec{r})\right]=0, \label{eq:sixcond-S1}\\
                                    &&  \vec{\nabla}\times\left[\varphi(\vec{r}),\
                                                \vec{\mathcal{A}}(\vec{r})\right]
                                          =2{\rm i} k\;\left[\vec{\mathcal{A}}(\vec{r})\times
                                                \vec{\mathcal{A}}(\vec{r})\right],\label{eq:sixcond-S2}\\
                                    &&  {\rm i} k\left[\vec{\nabla}\cdot \vec{\mathcal{A}}(\vec{r})\right]
                                          -{\nabla}^2\varphi(\vec{r})=0, \label{eq:sixcond-S3}\\
                                    &&\vec{\nabla}\cdot\left(\bigl[\varphi(\vec{r}),\
                                          \vec{\mathcal{A}}(\vec{r})\bigr]\right)=0,
                                                \label{eq:sixcond-S4}\\
                                    && -\vec{\nabla}\left(\vec{\nabla}\cdot\vec{\mathcal{A}}(\vec{r})
                                                \right)
                                          +\nabla^2 \vec{\mathcal{A}}(\vec{r})+k^2\;
                                                \vec{\mathcal{A}}(\vec{r})
                                          +{\rm i}\,k\;\vec{\nabla}\varphi(\vec{r})=0,
                                                \label{eq:sixcond-S5}\\
                                    && 2\,{\rm i}\,k\bigl[\varphi(\vec{r}),\ \vec{\mathcal{A}}(\vec{r})
                                                \bigr]
                                          +\vec{\nabla}\times\left[\vec{\mathcal{A}}(\vec{r})\times
                                                \vec{\mathcal{A}}(\vec{r})\right]=0.\label{eq:sixcond-S6}
                              \end{eqnarray}
                        \end{subequations}

                        \emph{Step 2.---}To determine the vector potential $\vec{\mathcal{A}}(\vec{r})$, we define its form as
                        \begin{eqnarray}
                              \vec{\mathcal{A}}(\vec{r}) =\vec{T}_{0} \openone +\vec{T}_{1} S_x
                                    +\vec{T}_{2} S_y +\vec{T}_{3} S_z,
                        \end{eqnarray}
                        with
                        \begin{eqnarray}
                              \vec{T}_{0} =(T_{0x}, T_{0y}, T_{0z}),\
                              \vec{T}_{1} =(T_{1x}, T_{1y}, T_{1z}),\
                              \vec{T}_{2} =(T_{2x}, T_{2y}, T_{2z}),\
                              \vec{T}_{3} =(T_{3x}, T_{3y}, T_{3z}),
                        \end{eqnarray}
                        and $S_x, S_y, S_z$ are three components of the angular-momentum operator
                        $\vec{S}=(S_x, S_y, S_z)$, which satisfies
                        \begin{eqnarray}
                              && \vec{S}\times\vec{S}={\rm i}\hbar\vec{S}.
                        \end{eqnarray}
                        Similarly, we define the scalar potential $\varphi(\vec{r})$ as
                        \begin{eqnarray}\label{eq:M-2}
                              && \varphi(\vec{r}) = \varphi_0 \openone + \varphi_1 S_x + \varphi_2 S_y
                                    + \varphi_3 S_z.
                        \end{eqnarray}
                        Here $\vec{T}_{\mu}\equiv\vec{T}_{\mu}(\vec{r})$, and $\varphi_{\mu}\equiv\varphi_{\mu}(\vec{r})$. In this work, for simplicity we have assumed that $\bigl[S_j ,\ \vec{r}\bigr]=0$, which means
                        $\bigl[S_j ,\ \vec{T}_{\ell}\bigr]=\bigl[S_j ,\ \varphi_\ell\bigr]=0$, with $j\in\{x,y,z\}$ and $\ell\in\{0,1,2,3\}$. For convenience, denote
                        \begin{equation}
                              \vec{\mathcal{M}}\equiv\vec{\mathcal{M}}(\vec{r})
                              =\vec{\mathcal{A}}(\vec{r})\times\vec{\mathcal{A}}(\vec{r}),
                        \end{equation}
                        which yields
                        \begin{eqnarray}
                              \vec{\mathcal{M}}={\rm i}\hbar\left[
                                    (\vec{T}_2 \times\vec{T}_3)S_x
                                    +(\vec{T}_3 \times\vec{T}_1)S_y
                                    +(\vec{T}_1 \times\vec{T}_2)S_z\right].
                        \end{eqnarray}
                        From \Eq{eq:sixcond-S6} we obtain
                        \begin{equation}
                              2{\rm i}k\,\vec{\nabla}\times[\varphi(\vec{r}),\ \vec{\mathcal{A}}(\vec{r})]
                              +\vec{\nabla}\times\biggl[\nabla
                                    \times\left(\vec{\mathcal{A}}(\vec{r})\times\vec{\mathcal{A}}(\vec{r})\right)\biggr]=0,
                        \end{equation}
                        i.e.
                        \begin{equation}
                              \dfrac{1}{2{\rm i}k} \vec{\nabla}\times\biggl[\nabla
                                    \times\left(\vec{\mathcal{A}}(\vec{r})\times\vec{\mathcal{A}}(\vec{r})\right)\biggr]
                              =-\vec{\nabla}\times[\varphi(\vec{r}),\
                                    \vec{\mathcal{A}}(\vec{r})],
                        \end{equation}
                        then combine Eq. (\ref{eq:sixcond-S2}) we immediately obtain
                        \begin{equation}
                              \dfrac{1}{2{\rm i}k} \vec{\nabla}\times\bigl(
                                    \nabla\times\vec{\mathcal{M}}\bigr)
                              =-2 {\rm i} k\,\vec{\mathcal{M}},
                        \end{equation}
                        viz.
                        \begin{equation}
                              \vec{\nabla}\times\Bigl(\vec{\nabla}\times\vec{\mathcal{M}}\Bigr)
                              =4\,k^2 \vec{\mathcal{M}},
                        \end{equation}
                        i.e.
                        \begin{equation}
                              \vec{\nabla}\times\Bigl(\vec{\nabla}\cdot\vec{\mathcal{M}}\Bigr)
                                    -\nabla^2 \vec{\mathcal{M}}=4\,k^2 \vec{\mathcal{M}}.
                        \end{equation}
                        Notice that \Eq{eq:sixcond-S1} indicates
                        \begin{equation}\label{eq:DivM0}
                              \vec{\nabla}\cdot\vec{\mathcal{M}}=0,
                        \end{equation}
                        then we attain the wave equation about $\vec{\mathcal{M}}$ as
                        \begin{equation}\label{eq:WaveM}
                              \nabla^2 \vec{\mathcal{M}}+4\,k^2 \vec{\mathcal{M}}=0,
                        \end{equation}
                        whose solution reads
                        \begin{equation}\label{eq:GeneM}
                              \vec{\mathcal{M}}=\vec{M}_1\,{\rm e}^{{\rm i}\,2\vec{k}\cdot\vec{r}}
                                    +\vec{M}_2\,{\rm e}^{-{\rm i}\,2\vec{k}\cdot\vec{r}},
                        \end{equation}
                        where $\vec{M}_1$ and $\vec{M}_2$ are two vectorial operators and independent of $\vec{r}$.

                        For simplicity, let
                        $\vec{M}_2 =0$, i.e.,
                        \begin{equation}
                              \vec{\mathcal{M}}=\vec{M}_1\,{\rm e}^{{\rm i}\,2\vec{k}\cdot\vec{r}},
                        \end{equation}
                        which implies that $\vec{T}_\ell =\vec{R}_\ell\,{\rm e}^{{\rm i}\vec{k}\cdot\vec{r}}$, $(\ell=0,1,2,3)$. We then obtain
                        \begin{equation}\label{eq:solu-A}
                              \vec{\mathcal{A}}(\vec{r})=\vec{\tau}\;{\rm e}^{
                                          {\rm i}\vec{k}\cdot\vec{r}},
                        \end{equation}
                        and
                        \begin{equation}\label{eq:M1}
                              \vec{M}_1 ={\rm i}\hbar \; \vec{\eta},
                        \end{equation}
                        with
                        \begin{eqnarray}\label{eq:solu-A-aa}
                        \vec{\tau} &=& \vec{R}_0 \openone+\vec{R}_1 S_x
                                    +\vec{R}_2 S_y +\vec{R}_3 S_z,\nonumber\\
                              \vec{\eta}&=& (\vec{R}_2 \times\vec{R}_3)S_x
                                    +(\vec{R}_3 \times\vec{R}_1)S_y +(\vec{R}_1 \times\vec{R}_2)S_z.
                        \end{eqnarray}
                        According to \Eq{eq:DivM0}, we arrive at
                        \begin{equation}
                              \vec{k}\cdot\vec{\eta} =0,
                        \end{equation}
                        namely
                        \begin{equation}
                              \vec{k}\cdot\bigl(\vec{R}_1 \times\vec{R}_2\bigr)
                              =\vec{k}\cdot\bigl(\vec{R}_2 \times\vec{R}_3\bigr)
                              =\vec{k}\cdot\bigl(\vec{R}_3 \times\vec{R}_1\bigr)=0,
                        \end{equation}
                        which indicates that four vectors $\{\vec{R}_1, \vec{R}_2, \vec{R}_3, \vec{k}\}$ locate on the same plane, and $\vec{R}_0$ is arbitrary.

                        \emph{Step 3.---}To determine the scalar potential $\varphi(\vec{r})$, similarly let
                        \begin{eqnarray}
                              \varphi(\vec{r}) =\left(\tilde{\varphi}_0 \openone
                                    +\tilde{\varphi}_1 S_x +\tilde{\varphi}_2 S_y
                                    +\tilde{\varphi}_3 S_z\right){\rm e}^{
                                          {\rm i}\vec{k}\cdot\vec{r}}.
                        \end{eqnarray}
                        From \Eq{eq:sixcond-S3}, we have
                        \begin{eqnarray}
                              && {\rm i} k\vec{\nabla}\cdot\left(\vec{\tau}\;{\rm e}^{
                                          {\rm i}\vec{k}\cdot\vec{r}}\right)
                                    -{\nabla}^2 \left[\left(\tilde{\varphi}_0 \openone
                                          +\tilde{\varphi}_1 S_x +\tilde{\varphi}_2 S_y
                                          +\tilde{\varphi}_3 S_z\right){\rm e}^{{\rm i}\vec{k}\cdot
                                                \vec{r}}\right]=0,
                        \end{eqnarray}
                        then we have
                        \begin{eqnarray}
                              && \hat{k}\cdot\vec{\tau}-\left(\tilde{\varphi}_0 \openone
                                    +\tilde{\varphi}_1 S_x +\tilde{\varphi}_2 S_y +\tilde{\varphi}_3 S_z
                                    \right)=0,
                        \end{eqnarray}
                        with $\hat{k}=\vec{k}/k$, which leads to
                        \begin{eqnarray}\label{eq:phiphi}
                              \tilde{\varphi}_0=\hat{k}\cdot\vec{R}_0,\quad
                              \tilde{\varphi}_1=\hat{k}\cdot\vec{R}_1,\quad
                              \tilde{\varphi}_2=\hat{k}\cdot\vec{R}_2,\quad
                              \tilde{\varphi}_3=\hat{k}\cdot\vec{R}_3.
                        \end{eqnarray}
                        This means that the scalar potential is determined by
                        \begin{eqnarray}\label{eq:PhiR}
                              \varphi(\vec{r})=\left(\hat{k}\cdot\vec{\tau}\right){\rm e}^{
                                    {\rm i}\vec{k}\cdot\vec{r}}.
                        \end{eqnarray}

                        \emph{Step 4.---}In step 2 and step 3, we have determined the forms of $\vec{\mathcal{A}}(\vec{r})$ and $\varphi(\vec{r})$, which are given in Eq. (\ref{eq:solu-A}) and Eq. (\ref{eq:PhiR}), respectively. Here we check whether they hold for all the six conditions Eqs. (\ref{eq:sixcond-S1})-(\ref{eq:sixcond-S6}). It is easy to verify that Eqs. (\ref{eq:sixcond-S1})-(\ref{eq:sixcond-S3}) are satisfied. We need to further check the rest three conditions (\ref{eq:sixcond-S4})-(\ref{eq:sixcond-S6}). Let us study Eq. (\ref{eq:sixcond-S6}) first. Because
                        \begin{eqnarray}
                              \bigl[\varphi(\vec{r}),\ \vec{\mathcal{A}}(\vec{r})\bigr]
                              &=& {\rm i}\hbar\;{\rm e}^{{\rm i}\,2\vec{k}\cdot\vec{r}} \biggl[
                                    \left(\tilde{\varphi}_2 \vec{R}_3
                                          -\tilde{\varphi}_3 \vec{R}_2\right)S_x
                                    +\left(\tilde{\varphi}_3 \vec{R}_1
                                          -\tilde{\varphi}_1 \vec{R}_3\right)S_y
                                    +\left(\tilde{\varphi}_1 \vec{R}_2
                                          -\tilde{\varphi}_2 \vec{R}_1\right)S_z\biggr],\\
                              \vec{\nabla}\times\vec{\mathcal{M}}&=&{\rm i}\,2\,{\rm e}^{
                                    {\rm i}\,2\vec{k}\cdot\vec{r}} \left(\vec{k}\times\vec{M}_1\right),
                        \end{eqnarray}
                        then \Eq{eq:sixcond-S6} can be recast as
                        \begin{eqnarray}\label{eq:Condit6Reduce}
                              {\rm i}\hbar\biggl[\left(\tilde{\varphi}_2 \vec{R}_3
                                          -\tilde{\varphi}_3 \vec{R}_2\right)S_x
                                    +\left(\tilde{\varphi}_3 \vec{R}_1
                                          -\tilde{\varphi}_1 \vec{R}_3\right)S_y
                                    +\left(\tilde{\varphi}_1 \vec{R}_2
                                          -\tilde{\varphi}_2 \vec{R}_1\right)S_z\biggr]
                                    +\hat{k}\times\vec{M}_1 =0.
                        \end{eqnarray}
                        By substituting Eq. (\ref{eq:M1}) and Eq. (\ref{eq:phiphi}) into Eq. (\ref{eq:Condit6Reduce}), after simplification, one finds that it is satisfied, thus Eq. (\ref{eq:sixcond-S6}) holds. Based on Eq.
                        (\ref{eq:sixcond-S6}), one can derive
                        \begin{eqnarray}
                              \vec{\nabla}\cdot \left[2{\rm i}k[\varphi(\vec{r}),\
                                          \vec{\mathcal{A}}(\vec{r})]
                                    +\nabla\times\left(\vec{\mathcal{A}}(\vec{r})\times
                                          \vec{\mathcal{A}}(\vec{r})\right)\right]
                              &=& 2{\rm i}k\; \vec{\nabla}\cdot \left([\varphi(\vec{r}),\
                                          \vec{\mathcal{A}}(\vec{r})]\right)
                                    +\vec{\nabla}\cdot\left(\nabla
                                          \times\vec{\mathcal{M}}(\vec{r})\right) \notag\\
                              &=& 2{\rm i}k\; \vec{\nabla}\cdot \left([\varphi(\vec{r}),\
                                    \vec{\mathcal{A}}(\vec{r})]\right)=0,
                        \end{eqnarray}
                        thus Eq. (\ref{eq:sixcond-S4}) is valid. For Eq. (\ref{eq:sixcond-S5}), because
                        \begin{eqnarray}\label{eq:DivAPhi}
                              \vec{\nabla}\cdot \vec{\mathcal{A}}(\vec{r}) &=& \vec{\nabla}\cdot\left(
                                    \vec{\tau}\;{\rm e}^{{\rm i}\vec{k}\cdot\vec{r}}\right)
                              ={\rm i}\left(\vec{k}\cdot\vec{\tau}\right){\rm e}^{
                                    {\rm i}\vec{k}\cdot\vec{r}}
                              ={\rm i}\,k\,\varphi(\vec{r}),
                        \end{eqnarray}
                        then Eq. (\ref{eq:sixcond-S5}) becomes
                        \begin{eqnarray}\label{eq:AA-2}
                        && \nabla^2 \vec{\mathcal{A}}(\vec{r})+k^2\; \vec{\mathcal{A}}(\vec{r})=0.
                        \end{eqnarray}
                        By substituting Eq. (\ref{eq:solu-A}) into Eq. (\ref{eq:AA-2}), one finds that it is satisfied, thus Eq. (\ref{eq:sixcond-S5}) holds.

                        In summary, from Eq. (\ref{eq:WaveAPhi}) we have the vector potential and the scalar potential as
                        \begin{eqnarray}\label{eq:WaveAPhi-1}
                              &&\vec{\mathcal{A}}(\vec{r}, t)=\vec{\tau}\;
                                    {\rm e}^{{\rm i}(\vec{k}\cdot\vec{r}-\omega t)},\nonumber\\
                              && \varphi(\vec{r}, t)=\left(\hat{k}\cdot\vec{\tau}\right)\;{\rm e}^{{\rm i}(\vec{k}\cdot\vec{r}-\omega t)}.
                        \end{eqnarray}

                        \emph{Step 5.---}We calculate the corresponding ``magnetic'' field $\vec{\mathcal{B}}$ and ``electric'' filed $\vec{\mathcal{E}}$. Based on Eq. (\ref{eq:WaveAPhi-1}), we can have
                        \begin{eqnarray}\label{eq:B-1-a}
                              \vec{\mathcal{B}}(\vec{r},t) &=& \vec{\nabla}\times
                                    \vec{\mathcal{A}}(\vec{r}, t)-{\rm i}\,g
                                          \left[\vec{\mathcal{A}}(\vec{r}, t)\times
                                                \vec{\mathcal{A}}(\vec{r}, t)\right]
                              =\vec{\nabla}\times\left[\vec{\tau}\;{\rm e}^{
                                          {\rm i}(\vec{k}\cdot\vec{r}-\omega t)}\right]
                                    -{\rm i}\,g \vec{\mathcal{M}}\,
                                          {\rm e}^{-2{\rm i}\omega t} \notag\\
                              &=& {\rm i}\left(\vec{k}\times\vec{\tau}\right){\rm e}^{
                                          {\rm i}(\vec{k}\cdot\vec{r}-\omega t)}
                                    +g \hbar\vec{\eta}\;{\rm e}^{2{\rm i}(\vec{k}\cdot\vec{r}-\omega t)}.
                        \end{eqnarray}
                        Due to Eq. (\ref{eq:sixcond-S6}), we have
                        \begin{equation}
                              2{\rm i}k[\varphi(\vec{r}),\ \vec{\mathcal{A}}(\vec{r})]
                              =-\vec{\nabla}\times\vec{\mathcal{M}}=-{\rm i}\,2{\rm e}^{
                                    {\rm i}\,2\vec{k}\cdot\vec{r}} \left(\vec{k}\times\vec{M}_1\right),
                        \end{equation}
                        thus
                        \begin{equation}
                              [\varphi(\vec{r}),\ \vec{\mathcal{A}}(\vec{r})]
                              =-\left(\hat{k}\times\vec{M}_1\right){\rm e}^{
                                    {\rm i}\,2\vec{k}\cdot\vec{r}}.
                        \end{equation}
                        From Eq. (\ref{eq:GenerSpinE}), we obtain the ``electric'' field as
                        \begin{eqnarray}\label{eq:SpinE-01}
                              \vec{\mathcal{E}}(\vec{r}, t)
                              &=&{\rm i}\,k\,\vec{\tau}\;{\rm e}^{
                                                {\rm i}(\vec{k}\cdot\vec{r}-\omega t)}
                                                -{\rm i} \vec{k}\left(\hat{k}\cdot\vec{\tau}\right)\;
                                          {\rm e}^{{\rm i}(\vec{k}\cdot\vec{r}-\omega t)}
                                    -g \hbar\left(\hat{k}\times \vec{\eta}\right){\rm e}^{
                                          {\rm i}\,2(\vec{k}\cdot\vec{r}-\omega t)} \notag\\
                              &=& -\hat{k}\times\Biggl[{\rm i}\,\left(\vec{k}\times \vec{\tau}\right)
                                    {\rm e}^{{\rm i}(\vec{k}\cdot\vec{r}-\omega t)}\Biggr]-g \hbar\left(\hat{k}\times \vec{\eta}\right){\rm e}^{{\rm i}\,2(\vec{k}\cdot\vec{r}-\omega t)}
                              =-\hat{k}\times \vec{\mathcal{B}}(\vec{r}, t).
                        \end{eqnarray}
                        This ends the proof.
                  \end{proof}

            \subsubsection{Lorentz Covariance of the Maxwell-Type Equations}\label{subsec:LorenzMaxwTyp}

                As mentioned in previous section, the Maxwell-type equations can be written in a concise form as
                  \begin{subequations}
                        \begin{eqnarray}
                              &&  \partial_\mu \mathcal{F}^{\mu\nu} = 0,
                                    \label{eq:MaxwTyp-1} \\
                              &&  \partial_\mu \mathcal{F}_{\nu\gamma}
                                    +\partial_\nu \mathcal{F}_{\gamma\mu}
                                    +\partial_\gamma \mathcal{F}_{\mu\nu} =0,
                                          \label{eq:MaxwTyp-2}
                        \end{eqnarray}
                  \end{subequations}
                which will be used to prove the Lorentz covariance. Assume there is another inertial reference system $(x',y',z',t')$ moves uniformly with the velocity $v$ along the $\hat{z}$-axis with respect to the former one $(x,y,z,t)$. Then we have the Lorentz transformation:
                  \begin{equation}\label{eq:XPrimX}
                        x'^\mu=c_{\mu\nu} x^\nu,
                  \end{equation}
                  with $c_{\mu\nu}$ signifying the elements of the Lorentz transformation matrix
                  \begin{equation}
                        (c_{\mu\nu})=C\equiv\begin{bmatrix}
                              \gamma & 0 & 0 & -\gamma\dfrac{v}{c} \\
                              0 & 1 & 0 & 0 \\
                              0 & 0 & 1 & 0 \\
                              -\gamma\dfrac{v}{c} & 0 & 0 & \gamma
                        \end{bmatrix}.
                  \end{equation}
                  According to the invariance of relativistic interval
                        $x'^\mu x'_\mu =c_{\mu\nu} x^\nu c^{\mu\alpha} x_\alpha
                        =c_{\mu\nu} c^{\mu\alpha} x^\nu x_\alpha =x^\nu x_\nu$, which leads to $c_{\mu\nu} c^{\mu\alpha} =\delta^\alpha_\nu$.

                 Note \Eq{eq:XPrimX} can be generalized to differential operations under the Lorentz transformation $C$. Because
                  \begin{equation}
                        x'^\mu=c_{\mu\nu} x^\nu \Rightarrow
                        \dfrac{\partial\,x'^\mu}{\partial\,x^\nu} =c_{\mu\nu}.
                  \end{equation}
                  After that,
                  \begin{equation}
                        \partial'_\mu=\dfrac{\partial}{\partial\,x'^\mu}
                        =\dfrac{\partial\,x^\nu}{\partial\,x'^\mu}
                              \dfrac{\partial}{\partial\,x^\nu}
                        =\dfrac{1}{\frac{\partial\,x'^\mu}{\partial\,x^\nu}}
                              \dfrac{\partial}{\partial\,x^\nu}
                        =\dfrac{1}{c_{\mu\nu}} \dfrac{\partial}{\partial\,x^\nu}
                        =\dfrac{\delta^\mu_\mu}{c_{\mu\nu}} \partial_\nu
                        =c^{\mu\nu} \partial_\nu.
                  \end{equation}
                  Similarly, we obtain
                  \begin{equation}
                        \mathbb{A}_\mu'=c^{\mu\nu} \mathbb{A}_\nu.
                  \end{equation}

                  After that, we arrive at
                  \begin{eqnarray}
                               \mathcal{F}'_{\mu\nu} &=&\partial'_\mu \mathbb{A}'_\nu
                                          -\partial'_\nu \mathbb{A}'_\mu
                                    +{\rm i}\,g[\mathbb{A}'_\mu,\ \mathbb{A}'_\nu]
                              =c^{\mu\alpha} c^{\nu\beta} F_{\alpha\beta}
                                    +{\rm i}\,g[
                                          c^{\mu\alpha} \mathbb{A}_\alpha,\
                                          c^{\nu\beta} \mathbb{A}_\beta]\nonumber\\
                              &=& c^{\mu\alpha} c^{\nu\beta} \biggl\{F_{\alpha\beta}
                                    +{\rm i}\,g[\mathbb{A}_\alpha,\
                                          \mathbb{A}_\beta]\biggr\}
                              =c^{\mu\alpha} c^{\nu\beta} \mathcal{F}_{\alpha\beta}.
                  \end{eqnarray}
                  Likewise, ${\mathcal{F}'}^{\mu\nu} =c_{\mu\alpha}\,c_{\nu\beta} \mathcal{F}^{\alpha\beta}$. Further,
                  \begin{equation}
                        \partial'_\mu {\mathcal{F}'}^{\mu\nu}
                        =\bigl(c^{\mu\alpha} \partial_\alpha\bigr)\bigl(
                              c_{\mu\beta}\,c_{\nu\xi} \mathcal{F}^{\beta\xi}\bigr)
                        =\bigl(c^{\mu\alpha} c_{\mu\beta}\,c_{\nu\xi}\bigr)
                              \partial_\alpha \mathcal{F}^{\beta\xi}
                        =\bigl(\delta^\alpha_\beta c_{\nu\xi}\bigr)
                              \partial_\alpha \mathcal{F}^{\beta\xi}
                        =c_{\nu\xi} \partial_\alpha \mathcal{F}^{\alpha\xi}=0.
                  \end{equation}
                  \begin{eqnarray}
                               \partial'_\mu \mathcal{F}'_{\nu\alpha}
                                    +\partial'_\nu \mathcal{F}'_{\alpha\mu}
                                    +\partial'_\alpha \mathcal{F}'_{\mu\nu}
                              &=& \bigl(c^{\mu\beta} \partial_\beta\bigr)\bigl(
                                          c^{\nu\zeta} c^{\alpha\xi} \mathcal{F}_{\zeta\xi}
                                          \bigr)
                                    +\bigl(c^{\nu\kappa} \partial_\kappa\bigr)
                                          \bigl(c^{\alpha\rho} c^{\mu\phi}
                                                \mathcal{F}_{\rho\phi}\bigr)
                                    +\bigl(c^{\alpha\psi} \partial_\psi\bigr)
                                          \bigl(c^{\mu\eta} c^{\nu\lambda}
                                                \mathcal{F}_{\eta\lambda}\bigr) \nonumber\\
                              &=& c^{\mu\beta} c^{\nu\zeta} c^{\alpha\xi}\,
                                          \partial_\beta \mathcal{F}_{\zeta\xi}
                                    +c^{\nu\kappa} c^{\alpha\rho} c^{\mu\phi}\,
                                          \partial_\kappa \mathcal{F}_{\rho\phi}
                                    +c^{\alpha\psi} c^{\mu\eta} c^{\nu\lambda}\,
                                          \partial_\psi \mathcal{F}_{\eta\lambda} \nonumber\\
                              &=& c^{\mu\beta} c^{\nu\zeta} c^{\alpha\xi}\,
                                          \partial_\beta \mathcal{F}_{\zeta\xi}
                                    +c^{\nu\zeta} c^{\alpha\xi} c^{\mu\beta}\,
                                          \partial_\zeta \mathcal{F}_{\xi\beta}
                                    +c^{\alpha\xi} c^{\mu\beta} c^{\nu\zeta}\,
                                          \partial_\xi \mathcal{F}_{\beta\zeta} \nonumber\\
                              &= & c^{\mu\beta} c^{\nu\zeta} c^{\alpha\xi}\bigl(
                                    \partial_\beta \mathcal{F}_{\zeta\xi}
                                    +\partial_\zeta \mathcal{F}_{\xi\beta}
                                    +\partial_\xi \mathcal{F}_{\beta\zeta}\bigr)=0.
                  \end{eqnarray}
                  Thus under the Lorentz transformation $C$, the forms of \Eq{eq:MaxwTyp-1} and \Eq{eq:MaxwTyp-2} are invariant.

\subsection{The Angular-Momentum Waves}\label{secAMW}
      Similar to the derivations of Eq. (\ref{eq:WavE-1}) and Eq. (\ref{eq:WavB-1}), based on the Maxwell-type equations one can \emph{strictly} derive the wave equations for the ``electric'' field $\vec{\mathcal{E}}(\vec{r}, t)$ and the ``magnetic'' field $\vec{\mathcal{B}}(\vec{r}, t)$. They are
      \begin{equation}\label{eq:WavE-2}
            \nabla^2 \vec{\mathcal{E}}-\dfrac{1}{c^2} \dfrac{\partial\,\vec{\mathcal{E}}}{\partial\,t^2} =0,
      \end{equation}
      \begin{equation}\label{eq:WavB-2}
            \nabla^2 \vec{\mathcal{B}}-\dfrac{1}{c^2} \dfrac{\partial\,\vec{\mathcal{B}}}{\partial\,t^2} =0.
      \end{equation}
      For the ``magnetic'' field in \Eq{eq:B-1-a} and the ``electric'' field
      \Eq{eq:SpinE-01}, i.e.,
      \begin{eqnarray}
          &&  \vec{\mathcal{B}}(\vec{r},t)={\rm i}\left(\vec{k}\times\vec{\tau}
                        \right){\rm e}^{{\rm i}(\vec{k}\cdot\vec{r}-\omega t)}
                  +g \hbar \vec{\eta}\;{\rm e}^{2{\rm i}(\vec{k}\cdot\vec{r}-\omega t)}, \\
          && \vec{\mathcal{E}}(\vec{r}, t)=-\hat{k}\times
                        \vec{\mathcal{B}}(\vec{r}, t),
            \end{eqnarray}
       one easily proves that
      \begin{eqnarray}
            \vec{k}\cdot \vec{\mathcal{B}}(\vec{r}, t) = 0,\quad
            \vec{k}\cdot \vec{\mathcal{E}}(\vec{r}, t) = 0,\quad
            \vec{\mathcal{B}}(\vec{r}, t) \cdot \vec{\mathcal{E}}(\vec{r}, t)=0,
      \end{eqnarray}
      that is the ``magnetic'' field $\vec{\mathcal{B}}(\vec{r}, t)$, the ``electric'' field $\vec{\mathcal{E}}(\vec{r}, t)$ and the propagation direction $\vec{k}$ are mutually perpendicular. The amplitudes of $\vec{\mathcal{B}}(\vec{r}, t)$ and
      $\vec{\mathcal{E}}(\vec{r}, t)$ contain the components of the $SU(2)$ angular-momentum operator $\vec{S}$, thus they are
      the $SU(2)$ angular-momentum waves.

      \newpage

       \subsection{The Comparison between Maxwell's Equations and Maxwell-Type Equations}

       The comparison between Maxwell's equations and Maxwell-type equations is listed in Table \ref{tab:Maxwell}.

            \begin{table}[htbp]
                  \caption{
                        The comparison between Maxwell's equations and the Maxwell-type equations (sourceless and in vacuum). The former predict the well-known electromagnetic waves $\{\vec{E}(\vec{r}, t), \vec{B}(\vec{r}, t)\}$, while the latter predict the $SU(2)$ angular-momentum waves
                        $\{\vec{\mathcal{E}}(\vec{r}, t), \vec{\mathcal{B}}(\vec{r}, t)\}$.}\label{tab:Maxwell}
                  \centering\begin{tabular}{|c|l|l|}
                  \hline
                  \cline{2-3}
                  & \quad \quad \quad Maxwell's Equations & \qquad\qquad\qquad
                        Maxwell-Type Equations \\
                  \hline
                  Expressions & $\partial_\mu F^{\mu\nu} = 0$\ \
                        & $ \partial_\mu \mathcal{F}^{\mu\nu}=0$ \\
                  & $\partial_\mu F_{\nu\gamma}+\partial_\nu F_{\gamma\mu}
                              +\partial_\gamma F_{\mu\nu} = 0$ &
                        $\partial_\mu \mathcal{F}_{\nu\gamma}
                                    +\partial_\nu \mathcal{F}_{\gamma\mu}
                                    +\partial_\gamma \mathcal{F}_{\mu\nu} =0$ \\
                  \hline
                  Tensors & $F_{\mu\nu} =\partial_\mu A_{\nu} -\partial_\nu A_{\mu}$\ \
                        & $\mathcal{F}_{\mu\nu}=\partial_\mu \mathbb{A}_{\nu}
                              -\partial_\nu \mathbb{A}_{\mu}
                              +{\rm i}\,g[\mathbb{A}_{\mu},\
                                    \mathbb{A}_{\nu}]$ \\
                  \hline
                  & $\vec{\nabla}\cdot\vec{E}=0$\qquad\qquad(1) & $\vec{\nabla}\cdot\vec{\mathcal{E}}=0$
                        \qquad\qquad(1) \\
                  Expressions & $\vec{\nabla}\times\vec{E}=-\dfrac{1}{c}
                        \dfrac{\partial\,\vec{B}}{\partial\,t}$\ \ (2)
                        & $\vec{\nabla}\times\vec{\mathcal{E}}=-\dfrac{1}{c}
                        \dfrac{\partial\,\vec{\mathcal{B}}}{\partial\,t}$\ \ \, (2) \\
                  & $\vec{\nabla}\cdot\vec{B}=0$\qquad\qquad(3) &
                        $\vec{\nabla}\cdot\vec{\mathcal{B}}=0$\qquad\qquad\ (3) \\
                  & $\vec{\nabla}\times\vec{B}=\dfrac{1}{c} \dfrac{\partial\,\vec{E}}{\partial\,t}$
                        \quad\,(4) & $\vec{\nabla}\times\vec{\mathcal{B}}=\dfrac{1}{c}
                        \dfrac{\partial\,\vec{\mathcal{E}}}{\partial\,t}$\quad\ \ \ (4) \\
                  \hline
                  Magnetic Field
                  & $\vec{B}=\vec{\nabla}\times\vec{A}$ & $\vec{\mathcal{B}}
                        =\vec{\nabla}\times\vec{\mathcal{A}} - {\rm i}\,g\,
                        \vec{\mathcal{A}}\times\vec{\mathcal{A}}$ \\
                  Electric Field & $\vec{E}=-\dfrac{1}{c} \dfrac{\partial\,\vec{A}}{\partial\,t} -\vec{\nabla}\varphi$
                        & $\vec{\mathcal{E}}=-\dfrac{1}{c} \dfrac{\partial\,\vec{\mathcal{A}}}{\partial\,t}
                              -\vec{\nabla}\varphi-{\rm i}\,g\, \left[
                                    \varphi,\ \vec{\mathcal{A}}\right]$ \\
                  \hline
                  Vector Potential & $\vec{A}(\vec{r}, t)=\vec{A}_{01}\;{\rm e}^{{\rm i}(\vec{k}\cdot\vec{r}
                        -\omega t)}$
                        & $\vec{\mathcal{A}}(\vec{r}, t)=\vec{\tau}\;{\rm e}^{
                              {\rm i}(\vec{k}\cdot\vec{r}-\omega t)}$\\
                  Scalar Potential & $\varphi(\vec{r}, t)=0$\;\;\; (Coulomb gauge)
                  & $\varphi(\vec{r}, t)
                        =\left(\hat{k}\cdot\vec{\tau}\right)\;{\rm e}^{
                              {\rm i}(\vec{k}\cdot\vec{r}-\omega t)}$ \\
                  & & $\vec{\tau}=\vec{R}_0 {\bf 1}
                        +\vec{R}_1 S_x+\vec{R}_2 S_y
                        +\vec{R}_3 S_z$ \\
                  \hline
                  Solutions of Fields& $\vec{B}(\vec{r}, t)={\rm i}(\vec{k}\times\vec{A}_{01})\,{\rm e}^{
                        {\rm i}(\vec{k}\cdot\vec{r}-\omega t)}$
                  & $\vec{\mathcal{B}}(\vec{r}, t) = {\rm i}\left(\vec{k}\times\vec{\tau}\right){\rm e}^{{\rm i}(\vec{k}\cdot\vec{r}-\omega t)}+g \hbar \; \vec{\eta}\;{\rm e}^{2{\rm i}(\vec{k}\cdot\vec{r}-\omega t)}$ \\
                  & $\vec{E}(\vec{r}, t)={\rm i}\,k\,\vec{A}_{01}\,{\rm e}^{
                        {\rm i}(\vec{k}\cdot\vec{r}-\omega t)}$ & $\vec{\mathcal{E}}(\vec{r}, t) = -\hat{k}\times \vec{\mathcal{B}}(\vec{r}, t)$\\
                  &  &  $\vec{\eta}=
                              (\vec{R}_2 \times\vec{R}_3)S_x
                              +(\vec{R}_3 \times\vec{R}_1)S_y
                              +(\vec{R}_1 \times\vec{R}_2)S_z$\\
                  \hline
                  Constraints & $\vec{k}\cdot \vec{A}_{01}=0$ &  $\vec{k}\cdot\bigl(\vec{R}_1 \times\vec{R}_2\bigr)
                  =\vec{k}\cdot\bigl(\vec{R}_2 \times\vec{R}_3\bigr)
                  =\vec{k}\cdot\bigl(\vec{R}_3 \times\vec{R}_1\bigr)=0$\\
                  \hline
                  Properties & $\vec{k}\perp\vec{B}, \; \vec{k}\perp\vec{E},\; \vec{B}\perp\vec{E},$
                        & $\vec{k}\perp\vec{\mathcal{B}}, \; \vec{k}\perp\vec{\mathcal{E}},\;\vec{\mathcal{B}}\perp\vec{\mathcal{E}},$ \\
                  & $\vec{B}=\hat{k}\times \vec{E}, \;\vec{E}=-\hat{k}\times \vec{B},\;\hat{k}\propto \vec{E}\times \vec{B}$
                        & $ \vec{\mathcal{B}}=\hat{k}\times \vec{\mathcal{E}},\; \vec{\mathcal{E}}=-\hat{k}\times \vec{\mathcal{B}}$, \; $\hat{k}\propto \vec{E}\times \vec{B}$, \\
                  &    & $ \vec{S}\times\vec{S}={\rm i} \hbar\vec{S}$, \; $\vec{\tau}\times \vec{\tau}={\rm i} \hbar \vec{\eta}$ \\
                  \hline
                  Predictions & \ Electromagnetic waves
                        & The $SU(2)$ angular-momentum waves  \\
                  & $\nabla^2 \vec{E}-\dfrac{1}{c^2} \dfrac{\partial\,\vec{E}}{\partial\,t^2} =0$
                        & $\nabla^2 \vec{\mathcal{E}}-\dfrac{1}{c^2} \dfrac{\partial\,\vec{\mathcal{E}}}{\partial\,t^2} =0$  \\
                  & $\nabla^2 \vec{B}-\dfrac{1}{c^2} \dfrac{\partial\,\vec{B}}{\partial\,t^2} =0$
                        & $\nabla^2 \vec{\mathcal{B}}-\dfrac{1}{c^2} \dfrac{\partial\,\vec{\mathcal{B}}}{\partial\,t^2} =0$  \\
                  \hline
                  Experimental Verification & Verified by German physicist Hertz (1886)
                        & Possibly be observed in spin-oscillation experiments, \\
                  &
                        & Waiting for experimental discovery   \\
                  \hline
                  \end{tabular}
                 \end{table}

                 \subsection{The Comparison between Yang-Mills Equations under Weak-Coupling Approximation and the Maxwell-Type Equations}

       The comparison between Yang-Mills equations under weak-coupling approximation and Yang-Mills equations under zero-coupling approximation (i.e., the Maxwell-type equations) is listed in Table \ref{tab:Maxwel2}.

            \begin{table}[htbp]
                  \caption{
                        The comparison between Yang-Mills equations under weak-coupling approximation and the Maxwell-type equations (sourceless and in vacuum). Both of them predict the $SU(2)$ angular-momentum waves.}\label{tab:Maxwel2}
                  \centering\begin{tabular}{|c|l|l|}
                  \hline
                  & \multicolumn{2}{c|}{Yang-Mills Equation: \;\; $D_\mu \mathcal{F}^{\mu\nu} =\partial_\mu \mathcal{F}^{\mu\nu}
                        -{\rm i}\,g\,[\mathbb{A}_\mu ,\ \mathcal{F}^{\mu\nu}]=0$,\;\;$D_\mu \mathcal{F}_{\nu\gamma} +D_\nu \mathcal{F}_{\gamma\mu}
                        +D_\gamma \mathcal{F}_{\mu\nu} =0$ } \\
                  \cline{2-3}
                  & \quad \quad \quad Weak-Coupling Approximation & \qquad\qquad\qquad
                        Maxwell-Type Equations \\
                  \hline
                  Expressions & Omitting the $g^2$ terms in YM equations \ \
                        &  Omitting the self-interaction terms in YM equations \\
                   &  \ \
                        & $ \partial_\mu \mathcal{F}^{\mu\nu}=0$ \\
                  &  &
                        $\partial_\mu \mathcal{F}_{\nu\gamma}
                                    +\partial_\nu \mathcal{F}_{\gamma\mu}
                                    +\partial_\gamma \mathcal{F}_{\mu\nu}=0$ \\
                  \hline
                  Tensors & \multicolumn{2}{c|}{$\mathcal{F}_{\mu\nu}=\partial_\mu
                              \mathbb{A}_{\nu} -\partial_\nu \mathbb{A}_{\mu}
                        +{\rm i}\,g[\mathbb{A}_{\mu},\ \mathbb{A}_{\nu}]$} \\
                  \hline
                  Expressions & $\vec{\nabla}\cdot\vec{\mathcal{E}}+ \textcolor{red}{\mathcal{W}_1}=0 $\qquad\qquad\qquad \;\;\;(1) & $\vec{\nabla}\cdot\vec{\mathcal{E}}=0$
                        \qquad\qquad(1) \\
                  & $-\dfrac{1}{c} \dfrac{\partial}{\partial\,t} \vec{\mathcal{B}}
                        -\vec{\nabla}\times\vec{\mathcal{E}}+ \textcolor{red}{\mathcal{W}_2}
                        =0$\ \ \;\; (2)
                        & $\vec{\nabla}\times\vec{\mathcal{E}}=-\dfrac{1}{c}
                        \dfrac{\partial\,\vec{\mathcal{B}}}{\partial\,t}$\ \ \, (2) \\
                  & $\vec{\nabla}\cdot\vec{\mathcal{B}}+ \textcolor{red}{\mathcal{W}_3}
                        =0$\qquad\qquad\qquad\;\; (3) &
                        $\vec{\nabla}\cdot\vec{\mathcal{B}}=0$\qquad\qquad\ (3) \\
                  & $-\dfrac{1}{c} \dfrac{\partial}{\partial\,t} \vec{\mathcal{E}}
                        +\vec{\nabla}\times\vec{\mathcal{B}}+\textcolor{red}{\mathcal{W}_4}
                        =0$
                        \quad\,(4) & $\vec{\nabla}\times\vec{\mathcal{B}}=\dfrac{1}{c}
                        \dfrac{\partial\,\vec{\mathcal{E}}}{\partial\,t}$\quad\ \ \ (4) \\
                  \hline
                  Magnetic Field
                  & \multicolumn{2}{c|}{$\vec{\mathcal{B}}
                        =\vec{\nabla}\times\vec{\mathcal{A}} -{\rm i}\,g\, \vec{\mathcal{A}}\times\vec{\mathcal{A}}$} \\
                  Electric Field & \multicolumn{2}{c|}{$\vec{\mathcal{E}}=-\dfrac{1}{c}
                        \dfrac{\partial\,\vec{\mathcal{A}}}{\partial\,t}
                              -\vec{\nabla}\varphi-{\rm i}\,g\, \left[
                                    \varphi,\ \vec{\mathcal{A}}\right]$} \\
                  \hline
                  Vector Potential & \multicolumn{2}{c|}{$\vec{\mathcal{A}}(\vec{r}, t)=\vec{\tau}\;{\rm e}^{
                              {\rm i}(\vec{k}\cdot\vec{r}-\omega t)}$} \\
                  Scalar Potential & \multicolumn{2}{c|}{$\varphi(\vec{r}, t)
                        =\left(\hat{k}\cdot\vec{\tau}\right)\;{\rm e}^{
                              {\rm i}(\vec{k}\cdot\vec{r}-\omega t)}$} \\
                  & \multicolumn{2}{c|}{$\vec{\tau}=\vec{R}_0 {\bf 1}
                        +\vec{R}_1 S_x+\vec{R}_2 S_y
                        +\vec{R}_3 S_z$} \\
                  \hline
                  Solutions of Fields & \multicolumn{2}{c|}{$\vec{\mathcal{B}}(\vec{r}, t) = {\rm i}\left(\vec{k}\times\vec{\tau}\right){\rm e}^{{\rm i}(\vec{k}\cdot\vec{r}-\omega t)}+g \hbar \; \vec{\eta}\;{\rm e}^{2{\rm i}(\vec{k}\cdot\vec{r}-\omega t)}$} \\
                  & \multicolumn{2}{c|}{$\vec{\mathcal{E}}(\vec{r}, t) = -\hat{k}\times \vec{\mathcal{B}}(\vec{r}, t)$} \\
                  & \multicolumn{2}{c|}{$\vec{\eta}=
                              (\vec{R}_2 \times\vec{R}_3)S_x
                              +(\vec{R}_3 \times\vec{R}_1)S_y
                              +(\vec{R}_1 \times\vec{R}_2)S_z$} \\
                  \hline
                  Constraints & \multicolumn{2}{c|}{$\vec{k}\cdot\bigl(\vec{R}_1 \times\vec{R}_2\bigr)
                  =\vec{k}\cdot\bigl(\vec{R}_2 \times\vec{R}_3\bigr)
                  =\vec{k}\cdot\bigl(\vec{R}_3 \times\vec{R}_1\bigr)=0$} \\
                  \hline
                  Properties & \multicolumn{2}{c|}{$\vec{k}\perp\vec{\mathcal{B}}, \; \vec{k}\perp\vec{\mathcal{E}},\;\vec{\mathcal{B}}\perp
                        \vec{\mathcal{E}},$} \\
                  & \multicolumn{2}{c|}{$\vec{\mathcal{B}}=\hat{k}\times \vec{\mathcal{E}},\; \vec{\mathcal{E}}=-\hat{k}\times \vec{\mathcal{B}}$, \; $\hat{k}\propto \vec{E}\times \vec{B}$,} \\
                  &  \multicolumn{2}{c|}{$ \vec{S}\times\vec{S}={\rm i} \hbar\vec{S}$, \; $\vec{\tau}\times \vec{\tau}={\rm i} \hbar \vec{\eta}$} \\
                  \hline
                  Predictions & \multicolumn{2}{c|}{The $SU(2)$ angular-momentum waves}\\
                  \cline{2-3}
                  & \multicolumn{2}{c|}{$\nabla^2 \vec{\mathcal{E}}-\dfrac{1}{c^2} \dfrac{\partial\,\vec{\mathcal{E}}}{\partial\,t^2} =0$} \\
                  & \multicolumn{2}{c|}{$\nabla^2 \vec{\mathcal{B}}-\dfrac{1}{c^2} \dfrac{\partial\,\vec{\mathcal{B}}}{\partial\,t^2} =0$}  \\
                  \hline
                  Experimental Verification & \multicolumn{2}{c|}{Possibly be observed in spin-oscillation experiments,} \\
                  & \multicolumn{2}{c|}{Waiting for experimental discovery} \\
                  \hline
                  \end{tabular}
                 \end{table}

                 Here
       \begin{eqnarray}
            &&\mathcal{W}_1=
                        -{\rm i}\,g\Biggl\{\bigg[\vec{\mathcal{A}}\cdot\Big(
                                          \dfrac{1}{c} \dfrac{\partial\,
                                                \vec{\mathcal{A}}}{\partial\,t}\Big)
                                    -\Big(\dfrac{1}{c} \dfrac{\partial\,
                                          \vec{\mathcal{A}}}{\partial\,t}\Big)\cdot
                                                \vec{\mathcal{A}}\bigg]
                              +\vec{\mathcal{A}}\cdot\bigl(\vec{\nabla}\varphi\bigr)
                              -\bigl(\vec{\nabla}\varphi\bigr)\cdot\vec{\mathcal{A}}
                              \Biggr\}, \nonumber\\
            &&\mathcal{W}_2={\rm i}\,g\Biggl\{\bigg[\vec{\mathcal{A}}
                                          \times\Big(\dfrac{1}{c}
                                                \dfrac{\partial\,\vec{\mathcal{A}}}{
                                                      \partial\,t}\Big)
                                    +\Big(\dfrac{1}{c} \dfrac{\partial\,
                                          \vec{\mathcal{A}}}{\partial\,t}\Big)\times
                                                \vec{\mathcal{A}}\bigg]
                              +\Bigl[\varphi,\ \big(\vec{\nabla}\times\vec{\mathcal{A}}
                                    \big)\Bigr]
                              +\bigl(\vec{\nabla}\varphi\bigr)\times\vec{\mathcal{A}}
                              +\vec{\mathcal{A}}\times\bigl(\vec{\nabla}\varphi\bigr)
                              \Biggr\}, \nonumber\\
            && \mathcal{W}_3=-{\rm i}\,g\,\vec{\nabla}\cdot\bigl(\vec{\mathcal{A}}
                              \times\vec{\mathcal{A}}\bigr), \nonumber\\
            && \mathcal{W}_4=-{\rm i}\,g\Biggl\{\bigg[\varphi,\
                                    \dfrac{1}{c} \dfrac{\partial\,\vec{\mathcal{A}}}{
                                          \partial\,t}\bigg]
                              +\Big[\varphi,\ \bigl(\vec{\nabla}\varphi\bigr)\Big]
                              -\vec{\mathcal{A}}\times\bigl(\vec{\nabla}\times
                                    \vec{\mathcal{A}}\bigr)
                              -\bigl(\vec{\nabla}\times\vec{\mathcal{A}}\bigr)\times
                                    \vec{\mathcal{A}}\Biggr\}.
      \end{eqnarray}
      Based on
      \begin{eqnarray}
       && \vec{\mathcal{A}}(\vec{r}, t)=\vec{\tau}\,{\rm e}^{
                        {\rm i}(\vec{k}\cdot\vec{r}-\omega\,t)}, \nonumber\\
                 && \varphi(\vec{r}, t) =\bigl(\vec{\tau}\cdot\hat{k}\bigr){\rm e}^{
                        {\rm i}(\vec{k}\cdot\vec{r}-\omega\,t)},
      \end{eqnarray}
      one may prove that
      \begin{eqnarray}\label{eq:W}
       && \mathcal{W}_1=\mathcal{W}_2=\mathcal{W}_3=\mathcal{W}_4=0.
      \end{eqnarray}
      \emph{Proof.---}Because
            \begin{equation}
                  \dfrac{1}{c} \dfrac{\partial\,\vec{\mathcal{A}}}{\partial\,t}
                  =-{\rm i}\,k\,\vec{\tau}\,{\rm e}^{{\rm i}(\vec{k}\cdot\vec{r}
                        -\omega\,t)} ,\quad
                  \vec{\nabla}\big[\varphi(\vec{r},t)\big]={\rm i}\,k\bigl(\vec{\tau}
                        \cdot\hat{k}\bigr)\hat{k}\,{\rm e}^{{\rm i}(\vec{k}\cdot\vec{r}
                              -\omega\,t)},\quad
                  \vec{\nabla}\times\vec{\mathcal{A}}={\rm i}\,k\bigl(\hat{k}\times
                        \vec{\tau}\bigr){\rm e}^{{\rm i}(\vec{k}\cdot\vec{r}
                              -\omega\,t)},
            \end{equation}
            then
            \begin{itemize}
                  \item [(i)]. For $\mathcal{W}_1$,
                        \begin{eqnarray}
                                    && \bigg[\vec{\mathcal{A}}\cdot\Big(
                                          \dfrac{1}{c} \dfrac{\partial\,
                                                \vec{\mathcal{A}}}{\partial\,t}\Big)
                                          -\Big(\dfrac{1}{c} \dfrac{\partial\,
                                                \vec{\mathcal{A}}}{\partial\,t}\Big)\cdot\vec{\mathcal{A}}\bigg]
                                          +\vec{\mathcal{A}}\cdot\bigl(\vec{\nabla}
                                                \varphi\bigr)
                                          -\bigl(\vec{\nabla}\varphi\bigr)\cdot\vec{
                                                \mathcal{A}} \nonumber\\
                                    &=& -{\rm i}\,k\Big[\bigl(\vec{\tau}\cdot\vec{\tau}
                                                \bigr)-\bigl(\vec{\tau}\cdot\vec{\tau}
                                                \bigr)\Big]{\rm e}^{{\rm i}\,2(\vec{k}
                                                      \cdot\vec{r}-\omega\,t)}
                                          +{\rm i}\,k\Big[\bigl(\vec{\tau}\cdot\hat{k}
                                                      \bigr)^2
                                                -\bigl(\vec{\tau}\cdot\hat{k}\bigr)^2
                                                \Big]{\rm e}^{{\rm i}\,2(\vec{k}\cdot
                                                      \vec{r}-\omega\,t)}
                                    =0.
                        \end{eqnarray}
                  \item [(ii)]. For $\mathcal{W}_2$,
                        \begin{eqnarray}
                                    && \bigg[\vec{\mathcal{A}}\times\Big(\dfrac{1}{c}
                                                \dfrac{\partial\,\vec{\mathcal{A}}}{
                                                      \partial\,t}\Big)
                                          +\Big(\dfrac{1}{c} \dfrac{\partial\,
                                                \vec{\mathcal{A}}}{\partial\,t}\Big)
                                                      \times\vec{\mathcal{A}}\bigg]
                                          +\Bigl[\varphi,\ \big(\vec{\nabla}\times
                                                \vec{\mathcal{A}}\big)\Bigr]
                                          +\bigl(\vec{\nabla}\varphi\bigr)\times\vec{
                                                \mathcal{A}}
                                          +\vec{\mathcal{A}}\times\bigl(\vec{\nabla}
                                                \varphi\bigr) \nonumber\\
                                    &=& -{\rm i}\,k\Big[\bigl(\vec{\tau}\times\vec{\tau}
                                                \bigr)+\bigl(\vec{\tau}\times\vec{\tau}
                                                \bigr)\Big]{\rm e}^{{\rm i}\,2(\vec{k}
                                                      \cdot\vec{r}-\omega\,t)}
                                          +{\rm i}\,k\Bigl[\bigl(\vec{\tau}\cdot\hat{k}
                                                \bigr),\ \bigl(\hat{k}\times\vec{\tau}
                                                \bigr)\Bigr]{\rm e}^{{\rm i}\,2(\vec{k}
                                                      \cdot\vec{r}-\omega\,t)}\nonumber\\
                                          &&+{\rm i}\,k\Bigl[\bigl(\vec{\tau}\cdot\hat{k}
                                                \bigr)\bigl(\hat{k}\times\vec{\tau}
                                                \bigr)+\bigl(\vec{\tau}\times\hat{k}
                                                \bigr)\bigl(\vec{\tau}\cdot\hat{k}
                                                \bigr)\Bigr]{\rm e}^{{\rm i}\,2(\vec{k}
                                                      \cdot\vec{r}-\omega\,t)} \nonumber\\
                                    &=& -{\rm i}\,2\,k\bigl(\vec{\tau}\times\vec{\tau}
                                                \bigr){\rm e}^{{\rm i}\,2(\vec{k}\cdot
                                                      \vec{r}-\omega\,t)}
                                          +{\rm i}\,k\,\hat{k}\times\Bigl[\bigl(
                                                \vec{\tau}\cdot\hat{k}\bigr),\
                                                \vec{\tau}\Bigr]{\rm e}^{{\rm i}\,2(
                                                      \vec{k}\cdot\vec{r}-\omega\,t)}
                                          +{\rm i}\,k\Bigl[\bigl(\vec{\tau}\cdot\hat{k}
                                                \bigr)\bigl(\hat{k}\times\vec{\tau}
                                                \bigr)-\bigl(\hat{k}\times\vec{\tau}
                                                \bigr)\bigl(\vec{\tau}\cdot\hat{k}
                                                \bigr)\Bigr]{\rm e}^{{\rm i}\,2(\vec{k}
                                                      \cdot\vec{r}-\omega\,t)} \nonumber\\
                                    &=& -{\rm i}\,2\,k\bigl(\vec{\tau}\times\vec{\tau}
                                                \bigr){\rm e}^{{\rm i}\,2(\vec{k}\cdot
                                                      \vec{r}-\omega\,t)}
                                          +{\rm i}\,2\,k\,\hat{k}\times\Bigl[\bigl(
                                                \vec{\tau}\cdot\hat{k}\bigr),\
                                                \vec{\tau}\Bigr]{\rm e}^{{\rm i}\,2(
                                                      \vec{k}\cdot\vec{r}-\omega\,t)}
                                    ={\rm i}\,2\,k\,\biggl\{\hat{k}\times\Bigl[\bigl(
                                                \vec{\tau}\cdot\hat{k}\bigr),\
                                                \vec{\tau}\Bigr]
                                          -\bigl(\vec{\tau}\times\vec{\tau}\bigr)
                                          \biggr\}{\rm e}^{{\rm i}\,2(\vec{k}\cdot
                                                \vec{r}-\omega\,t)} \nonumber\\
                                    &=& -2\,k\,\hbar\bigl(\hat{k}\times\vec{\xi}
                                          -\vec{\eta}\bigr){\rm e}^{{\rm i}\,2(
                                                \vec{k}\cdot\vec{r}-\omega\,t)}
                                    =0.
                        \end{eqnarray}
                  \item [(iii)]. For $\mathcal{W}_3$,
                        \begin{equation}
                              \begin{split}
                                    & \vec{\nabla}\cdot\bigl(\vec{\mathcal{A}}\times
                                          \vec{\mathcal{A}}\bigr)
                                    =\vec{\nabla}\cdot\Bigl[\bigl(\vec{\tau}\times
                                          \vec{\tau}\bigr){\rm e}^{{\rm i}\,2(
                                                \vec{k}\cdot\vec{r}-\omega\,t)}\Bigr]
                                    ={\rm i}\hbar\,\vec{\nabla}\cdot\Bigl[\vec{\eta}\,
                                          {\rm e}^{{\rm i}\,2(\vec{k}\cdot\vec{r}
                                                -\omega\,t)}\Bigr]
                                    =-2\,\hbar\,k\bigl(\hat{k}\cdot\vec{\eta}\bigr)
                                          {\rm e}^{{\rm i}\,2(\vec{k}\cdot\vec{r}
                                                -\omega\,t)}
                                    =0.
                              \end{split}
                        \end{equation}
                  \item [(iv)]. For $\mathcal{W}_4$,
                        \begin{eqnarray}
                                    && \bigg[\varphi,\ \dfrac{1}{c}
                                          \dfrac{\partial\,\vec{\mathcal{A}}}{
                                                \partial\,t}\bigg]
                                          +\Big[\varphi,\ \bigl(\vec{\nabla}\varphi
                                                \bigr)\Big]
                                          -\vec{\mathcal{A}}\times\bigl(\vec{\nabla}
                                                \times\vec{\mathcal{A}}\bigr)
                                          -\bigl(\vec{\nabla}\times\vec{\mathcal{A}}
                                                \bigr)\times\vec{\mathcal{A}} \nonumber\\
                                    &=& -{\rm i}\,k\Big[\bigl(\vec{\tau}\cdot\hat{k}
                                                \bigr),\ \vec{\tau}\Big]{\rm e}^{
                                                      {\rm i}\,2(\vec{k}\cdot\vec{r}
                                                            -\omega\,t)}
                                          +{\rm i}\,k\Big[\bigl(\vec{\tau}\cdot\hat{k}
                                                \bigr),\ \bigl(\vec{\tau}\cdot\hat{k}
                                                \bigr)\hat{k}\Big]{\rm e}^{{\rm i}\,2(
                                                      \vec{k}\cdot\vec{r}-\omega\,t)}
                                          -{\rm i}\,k\Big[\vec{\tau}\times\bigl(\hat{k}
                                                      \times\vec{\tau}\bigr)
                                                +\bigl(\hat{k}\times\vec{\tau}\bigr)
                                                      \times\vec{\tau}\Big]{\rm e}^{
                                                            {\rm i}\,2(\vec{k}\cdot
                                                                  \vec{r}-\omega\,t)} \nonumber\\
                                    &=& -{\rm i}\,k\Big[\bigl(\vec{\tau}\cdot\hat{k}
                                                \bigr),\ \vec{\tau}\Big]{\rm e}^{
                                                      {\rm i}\,2(\vec{k}\cdot\vec{r}
                                                            -\omega\,t)}
                                          -{\rm i}\,k\Big[\vec{\tau}\times\bigl(\hat{k}
                                                      \times\vec{\tau}\bigr)
                                                -\bigl(\vec{\tau}\times\hat{k}\bigr)
                                                      \times\vec{\tau}\Big]{\rm e}^{
                                                            {\rm i}\,2(\vec{k}\cdot
                                                                  \vec{r}-\omega\,t)} \nonumber\\
                                    &=& -{\rm i}\,k\Big[\bigl(\vec{\tau}\cdot\hat{k}
                                                \bigr),\ \vec{\tau}\Big]{\rm e}^{
                                                      {\rm i}\,2(\vec{k}\cdot\vec{r}
                                                            -\omega\,t)}
                                          -{\rm i}\,k\Big[\bigl(\vec{\tau}\cdot
                                                      \vec{\tau}\bigr)\hat{k}
                                                -\bigl(\vec{\tau}\cdot\hat{k}\bigr)
                                                      \vec{\tau}
                                                -\bigl(\vec{\tau}\cdot\vec{\tau}\bigr)
                                                      \hat{k}
                                                +\vec{\tau}\bigl(\vec{\tau}\cdot\hat{k}
                                                      \bigr)\Big]{\rm e}^{{\rm i}\,2(
                                                            \vec{k}\cdot\vec{r}
                                                            -\omega\,t)} \nonumber\\
                                    &=& -{\rm i}\,k\Big[\bigl(\vec{\tau}\cdot\hat{k}
                                                \bigr),\ \vec{\tau}\Big]{\rm e}^{
                                                      {\rm i}\,2(\vec{k}\cdot\vec{r}
                                                            -\omega\,t)}
                                          +{\rm i}\,k\Big[\bigl(\vec{\tau}\cdot\hat{k}
                                                \bigr),\ \vec{\tau}\Big]{\rm e}^{
                                                      {\rm i}\,2(\vec{k}\cdot\vec{r}
                                                            -\omega\,t)}
                                    =0.
                        \end{eqnarray}
            \end{itemize}
            Based on Eq. (\ref{eq:W}), from Table \ref{tab:Maxwel2} one finds that the Yang-Mills equations under the weak-coupling approximation turns to the Maxwell-type equations, thus it is the reason why they share the same $SU(2)$ angular-momentum waves.


\newpage
\newpage
\section{Possible Detection of Angular-Momentum Waves with the ``Zitterbewegung'' of Electron}

                \subsection{The ``Position Zitterbewegung'' of Electron}

            When Schr\"odinger first examined the expectation value of the position operator of a relativistic free electron, he discovered a kind of surprising highly frequent oscillation, which was called the
            ``position Zitterbewegung'' (PZ) \cite{1931Zitter}. In the following, we make a brief review.

            In relativistic quantum mechanics, the Dirac equation of a free electron reads
            \begin{equation}
                  H_{\rm e}\,\ket{\Psi}=\left(c\,\vec{\alpha}\cdot\vec{p}
                        +\beta\,m\,c^2\right)\ket{\Psi}
                  =E\,\ket{\Psi},
            \end{equation}
            where
            \begin{equation}\label{eq:Dirac}
                  H_{\rm e}= c\,\vec{\alpha}\cdot\vec{p}+\beta\,m\,c^2
            \end{equation}
            is the Hamiltonian operator of the electron, $m$ is the rest mass, $c$ is the speed of light in vacuum, $\vec{p}=(p_x, p_y, p_z)$ is the linear momentum, $E$ is the energy, and $\ket{\Psi}$ is the wavefunction. In Eq. (\ref{eq:Dirac}), $\vec{\alpha}$ and $\beta$ are Dirac's matrices, whose explicit forms are given by
            \begin{equation}\label{eq:EP3}
            \vec{\alpha}=\begin{bmatrix}
                  0 & \vec{\sigma} \\
                  \vec{\sigma} & 0
            \end{bmatrix},
            \quad \beta=\begin{bmatrix}
                  \openone & 0 \\
                  0 & -\openone
            \end{bmatrix},
            \end{equation}
            with $\vec{\sigma}=(\sigma_x,\sigma_y,\sigma_z)$ indicating the vector of Pauli matrices, and $\openone$ expressing the the $2\times 2$ identity matrix. It is easy to check that
            \begin{equation}
                  \alpha_i^2=\beta^2={\bf{1}}, \;\;\; \left\{\alpha_i,\beta\right\}=\alpha_i\beta+\beta\alpha_i=0,\;\;\;  \left\{\alpha_i,\alpha_j\right\}=2\,\delta_{ij}\,{\bf{1}},
            \end{equation}
            where ${\bf{1}}=\openone\otimes\openone$ is the $4\times 4$ identity matrix, and $\delta_{ij}$ is the Kronecker delta symbold (with $\delta_{ij}$ equaling to 1 for $i=j$, and 0 otherwise).

            One may introduce the helicity operator
            \begin{equation}
                  \hat{\Lambda}= \vec{S}\cdot\hat{p},
            \end{equation}
            where
            \begin{equation}\label{eq:spin}
            \vec{S}=\dfrac{\hbar}{2} \vec{\Sigma}
            \end{equation}
            is the spin operator,
            \begin{equation}
            \vec{\Sigma}\equiv\begin{bmatrix}
                              \vec{\sigma} & 0 \\
                              0 & \vec{\sigma}
                        \end{bmatrix},
            \end{equation}
            $\hat{p}=\vec{p}/|\vec{p}|$, $\hbar=h/2\pi$, and $h$ is Planck's constant. It is easy to prove that the helicity operator $\hat{\Lambda}$ commutes with the Hamiltonian $ H_{\rm e}$, i.e.,
            \begin{equation}
            [H_{\rm e}, \hat{\Lambda}]=H_{\rm e} \hat{\Lambda}-\hat{\Lambda} H_{\rm e}=0.
            \end{equation}
            Thereby, in the momentum space, one may express the wavefunction $\ket{\Psi}$ as the common eigenstates of the set $\{H_{\rm e}, \hat{\Lambda}\}$. Since the Hamiltonian operator is a $4\times 4$ matrix, it accordingly has four eigen-energy and four eigenfunctions. Explicitly they are
            \begin{equation}\label{eq:mom}
                  \begin{split}
                        & E_{\rm +} \begin{cases}
                              & |{\Psi}_1\rangle
                              =\dfrac{1}{\sqrt{2\,E_{\rm p}\,p(p+p_z)}} \left[
                                    u_+ \Bigl(p+p_z ,\ p_+\Bigr)
                                    ,\dfrac{c\,p}{u_+}\Bigl(p+p_z,\ p_+ \Bigr)\right]^{
                                          \rm T}, \\
                              & |{\Psi}_2\rangle
                              =\dfrac{1}{\sqrt{2\,E_{\rm p}\,p(p+p_z)}}\left[
                                    u_+ \Bigl(-p_-, p+p_z\Bigr),-\dfrac{c\,p}{u_+}\Bigl(
                                          -p_-, p+p_z \Bigr)\right]^{\rm T},
                        \end{cases} \\
                        & E_{\rm -} \begin{cases}
                              & |{\Psi}_3\rangle=\dfrac{1}{\sqrt{2\,E_{\rm p}\,
                                    p(p+p_z)}} \left[u_- \Bigl(p+p_z ,\ p_+\Bigr)
                                          ,-\dfrac{c\,p}{u_-}\Bigl(
                                                p+p_z,\ p_+ \Bigr)\right]^{\rm T}, \\
                              & |{\Psi}_4\rangle
                              =\dfrac{1}{\sqrt{2\,E_{\rm p}\,p(p+p_z)}}\left[
                                    u_- \Bigl(-p_-,p+p_z\Bigr)
                                    ,\dfrac{c\,p}{u_-}\Bigl(-p_-, p+p_z \Bigr)\right]^{
                                          \rm T},
                        \end{cases}
                  \end{split}
            \end{equation}
            where
            \begin{eqnarray}
                  &&  E_{\rm p}=\sqrt{p^2c^2+m^2c^4}=E_{\rm +}, \;\;\;\;\;\; -E_{\rm p}=-\sqrt{p^2c^2+m^2c^4}=E_-,
            \end{eqnarray}
            and we have denoted
            \begin{eqnarray}
                  p_{\pm}=p_x \pm\mathrm{i}\,p_y,\qquad p=\sqrt{p_x^2+p_y^2+p_z^2}, \qquad
                  u_{\pm}=\sqrt{E_{\rm p}\pm m\,c^2}.
            \end{eqnarray}
            For the four eigenfunctions $\{|{\Psi}_1\rangle,|{\Psi}_2\rangle,|{\Psi}_3\rangle,|{\Psi}_4\rangle\}$ in Eq. (\ref{eq:mom}), the first two eigenfunctions (i.e., $\{|{\Psi}_1\rangle, |{\Psi}_2\rangle\}$) correspond to the positive energy $E_+$, and the last two eigenfunctions (i.e., $\{|{\Psi}_3\rangle, |{\Psi}_4\rangle\}$) correspond to the negative energy $E_-$. Similarly, for the helicity operator, we have
            \begin{equation}
            \hat{\Lambda}\,|{\Psi}\rangle=\lambda |{\Psi}\rangle,
            \end{equation}
            and explicitly
            \begin{equation}
                  \hat{\Lambda}\,|{\Psi}_1\rangle=+\dfrac{\hbar}{2} |{\Psi}_1\rangle,
                  \quad
                  \hat{\Lambda}\,|{\Psi}_2\rangle=-\dfrac{\hbar}{2} |{\Psi}_2\rangle,
                  \quad
                  \hat{\Lambda}\,|{\Psi}_3\rangle=+\dfrac{\hbar}{2} |{\Psi}_3\rangle,
                  \quad
                  \hat{\Lambda}\,|{\Psi}_4\rangle=-\dfrac{\hbar}{2} |{\Psi}_4\rangle,
            \end{equation}
            which means that $\{|{\Psi}_1\rangle, |{\Psi}_3\rangle\}$ correspond to the positive helicity $\lambda_+=+\hbar/2$, and $\{|{\Psi}_2\rangle, |{\Psi}_4\rangle\}$ correspond to the negative helicity $\lambda_-=-\hbar/2$.

            Schr\"odinger has made the following observation.

            \emph{Observation 1.---}Only when Dirac's electron is in a superposition state of positive-energy and negative-energy (i.e., $|\Psi\rangle=\cos\theta |\psi_+\rangle +\sin\theta |\psi_-\rangle$), is there a PZ phenomenon.
            \begin{proof}
                  Here we show a simple proof. From the Heisenberg equation, one has
                  \begin{eqnarray}\label{eq:D-3}
                        && \frac{{\rm d}\,\vec{p}}{{\rm d}\,t}=\frac{1}{{\rm i}\hbar}[\vec{p}, H_{\rm e}]=0, \nonumber\\
                        && \frac{{\rm d}\,H_{\rm e}}{{\rm d}\,t}=\frac{1}{{\rm i}\hbar}[H_{\rm e}, H_{\rm e}]=0,
                  \end{eqnarray}
                  which means that $\vec{p}$ and $H_{\rm e}$ are conservation quantities. For the position operator $\vec{r}$, we have
                  \begin{eqnarray}\label{eq:D-4}
                        \frac{{\rm d}\,\vec{r}}{{\rm d}\,t}=\frac{1}{{\rm i}\hbar}[\vec{r}, H_{\rm e}]=\frac{1}{{\rm i}\hbar}[\vec{r} \vec{\alpha}\cdot\vec{p}]=c\vec{\alpha}.
                  \end{eqnarray}
                  For the operator $\vec{\alpha}$, we obtain
                  \begin{eqnarray}\label{eq:D-5}
                        \frac{{\rm d}\,\vec{\alpha}}{{\rm d}\,t}=\frac{1}{{\rm i}\hbar}[\vec{\alpha}, H_{\rm e}]
                        =\frac{1}{{\rm i}\hbar}\left(\{\vec{\alpha}, H_{\rm e}\}-2\,H_{\rm e}\,\vec{\alpha}\right)
                        =\frac{1}{{\rm i}\hbar}\left(2c\vec{p}-2\,H_{\rm e}\,\vec{\alpha}\right)=-\frac{2\,H_{\rm e}}{{\rm i}\hbar}\left(\vec{\alpha}-c H^{-1}_{\rm e}\vec{p}\right),
                  \end{eqnarray}
                  which yields
                  \begin{eqnarray}\label{eq:D-7}
                        \vec{\alpha}(t)=c H^{-1}_{\rm e}\vec{p}+{\rm e}^{
                              \frac{{\rm i}\,2\,H_{\rm e}t}{\hbar}}\left[\vec{\alpha}(0)
                                    -c H^{-1}_{\rm e}\vec{p}\right].
                  \end{eqnarray}
                  One can check the following anti-commutative relation
                  \begin{eqnarray}\label{eq:D-8a}
                        \Bigl\{H_{\rm e},\ \left[\vec{\alpha}(0)-c H^{-1}_{\rm e}\vec{p}\right]\Bigr\}=0,
                  \end{eqnarray}
                  thus
                  \begin{eqnarray}\label{eq:D-8b}
                        H_{\rm e} \left[\vec{\alpha}(0)-c H^{-1}_{\rm e}\vec{p}\right]=-\left[\vec{\alpha}(0)-c H^{-1}_{\rm e}\vec{p}\right]H_{\rm e},
                  \end{eqnarray}
                  which leads to
                  \begin{eqnarray}\label{eq:D-8c}
                        {\rm e}^{\frac{{\rm i}\,2\,H_{\rm e}t}{\hbar}}\left[\vec{\alpha}(0)-c H^{-1}_{\rm e}\vec{p}\right]=
                        \left[\vec{\alpha}(0)-c H^{-1}_{\rm e}\vec{p}\right] {\rm e}^{\frac{-{\rm i}\,2\,H_{\rm e}t}{\hbar}}.
                  \end{eqnarray}
                  By substituting Eq. (\ref{eq:D-7}) and Eq. (\ref{eq:D-8c}) into Eq. (\ref{eq:D-4}), we have
                  \begin{eqnarray}\label{eq:D-9}
                        \frac{{\rm d}\,\vec{r}}{{\rm d}\,t}=c\vec{\alpha}(t)=c^2 H^{-1}_{\rm e}\vec{p}+\left[c\vec{\alpha}(0)-c^2 H^{-1}_{\rm e}\vec{p}\right]{\rm e}^{\frac{-{\rm i}\,2\,H_{\rm e}t}{\hbar}},
                  \end{eqnarray}
                  then we have
                  \begin{eqnarray}
                        \vec{r}(t)&=&\vec{r}(0)+c^2 H^{-1}_{\rm e}\vec{p}\;t+\int \left[c\vec{\alpha}(0)-c^2 H^{-1}_{\rm e}\vec{p}\right]{\rm e}^{\frac{-{\rm i}\,2\,H_{\rm e}t}{\hbar}}{\rm d}t\nonumber\\
                        &=&\vec{r}(0)+c^2 H^{-1}_{\rm e}\vec{p}\;t+\frac{{\rm i}\hbar c}{2}\left[
                              \vec{\alpha}(0)-c H^{-1}_{\rm e}\vec{p}\right] H^{-1}_{\rm e} \left(
                                    {\rm e}^{\frac{-{\rm i}\,2\,H_{\rm e}t}{\hbar}}+C\right).
                  \end{eqnarray}
                  To ensure $\vec{r}(t)\big|_{t=0} =\vec{r}(0)$, we have to choose $C=-1$, namely
                  \begin{equation}\label{eq:D-10}
                        \vec{r}(t)=\vec{r}(0)+c^2 H^{-1}_{\rm e} \vec{p}\,t
                              +\dfrac{{\rm i}\hbar c}{2}\left[\vec{\alpha}(0)
                                    -c\,H^{-1}_{\rm e} \vec{p}\right] H^{-1}_{\rm e} \left(
                                          {\rm e}^{\frac{-{\rm i}\,2\,H_{\rm e}t}{\hbar}} -1\right).
                  \end{equation}

                  For the time-dependent position operator $\vec{r}(t)$ in Eq. (\ref{eq:D-10}), one may notice that the third term refers to oscillation, namely the PZ. Based on Eq. (\ref{eq:D-10}), the PZ operator is defined as
                  \begin{equation}
                        \hat{\mathcal{Z}}_r=\left\{\frac{{\rm i}\hbar c}{2}\left[\vec{\alpha}(0)-c H^{-1}_{\rm e}\vec{p}\right] H^{-1}_{\rm e} \left(
                              {\rm e}^{\frac{-{\rm i}\,2\,H_{\rm e} t}{\hbar}} -1\right)\right\},
                  \end{equation}
                  which is a Hermitian operator. One needs to calculate the following expectation value of the PZ operator,
                  \begin{eqnarray}\label{eq:D-12a}
                        \mathcal{Z}_r=\langle \Psi| \hat{\mathcal{Z}}_r |\Psi\rangle,
                  \end{eqnarray}
                  where $|\Psi\rangle$ is the quantum state of Dirac's electron.

                  After introducing the projective operators
                  \begin{eqnarray}\label{eq:D-14}
                        \Pi_\pm=\frac{1}{2}\left({\bf 1}\pm \frac{H_{\rm e}}{\sqrt{p^2c^2+m^2c^4}}\right), \;\;\;\;\; \Pi_\pm^2=\Pi_\pm,
                  \end{eqnarray}
                  we easily have
                  \begin{eqnarray}\label{eq:D-14a}
                        && \Pi_+ |\Psi_1\rangle = |\Psi_1\rangle, \;\; \Pi_+ |\Psi_2\rangle = |\Psi_2\rangle, \;\; \Pi_+ |\Psi_3\rangle = 0, \;\; \Pi_+ |\Psi_4\rangle = 0, \nonumber\\
                        && \Pi_- |\Psi_1\rangle = 0, \;\; \Pi_- |\Psi_2\rangle = 0, \;\; \Pi_- |\Psi_3\rangle = |\Psi_3\rangle, \;\; \Pi_- |\Psi_4\rangle = |\Psi_4\rangle.
                  \end{eqnarray}
                  Based on Eq. (\ref{eq:D-8b}), we then obtain
                  \begin{eqnarray}\label{eq:D-17}
                        \Pi_+ \left[\vec{\alpha}(0)-c H^{-1}_{\rm e}\vec{p}\right]\Pi_+
                        =\frac{1}{4} \left({\bf 1}+\frac{H_{\rm e}}{\sqrt{p^2c^2+m^2c^4}}
                              \right)\left[\vec{\alpha}(0)-c H^{-1}_{\rm e}\vec{p}\right]
                                    \left({\bf 1}+\frac{H_{\rm e}}{\sqrt{p^2c^2+m^2c^4}}
                                    \right)=0,
                  \end{eqnarray}
                  and similarly,
                  \begin{eqnarray}\label{eq:D-18}
                        \Pi_- \left[\vec{\alpha}(0)-c H^{-1}_{\rm e}\vec{p}\right]\Pi_- =0.
                  \end{eqnarray}
                  The above results imply that
                  \begin{eqnarray}\label{eq:D-19}
                        \Pi_+ \hat{\mathcal{Z}}_r \Pi_+ =0,\;\;\;\;\;
                        \Pi_- \hat{\mathcal{Z}}_r \Pi_- =0.
                  \end{eqnarray}

                  Suppose the quantum state $|\Psi\rangle$ is in a superposition of only positive-energy states, i.e.,
                  \begin{eqnarray}\label{eq:D-20}
                        |\Psi\rangle\equiv|\psi_+\rangle = c_1 |\Psi_1\rangle
                              + c_2 |\Psi_2\rangle,
                  \end{eqnarray}
                  therefore
                  \begin{eqnarray}\label{eq:D-20b}
                        \mathcal{Z}_r=\langle\psi_+ | \hat{\mathcal{Z}}_r |\psi_+\rangle
                        =\langle\psi_+ | \left(\Pi_+ \; \hat{\mathcal{Z}}_r\;\Pi_+ \right)
                              |\psi_+\rangle=0.
                  \end{eqnarray}
                  Similarly, one has
                  \begin{eqnarray}\label{eq:D-20d}
                        \mathcal{Z}_r= \langle\psi_- | \hat{\mathcal{Z}}_r |\psi_-\rangle
                        = \langle\psi_- | \left(\Pi_- \; \hat{\mathcal{Z}}_r\;\Pi_- \right)|
                              \psi_-\rangle=0.
                  \end{eqnarray}
                  Hence, as Schr\"odinger has pointed out, if Dirac's electron is in a superposition of only positive- (or negative-) energy, then
                  $\mathcal{Z}_r=0$, and there is no any PZ phenomenon.

                  Let us now choose the state $|\Psi\rangle$ as a superposition of the positive-energy state $|\psi_+\rangle$ and the negative-energy state $|\psi_-\rangle$, i.e.,
                  \begin{eqnarray}\label{eq:D-21}
                        |\Psi\rangle=\cos\theta |\psi_+\rangle
                              +\sin\theta |\psi_-\rangle
                        =\cos\theta \left( c_1 |\Psi_1\rangle+ c_2 |\Psi_2\rangle \right)
                              +\sin\theta \left(c_3 |\Psi_3\rangle+c_4 |\Psi_4\rangle \right),
                  \end{eqnarray}
                  with $\theta\in[0,\pi/2]$. We then have
                  \begin{eqnarray}\label{eq:D-21aa}
                  && \mathcal{Z}_r = \langle \Psi| \hat{\mathcal{Z}}_r |\Psi\rangle
                        =\Bigl(\cos\theta\Bra{\psi_+}+\sin\theta\Bra{\psi_-}\Bigr)
                              \hat{\mathcal{Z}}_r\Bigl(\cos\theta\Ket{\psi_+}
                                    +\sin\theta\Ket{\psi_-}\Bigr) \nonumber\\
                        &&= \sin(2\theta) \Bigl[{\rm Re}\left( c_1^*c_3\langle \Psi_1|
                                    \hat{\mathcal{Z}}_r|\Psi_3 \rangle \right)
                              +{\rm Re}\left( c_1^*c_4\langle \Psi_1|
                                    \hat{\mathcal{Z}}_r|\Psi_4 \rangle\right)
                              +{\rm Re}\left( c_2^*c_3\langle \Psi_2|
                                    \hat{\mathcal{Z}}_r|\Psi_3 \rangle \right)
                              +{\rm Re}\left( c_2^*c_4\langle \Psi_2|
                                    \hat{\mathcal{Z}}_r|\Psi_4 \rangle \right)\Bigr].
                  \end{eqnarray}
                  Thus, we need to calculate the following four terms:
                  $\langle\Psi_1|\hat{\mathcal{Z}}|\Psi_3 \rangle,
                  \langle\Psi_1|\hat{\mathcal{Z}}|\Psi_4 \rangle,
                  \langle\Psi_2|\hat{\mathcal{Z}}|\Psi_3 \rangle,
                  \langle\Psi_2|\hat{\mathcal{Z}}|\Psi_4 \rangle$. In general, these four terms are not zeros.

                  For simplicity, we only consider the case with $c_1=c_3=1, c_2=c_4=0$, namely
                  \begin{eqnarray}\label{eq:D-21-ma}
                        |\Psi\rangle=\cos\theta |\Psi_1\rangle +\sin\theta |\Psi_3\rangle.
                  \end{eqnarray}
                  In this  simple case, we have
                  \begin{eqnarray}\label{eq:D-21-mb}
                        \mathcal{Z}_r=\langle \Psi| \hat{\mathcal{Z}}_r |\Psi\rangle
                        =\sin(2\theta) {\rm Re}\left[
                              \langle \Psi_1|\hat{\mathcal{Z}}_r |\Psi_3 \rangle \right],
                  \end{eqnarray}
                  where
                  \begin{eqnarray}\label{eq:D-22-a}
                        \langle\Psi_1|\hat{\mathcal{Z}}_r|\Psi_3 \rangle &=& \langle\Psi_1|
                              \left[\frac{{\rm i}\hbar c}{2}\left[\vec{\alpha}(0)
                                    -c H^{-1}_{\rm e}\vec{p}\right] H^{-1}_{\rm e} \left(
                                          {\rm e}^{\frac{-{\rm i}\,2\,H_{\rm e} t}{\hbar}}
                                                -1\right)\right]|\Psi_3\rangle \notag\\
                        &=& \frac{-{\rm i}\hbar c}{2 E_{\rm p}} \left({\rm e}^{
                              \frac{{\rm i}\,2\,E_{\rm p} t}{\hbar}} -1\right)\langle\Psi_1|
                                    \vec{\alpha}(0)|\Psi_3\rangle,
                  \end{eqnarray}
                  here in the last step we have used the orthogonal relation $\langle\Psi_1|\Psi_3 \rangle=0$.
                  Based on the eigenfunctions $\{|\Psi_1\rangle, |\Psi_3\rangle\}$ in Eq. (\ref{eq:mom}), after some careful calculations we have
                  \begin{eqnarray}
                        \mathcal{Z}_r &=& \langle \Psi| \hat{\mathcal{Z}}_r |\Psi\rangle
                        =\sin(2\theta){\rm Re}\left[
                              \langle \Psi_1|\hat{\mathcal{Z}}_r|\Psi_3 \rangle \right]
                        =\sin(2\theta){\rm Re}\left[
                              -\frac{-{\rm i}\,\hbar\,c}{2\,E_{\rm p}} \left(
                                    {\rm e}^{\frac{{\rm i}\,2\,E_{\rm p} t}{\hbar}} -1
                                          \right)
                                    \dfrac{mc^2}{E_{\rm p}} \dfrac{\vec{p}}{p}\right] \nonumber \\
                        &=& \sin(2\theta)\dfrac{\hbar c}{2 E_{\rm p}} \dfrac{mc^2}{E_{\rm p}}
                              \dfrac{\vec{p}}{p} {\rm Re}\left[
                                    {\rm i} \left({\rm e}^{\frac{{\rm i}\,2\,E_{\rm p} t}{\hbar}} -1 \right)\right]
                        =\hat{p}\;\sin(2\theta)
                              \dfrac{\hbar c \times mc^2}{2 E^2_{\rm p}} {\rm Re}\left[
                                    {\rm i} \left({\rm e}^{
                                          \frac{{\rm i}\,2\,E_{\rm p} t}{\hbar}} -1 \right) \right] \nonumber \\
                        &=& \hat{p}\;\sin(2\theta)\dfrac{\hbar c \times mc^2}{2 E^2_{\rm p}}
                              \left[-\sin\left(\frac{2\,E_{\rm p} t}{\hbar}\right)\right]
                        =-\hat{p}\;A \sin(\omega t).
                  \end{eqnarray}
                  Based on which, we have the amplitude and frequency as
                  \begin{eqnarray}\label{eq:D-21-e}
                        A=\sin(2\theta) \frac{\hbar c}{2 E_{\rm p}}
                              \dfrac{mc^2}{{E_{\rm p}}},\qquad
                        \omega=\frac{2\,E_{\rm p}}{\hbar}.
                  \end{eqnarray}
                  One may estimate the amplitude as
                  \begin{eqnarray}\label{eq:D-21-f}
                        A &=& \sin(2\theta) \frac{\hbar c}{2 E_{\rm p}} \dfrac{mc^2}{{E_{\rm p}}} \nonumber\\
                        &\leq &  \frac{\hbar c}{2 E_{\rm p}} \dfrac{mc^2}{{E_{\rm p}}}=\dfrac{\hbar c \times mc^2}{2E^2_{\rm p}}
                        =\dfrac{\hbar c \times mc^2}{2 (m^2c^4+p^2c^2)} \nonumber\\
                        &\leq &  \dfrac{\hbar c \times mc^2}{2 (m^2c^4)}=\frac{1}{4\pi} \dfrac{h}{mc}=\frac{1}{4\pi} \lambda_{\rm e}\nonumber\\
                        &\approx&1.9308 \times 10^{-13},
                  \end{eqnarray}
                  where
                  \begin{eqnarray}\label{eq:D-21-g}
                              \lambda_{\rm e} =\dfrac{h}{mc}\approx 2.42631\times 10^{-12} \;{\rm m},
                  \end{eqnarray}
                  is the Compton wavelength of an electron. This means that the amplitude $A$ of oscillation is less than $ \lambda_{\rm e}/(4\pi)\approx 1.9308 \times 10^{-13}$. Here some constants are
                  \begin{eqnarray}\label{eq:D-21-h}
                              h &=& 6.62607015\times 10^{-34} \; {\rm J\cdot s}, \nonumber\\
                              c &=& 2.99792458 \times 10^8 \;{\rm m/s}, \nonumber\\
                              m &=& 9.10938188\times 10^{-31} \; {\rm kg}\nonumber\\
                              \pi &=& 3.141592654.
                  \end{eqnarray}
                  By using the data of Eq. (\ref{eq:D-21-h}), one may easily obtain Eq. (
                        \ref{eq:D-21-g}). Further, one could estimate the frequency as
                  \begin{eqnarray}\label{eq:D-21-i}
                        \omega=\frac{2\,E_{\rm p}}{\hbar}
                        =\frac{2\,\sqrt{m^2c^4+p^2c^2}}{\hbar}\geq \frac{2\,mc^2}{\hbar}
                        =\frac{4\pi c}{\lambda_{\rm e}}
                        \approx {1.55269 \times 10^{21}\; {\rm s^{-1}}}.
                  \end{eqnarray}

                  Similarly, if $|\Psi\rangle$ is chosen as the superposition states of
                  $\{|\Psi_1\rangle,|\Psi_4\rangle\}$, $\{|\Psi_2\rangle,|\Psi_3\rangle\}$, or $\{|\Psi_2\rangle,|\Psi_4\rangle\}$, one generally has the Zitterbewegung
                  $\mathcal{Z}_r\neq 0$. In summary, only when Dirac's electron is in a superposition of positive-energy states and negative-energy states, is there a PZ phenomenon. This ends the proof.
            \end{proof}
      \subsection{The ``Spin Zitterbewegung'' of Electron}
            The spin operator has been introduced in Eq. (\ref{eq:spin}), i.e.,
            \begin{equation}
                  \vec{S}=\dfrac{\hbar}{2} \vec{\Sigma}\equiv\dfrac{\hbar}{2}
                        \begin{bmatrix}
                              \vec{\sigma} & 0 \\
                              0 & \vec{\sigma}
                        \end{bmatrix}.
            \end{equation}
            There are at least three different methods to derive the ``spin Zitterbewegung'' (SZ) operator. Here we introduce the first method, which is very simple and shows a direct connection with the PZ operator.

            Let us introduce the total angular momentum operator as
            \begin{equation}
            \vec{J}=\vec{L}+\vec{S},
            \end{equation}
            with
            \begin{equation}
            \vec{L}=\vec{r}\times \vec{p}
            \end{equation}
            being the orbital angular momentum operator. It is well-known that the operator $\vec{J}$ is conserved for a free electron, i.e.,
            \begin{equation}
            [\vec{J}, H_{\rm e}]=0,
            \end{equation}
            which leads to
            \begin{equation}\label{eq:total-1}
            \vec{J}(t)=\vec{J}(0)
            \end{equation}
            for any time $t$. Based on Eq. (\ref{eq:total-1}) one obtains
            \begin{equation}\label{eq:total-2}
            \vec{L}(t)+\vec{S}(t)=\vec{L}(0)+\vec{S}(0),
            \end{equation}
            i.e., the time evolution of the spin operator $\vec{S}(t)$ is given by
            \begin{eqnarray}
                        \vec{S}(t)&=&\vec{L}(0)+\vec{S}(0)-\vec{L}(t)=\vec{r}(0)\times\vec{p}(0)+\vec{S}(0)-\vec{r}(t)\times\vec{p}(t)= \vec{r}(0)\times\vec{p}+\vec{S}(0)-\vec{r}(t)\times\vec{p}\nonumber\\
                        &=& \vec{r}(0)\times\vec{p}+\vec{S}(0)-\biggl\{
                              \vec{r}(0)+c^2 H^{-1}_{\rm e} \vec{p}\,t
                              +\dfrac{{\rm i}\hbar c}{2}\left[\vec{\alpha}(0)
                                    -c\,H^{-1}_{\rm e} \vec{p}\right] H^{-1}_{\rm e}
                                          \left({\rm e}^{
                                                \frac{-{\rm i}\,2\,H_{\rm e}t}{\hbar}}
                                                -1\right)\biggr\}\times\vec{p} \nonumber\\
                        &=& \vec{r}(0)\times\vec{p}+\vec{S}(0)-\biggl\{
                              \vec{r}(0)+c^2 H^{-1}_{\rm e} \vec{p}\,t
                              +\hat{\mathcal{Z}}_r\biggr\}\times\vec{p} = \vec{r}(0)\times\vec{p}+\vec{S}(0)-\biggl\{
                              \vec{r}(0)\times\vec{p}+\hat{\mathcal{Z}}_r \times\vec{p}\biggr\}\nonumber\\
                        &=& \vec{S}(0)-\hat{\mathcal{Z}}_r \times\vec{p}.
            \end{eqnarray}
            Based on $\vec{S}(t)$, we immediately extract out its second term as the SZ operator, i.e.,
            \begin{eqnarray}
            \hat{\mathcal{Z}}_s=-\hat{\mathcal{Z}}_r \times\vec{p},
            \end{eqnarray}
            or explicitly
            \begin{eqnarray}
            \hat{\mathcal{Z}}_s=-\hat{\mathcal{Z}}_r \times\vec{p}=-\dfrac{{\rm i}\hbar c}{2}\left[
                                    \vec{\alpha}(0)\times\vec{p}\right] H^{-1}_{\rm e}
                                          \left({\rm e}^{\frac{-{\rm i}\,2\,H_{\rm e}t}{
                                                \hbar}} -1\right).
            \end{eqnarray}
            When Dirac's electron is in a quantum state $|\Psi\rangle$, the expectation value of the SZ operator reads
            \begin{eqnarray}\label{eq:sr-1}
                  \mathcal{Z}_s=\langle \Psi| \hat{\mathcal{Z}}_s |\Psi\rangle= -\langle \Psi| \hat{\mathcal{Z}}_r |\Psi\rangle \times\vec{p}.
            \end{eqnarray}
            If $\mathcal{Z}_s\neq 0$, we then say that there is a SZ phenomenon for Dirac's electron.

            \begin{remark}\emph{The Second Method of Deriving $\vec{S}(t)$}.
            One can check that
                  \begin{equation}\label{eq:SAlpha}
                        \begin{split}
                              \vec{S}=-{\rm i}\dfrac{\hbar}{4} \vec{\alpha}\times\vec{\alpha}.
                        \end{split}
                  \end{equation}
                  Then by combining \Eq{eq:D-7} with \Eq{eq:SAlpha}, we get
                  \begin{equation}
                        \begin{split}
                              & \vec{S}(t)=-{\rm i}\dfrac{\hbar}{4} \vec{\alpha}(t)
                                    \times\vec{\alpha}(t)
                              =-{\rm i}\dfrac{\hbar}{4} \biggl\{
                                          c\,H_{\rm e}^{-1} \vec{p}+{\rm e}^{
                                                \frac{{\rm i}\,2\,H_{\rm e}t}{\hbar}}
                                                      \left[\vec{\alpha}(0)
                                                            -c\,H_{\rm e}^{-1} \vec{p}
                                                            \right]\biggr\}
                                    \times\biggl\{c\,H_{\rm e}^{-1} \vec{p}
                                          +{\rm e}^{\frac{{\rm i}\,2\,H_{\rm e}t}{\hbar}}
                                                \left[\vec{\alpha}(0)-c\,H_{\rm e}^{-1}
                                                      \vec{p}\right]\biggr\} \\
                              =& -{\rm i}\dfrac{\hbar}{4} \biggl\{
                                          c\,H_{\rm e}^{-1} \vec{p}
                                          +\left[\vec{\alpha}(0)-c\,H_{\rm e}^{-1}
                                                \vec{p}\right]{\rm e}^{
                                                      -\frac{{\rm i}\,2\,H_{\rm e}t}{
                                                            \hbar}}\biggr\}
                                    \times\biggl\{c\,H_{\rm e}^{-1} \vec{p}
                                          +\left[\vec{\alpha}(0)-c\,H_{\rm e}^{-1}
                                                \vec{p}\right]{\rm e}^{
                                                      -\frac{{\rm i}\,2\,H_{\rm e}t}{\hbar}}
                                                      \biggr\} \\
                              =& -{\rm i}\dfrac{\hbar}{4} \biggl\{
                                          c\,H_{\rm e}^{-1} \vec{p}\times\left[
                                                \vec{\alpha}(0)-c\,H_{\rm e}^{-1}
                                                      \vec{p}\right]{\rm e}^{
                                                            -\frac{{\rm i}\,2\,H_{\rm e}
                                                                  t}{\hbar}}
                                    +c\left[\vec{\alpha}(0)-c\,H_{\rm e}^{-1} \vec{p}
                                          \right]\times\vec{p}\,H_{\rm e}^{-1} {\rm e}^{
                                                -\frac{{\rm i}\,2\,H_{\rm e}t}{\hbar}}
                                                \\
                                    &\qquad +\left[\vec{\alpha}(0)-c\,H_{\rm e}^{-1}
                                                \vec{p}\right]{\rm e}^{
                                                      -\frac{{\rm i}\,2\,H_{\rm e}t}{
                                                            \hbar}}
                                          \times\left[\vec{\alpha}(0)-c\,H_{\rm e}^{-1}
                                                \vec{p}\right]{\rm e}^{
                                                      -\frac{{\rm i}\,2\,H_{\rm e}t}{
                                                            \hbar}}\biggr\} \\
                              =& -{\rm i}\dfrac{\hbar}{4} \biggl\{c\,H_{\rm e}^{-1}
                                          \left[\vec{p}\times\vec{\alpha}(0)\right]
                                                {\rm e}^{-\frac{{\rm i}\,2\,H_{\rm e} t}
                                                      {\hbar}}
                                    +c\left[\vec{\alpha}(0)\times\vec{p}\right]
                                          H_{\rm e}^{-1} {\rm e}^{
                                                -\frac{{\rm i}\,2\,H_{\rm e}t}{\hbar}}
                                    +\left[\vec{\alpha}(0)-c\,H_{\rm e}^{-1}
                                                \vec{p}\right]{\rm e}^{
                                                      -\frac{{\rm i}\,2\,H_{\rm e}t}{
                                                            \hbar}}
                                          \times{\rm e}^{\frac{{\rm i}\,2\,H_{\rm e}t}{
                                                \hbar}}\left[\vec{\alpha}(0)
                                                      -c\,H_{\rm e}^{-1} \vec{p}\right]
                                                      \biggr\} \\
                              =& -{\rm i}\dfrac{\hbar}{4} \biggl\{-c\left[
                                          \vec{p}\times\vec{\alpha}(0)\right]
                                                H_{\rm e}^{-1}{\rm e}^{
                                                      -\frac{{\rm i}\,2\,H_{\rm e} t}{
                                                            \hbar}}
                                    +c\left[\vec{\alpha}(0)\times\vec{p}\right]
                                          H_{\rm e}^{-1} {\rm e}^{
                                                -\frac{{\rm i}\,2\,H_{\rm e}t}{\hbar}}
                                    +\left[\vec{\alpha}(0)-c\,H_{\rm e}^{-1} \vec{p}
                                          \right]\times\left[\vec{\alpha}(0)
                                                -c\,H_{\rm e}^{-1} \vec{p}\right]
                                                \biggr\} \\
                              =& -{\rm i}\dfrac{\hbar}{4} \biggl\{
                                    \vec{\alpha}(0)\times\vec{\alpha}(0)
                                    +c\bigl[\vec{\alpha}(0)\times\vec{p}
                                          -\vec{p}\times\vec{\alpha}(0)\bigr]
                                                H_{\rm e}^{-1} {\rm e}^{
                                                      -\frac{{\rm i}\,2\,H_{\rm e}t}{
                                                            \hbar}}
                                    -c\bigl[\vec{\alpha}(0)\times\vec{p}
                                          -\vec{p}\times\vec{\alpha}(0)\bigr]
                                                H_{\rm e}^{-1}\biggr\} \\
                              =& -{\rm i}\dfrac{\hbar}{4} \biggl\{
                                    \vec{\alpha}(0)\times\vec{\alpha}(0)
                                    +2\,c\bigl[\vec{\alpha}(0)\times\vec{p}\bigr]
                                          H_{\rm e}^{-1} {\rm e}^{
                                                -\frac{{\rm i}\,2\,H_{\rm e}t}{\hbar}}
                                    -2\,c\bigl[\vec{\alpha}(0)\times\vec{p}\bigr]
                                          H_{\rm e}^{-1}\biggr\} \\
                              =& \vec{S}(0)-\dfrac{{\rm i}\hbar c}{2}\left[
                                    \vec{\alpha}(0)\times\vec{p}\right] H^{-1}_{\rm e}
                                          \left({\rm e}^{\frac{-{\rm i}\,2\,H_{\rm e}t}{
                                                \hbar}} -1\right).
                        \end{split}
                  \end{equation}
            \end{remark}
            \begin{remark}\emph{The Third Method of Deriving $\vec{S}(t)$}.
                  Since
                  \begin{equation}
                        \begin{split}
                              & \left[\vec{S},\ \vec{\alpha}\cdot\vec{p}\right]_x
                              =\left[S_x,\ \vec{\alpha}\cdot\vec{p}\right]
                              =S_x (\vec{\alpha}\cdot\vec{p})-(\vec{\alpha}\cdot\vec{p})S_x \\
                              =& \dfrac{\hbar}{2} \begin{bmatrix}
                                                \sigma_x & 0 \\
                                                0 & \sigma_x
                                          \end{bmatrix}\begin{bmatrix}
                                                0 & \sigma_j\,p_j \\
                                                \sigma_j\,p_j & 0
                                          \end{bmatrix}
                                    -\dfrac{\hbar}{2} \begin{bmatrix}
                                                0 & \sigma_j\,p_j \\
                                                \sigma_j\,p_j & 0
                                          \end{bmatrix}\begin{bmatrix}
                                                \sigma_x & 0 \\
                                                0 & \sigma_x
                                          \end{bmatrix}
                              =\dfrac{\hbar}{2} \begin{bmatrix}
                                          0 & \sigma_x (\sigma_j\,p_j)
                                                -(\sigma_j\,p_j)\sigma_x  \\
                                          \sigma_x (\sigma_j\,p_j)
                                                -(\sigma_j\,p_j)\sigma_x & 0
                                    \end{bmatrix} \\
                              =& \dfrac{\hbar}{2} \begin{bmatrix}
                                          0 & \bigl[\openone\,p_x -{\rm i}(\sigma_y\,p_z
                                                -\sigma_z\,p_y)\bigr]-\bigl[
                                                      \openone\,p_x +{\rm i}(\sigma_y\,p_z
                                                            -\sigma_z\,p_y)\bigr] \\
                                          \bigl[\openone\,p_x -{\rm i}(\sigma_y\,p_z
                                                -\sigma_z\,p_y)\bigr]-\bigl[
                                                      \openone\,p_x +{\rm i}(\sigma_y\,p_z
                                                            -\sigma_z\,p_y)\bigr] & 0
                                    \end{bmatrix} \\
                              =& -{\rm i}\hbar\begin{bmatrix}
                                          0 & (\vec{\sigma}\times\vec{p})_x \\
                                          (\vec{\sigma}\times\vec{p})_x & 0
                                    \end{bmatrix}
                              =-{\rm i}\hbar(\vec{\alpha}\times\vec{p})_x,
                        \end{split}
                  \end{equation}
                  where $j=x,y,z$, and Einstein's summation convention is adopted. Then we have
                  \begin{equation}
                        \left[\vec{S},\ \vec{\alpha}\cdot\vec{p}\right]
                        =-{\rm i}\hbar\,\vec{\alpha}\times\vec{p}.
                  \end{equation}
                  Besides,
                  \begin{equation}
                        \left[\vec{S},\ \beta\right]=\vec{S}\,\beta-\beta\,\vec{S}
                        =\dfrac{\hbar}{2} \begin{bmatrix}
                                          \vec{\sigma} & 0 \\
                                          0 & \vec{\sigma}
                                    \end{bmatrix}\begin{bmatrix}
                                          \openone & 0 \\
                                          0 & -\openone
                                    \end{bmatrix}
                              -\dfrac{\hbar}{2} \begin{bmatrix}
                                          \openone & 0 \\
                                          0 & -\openone
                                    \end{bmatrix}\begin{bmatrix}
                                          \vec{\sigma} & 0 \\
                                          0 & \vec{\sigma}
                                    \end{bmatrix}
                        =\dfrac{\hbar}{2} \begin{bmatrix}
                                          \vec{\sigma} & 0 \\
                                          0 & -\vec{\sigma}
                                    \end{bmatrix}
                              -\dfrac{\hbar}{2} \begin{bmatrix}
                                          \vec{\sigma} & 0 \\
                                          0 & -\vec{\sigma}
                                    \end{bmatrix}
                        =0.
                  \end{equation}
                  After that,
                  \begin{eqnarray}
                        && \dfrac{{\rm d}\,\vec{S}}{{\rm d}\,t}
                        =\dfrac{1}{{\rm i}\hbar} \left[\vec{S}, H_{\rm e}\right]
                        =\dfrac{1}{{\rm i}\hbar} \left[\vec{S},\
                              c\,\vec{\alpha}\cdot\vec{p}+\beta\,m\,c^2\right]
                        =\dfrac{c}{{\rm i}\hbar} \left[\vec{S},\ \vec{\alpha}\cdot\vec{p}
                              \right]
                        =-c\,\vec{\alpha}\times\vec{p} \notag\\
                        &=& -c\Bigl\{c\,H_{\rm e}^{-1} \vec{p}+{\rm e}^{
                              \frac{{\rm i}\,2\,H_{\rm e}t}{\hbar}} \left[\vec{\alpha}(0)
                                    -c\,H_{\rm e}^{-1} \vec{p}\right]\Bigr\}\times\vec{p}
                        =-c\,{\rm e}^{\frac{{\rm i}\,2\,H_{\rm e}t}{\hbar}} \left[
                              \vec{\alpha}(0)\times\vec{p}\right],
                  \end{eqnarray}
                  then
                  \begin{equation}
                        \vec{S}(t)=-c\int\Bigl\{{\rm e}^{\frac{{\rm i}\,2\,H_{\rm e}t}{
                              \hbar}} \left[\vec{\alpha}(0)\times\vec{p}\right]\Bigr\}{\rm d}t
                        =-c\int\Bigl\{{\rm e}^{
                              \frac{{\rm i}\,2\,H_{\rm e}t}{\hbar}} {\rm d}t\Bigr\}\left[
                                    \vec{\alpha}(0)\times\vec{p}\right]
                        =-\dfrac{c\,\hbar}{{\rm i}\,2\,H_{\rm e}}\,{\rm e}^{
                              \frac{{\rm i}\,2\,H_{\rm e}t}{\hbar}}\left[
                                    \vec{\alpha}(0)\times\vec{p}\right]+C.
                  \end{equation}
                  Because
                  \begin{eqnarray}
                        \Bigl\{H_{\rm e},\ \left[\vec{\alpha}(0)\times\vec{p}\right]
                                    \Bigr\}_x
                              &=&\Bigl\{H_{\rm e},\ (\alpha_y\,p_z -\alpha_z\,p_y)\Bigr\}
                              =c\Bigl\{\vec{\alpha}\cdot\vec{p},\
                                    (\alpha_y\,p_z -\alpha_z\,p_y)\Bigr\}\nonumber\\
                              &=& c\Bigl\{\vec{\alpha}\cdot\vec{p},\ \alpha_y\,p_z\Bigr\}
                                    -c\Bigl\{\vec{\alpha}\cdot\vec{p},\ \alpha_z\,p_y\Bigr\}
                              = c\Bigl\{\vec{\alpha}\cdot\vec{p},\ \alpha_y\Bigr\}p_z
                                    -c\Bigl\{\vec{\alpha}\cdot\vec{p},\ \alpha_z\Bigr\}p_y\nonumber\\
                              &=&c\,{\bf 1}\bigl(2\,p_y\bigr)p_z
                                    -c\,{\bf 1}\bigl(2\,p_z\bigr)p_y
                              =0,
                  \end{eqnarray}
                  i.e.
                  $\bigl\{H_{\rm e},\ \left[\vec{\alpha}(0)\times\vec{p}\right]\bigr\}=0$,
                  thus
                  \begin{eqnarray}
                        H_{\rm e} \left[\vec{\alpha}(0)\times\vec{p}\right]
                        =-\left[\vec{\alpha}(0)\times\vec{p}\right]H_{\rm e},
                  \end{eqnarray}
                  which leads to
                  \begin{eqnarray}
                        {\rm e}^{\frac{{\rm i}\,2\,H_{\rm e}t}{\hbar}} \left[
                              \vec{\alpha}(0)\times\vec{p}\right]
                        =\left[\vec{\alpha}(0)\times\vec{p}\right] {\rm e}^{
                              \frac{-{\rm i}\,2\,H_{\rm e}t}{\hbar}}.
                  \end{eqnarray}
                  For the sake of $\lim_{t\to 0} \vec{S}(t)=\vec{S}(0)$, we obtain
                  \begin{equation}
                        -\dfrac{c\,\hbar}{{\rm i}\,2\,H_{\rm e}} \left[
                              \vec{\alpha}(0)\times\vec{p}\right]+C=\vec{S}(0),
                  \end{equation}
                  which leads to
                  \begin{equation}
                        C=\vec{S}(0)+\dfrac{c\,\hbar}{{\rm i}\,2\,H_{\rm e}} \left[
                              \vec{\alpha}(0)\times\vec{p}\right]
                        =\vec{S}(0)-\dfrac{{\rm i}\,c\,\hbar}{2\,H_{\rm e}} \left[
                              \vec{\alpha}(0)\times\vec{p}\right]
                        =\vec{S}(0)+\dfrac{{\rm i}\,c\,\hbar}{2\,} \left[
                              \vec{\alpha}(0)\times\vec{p}\right]H_{\rm e}^{-1},
                  \end{equation}
                  then
                  \begin{equation}
                        \begin{split}
                              & \vec{S}(t)=-\dfrac{c\,\hbar}{{\rm i}\,2\,H_{\rm e}}\,
                                    {\rm e}^{\frac{{\rm i}\,2\,H_{\rm e}t}{\hbar}} \left[
                                          \vec{\alpha}(0)\times\vec{p}\right]+C
                              =\dfrac{{\rm i}\,c\,\hbar}{2\,H_{\rm e}}\,
                                    {\rm e}^{\frac{{\rm i}\,2\,H_{\rm e}t}{\hbar}} \left[
                                          \vec{\alpha}(0)\times\vec{p}\right]+C
                              =\dfrac{{\rm i}\,c\,\hbar}{2\,H_{\rm e}} \left[
                                    \vec{\alpha}(0)\times\vec{p}\right]{\rm e}^{
                                          \frac{-{\rm i}\,2\,H_{\rm e}t}{\hbar}} +C \\
                              =& -\dfrac{{\rm i}\,c\,\hbar}{2} \left[
                                    \vec{\alpha}(0)\times\vec{p}\right]H_{\rm e}^{-1}
                                          {\rm e}^{\frac{-{\rm i}\,2\,H_{\rm e}t}{\hbar}} +C
                              =\vec{S}(0)-\dfrac{{\rm i}\hbar c}{2}\left[
                                    \vec{\alpha}(0)\times\vec{p}\right] H^{-1}_{\rm e}
                                          \left({\rm e}^{\frac{-{\rm i}\,2\,H_{\rm e}t}{
                                                \hbar}} -1\right).
                        \end{split}
                  \end{equation}
            \end{remark}

            Similar to Schr\"odinger, for the SZ we have the following observation.

            \emph{Observation 2.---}Only when Dirac's electron is in a superposition state with opposite energy and opposite helicity (i.e.,
            $|\Psi\rangle=\cos\theta |\psi_{+,+\lambda}\rangle +\sin\theta |\psi_{-,-\lambda}\rangle$, $\lambda=\hbar/2$), is there a SZ phenomenon.

            \begin{proof} We need to calculate the expectation values of the SZ operator. From Eq. (\ref{eq:sr-1}) we have known that
            \begin{eqnarray}\label{eq:sr-2}
                  \mathcal{Z}_s= -\langle \Psi| \hat{\mathcal{Z}}_r |\Psi\rangle \times\vec{p}.
            \end{eqnarray}
            Therefore, if $\mathcal{Z}_r= 0$, one must have $\mathcal{Z}_s= 0$. Based on the \emph{observation 1}, to obtain the non-zero expectation value of $\mathcal{Z}_s$, at least Dirac's electron is in a superposition state with opposite energy. In the following, we further show that $\mathcal{Z}_s= 0$ if Dirac's electron is in a superposition state with the same helicity.

            Let us introduce the projective operators
            \begin{eqnarray}
                  \Pi^s_\pm =\frac{1}{2}\left({\bf 1}\pm \dfrac{2}{\hbar} \hat{\Lambda}
                        \right),\qquad
                  (\Pi^s_\pm)^2 =\Pi^s_\pm,
            \end{eqnarray}
            we easily have
            \begin{eqnarray}
                  && \Pi^S_+ |\Psi_1\rangle = |\Psi_1\rangle, \;\;
                  \Pi^S_+ |\Psi_2\rangle = 0, \;\;
                  \Pi^S_+ |\Psi_3\rangle = |\Psi_3\rangle, \;\;
                  \Pi^S_+ |\Psi_4\rangle = 0, \nonumber\\
                  && \Pi^S_- |\Psi_1\rangle = 0, \;\;
                  \Pi^S_- |\Psi_2\rangle = |\Psi_2\rangle, \;\;
                  \Pi^S_- |\Psi_3\rangle = 0, \;\;
                  \Pi^S_- |\Psi_4\rangle = |\Psi_4\rangle.
            \end{eqnarray}
            Because
            \begin{equation}
                  \begin{split}
                        & \bigl\{\hat{\Lambda},\ \left[\vec{\alpha}(0)\times\vec{p}
                              \right]\bigr\}
                        =\dfrac{\hbar}{2} \Bigl\{\vec{\Sigma}\cdot\hat{p},\
                              \vec{\alpha}\times\vec{p}\Bigr\}
                        =\dfrac{\hbar}{2} \Bigl\{\vec{\Sigma}\cdot\hat{p},\
                              \vec{\alpha}\Bigr\}\times\vec{p} \\
                        =& \dfrac{\hbar}{2} \Biggl(\begin{bmatrix}
                                          \vec{\sigma}\cdot\hat{p} & 0 \\
                                          0 & \vec{\sigma}\cdot\hat{p}
                                    \end{bmatrix}\begin{bmatrix}
                                          0 & \vec{\sigma} \\
                                          \vec{\sigma} & 0
                                    \end{bmatrix}
                              +\begin{bmatrix}
                                          0 & \vec{\sigma} \\
                                          \vec{\sigma} & 0
                                    \end{bmatrix}\begin{bmatrix}
                                          \vec{\sigma}\cdot\hat{p} & 0 \\
                                          0 & \vec{\sigma}\cdot\hat{p}
                                    \end{bmatrix}\Biggr)\times\vec{p}\\
                        =&\dfrac{\hbar}{2} \Biggl\{\begin{bmatrix}
                                          0 & (\vec{\sigma}\cdot\hat{p})
                                                \vec{\sigma} \\
                                          (\vec{\sigma}\cdot\hat{p})
                                                \vec{\sigma} & 0
                                    \end{bmatrix}
                              +\begin{bmatrix}
                                          0 & \vec{\sigma}(
                                                \vec{\sigma}\cdot\hat{p}) \\
                                          \vec{\sigma}(\vec{\sigma}\cdot\hat{p})
                                                & 0
                                    \end{bmatrix}\Biggr\}\times\vec{p} \\
                        =& \dfrac{\hbar}{2} \begin{bmatrix}
                                    0 & (\vec{\sigma}\cdot\hat{p})\vec{\sigma}
                                          +\vec{\sigma}(\vec{\sigma}\cdot\hat{p})
                                          \\
                                    (\vec{\sigma}\cdot\hat{p})\vec{\sigma}
                                          +\vec{\sigma}(\vec{\sigma}\cdot
                                                \hat{p}) & 0
                              \end{bmatrix}\times\vec{p}
                        =\dfrac{\hbar}{2} \begin{bmatrix}
                                    0 & 2\,\hat{p} \\
                                    2\,\hat{p} & 0
                              \end{bmatrix}\times\vec{p}
                        =0.
                  \end{split}
            \end{equation}
            We then have
            \begin{eqnarray}
                  && \Pi^s_+ \left[\vec{\alpha}(0)\times\vec{p}\right]\Pi^s_+
                  =\frac{1}{4} \left({\bf 1}+\dfrac{2}{\hbar} \hat{\Lambda}\right)\left[
                        \vec{\alpha}(0)\times\vec{p}\right]\left({\bf 1}
                              +\dfrac{2}{\hbar} \hat{\Lambda}\right) \notag\\
                  &=& \frac{1}{4} \left(\left[\vec{\alpha}(0)\times\vec{p}\right]
                        +\frac{2}{\hbar} \bigl\{\hat{\Lambda},\
                              \left[\vec{\alpha}(0)\times\vec{p}\right]\bigr\}
                        +\frac{4}{\hbar^4} \hat{\Lambda} \left[
                              \vec{\alpha}(0)\times\vec{p}\right]\hat{\Lambda}\right)
                              \nonumber\\
                  &=& \frac{1}{4} \left[\left[\vec{\alpha}(0)\times\vec{p}\right]
                        -\frac{4}{\hbar^2} \hat{\Lambda}^2 \left[
                              \vec{\alpha}(0)\times\vec{p}\right] \right]
                  =\frac{1}{4} \Bigl\{\left[\vec{\alpha}(0)\times\vec{p}\right]
                        -\left[\vec{\alpha}(0)\times\vec{p}\right]\Bigr\}
                  =0,
            \end{eqnarray}
            similarly, we have
            \begin{eqnarray}
                  \Pi^s_- \left[\vec{\alpha}(0)\times\vec{p}\right]\Pi^s_- =0.
            \end{eqnarray}
            The above results lead to
            \begin{eqnarray}
                  \Pi^s_+ \hat{\mathcal{Z}}_s \Pi^s_+ =0,\;\;\;\;\;
                  \Pi^s_- \hat{\mathcal{Z}}_s \Pi^s_- =0.
            \end{eqnarray}
            Thus, if $|\Psi\rangle$ is in a superposition state of the same helicity, one will have
            \begin{eqnarray}
                  \mathcal{Z}_s=\langle \Psi|\hat{\mathcal{Z}}_s|\Psi\rangle=  \langle \Psi|\Pi^s_+ \hat{\mathcal{Z}}_s \Pi^s_+|\Psi\rangle =0,\;\;\;{\rm or}\;\;\;
                  \mathcal{Z}_s=\langle \Psi|\hat{\mathcal{Z}}_s|\Psi\rangle=  \langle \Psi|\Pi^s_- \hat{\mathcal{Z}}_s \Pi^s_-|\Psi\rangle =0.
            \end{eqnarray}
            In summary, only when Dirac's electron is in a superposition state with opposite energy and opposite helicity, is there a SZ phenomenon. This ends the proof.
            \end{proof}

            \emph{Example 2.---}Dirac's electron is in the following superposition state
            \begin{eqnarray}
            |\Psi\rangle=\cos\theta |\Psi_1\rangle +\sin\theta |\Psi_4\rangle,
            \end{eqnarray}
            where $|\Psi_1\rangle$ is the eigenfunction with positive energy and positive helicity, while $|\Psi_3\rangle$ is the eigenfunction with negative energy and negative helicity. We then have
            \begin{eqnarray}
                  \mathcal{Z}_s &=& \langle\Psi| \hat{\mathcal{Z}}_s |\Psi\rangle
                  =\Bigl(\cos\theta\Bra{\Psi_1}+\sin\theta\Bra{\Psi_4}\Bigr)
                        \hat{\mathcal{Z}}_s\Bigl(\cos\theta\Ket{\Psi_1}
                              +\sin\theta\Ket{\Psi_4}\Bigr) = \sin(2\theta) \;{\rm Re}\left[\langle \Psi_1|
                              \hat{\mathcal{Z}}_s|\Psi_4 \rangle \right].
            \end{eqnarray}
            where
            \begin{eqnarray}
                        \langle\Psi_1|\hat{\mathcal{Z}}_S|\Psi_4\rangle
                  &=&-\dfrac{{\rm i}\hbar c}{2} \langle\Psi_1|\biggl\{\left[
                        \vec{\alpha}(0)\times\vec{p}\right] H^{-1}_{\rm e} \left(
                              {\rm e}^{\frac{-{\rm i}\,2\,H_{\rm e}t}{\hbar}} -1
                              \right)\biggr\}|\Psi_4\rangle\nonumber\\
                  &=&-\dfrac{{\rm i}\hbar c}{2} \left({\rm e}^{
                        \frac{{\rm i}\,2\,E_{\rm p} t}{\hbar}} -1\right)\langle\Psi_1|
                              \biggl\{\left[\vec{\alpha}(0)\times\vec{p}\right]
                                    H^{-1}_{\rm e}\biggr\}|\Psi_4\rangle \notag\\
                  &=& \dfrac{{\rm i}\hbar c}{2\,E_{\rm p}} \left({\rm e}^{
                        \frac{{\rm i}\,2\,E_{\rm p} t}{\hbar}} -1\right)\langle\Psi_1|
                              \left[\vec{\alpha}(0)\times\vec{p}\right]|\Psi_4\rangle.
            \end{eqnarray}
            Using the eigenstates in Eq. (\ref{eq:mom}), the direct calculation shows that
            \begin{eqnarray}
                  \langle\Psi_1|\vec{\alpha}|\Psi_4\rangle
                  =\left[1-\dfrac{p_x (p_x -{\rm i}\,p_y)}{p(p+p_z)}\right]
                              \vec{e}_x
                        -\left[{\rm i}+\dfrac{p_y (p_x -{\rm i}\,p_y)}{p(p+p_z)}
                              \right]\vec{e}_y
                        -\dfrac{(p_x -{\rm i}\,p_y)}{p} \vec{e}_z,
            \end{eqnarray}
            then one obtains
            \begin{equation}
                  \begin{split}
                        & \langle\Psi_1|\left[\vec{\alpha}(0)\times\vec{p}\right]|
                              \Psi_4\rangle
                        =\langle\Psi_1|\vec{\alpha}|\Psi_4\rangle\times\vec{p}\\
                        =&\Biggl\{\left[1-\dfrac{p_x (p_x -{\rm i}\,p_y)}{p(p+p_z)}
                                    \right]\vec{e}_x
                              -\left[{\rm i}+\dfrac{p_y (p_x -{\rm i}\,p_y)}{p(p+p_z)}
                                    \right]\vec{e}_y
                              -\dfrac{(p_x -{\rm i}\,p_y)}{p} \vec{e}_z\Biggr\}
                              \times\vec{p} \\
                        =& \Biggl\{-\left[{\rm i}
                                    +\dfrac{p_y (p_x -{\rm i}\,p_y)}{p(p+p_z)}\right]p_z
                                    +\dfrac{(p_x -{\rm i}\,p_y)}{p} p_y\Biggr\}
                                    \vec{e}_x
                              +\Biggl\{-\dfrac{(p_x -{\rm i}\,p_y)}{p} p_x
                                    -\left[1-\dfrac{p_x (p_x -{\rm i}\,p_y)}{p(p+p_z)}
                                          \right]p_z\Biggr\}\vec{e}_y \\
                              & +\Biggl\{\left[1
                                          -\dfrac{p_x (p_x -{\rm i}\,p_y)}{p(p+p_z)}
                                          \right]p_y
                                    +\left[{\rm i}
                                          +\dfrac{p_y (p_x -{\rm i}\,p_y)}{p(p+p_z)}
                                          \right]p_x\Biggr\}\vec{e}_z \\
                        =& \left[\dfrac{p_y (p_x -{\rm i}\,p_y)}{p+p_z}
                                    -{\rm i}\,p_z\right]\vec{e}_x
                              -\left[\dfrac{p_x (p_x -{\rm i}\,p_y)}{p+p_z} +p_z\right]
                                    \vec{e}_y
                              +({\rm i}\,p_x +p_y)\vec{e}_z \\
                        =& \Biggl[\dfrac{p_x p_y}{p+p_z} \vec{e}_x
                                    -\biggl(\dfrac{p_x^2}{p+p_z} +p_z\biggr) \vec{e}_y
                                    +p_y \vec{e}_z\Biggr]
                              +{\rm i}\Biggl[-\biggl(\dfrac{p_y^2}{p+p_z} +p_z\biggr)
                                          \vec{e}_x
                                    +\dfrac{p_x p_y}{p+p_z} \vec{e}_y +p_x \vec{e}_z
                                    \Biggr].
                  \end{split}
            \end{equation}
            Based on the above results, we have
            \begin{eqnarray}
                  && \langle\Psi_1|\hat{\mathcal{Z}}_S|\Psi_4\rangle
                  =\dfrac{{\rm i}\hbar c}{2\,E_{\rm p}} \left({\rm e}^{
                        \frac{{\rm i}\,2\,E_{\rm p} t}{\hbar}} -1\right)\langle\Psi_1|
                              \left[\vec{\alpha}(0)\times\vec{p}\right]|\Psi_4\rangle
                              \notag\\
                  &=& \dfrac{{\rm i}\hbar c}{2\,E_{\rm p}} \left({\rm e}^{
                        \frac{{\rm i}\,2\,E_{\rm p} t}{\hbar}} -1\right)\Biggl\{
                              \biggl[\dfrac{p_x p_y}{p+p_z} \vec{e}_x
                                    -\Bigl(\dfrac{p_x^2}{p+p_z} +p_z\Bigr) \vec{e}_y
                                    +p_y \vec{e}_z\biggr]
                              +{\rm i}\biggl[-\Bigl(\dfrac{p_y^2}{p+p_z} +p_z\Bigr)
                                          \vec{e}_x
                                    +\dfrac{p_x p_y}{p+p_z} \vec{e}_y +p_x \vec{e}_z
                                    \biggr]\Biggr\}  \notag\\
                  &=& \dfrac{\hbar c}{2\,E_{\rm p}} \Biggl\{{\rm i}\biggl[\cos\left(
                                    \frac{2\,E_{\rm p} t}{\hbar}\right)-1\Biggr]
                              -\sin\left(
                                    \frac{2\,E_{\rm p} t}{\hbar}\right)\Biggr\}\times \notag\\
                        && \Biggl\{\biggl[\dfrac{p_x p_y}{p+p_z} \vec{e}_x
                                    -\Bigl(\dfrac{p_x^2}{p+p_z} +p_z\Bigr) \vec{e}_y
                                    +p_y \vec{e}_z\biggr]
                              +{\rm i}\biggl[-\Bigl(\dfrac{p_y^2}{p+p_z} +p_z\Bigr)
                                          \vec{e}_x
                                    +\dfrac{p_x p_y}{p+p_z} \vec{e}_y +p_x \vec{e}_z
                                    \biggr]\Biggr\},
            \end{eqnarray}
            which means
            \begin{equation}
                  \begin{split}
                              {\rm Re}\bigl(\bra{\Psi_1}\hat{\mathcal{Z}}_s\ket{\Psi_4}
                              \bigr) & = -\dfrac{\hbar c}{2\,E_{\rm p}} \Biggl\{\biggl[\cos\left(
                                    \frac{2\,E_{\rm p} t}{\hbar}\right)-1\biggr]\biggl[
                                          -\Bigl(\dfrac{p_y^2}{p+p_z} +p_z\Bigr)
                                                \vec{e}_x
                                          +\dfrac{p_x p_y}{p+p_z} \vec{e}_y
                                          +p_x \vec{e}_z\biggr]\\
                              & \qquad\qquad+\sin\left(\frac{2\,E_{\rm p} t}{\hbar}\right)\biggl[
                                    \dfrac{p_x p_y}{p+p_z} \vec{e}_x
                                    -\Bigl(\dfrac{p_x^2}{p+p_z} +p_z\Bigr)\vec{e}_y
                                    +p_y \vec{e}_z\biggr]\Biggr\} \\
                        =& -\dfrac{\hbar c}{2\,E_{\rm p}} \Biggl(
                              \biggl\{-\Bigl[\cos\Bigl(\frac{2\,E_{\rm p} t}{\hbar}
                                          \Bigr)-1\Bigr]\Bigl(\dfrac{p_y^2}{p+p_z}
                                                +p_z\Bigr)
                                    +\sin\Bigl(\frac{2\,E_{\rm p} t}{\hbar}\Bigr)
                                          \dfrac{p_x p_y}{p+p_z}\biggr\}\vec{e}_x \\
                              &\qquad\qquad +\biggl\{\Bigl[\cos\Bigl(\frac{2\,E_{\rm p} t}{
                                          \hbar}\Bigr)-1\Bigr]\dfrac{p_x p_y}{p+p_z}
                                    -\sin\Bigl(\frac{2\,E_{\rm p} t}{\hbar}\Bigr)\Bigl(
                                          \dfrac{p_x^2}{p+p_z} +p_z\Bigr)\biggr\}
                                                \vec{e}_y\\
                              &\qquad\qquad +\biggl\{\Bigl[\cos\Bigl(\frac{2\,E_{\rm p} t}{\hbar}
                                          \Bigr)-1\Bigr]p_x
                                    +\sin\Bigl(\frac{2\,E_{\rm p} t}{\hbar}\Bigr)
                                          p_y\biggr\}\vec{e}_z\Biggr).
                  \end{split}
            \end{equation}
            Thus one obtains
            \begin{eqnarray}\label{eq:ZSPsi14}
                        \mathcal{Z}_s &=&\langle\Psi|\hat{\mathcal{Z}}_s|\Psi\rangle
                        =\sin(2\theta){\rm Re}\left[\langle \Psi_1|\hat{\mathcal{Z}}_s|
                              \Psi_4\rangle\right] \nonumber\\
                        &=& -\sin(2\theta)\dfrac{\hbar c}{2\,E_{\rm p}} \Biggl(
                              \biggl\{-\Bigl[\cos\Bigl(\frac{2\,E_{\rm p} t}{\hbar}
                                          \Bigr)-1\Bigr]\Bigl(\dfrac{p_y^2}{p+p_z}
                                                +p_z\Bigr)
                                    +\sin\Bigl(\frac{2\,E_{\rm p} t}{\hbar}\Bigr)
                                          \dfrac{p_x p_y}{p+p_z}\biggr\}\vec{e}_x \nonumber\\
                              &&\qquad\qquad\qquad\quad +\biggl\{\Bigl[\cos\Bigl(\frac{2\,E_{\rm p} t}{
                                          \hbar}\Bigr)-1\Bigr]\dfrac{p_x p_y}{p+p_z}
                                    -\sin\Bigl(\frac{2\,E_{\rm p} t}{\hbar}\Bigr)\Bigl(
                                          \dfrac{p_x^2}{p+p_z} +p_z\Bigr)\biggr\}
                                                \vec{e}_y \nonumber\\
                              &&\qquad\qquad\qquad\quad +\biggl\{
                                    \Bigl[\cos\Bigl(\frac{2\,E_{\rm p} t}{\hbar}
                                          \Bigr)-1\Bigr]p_x
                                    +\sin\Bigl(\frac{2\,E_{\rm p} t}{\hbar}\Bigr)
                                          p_y\biggr\}\vec{e}_z\Biggr).
            \end{eqnarray}

            For the special case of $\vec{p}=p\,\vec{e}_z$, viz.
                  $p_x =p_y =0$, and $p_z =p$, \Eq{eq:ZSPsi14} degenerates into
                  \begin{equation}\label{eq:ZSPsi14Z}
                        \begin{split}
                                    \mathcal{Z}_s =&\sin(2\theta)\dfrac{c\,\hbar\,p}{
                                    2\,E_{\rm p}} \Biggl\{
                                          \biggl[\cos\Bigl(\frac{2\,E_{\rm p} t}{\hbar}
                                                \Bigr)-1\biggr]\vec{e}_x
                                          +\sin\Bigl(\frac{2\,E_{\rm p} t}{\hbar}\Bigr)
                                                \vec{e}_y\Biggr\} \\
                              =& \sin(2\theta)\dfrac{c\,\hbar\,p}{2\,E_{\rm p}}
                                    \Biggl[2\,\sin\Bigl(\frac{E_{\rm p} t}{\hbar}\Bigr)
                                                \cos\Bigl(\frac{E_{\rm p} t}{\hbar}
                                                      \Bigr)\vec{e}_y
                                          -2\,{\sin^2}\Bigl(\frac{E_{\rm p} t}{
                                                \hbar}\Bigr)\vec{e}_x\Biggr] \\
                              =& 2\,\sin(2\theta)\dfrac{c\,\hbar\,p}{E_{\rm p}}
                                    \sin\Bigl(\frac{E_{\rm p} t}{\hbar}\Bigr)\Biggl[
                                          \cos\left(\frac{E_{\rm p} t}{\hbar}\right)
                                                \vec{e}_y
                                          -\sin\left(\frac{E_{\rm p} t}{\hbar}\right)
                                                \vec{e}_x\Biggr],
                        \end{split}
                  \end{equation}
            thus in general one has $\mathcal{Z}_s \neq 0$. In the next section, the Zitterbewegung of spin can be used as a source to produce the spin angular-momentum waves.
      \subsection{Thought Experiment to Detect Angular-Momentum Waves}
            Electromagnetic waves are produced via electromagnetic oscillation in spacetime. which was first verified experimentally by H. R. Hertz. Later he measured the velocity of wave, found its equivalence to the velocity of light \cite{1888hertz}, and confirmed that the light is a kind of electromagnetic waves.

\begin{figure}[t]
                  \centering
                  \includegraphics[width=150mm]{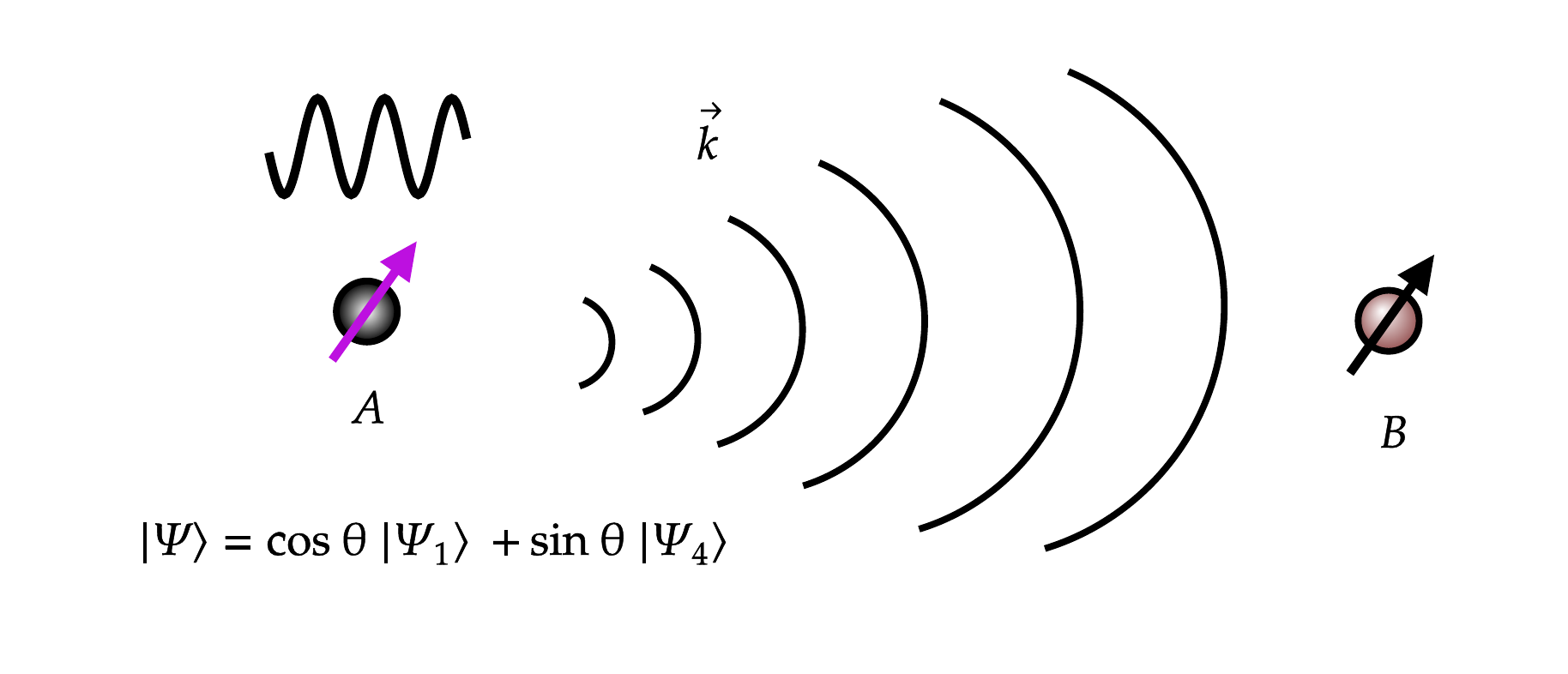}
                  \caption{Illustration of a thought experiment on detecting the spin angular momentum waves (AMW). Particle $A$ is a free Dirac's electron, which is prepared in a superposition state of $|\Psi_1\rangle$ (positive energy and positive helicity) and $|\Psi_4\rangle$ (negative energy and negative helicity). In such a superposition state, the spin of the electron oscillates rapidly due to the relativistic phenomenon of ``spin Zitterbewegung''. Because of the spin oscillation, particle $A$ emits the spin AMW along the direction $\vec{k}$. Due to the Pauli-like interaction $\vec{S}_b \cdot\vec{\mathcal{B}}$, the distanced particle $B$ can experience the existence of the spin AMW emitted by particle $A$.}
                  \label{fig:AMWave}
            \end{figure}

            In general, an oscillating charged particle can emit (or radiate) electromagnetic waves \cite{EMWave}. Analogously, the ``\emph{oscillation of angular momentum}'' can be regarded as a mechanism to emit the angular-momentum waves. The SZ, as a phenomenon of spin oscillation, can be used to radiate the spin angular-momentum waves. In Fig. \ref{fig:AMWave}, we illustrate a possible way to detect the spin angular-momentum waves. A Dirac's electron (particle $A$) is prepared in the superposition of different energy and different helicity, i.e.
            $|\Psi\rangle=\cos\theta|\Psi_1\rangle +\sin\theta|\Psi_4\rangle$. In such a state, there exists a phenomenon of SZ, whose expectation value $\mathcal{Z}_s \neq 0$ in general. The oscillation of the spin can be viewed as a source to emit the angular-momentum waves (i.e., the ``magnetic'' field \eqref{eq:BB} and the ``electric'' field \eqref{eq:EE}) along the direction of the wave ket $\vec{k}$. Particle $B$ with spin $\vec{S}_b$ is used to detect the spin angular-momentum waves. Due to the Pauli-like interaction $\vec{S}_b\cdot\vec{\mathcal{B}}$, the spin of particle $B$ can experience the existence of the spin angular-momentum waves emitted by particle $A$.

\newpage
\section{Some Other Calculations}

       \subsection{$SU(N)$ Operator Solutions of the Maxwell-Type Equations}

       Electromagnetic waves are the $U(1)$ solutions of the Maxwell-type equations (or Maxwell's equations). The $SU(2)$ angular-momentum waves are $SU(2)$ operator solutions of the Maxwell-type equations. Besides these solutions, there are generally $SU(N)$ operator solutions of the Maxwell-type equations, which can be viewed as the $SU(N)$ waves.

       \emph{Proof.---}For the generators of $SU(N)$ group, the commutation and anti-commutation relations satisfy
                        \begin{eqnarray}\label{eq:SUNComut}
                              \left[\hat{G}_a,\ \hat{G}_b\right]=2\,{\rm i} \sum_c f^{abc}
                                    \hat{G}_c,\qquad
                              \{\hat{G}_a,\ \hat{G}_b\}=\dfrac{4}{N} \delta_{ab} \openone
                                    +2\sum_c d^{abc} \hat{G}_c.
                        \end{eqnarray}
                        When $N=3$, the eight generators (Gell-Mann matrices) of $SU(3)$ group read (see, e.g. \cite{GreinerSym})
                        \begin{eqnarray}\label{eq-su3gen}
                              \hat{G}_1 &=& \begin{pmatrix}
                                          0 & 1 & 0 \\ 1 & 0 & 0 \\ 0 & 0 & 0
                                    \end{pmatrix},\
                              \hat{G}_2 \,= \begin{pmatrix}
                                          0 & -{\rm i} & 0 \\ {\rm i} & 0 & 0 \\ 0 & 0 & 0
                                    \end{pmatrix},\
                              \hat{G}_3 \,= \begin{pmatrix}
                                          1 & 0 & 0 \\ 0 & -1 & 0 \\ 0 & 0 & 0
                                    \end{pmatrix},\
                              \hat{G}_4 =\begin{pmatrix}
                                          0 & 0 & 1 \\ 0 & 0 & 0 \\ 1 & 0 & 0
                                    \end{pmatrix}, \notag\\
                              \hat{G}_5 &=& \begin{pmatrix}
                                          0 & 0 & -{\rm i} \\ 0 & 0 & 0 \\ {\rm i} & 0 & 0
                                    \end{pmatrix},\
                              \hat{G}_6 =\begin{pmatrix}
                                          0 & 0 & 0 \\ 0 & 0 & 1 \\ 0 & 1 & 0
                                    \end{pmatrix},\
                              \hat{G}_7 =\begin{pmatrix}
                                          0 & 0 & 0 \\ 0 & 0 & -{\rm i} \\ 0 & {\rm i} & 0
                                    \end{pmatrix},\
                              \hat{G}_8 =\frac{1}{\sqrt{3}} \begin{pmatrix}
                                          1 & 0 & 0 \\ 0 & 1 & 0 \\ 0 & 0 & -2
                                    \end{pmatrix}.
                        \end{eqnarray}
                        Here the structure constants
                        \begin{eqnarray}
                              f^{abc} =-{\rm i}\dfrac{1}{4} \operatorname{tr}\bigl(\hat{G}_a
                                    [\hat{G}_b,\ \hat{G}_c]\bigr),\qquad
                              d^{abc} =\dfrac{1}{4} \operatorname{tr}\bigl(\hat{G}_a \{
                                    \hat{G}_b,\ \hat{G}_c\}\bigr),
                        \end{eqnarray}
                        which are completely antisymmetric and symmetric for the three indices respectively, i.e.,
                        \begin{eqnarray}\label{eq:AntSymFD}
                              f^{abc}=-f^{acb}=-f^{bac} ,\qquad d^{abc}=d^{acb}=d^{bac}.
                        \end{eqnarray}
                        Combine the equations in \eqref{eq:SUNComut} together, then we get the product of two generators in the following form,
                        \begin{eqnarray}\label{eq:SUNproduct}
                              \hat{G}_a \hat{G}_b =\dfrac{1}{2} \bigl([\hat{G}_a,\hat{G}_b]
                                    + \{\hat{G}_a,\hat{G}_b\}\bigr)
                              =\dfrac{2}{N} \delta_{ab} \openone + \sum_c \left(d^{abc}
                                    + \mathrm{i}\,f^{abc}\right) \hat{G}_c.
                        \end{eqnarray}

                        Assume the vector potential and scalar potential are expanded as
                        \begin{eqnarray}
                              \mathcal{\vec{A}}(\vec{r})=\vec{T}_0 \openone
                                    +\sum_{l=1}^{N^2 -1} \vec{T}_l\,\hat{G}_l,\qquad
                              \varphi(\vec{r})=\varphi_0 \openone +\sum_{l=1}^{N^2 -1}
                                    \varphi_l\,\hat{G}_l,
                        \end{eqnarray}
                        where $\hat{G}_l$ are the generators of $SU(N)$ groups. After that, we start to find the restrictions of $\vec{T}_i$ and $\varphi_i$ based on \Eq{eq:sixcond-S1} to \Eq{eq:sixcond-S6}. For convenience, denote
                        \begin{eqnarray}\label{eq:SUNM}
                              \vec{\mathcal{M}}(\vec{r})=\vec{\mathcal{A}}(\vec{r})\times
                                    \vec{\mathcal{A}}(\vec{r})
                              =\sum_{l,m}^{N^2 -1} \bigl(\vec{T}_l\times\vec{T}_m\bigr)
                                    \hat{G}_l\,\hat{G}_m.
                        \end{eqnarray}
                        By substituting \Eq{eq:SUNproduct} into \Eq{eq:SUNM}, we obtain
                        \begin{eqnarray}
                              \vec{\mathcal{M}}(\vec{r})=\sum_{l,m,n}^{N^2 -1} \left(
                                    \vec{T}_l \times\vec{T}_m\right)\left(d^{lmn}
                                          + \mathrm{i}\,f^{lmn}\right)\hat{G}_n
                              ={\rm i}\sum_{l,m,n}^{N^2 -1} \left(\vec{T}_l \times\vec{T}_m\right)
                                    f^{lmn}\hat{G}_n.
                        \end{eqnarray}
                        Based on \Eq{eq:sixcond-S2} and \Eq{eq:sixcond-S6}, the solution of
                        $\vec{\mathcal{M}}(\vec{r})$ is still given by
                        \begin{equation}
                              \vec{\mathcal{M}}(\vec{r})=\vec{M}_1\,{\rm e}^{
                                          {\rm i}\,2\vec{k}\cdot\vec{r}}
                                    +\vec{M}_2\,{\rm e}^{-{\rm i}\,2\vec{k}\cdot\vec{r}},
                        \end{equation}
                        Take $\vec{M}_2=0$ for simplicity, then the form of $\vec{T}_l$ must be
                        \begin{eqnarray}
                              \vec{T}_l =\vec{R}_l\,{\rm e}^{{\rm i}\,\vec{k}\cdot\vec{r}},
                        \end{eqnarray}
                        and
                        \begin{eqnarray}
                              \vec{M}_1 ={\rm i}\sum_{l,m,n}^{N^2 -1} \left(
                                    \vec{R}_l \times\vec{R}_m\right)\,f^{lmn} \hat{G}_n
                              \equiv{\rm i}\,\vec{\eta}.
                        \end{eqnarray}
                        After that, from \Eq{eq:sixcond-S1} we have
                        \begin{equation}
                              \begin{split}
                                    & \vec{\nabla}\cdot\vec{\mathcal{M}}(\vec{r})=\vec{\nabla}
                                          \cdot\left[\sum_{l,m,n}^{N^2 -1} \left(
                                                \vec{T}_l \times\vec{T}_m\right)f^{lmn}\hat{G}_n\right]
                                    =\vec{\nabla}\cdot\left[{\rm e}^{{\rm i}\,2\,\vec{k}\cdot\vec{r}}
                                          \sum_{l,m,n}^{N^2 -1} \left(\vec{R}_l \times\vec{R}_m\right)
                                                f^{lmn}\hat{G}_n\right] \\
                                    =& \Big[\nabla\bigl({\rm e}^{{\rm i}\,2\,\vec{k}\cdot\vec{r}}\bigr)
                                          \Big]\cdot\biggl[\sum_{l,m,n}^{N^2 -1} \left(
                                                \vec{R}_l \times\vec{R}_m\right)f^{lmn}\hat{G}_n\biggr]
                                    ={\rm i}\,2\,\vec{k}\cdot\biggl[\sum_{l,m,n}^{N^2 -1} \left(
                                          \vec{R}_l \times\vec{R}_m\right)f^{lmn}\hat{G}_n\biggr]{\rm e}^{
                                                {\rm i}\,2\,\vec{k}\cdot\vec{r}} \\
                                    =& 0,
                              \end{split}
                        \end{equation}
                        which alludes to
                        \begin{eqnarray}
                              \vec{k}\cdot\biggl[\sum_{l,m,n}^{N^2 -1} \left(
                                    \vec{R}_l \times\vec{R}_m\right)f^{lmn}\hat{G}_n\biggr]=0.
                        \end{eqnarray}
                        As $\hat{G}_n$, $(n\in\{1,2,...,N^2 -1\})$, are linear independent, thus for any wave vector $\vec{k}$, we attain
                        \begin{eqnarray}\label{eq:ConstrR}
                              \vec{k}\cdot\biggl[\sum_{l,m}^{N^2 -1} \left(
                                    \vec{R}_l \times\vec{R}_m\right)f^{lmn}\biggr]\hat{
                                          G}_n =0,\qquad n\in\{1,2,...,N^2 -1\}.
                        \end{eqnarray}
                        Give $n$, there are $(N^2 -1)^2$ terms in \Eq{eq:ConstrR}, while the condition of antisymmetry in \Eq{eq:AntSymFD} reduces the number of terms to $\binom{N^2 -1}{2} =(N^2 -1)(N^2 -2)/2$.
                        \Eq{eq:ConstrR} holds for arbitrary wave vector $\vec{k}$, hence
                        \begin{eqnarray}
                              \vec{k}\cdot\left(\vec{R}_l\times\vec{R}_m\right)=0,\qquad
                                    f^{lmn} \neq 0,\ \forall\,n\in\{1,2,...,N^2 -1\},
                        \end{eqnarray}
                        namely the vectors $\vec{R}_l,\ (i=1,\dots,N^2 -1)$, and the wave vector $\vec{k}$ locate on the same plane, while
                        $\vec{R}_0$ is arbitrary.
                        \begin{proposition}
                              In the case of $N=3$, the exclusive nonvanishing structure constants read \cite{GreinerSym}:
                              \begin{equation}
                                    f^{123} =1,\quad
                                    f^{147} =f^{246} =f^{257} =f^{345} =\dfrac{1}{2} ,
                                          \quad
                                    f^{156} =f^{367} =-\dfrac{1}{2} ,\quad
                                    f^{458} =f^{678} =\dfrac{\sqrt{3}}{2}.
                              \end{equation}
                              Thus
                              \begin{itemize}
                                    \item [(i)]. For $n=1$,
                                          \begin{equation}
                                                \vec{k}\cdot\left(\vec{R}_2 \times
                                                            \vec{R}_3\right)f^{231}
                                                      +\vec{k}\cdot\left(\vec{R}_4
                                                            \times\vec{R}_7\right)f^{
                                                                  471}
                                                      +\vec{k}\cdot\left(\vec{R}_5
                                                            \times\vec{R}_6\right)f^{
                                                                  561} =0,
                                          \end{equation}
                                          i.e.
                                          \begin{equation}
                                                \vec{k}\cdot\left(\vec{R}_2 \times
                                                            \vec{R}_3\right)f^{123}
                                                      +\vec{k}\cdot\left(\vec{R}_4
                                                            \times\vec{R}_7\right)f^{
                                                                  147}
                                                      +\vec{k}\cdot\left(\vec{R}_5
                                                            \times\vec{R}_6\right)f^{
                                                                  156} =0,
                                          \end{equation}
                                          i.e.
                                          \begin{equation}
                                                \vec{k}\cdot\left(\vec{R}_2 \times
                                                            \vec{R}_3\right)
                                                      +\dfrac{1}{2} \vec{k}\cdot\left(
                                                            \vec{R}_4 \times\vec{R}_7
                                                            \right)
                                                      -\dfrac{1}{2} \vec{k}\cdot\left(
                                                            \vec{R}_5 \times\vec{R}_6
                                                            \right)=0.
                                          \end{equation}
                                    \item [(ii)]. For $n=2$,
                                          \begin{equation}
                                                \vec{k}\cdot\left(\vec{R}_1 \times
                                                            \vec{R}_3\right)f^{132}
                                                      +\vec{k}\cdot\left(\vec{R}_4
                                                            \times\vec{R}_6\right)f^{
                                                                  462}
                                                      +\vec{k}\cdot\left(\vec{R}_5
                                                            \times\vec{R}_7\right)f^{
                                                                  572} =0,
                                          \end{equation}
                                          i.e.
                                          \begin{equation}
                                                -\vec{k}\cdot\left(\vec{R}_1 \times
                                                            \vec{R}_3\right)f^{123}
                                                      +\vec{k}\cdot\left(\vec{R}_4
                                                            \times\vec{R}_6\right)f^{
                                                                  246}
                                                      +\vec{k}\cdot\left(\vec{R}_5
                                                            \times\vec{R}_7\right)f^{
                                                                  257} =0,
                                          \end{equation}
                                          i.e.
                                          \begin{equation}
                                                -\vec{k}\cdot\left(\vec{R}_1 \times
                                                            \vec{R}_3\right)
                                                      +\dfrac{1}{2} \vec{k}\cdot\left(
                                                            \vec{R}_4 \times\vec{R}_6
                                                            \right)
                                                      -\dfrac{1}{2} \vec{k}\cdot\left(
                                                            \vec{R}_5 \times\vec{R}_7
                                                            \right)=0.
                                          \end{equation}
                                    \item [(iii)]. For $n=3$,
                                          \begin{equation}
                                                \vec{k}\cdot\left(\vec{R}_1 \times
                                                            \vec{R}_2\right)f^{123}
                                                      +\vec{k}\cdot\left(\vec{R}_4
                                                            \times\vec{R}_5\right)f^{
                                                                  453}
                                                      +\vec{k}\cdot\left(\vec{R}_6
                                                            \times\vec{R}_7\right)f^{
                                                                  673} =0,
                                          \end{equation}
                                          i.e.
                                          \begin{equation}
                                                \vec{k}\cdot\left(\vec{R}_1 \times
                                                            \vec{R}_2\right)f^{123}
                                                      +\vec{k}\cdot\left(\vec{R}_4
                                                            \times\vec{R}_5\right)f^{
                                                                  345}
                                                      +\vec{k}\cdot\left(\vec{R}_6
                                                            \times\vec{R}_7\right)f^{
                                                                  367} =0,
                                          \end{equation}
                                          i.e.
                                          \begin{equation}
                                                \vec{k}\cdot\left(\vec{R}_1 \times
                                                            \vec{R}_2\right)
                                                      +\dfrac{1}{2} \vec{k}\cdot\left(
                                                            \vec{R}_4 \times\vec{R}_5
                                                            \right)
                                                      -\dfrac{1}{2} \vec{k}\cdot\left(
                                                            \vec{R}_6 \times\vec{R}_7
                                                            \right)=0.
                                          \end{equation}
                                    \item [(iv)]. For $n=4$,
                                          \begin{equation}
                                                \vec{k}\cdot\left(\vec{R}_1 \times
                                                            \vec{R}_7\right)f^{174}
                                                      +\vec{k}\cdot\left(\vec{R}_2
                                                            \times\vec{R}_6\right)f^{
                                                                  264}
                                                      +\vec{k}\cdot\left(\vec{R}_3
                                                            \times\vec{R}_5\right)f^{
                                                                  354}
                                                      +\vec{k}\cdot\left(\vec{R}_5
                                                            \times\vec{R}_8\right)f^{
                                                                  584} =0,
                                          \end{equation}
                                          i.e.
                                          \begin{equation}
                                                -\vec{k}\cdot\left(\vec{R}_1 \times
                                                            \vec{R}_7\right)f^{147}
                                                      -\vec{k}\cdot\left(\vec{R}_2
                                                            \times\vec{R}_6\right)f^{
                                                                  246}
                                                      -\vec{k}\cdot\left(\vec{R}_3
                                                            \times\vec{R}_5\right)f^{
                                                                  345}
                                                      +\vec{k}\cdot\left(\vec{R}_5
                                                            \times\vec{R}_8\right)f^{
                                                                  458} =0,
                                          \end{equation}
                                          i.e.
                                          \begin{equation}
                                                -\dfrac{1}{2} \vec{k}\cdot\left(
                                                      \vec{R}_1 \times\vec{R}_7\right)
                                                -\dfrac{1}{2} \vec{k}\cdot\left(
                                                      \vec{R}_2 \times\vec{R}_6\right)
                                                +\dfrac{1}{2} \vec{k}\cdot\left(
                                                      \vec{R}_3 \times\vec{R}_5\right)
                                                +\dfrac{\sqrt{3}}{2} \vec{k}\cdot\left(
                                                      \vec{R}_5 \times\vec{R}_8\right)
                                                =0.
                                          \end{equation}
                                    \item [(v)]. For $n=5$,
                                          \begin{equation}
                                                \vec{k}\cdot\left(\vec{R}_2 \times
                                                      \vec{R}_7\right)f^{275}
                                                +\vec{k}\cdot\left(\vec{R}_3 \times
                                                      \vec{R}_4\right)f^{345}
                                                +\vec{k}\cdot\left(\vec{R}_1 \times
                                                      \vec{R}_6\right)f^{165}
                                                +\vec{k}\cdot\left(\vec{R}_4 \times
                                                      \vec{R}_8\right)f^{485} =0,
                                          \end{equation}
                                          i.e.
                                          \begin{equation}
                                                -\vec{k}\cdot\left(\vec{R}_2 \times
                                                      \vec{R}_7\right)f^{257}
                                                +\vec{k}\cdot\left(\vec{R}_3 \times
                                                      \vec{R}_4\right)f^{345}
                                                -\vec{k}\cdot\left(\vec{R}_1 \times
                                                      \vec{R}_6\right)f^{156}
                                                -\vec{k}\cdot\left(\vec{R}_4 \times
                                                      \vec{R}_8\right)f^{458} =0,
                                          \end{equation}
                                          i.e.
                                          \begin{equation}
                                                -\dfrac{1}{2} \vec{k}\cdot\left(
                                                      \vec{R}_2 \times\vec{R}_7\right)
                                                +\dfrac{1}{2} \vec{k}\cdot\left(
                                                      \vec{R}_3 \times\vec{R}_4\right)
                                                +\dfrac{1}{2} \vec{k}\cdot\left(
                                                      \vec{R}_1 \times\vec{R}_6\right)
                                                -\dfrac{\sqrt{3}}{2} \vec{k}\cdot\left(
                                                      \vec{R}_4 \times\vec{R}_8\right)
                                                =0.
                                          \end{equation}
                                    \item [(vi)]. For $n=6$,
                                          \begin{equation}
                                                \vec{k}\cdot\left(\vec{R}_2 \times
                                                      \vec{R}_4\right)f^{246}
                                                +\vec{k}\cdot\left(\vec{R}_1 \times
                                                      \vec{R}_5\right)f^{156}
                                                +\vec{k}\cdot\left(\vec{R}_3 \times
                                                      \vec{R}_7\right)f^{376}
                                                +\vec{k}\cdot\left(\vec{R}_7 \times
                                                      \vec{R}_8\right)f^{786} =0,
                                          \end{equation}
                                          i.e.
                                          \begin{equation}
                                                \vec{k}\cdot\left(\vec{R}_2 \times
                                                      \vec{R}_4\right)f^{246}
                                                +\vec{k}\cdot\left(\vec{R}_1 \times
                                                      \vec{R}_5\right)f^{156}
                                                -\vec{k}\cdot\left(\vec{R}_3 \times
                                                      \vec{R}_7\right)f^{367}
                                                +\vec{k}\cdot\left(\vec{R}_7 \times
                                                      \vec{R}_8\right)f^{678} =0,
                                          \end{equation}
                                          i.e.
                                          \begin{equation}
                                                \dfrac{1}{2} \vec{k}\cdot\left(
                                                      \vec{R}_2 \times\vec{R}_4\right)
                                                -\dfrac{1}{2} \vec{k}\cdot\left(
                                                      \vec{R}_1 \times\vec{R}_5\right)
                                                +\dfrac{1}{2} \vec{k}\cdot\left(
                                                      \vec{R}_3 \times\vec{R}_7\right)
                                                +\dfrac{\sqrt{3}}{2} \vec{k}\cdot\left(
                                                      \vec{R}_7 \times\vec{R}_8\right)
                                                =0.
                                          \end{equation}
                                    \item [(vii)]. For $n=7$,
                                          \begin{equation}
                                                \vec{k}\cdot\left(\vec{R}_1 \times
                                                      \vec{R}_4\right)f^{147}
                                                +\vec{k}\cdot\left(\vec{R}_2 \times
                                                      \vec{R}_5\right)f^{257}
                                                +\vec{k}\cdot\left(\vec{R}_3 \times
                                                      \vec{R}_6\right)f^{367}
                                                +\vec{k}\cdot\left(\vec{R}_6 \times
                                                      \vec{R}_8\right)f^{687} =0,
                                          \end{equation}
                                          i.e.
                                          \begin{equation}
                                                \vec{k}\cdot\left(\vec{R}_1 \times
                                                      \vec{R}_4\right)f^{147}
                                                +\vec{k}\cdot\left(\vec{R}_2 \times
                                                      \vec{R}_5\right)f^{257}
                                                +\vec{k}\cdot\left(\vec{R}_3 \times
                                                      \vec{R}_6\right)f^{367}
                                                -\vec{k}\cdot\left(\vec{R}_6 \times
                                                      \vec{R}_8\right)f^{678} =0,
                                          \end{equation}
                                          i.e.
                                          \begin{equation}
                                                \dfrac{1}{2} \vec{k}\cdot\left(
                                                      \vec{R}_1 \times\vec{R}_4\right)
                                                +\dfrac{1}{2} \vec{k}\cdot\left(
                                                      \vec{R}_2 \times\vec{R}_5\right)
                                                -\dfrac{1}{2} \vec{k}\cdot\left(
                                                      \vec{R}_3 \times\vec{R}_6\right)
                                                -\dfrac{\sqrt{3}}{2} \vec{k}\cdot\left(
                                                      \vec{R}_6 \times\vec{R}_8\right)
                                                =0.
                                          \end{equation}
                                    \item [(viii)]. For $n=8$,
                                          \begin{equation}
                                                \vec{k}\cdot\left(\vec{R}_4 \times
                                                            \vec{R}_5\right)f^{458}
                                                      +\vec{k}\cdot\left(\vec{R}_6
                                                            \times\vec{R}_7\right)f^{
                                                                  678} =0,
                                          \end{equation}
                                          i.e.
                                          \begin{equation}
                                                \vec{k}\cdot\left(\vec{R}_4 \times
                                                            \vec{R}_5\right)
                                                      +\vec{k}\cdot\left(\vec{R}_6
                                                            \times\vec{R}_7\right)=0.
                                          \end{equation}
                              \end{itemize}
                              In summary, we attain the set of equations as follows.
                              \begin{align}
                                    & \vec{k}\cdot\left(\vec{R}_2 \times\vec{R}_3\right)
                                          +\dfrac{1}{2} \vec{k}\cdot\left(\vec{R}_4
                                                \times\vec{R}_7\right)
                                          -\dfrac{1}{2} \vec{k}\cdot\left(\vec{R}_5
                                                \times\vec{R}_6\right)=0, \label{eq:KREq1} \\
                                    & -\vec{k}\cdot\left(\vec{R}_1 \times\vec{R}_3
                                                \right)
                                          +\dfrac{1}{2} \vec{k}\cdot\left(\vec{R}_4
                                                \times\vec{R}_6\right)
                                          -\dfrac{1}{2} \vec{k}\cdot\left(\vec{R}_5
                                                \times\vec{R}_7\right)=0, \\
                                    & \vec{k}\cdot\left(\vec{R}_1 \times\vec{R}_2\right)
                                          +\dfrac{1}{2} \vec{k}\cdot\left(\vec{R}_4
                                                \times\vec{R}_5\right)
                                          -\dfrac{1}{2} \vec{k}\cdot\left(\vec{R}_6
                                                \times\vec{R}_7\right)=0, \\
                                    & -\dfrac{1}{2} \vec{k}\cdot\left(\vec{R}_1
                                                \times\vec{R}_7\right)
                                          -\dfrac{1}{2} \vec{k}\cdot\left(\vec{R}_2
                                                \times\vec{R}_6\right)
                                          +\dfrac{1}{2} \vec{k}\cdot\left(\vec{R}_3
                                                \times\vec{R}_5\right)
                                          +\dfrac{\sqrt{3}}{2} \vec{k}\cdot\left(
                                                \vec{R}_5 \times\vec{R}_8\right)=0, \\
                                    & -\dfrac{1}{2} \vec{k}\cdot\left(\vec{R}_2
                                                \times\vec{R}_7\right)
                                          +\dfrac{1}{2} \vec{k}\cdot\left(\vec{R}_3
                                                \times\vec{R}_4\right)
                                          +\dfrac{1}{2} \vec{k}\cdot\left(\vec{R}_1
                                                \times\vec{R}_6\right)
                                          -\dfrac{\sqrt{3}}{2} \vec{k}\cdot\left(
                                                \vec{R}_4 \times\vec{R}_8\right)=0, \\
                                    & \dfrac{1}{2} \vec{k}\cdot\left(\vec{R}_2
                                                \times\vec{R}_4\right)
                                          -\dfrac{1}{2} \vec{k}\cdot\left(
                                                \vec{R}_1 \times\vec{R}_5\right)
                                          +\dfrac{1}{2} \vec{k}\cdot\left(
                                                \vec{R}_3 \times\vec{R}_7\right)
                                          +\dfrac{\sqrt{3}}{2} \vec{k}\cdot\left(
                                                \vec{R}_7 \times\vec{R}_8\right)=0, \\
                                    & \dfrac{1}{2} \vec{k}\cdot\left(\vec{R}_1
                                                \times\vec{R}_4\right)
                                          +\dfrac{1}{2} \vec{k}\cdot\left(
                                                \vec{R}_2 \times\vec{R}_5\right)
                                          -\dfrac{1}{2} \vec{k}\cdot\left(
                                                \vec{R}_3 \times\vec{R}_6\right)
                                          -\dfrac{\sqrt{3}}{2} \vec{k}\cdot\left(
                                                \vec{R}_6 \times\vec{R}_8\right)=0, \\
                                    & \vec{k}\cdot\left(\vec{R}_4 \times\vec{R}_5\right)
                                          +\vec{k}\cdot\left(\vec{R}_6 \times\vec{R}_7
                                          \right)=0. \label{eq:KREq8}
                              \end{align}
                        \end{proposition}
                        The equations \eqref{eq:KREq1}-\eqref{eq:KREq8} holds for any $\vec{k}$, which leads to
                        \begin{equation}
                              \begin{split}
                                    & \vec{k}\cdot\left(\vec{R}_2 \times\vec{R}_3\right)
                                    =\vec{k}\cdot\left(\vec{R}_4 \times\vec{R}_7\right)
                                    =\vec{k}\cdot\left(\vec{R}_5 \times\vec{R}_6\right)
                                    =\vec{k}\cdot\left(\vec{R}_1 \times\vec{R}_3\right)
                                    =\vec{k}\cdot\left(\vec{R}_4 \times\vec{R}_6\right)
                                    =\vec{k}\cdot\left(\vec{R}_5 \times\vec{R}_7\right)
                                          \\
                                    =& \vec{k}\cdot\left(\vec{R}_1 \times\vec{R}_2
                                          \right)
                                    =\vec{k}\cdot\left(\vec{R}_4 \times\vec{R}_5\right)
                                    =\vec{k}\cdot\left(\vec{R}_6 \times\vec{R}_7\right)
                                    =\vec{k}\cdot\left(\vec{R}_1 \times\vec{R}_7\right)
                                    =\vec{k}\cdot\left(\vec{R}_2 \times\vec{R}_6\right)
                                    =\vec{k}\cdot\left(\vec{R}_3 \times\vec{R}_5\right)
                                          \\
                                    =& \vec{k}\cdot\left(\vec{R}_5 \times\vec{R}_8
                                          \right)
                                    =\vec{k}\cdot\left(\vec{R}_2 \times\vec{R}_7\right)
                                    =\vec{k}\cdot\left(\vec{R}_3 \times\vec{R}_4\right)
                                    =\vec{k}\cdot\left(\vec{R}_1 \times\vec{R}_6\right)
                                    =\vec{k}\cdot\left(\vec{R}_4 \times\vec{R}_8\right)
                                    =\vec{k}\cdot\left(\vec{R}_2 \times\vec{R}_4\right)
                                          \\
                                    =& \vec{k}\cdot\left(\vec{R}_1 \times\vec{R}_5
                                          \right)
                                    =\vec{k}\cdot\left(\vec{R}_3 \times\vec{R}_7\right)
                                    =\vec{k}\cdot\left(\vec{R}_7 \times\vec{R}_8\right)
                                    =\vec{k}\cdot\left(\vec{R}_1 \times\vec{R}_4\right)
                                    =\vec{k}\cdot\left(\vec{R}_2 \times\vec{R}_5\right)
                                    =\vec{k}\cdot\left(\vec{R}_3 \times\vec{R}_6\right)
                                          \\
                                    =& \vec{k}\cdot\left(\vec{R}_6 \times\vec{R}_8
                                          \right)=0,
                              \end{split}
                        \end{equation}
                        i.e.
                        \begin{eqnarray}
                              \vec{k}\cdot\left(\vec{R}_l\times\vec{R}_m\right)=0,\qquad
                                    f^{lmn} \neq 0,\ \forall\,n\in\{1,2,...,8\}.
                        \end{eqnarray}

                        For convenience, cite the notations
                        \begin{eqnarray}
                              \vec{\tau} &=& \vec{R}_0 \openone+\sum_{l=1}^{N^2 -1} \vec{R}_l\,\hat{G}_l,
                        \end{eqnarray}
                        then
                        \begin{equation}\label{eq:SU3A}
                              \mathcal{\vec{A}}(\vec{r})=\vec{T}_0 \openone
                                    +\sum_{l=1}^{N^2 -1} \vec{T}_l\,\hat{G}_l
                              =\Bigl(\vec{R}_0 \openone+\sum_{l=1}^{N^2 -1} \vec{R}_l\,\hat{G}_l\Bigr)
                                    {\rm e}^{{\rm i}\vec{k}\cdot\vec{r}}
                              =\vec{\tau}\,{\rm e}^{{\rm i}\vec{k}\cdot\vec{r}}.
                        \end{equation}
                        To determine the scalar potential $\varphi(\vec{r})$, similarly let
                        \begin{eqnarray}
                              \varphi(\vec{r})=\Bigl(\tilde{\varphi}_0 \openone
                                    +\sum_{l=1}^{N^2 -1} \tilde{\varphi}_l\,\hat{G}_l\Bigr){\rm e}^{
                                          {\rm i}\vec{k}\cdot\vec{r}}.
                        \end{eqnarray}
                        From \Eq{eq:sixcond-S3}, we have
                        \begin{eqnarray}
                              {\rm i} k\vec{\nabla}\cdot\left(\vec{\tau}\;{\rm e}^{
                                          {\rm i}\vec{k}\cdot\vec{r}}\right)
                                    -{\nabla}^2 \left[\Bigl(\tilde{\varphi}_0 \openone
                                    +\sum_{l=1}^{N^2 -1} \tilde{\varphi}_l\,\hat{G}_l\Bigr){\rm e}^{
                                          {\rm i}\vec{k}\cdot\vec{r}}\right]=0,
                        \end{eqnarray}
                        then we have
                        \begin{eqnarray}
                              \hat{k}\cdot\vec{\tau}-\Bigl(\tilde{\varphi}_0 \openone
                                    +\sum_{l=1}^{N^2 -1} \tilde{\varphi}_l\,\hat{G}_l\Bigr)=0,
                        \end{eqnarray}
                        with $\hat{k}=\vec{k}/k$, which leads to
                        \begin{eqnarray}\label{eq:phi0phil}
                              \tilde{\varphi}_0 =\hat{k}\cdot\vec{R}_0,\quad
                              \tilde{\varphi}_l =\hat{k}\cdot\vec{R}_l,\quad
                              l\in\{1,2,...,N^2 -1\}.
                        \end{eqnarray}
                        This means that the scalar potential is determined by
                        \begin{eqnarray}
                              \varphi(\vec{r})=\left(\hat{k}\cdot\vec{\tau}\right){\rm e}^{
                                    {\rm i}\vec{k}\cdot\vec{r}}.
                        \end{eqnarray}

                        Next we check whether they hold for all the six conditions Eqs.
                        (\ref{eq:sixcond-S1})-(\ref{eq:sixcond-S6}). It is easy to verify that Eqs. (
                              \ref{eq:sixcond-S1})-(\ref{eq:sixcond-S3}) are satisfied. We need
                        to further check the rest three conditions (\ref{eq:sixcond-S4})-(
                              \ref{eq:sixcond-S6}). Let us study Eq. (\ref{eq:sixcond-S6}) first.
                        Because
                        \begin{equation}
                              \begin{split}
                                    & \bigl[\varphi(\vec{r}),\ \vec{\mathcal{A}}(\vec{r})\bigr]
                                    =\bigg[\Bigl(\tilde{\varphi}_0 \openone
                                                +\sum_{l=1}^{N^2 -1} \tilde{\varphi}_l\,\hat{G}_l\Bigr),\
                                          \Bigl(\vec{R}_0 \openone+\sum_{m=1}^{N^2 -1} \vec{R}_m\,\hat{G}_m
                                          \Bigr)\bigg]{\rm e}^{{\rm i}2\vec{k}\cdot\vec{r}}
                                    =\bigg[\sum_{l=1}^{N^2 -1} \tilde{\varphi}_l\,\hat{G}_l,\
                                          \sum_{m=1}^{N^2 -1} \vec{R}_m\,\hat{G}_m\bigg]{\rm e}^{
                                                {\rm i}2\vec{k}\cdot\vec{r}} \\
                                    =& \sum_{l,m=1}^{N^2 -1} \tilde{\varphi}_l\,\vec{R}_m\,\big[\hat{G}_l,\
                                          \hat{G}_m\big]{\rm e}^{{\rm i}2\vec{k}\cdot\vec{r}}
                                    =2\,{\rm i}\left(\sum_{l,m,n=1}^{N^2 -1} \tilde{\varphi}_l\,\vec{R}_m\,
                                          f^{lmn} \hat{G}_n\right){\rm e}^{
                                                {\rm i}2\vec{k}\cdot\vec{r}}
                                    =2\,{\rm i}\,\left[\sum_{l,m,n=1}^{N^2 -1} \bigl(
                                          \hat{k}\cdot\vec{R}_l\bigr)\vec{R}_m\,f^{lmn} \hat{G}_n
                                          \right]{\rm e}^{{\rm i}2\vec{k}\cdot\vec{r}},
                              \end{split}
                        \end{equation}
                        while
                        \begin{equation}
                              \begin{split}
                                    & \vec{\nabla}\times\vec{\mathcal{M}}(\vec{r})
                                    =\vec{\nabla}\times\left[{\rm e}^{
                                          {\rm i}\,2\,\vec{k}\cdot\vec{r}} \sum_{l,m,n}^{N^2 -1} \left(
                                                \vec{R}_l \times\vec{R}_m\right){\rm i}\,f^{lmn}
                                                      \hat{G}_n\right]
                                    =\left(\vec{\nabla}\,{\rm e}^{{\rm i}\,2\,\vec{k}\cdot\vec{r}}
                                          \right)\times\left[\sum_{l,m,n}^{N^2 -1} \left(
                                                      \vec{R}_l \times\vec{R}_m\right){\rm i}\,
                                                            f^{lmn}\hat{G}_n\right] \\
                                    =& {\rm i}\,2\left[\sum_{l,m,n}^{N^2 -1} \vec{k}\times\left(
                                                \vec{R}_l \times\vec{R}_m\right){\rm i}\,f^{lmn}
                                                      \hat{G}_n\right]{\rm e}^{
                                                            {\rm i}\,2\,\vec{k}\cdot\vec{r}}
                                    ={\rm i}\,2\bigg\{\sum_{l,m,n}^{N^2 -1} \Big[
                                          \bigl(\vec{k}\cdot\vec{R}_m\bigr)\vec{R}_l
                                          -\bigl(\vec{k}\cdot\vec{R}_l\bigr)\vec{R}_m\Big]{\rm i}\,
                                                f^{lmn} \hat{G}_n\bigg\}{\rm e}^{
                                                      {\rm i}\,2\,\vec{k}\cdot\vec{r}}.
                              \end{split}
                        \end{equation}
                        Then \Eq{eq:sixcond-S6} can be recast as
                        \begin{equation}
                              2\,{\rm i}\,\left[\sum_{l,m,n=1}^{N^2 -1} \bigl(\vec{k}\cdot\vec{R}_l\bigr)
                                    \vec{R}_m\,f^{lmn} \hat{G}_n\right]{\rm e}^{
                                          {\rm i}2\vec{k}\cdot\vec{r}}
                              +\bigg\{\sum_{l,m,n}^{N^2 -1} \Big[\bigl(\vec{k}\cdot\vec{R}_m\bigr)
                                          \vec{R}_l -\bigl(\vec{k}\cdot\vec{R}_l\bigr)\vec{R}_m
                                          \Big]{\rm i}\,f^{lmn} \hat{G}_n\bigg\}{\rm e}^{
                                                {\rm i}\,2\,\vec{k}\cdot\vec{r}} =0,
                        \end{equation}
                        i.e.
                        \begin{equation}
                              2\,{\rm i}\,\left[\sum_{l,m,n=1}^{N^2 -1} \bigl(\vec{k}\cdot\vec{R}_l\bigr)
                                          \vec{R}_m\,f^{lmn} \hat{G}_n\right]
                              +{\rm i}\bigg\{\sum_{l,m,n}^{N^2 -1} \Big[
                                    -\bigl(\vec{k}\cdot\vec{R}_m\bigr)\vec{R}_l f^{mln}
                                    -\bigl(\vec{k}\cdot\vec{R}_l\bigr)\vec{R}_m f^{lmn}\Big]
                                          \hat{G}_n\bigg\} =0,
                        \end{equation}
                        i.e.
                        \begin{equation}
                              2\,{\rm i}\,\bigg[\sum_{l,m,n=1}^{N^2 -1} \bigl(\vec{k}\cdot\vec{R}_l\bigr)
                                    \vec{R}_m\,f^{lmn} \hat{G}_n\bigg]
                              -2\,{\rm i}\bigg[\sum_{l,m,n}^{N^2 -1} \bigl(\vec{k}\cdot\vec{R}_l\bigr)
                                    \vec{R}_m f^{lmn}\hat{G}_n\bigg]=0,
                        \end{equation}
                        which holds naturally.

                        For Eq. (\ref{eq:sixcond-S5}), because
                        \begin{eqnarray}
                              \vec{\nabla}\cdot \vec{\mathcal{A}}(\vec{r})
                              &=& \vec{\nabla}\cdot\left(\vec{\tau}\;{\rm e}^{
                                    {\rm i}\vec{k}\cdot\vec{r}}\right)
                              ={\rm i}\left(\vec{k}\cdot\vec{\tau}\right){\rm e}^{
                                    {\rm i}\vec{k}\cdot\vec{r}}
                              ={\rm i}\,k\,\varphi(\vec{r}),
                        \end{eqnarray}
                        then Eq. (\ref{eq:sixcond-S5}) becomes
                        \begin{eqnarray}\label{eq:WavEqSU3A}
                              \nabla^2 \vec{\mathcal{A}}(\vec{r})+k^2\;\vec{\mathcal{A}}(\vec{r})=0.
                        \end{eqnarray}
                        By substituting Eq. (\ref{eq:SU3A}) into Eq. (\ref{eq:WavEqSU3A}), one finds that it is satisfied, thus Eq. (\ref{eq:sixcond-S5}) holds.

                        In summary, we have the vector potential and the scalar potential as
                        \begin{eqnarray}
                              &&\vec{\mathcal{A}}(\vec{r}, t)=\vec{\tau}\;
                                    {\rm e}^{{\rm i}(\vec{k}\cdot\vec{r}-\omega t)},\nonumber\\
                              && \varphi(\vec{r}, t)=\left(\hat{k}\cdot\vec{\tau}\right)\;{\rm e}^{{\rm i}(\vec{k}\cdot\vec{r}-\omega t)},
                        \end{eqnarray}
                        with
                        \begin{eqnarray}
                              \vec{\tau}=\vec{R}_0 \openone+\sum_{l=1}^{N^2 -1} \vec{R}_l\,\hat{G}_l.
                        \end{eqnarray}
                        Note via \Eq{eq:PhiAM1}, we have
                        \begin{equation}
                              \Big[\bigl(\vec{\tau}\cdot\hat{k}\bigr),\ \vec{\tau}\Big]
                              =-\left(\hat{k}\times\vec{M}_1\right)
                              =-\left[\hat{k}\times\bigl(\vec{\tau}\times\vec{\tau}
                                    \bigr)\right],
                        \end{equation}
                        i.e.
                        \begin{equation}
                              \hat{k}\times\Big[\bigl(\vec{\tau}\cdot\hat{k}\bigr),\
                                    \vec{\tau}\Big]
                              =-\hat{k}\times\left[\hat{k}\times\bigl(\vec{\tau}
                                    \times\vec{\tau}\bigr)\right]
                              =-\left[\hat{k}\cdot\bigl(\vec{\tau}\times\vec{\tau}\bigr)
                                    \hat{k}-\bigl(\vec{\tau}\times\vec{\tau}\bigr)
                                    \right]=\bigl(\vec{\tau}\times\vec{\tau}\bigr).
                        \end{equation}

                        Finally, we calculate the corresponding ``magnetic'' field
                        $\vec{\mathcal{B}}$ and ``electric'' filed $\vec{\mathcal{E}}$ as follows.
                        \begin{eqnarray}
                              \vec{\mathcal{B}}(\vec{r},t) &=& \vec{\nabla}\times
                                    \vec{\mathcal{A}}(\vec{r}, t)-{\rm i}\,g
                                          \left[\vec{\mathcal{A}}(\vec{r}, t)\times
                                                \vec{\mathcal{A}}(\vec{r}, t)\right]
                              =\vec{\nabla}\times\left[\vec{\tau}\;{\rm e}^{
                                          {\rm i}(\vec{k}\cdot\vec{r}-\omega t)}\right]
                                    -{\rm i}\,g \vec{\mathcal{M}}(\vec{r})\,
                                          {\rm e}^{-2{\rm i}\omega t} \notag\\
                              &=& {\rm i}\left(\vec{k}\times\vec{\tau}\right){\rm e}^{
                                          {\rm i}(\vec{k}\cdot\vec{r}-\omega t)}
                                    +g\,\vec{\eta}\;{\rm e}^{
                                          2{\rm i}(\vec{k}\cdot\vec{r}-\omega t)}.
                        \end{eqnarray}
                        Due to Eq. (\ref{eq:sixcond-S6}), we have
                        \begin{equation}
                              2{\rm i}k[\varphi(\vec{r}),\ \vec{\mathcal{A}}(\vec{r})]
                              =-\vec{\nabla}\times\vec{\mathcal{M}}=-{\rm i}\,2{\rm e}^{
                                    {\rm i}\,2\vec{k}\cdot\vec{r}} \left(\vec{k}\times\vec{M}_1\right),
                        \end{equation}
                        thus
                        \begin{equation}\label{eq:PhiAM1}
                              [\varphi(\vec{r}),\ \vec{\mathcal{A}}(\vec{r})]
                              =-\left(\hat{k}\times\vec{M}_1\right){\rm e}^{
                                    {\rm i}\,2\vec{k}\cdot\vec{r}}.
                        \end{equation}
                        From Eq. (\ref{eq:GenerSpinE}), we obtain the ``electric'' field as
                        \begin{eqnarray}
                              \vec{\mathcal{E}}(\vec{r}, t)
                              &=&{\rm i}\,k\,\vec{\tau}\;{\rm e}^{
                                                {\rm i}(\vec{k}\cdot\vec{r}-\omega t)}
                                                -{\rm i} \vec{k}\left(\hat{k}\cdot\vec{\tau}\right)\;
                                          {\rm e}^{{\rm i}(\vec{k}\cdot\vec{r}-\omega t)}
                                    -g\left(\hat{k}\times \vec{\eta}\right){\rm e}^{
                                          {\rm i}\,2(\vec{k}\cdot\vec{r}-\omega t)} \notag\\
                              &=& -\hat{k}\times\Biggl[{\rm i}\,\left(\vec{k}\times \vec{\tau}
                                          \right){\rm e}^{{\rm i}(\vec{k}\cdot\vec{r}-\omega t)}
                                          \Biggr]
                                    -g\left(\hat{k}\times \vec{\eta}\right){\rm e}^{
                                          {\rm i}\,2(\vec{k}\cdot\vec{r}-\omega t)}
                              =-\hat{k}\times \vec{\mathcal{B}}(\vec{r}, t).
                        \end{eqnarray}

      \subsection{$SU(N)$ Operator Solutions of Yang-Mills Equations under Weak-Coupling Approximation}
            Note the difference of Maxwell-type equations and the Yang-Mills equations under weak-coupling approximation is nothing but the following four terms:
            \begin{eqnarray}
                  &&\mathcal{W}_1=
                              -{\rm i}\,g\Biggl\{\bigg[\vec{\mathcal{A}}\cdot\Big(
                                                \dfrac{1}{c} \dfrac{\partial\,
                                                      \vec{\mathcal{A}}}{\partial\,t}\Big)
                                          -\Big(\dfrac{1}{c} \dfrac{\partial\,
                                                \vec{\mathcal{A}}}{\partial\,t}\Big)\cdot
                                                      \vec{\mathcal{A}}\bigg]
                                    +\vec{\mathcal{A}}\cdot\bigl(\vec{\nabla}\varphi\bigr)
                                    -\bigl(\vec{\nabla}\varphi\bigr)\cdot\vec{\mathcal{A}}
                                    \Biggr\}, \nonumber\\
                  &&\mathcal{W}_2={\rm i}\,g\Biggl\{\bigg[\vec{\mathcal{A}}
                                                \times\Big(\dfrac{1}{c}
                                                      \dfrac{\partial\,\vec{\mathcal{A}}}{
                                                            \partial\,t}\Big)
                                          +\Big(\dfrac{1}{c} \dfrac{\partial\,
                                                \vec{\mathcal{A}}}{\partial\,t}\Big)\times
                                                      \vec{\mathcal{A}}\bigg]
                                    +\Bigl[\varphi,\ \big(\vec{\nabla}\times\vec{\mathcal{A}}
                                          \big)\Bigr]
                                    +\bigl(\vec{\nabla}\varphi\bigr)\times\vec{\mathcal{A}}
                                    +\vec{\mathcal{A}}\times\bigl(\vec{\nabla}\varphi\bigr)
                                    \Biggr\}, \nonumber\\
                  && \mathcal{W}_3=-{\rm i}\,g\,\vec{\nabla}\cdot\bigl(\vec{\mathcal{A}}
                                    \times\vec{\mathcal{A}}\bigr), \nonumber\\
                  && \mathcal{W}_4=-{\rm i}\,g\Biggl\{\bigg[\varphi,\
                                          \dfrac{1}{c} \dfrac{\partial\,\vec{\mathcal{A}}}{
                                                \partial\,t}\bigg]
                                    +\Big[\varphi,\ \bigl(\vec{\nabla}\varphi\bigr)\Big]
                                    -\vec{\mathcal{A}}\times\bigl(\vec{\nabla}\times
                                          \vec{\mathcal{A}}\bigr)
                                    -\bigl(\vec{\nabla}\times\vec{\mathcal{A}}\bigr)\times
                                          \vec{\mathcal{A}}\Biggr\}.
            \end{eqnarray}
      Based on
      \begin{eqnarray}
       && \vec{\mathcal{A}}(\vec{r}, t)=\vec{\tau}\,{\rm e}^{
                        {\rm i}(\vec{k}\cdot\vec{r}-\omega\,t)}, \nonumber\\
                 && \varphi(\vec{r}, t) =\bigl(\vec{\tau}\cdot\hat{k}\bigr){\rm e}^{
                        {\rm i}(\vec{k}\cdot\vec{r}-\omega\,t)},
      \end{eqnarray}
      one may prove that
      \begin{eqnarray}
       && \mathcal{W}_1=\mathcal{W}_2=\mathcal{W}_3=\mathcal{W}_4=0.
      \end{eqnarray}
      \emph{Proof.---}Because
            \begin{equation}
                  \dfrac{1}{c} \dfrac{\partial\,\vec{\mathcal{A}}}{\partial\,t}
                  =-{\rm i}\,k\,\vec{\tau}\,{\rm e}^{{\rm i}(\vec{k}\cdot\vec{r}
                        -\omega\,t)} ,\quad
                  \vec{\nabla}\big[\varphi(\vec{r},t)\big]={\rm i}\,k\bigl(\vec{\tau}
                        \cdot\hat{k}\bigr)\hat{k}\,{\rm e}^{{\rm i}(\vec{k}\cdot\vec{r}
                              -\omega\,t)},\quad
                  \vec{\nabla}\times\vec{\mathcal{A}}={\rm i}\,k\bigl(\hat{k}\times
                        \vec{\tau}\bigr){\rm e}^{{\rm i}(\vec{k}\cdot\vec{r}
                              -\omega\,t)},
            \end{equation}
            then
            \begin{itemize}
                  \item [(i)]. For $\mathcal{W}_1$,
                        \begin{eqnarray}
                                    && \bigg[\vec{\mathcal{A}}\cdot\Big(
                                          \dfrac{1}{c} \dfrac{\partial\,
                                                \vec{\mathcal{A}}}{\partial\,t}\Big)
                                          -\Big(\dfrac{1}{c} \dfrac{\partial\,
                                                \vec{\mathcal{A}}}{\partial\,t}\Big)\cdot\vec{\mathcal{A}}\bigg]
                                          +\vec{\mathcal{A}}\cdot\bigl(\vec{\nabla}
                                                \varphi\bigr)
                                          -\bigl(\vec{\nabla}\varphi\bigr)\cdot\vec{
                                                \mathcal{A}} \nonumber\\
                                    &=& -{\rm i}\,k\Big[\bigl(\vec{\tau}\cdot\vec{\tau}
                                                \bigr)-\bigl(\vec{\tau}\cdot\vec{\tau}
                                                \bigr)\Big]{\rm e}^{{\rm i}\,2(\vec{k}
                                                      \cdot\vec{r}-\omega\,t)}
                                          +{\rm i}\,k\Big[\bigl(\vec{\tau}\cdot\hat{k}
                                                      \bigr)^2
                                                -\bigl(\vec{\tau}\cdot\hat{k}\bigr)^2
                                                \Big]{\rm e}^{{\rm i}\,2(\vec{k}\cdot
                                                      \vec{r}-\omega\,t)}
                                    =0.
                        \end{eqnarray}
                  \item [(ii)]. For $\mathcal{W}_2$,
                        \begin{eqnarray}
                                    && \bigg[\vec{\mathcal{A}}\times\Big(\dfrac{1}{c}
                                                \dfrac{\partial\,\vec{\mathcal{A}}}{
                                                      \partial\,t}\Big)
                                          +\Big(\dfrac{1}{c} \dfrac{\partial\,
                                                \vec{\mathcal{A}}}{\partial\,t}\Big)
                                                      \times\vec{\mathcal{A}}\bigg]
                                          +\Bigl[\varphi,\ \big(\vec{\nabla}\times
                                                \vec{\mathcal{A}}\big)\Bigr]
                                          +\bigl(\vec{\nabla}\varphi\bigr)\times\vec{
                                                \mathcal{A}}
                                          +\vec{\mathcal{A}}\times\bigl(\vec{\nabla}
                                                \varphi\bigr) \nonumber\\
                                    &=& -{\rm i}\,k\Big[\bigl(\vec{\tau}\times\vec{\tau}
                                                \bigr)+\bigl(\vec{\tau}\times\vec{\tau}
                                                \bigr)\Big]{\rm e}^{{\rm i}\,2(\vec{k}
                                                      \cdot\vec{r}-\omega\,t)}
                                          +{\rm i}\,k\Bigl[\bigl(\vec{\tau}\cdot\hat{k}
                                                \bigr),\ \bigl(\hat{k}\times\vec{\tau}
                                                \bigr)\Bigr]{\rm e}^{{\rm i}\,2(\vec{k}
                                                      \cdot\vec{r}-\omega\,t)}\nonumber\\
                                          &&+{\rm i}\,k\Bigl[\bigl(\vec{\tau}\cdot\hat{k}
                                                \bigr)\bigl(\hat{k}\times\vec{\tau}
                                                \bigr)+\bigl(\vec{\tau}\times\hat{k}
                                                \bigr)\bigl(\vec{\tau}\cdot\hat{k}
                                                \bigr)\Bigr]{\rm e}^{{\rm i}\,2(\vec{k}
                                                      \cdot\vec{r}-\omega\,t)} \nonumber\\
                                    &=& -{\rm i}\,2\,k\bigl(\vec{\tau}\times\vec{\tau}
                                                \bigr){\rm e}^{{\rm i}\,2(\vec{k}\cdot
                                                      \vec{r}-\omega\,t)}
                                          +{\rm i}\,k\,\hat{k}\times\Bigl[\bigl(
                                                \vec{\tau}\cdot\hat{k}\bigr),\
                                                \vec{\tau}\Bigr]{\rm e}^{{\rm i}\,2(
                                                      \vec{k}\cdot\vec{r}-\omega\,t)}
                                          +{\rm i}\,k\Bigl[\bigl(\vec{\tau}\cdot\hat{k}
                                                \bigr)\bigl(\hat{k}\times\vec{\tau}
                                                \bigr)-\bigl(\hat{k}\times\vec{\tau}
                                                \bigr)\bigl(\vec{\tau}\cdot\hat{k}
                                                \bigr)\Bigr]{\rm e}^{{\rm i}\,2(\vec{k}
                                                      \cdot\vec{r}-\omega\,t)} \nonumber\\
                                    &=& -{\rm i}\,2\,k\bigl(\vec{\tau}\times\vec{\tau}
                                                \bigr){\rm e}^{{\rm i}\,2(\vec{k}\cdot
                                                      \vec{r}-\omega\,t)}
                                          +{\rm i}\,2\,k\,\hat{k}\times\Bigl[\bigl(
                                                \vec{\tau}\cdot\hat{k}\bigr),\
                                                \vec{\tau}\Bigr]{\rm e}^{{\rm i}\,2(
                                                      \vec{k}\cdot\vec{r}-\omega\,t)}
                                    ={\rm i}\,2\,k\,\biggl\{\hat{k}\times\Bigl[\bigl(
                                                \vec{\tau}\cdot\hat{k}\bigr),\
                                                \vec{\tau}\Bigr]
                                          -\bigl(\vec{\tau}\times\vec{\tau}\bigr)
                                          \biggr\}{\rm e}^{{\rm i}\,2(\vec{k}\cdot
                                                \vec{r}-\omega\,t)} \nonumber \\
                                    &=& 0.
                        \end{eqnarray}
                  \item [(iii)]. For $\mathcal{W}_3$,
                        \begin{equation}
                              \begin{split}
                                    & \vec{\nabla}\cdot\bigl(\vec{\mathcal{A}}\times
                                          \vec{\mathcal{A}}\bigr)
                                    =\vec{\nabla}\cdot\Bigl[\bigl(\vec{\tau}\times
                                          \vec{\tau}\bigr){\rm e}^{{\rm i}\,2(
                                                \vec{k}\cdot\vec{r}-\omega\,t)}\Bigr]
                                    =\vec{\nabla}\cdot\Bigl[\vec{M}_1 {\rm e}^{{\rm i}\,
                                          2(\vec{k}\cdot\vec{r}-\omega\,t)}\Bigr]
                                    =0.
                              \end{split}
                        \end{equation}
                  \item [(iv)]. For $\mathcal{W}_4$,
                        \begin{eqnarray}
                                    && \bigg[\varphi,\ \dfrac{1}{c}
                                          \dfrac{\partial\,\vec{\mathcal{A}}}{
                                                \partial\,t}\bigg]
                                          +\Big[\varphi,\ \bigl(\vec{\nabla}\varphi
                                                \bigr)\Big]
                                          -\vec{\mathcal{A}}\times\bigl(\vec{\nabla}
                                                \times\vec{\mathcal{A}}\bigr)
                                          -\bigl(\vec{\nabla}\times\vec{\mathcal{A}}
                                                \bigr)\times\vec{\mathcal{A}} \nonumber\\
                                    &=& -{\rm i}\,k\Big[\bigl(\vec{\tau}\cdot\hat{k}
                                                \bigr),\ \vec{\tau}\Big]{\rm e}^{
                                                      {\rm i}\,2(\vec{k}\cdot\vec{r}
                                                            -\omega\,t)}
                                          +{\rm i}\,k\Big[\bigl(\vec{\tau}\cdot\hat{k}
                                                \bigr),\ \bigl(\vec{\tau}\cdot\hat{k}
                                                \bigr)\hat{k}\Big]{\rm e}^{{\rm i}\,2(
                                                      \vec{k}\cdot\vec{r}-\omega\,t)}
                                          -{\rm i}\,k\Big[\vec{\tau}\times\bigl(\hat{k}
                                                      \times\vec{\tau}\bigr)
                                                +\bigl(\hat{k}\times\vec{\tau}\bigr)
                                                      \times\vec{\tau}\Big]{\rm e}^{
                                                            {\rm i}\,2(\vec{k}\cdot
                                                                  \vec{r}-\omega\,t)} \nonumber\\
                                    &=& -{\rm i}\,k\Big[\bigl(\vec{\tau}\cdot\hat{k}
                                                \bigr),\ \vec{\tau}\Big]{\rm e}^{
                                                      {\rm i}\,2(\vec{k}\cdot\vec{r}
                                                            -\omega\,t)}
                                          -{\rm i}\,k\Big[\vec{\tau}\times\bigl(\hat{k}
                                                      \times\vec{\tau}\bigr)
                                                -\bigl(\vec{\tau}\times\hat{k}\bigr)
                                                      \times\vec{\tau}\Big]{\rm e}^{
                                                            {\rm i}\,2(\vec{k}\cdot
                                                                  \vec{r}-\omega\,t)} \nonumber\\
                                    &=& -{\rm i}\,k\Big[\bigl(\vec{\tau}\cdot\hat{k}
                                                \bigr),\ \vec{\tau}\Big]{\rm e}^{
                                                      {\rm i}\,2(\vec{k}\cdot\vec{r}
                                                            -\omega\,t)}
                                          -{\rm i}\,k\Big[\bigl(\vec{\tau}\cdot
                                                      \vec{\tau}\bigr)\hat{k}
                                                -\bigl(\vec{\tau}\cdot\hat{k}\bigr)
                                                      \vec{\tau}
                                                -\bigl(\vec{\tau}\cdot\vec{\tau}\bigr)
                                                      \hat{k}
                                                +\vec{\tau}\bigl(\vec{\tau}\cdot\hat{k}
                                                      \bigr)\Big]{\rm e}^{{\rm i}\,2(
                                                            \vec{k}\cdot\vec{r}
                                                            -\omega\,t)} \nonumber\\
                                    &=& -{\rm i}\,k\Big[\bigl(\vec{\tau}\cdot\hat{k}
                                                \bigr),\ \vec{\tau}\Big]{\rm e}^{
                                                      {\rm i}\,2(\vec{k}\cdot\vec{r}
                                                            -\omega\,t)}
                                          +{\rm i}\,k\Big[\bigl(\vec{\tau}\cdot\hat{k}
                                                \bigr),\ \vec{\tau}\Big]{\rm e}^{
                                                      {\rm i}\,2(\vec{k}\cdot\vec{r}
                                                            -\omega\,t)}
                                    =0.
                        \end{eqnarray}
            \end{itemize}
            Thus the $SU(N)$ operator solutions of the Yang-Mills equations under weak-coupling approximation are the same as those of Maxwell-type equations.
\newpage
\section{Discussion}

      In this section, we would like to discuss some possible comments.
      
      \begin{proof}[Comment (1).] Discussion about the coupling parameter $g$.

            \emph{Answer}:
            From the Yang-Mills equations in \Eq{eq:YM1a} and \Eq{eq:YM1b}, i.e.,
            \begin{subequations}
                  \begin{eqnarray}
                        && D_\mu \mathcal{F}^{\mu\nu} =\partial_\mu \mathcal{F}^{\mu\nu}
                              -{\rm i}\,g\Bigl[\mathbb{A}_\mu ,\ \mathcal{F}^{\mu\nu}\Bigr]=0, \label{eq:E-0a} \\
                        && D_\mu \mathcal{F}_{\nu\gamma} +D_\nu \mathcal{F}_{\gamma\mu}
                              +D_\gamma \mathcal{F}_{\mu\nu} =0, \label{eq:E-0b}
                  \end{eqnarray}
           \end{subequations}
            we have known that the parameter $g$ characterizes the coupling between the potential $\mathbb{A}_\mu$ and the field $\mathcal{F}^{\mu\nu}$. In our work, we have considered the weak coupling approximation, namely, $g$ is very small, so that $g^2 \approx 0$. People might ask a question: For what energy scale, can the coupling $g$ be regarded as small? We have searched relevant references. However, to our knowledge, it seems that no one has explicitly answered this question. Nonetheless, we have found some weak coupling cases in QCD and nuclear physics, there are many weak coupling cases in QCD, nuclear physics \cite{1985NPBGasser}, and artificial gauge fields \cite{2011RMPJean,2011NatureLin} (where $g$ embodies the coupling interaction strength, rather than the energy scale). Hence, in this work we currently treat $g$ as an adjustable parameter to calibrate the non-Abelian property of the given system, i.e., $g=0$ (Abelian), $g\neq 0\ \&\ g^2 \approx 0$ (weak non-Abelian), and non-Abelian for the general nonzero $g$. It is important to consider the significant problem that for what energy scale the coupling $g$ can be regarded as small. We shall study the problem in our next work.
      \end{proof}
      \begin{proof}[Comment (2).] Discussion about the stability of solutions.

            \emph{Answer}:
            In this work, we have sought a plane-wave solution of the linearized Yang-Mills equations in vacuum. People might ask: Is this solution stable? There exists no decay factor in our solutions (see (8), (9), and (14) in the main text) of the linearized Yang-Mills equations, so they are stable. Explicitly, the propagation part $e^{i(\vec{k}\cdot\vec{r}-\omega t)}$ is stable (since no decay factor), and the amplitude part, as the expectation of the spin Zitterbewegung operator when observed is stable too, see Eqs. \eqref{eq:ZSPsi14} and \eqref{eq:ZSPsi14Z}.
      \end{proof}
      \begin{proof}[Comment (3).]
            Discussion about distinguishing electromagnetic waves and angular-momentum waves.

            \emph{Answer}:
            In this work, we have devised a thought experiment based on spin ``Zitterbewegung''. However, the spin oscillation of a Dirac electron can emit spin angular-momentum wave together with electromagnetic waves. People might ask: How can we distinguish the contribution from electromagnetic waves and (spin) angular-momentum waves in our thought experiments? In our thought experiment (see Fig.\ref{fig:AMWave}), a Dirac's electron (particle $A$) is prepared in the superposition of different energy and different helicity, i.e., $|\Psi\rangle=\cos\theta|\Psi_1\rangle +\sin\theta|\Psi_4\rangle$. In such a state, there exist both ``position Zitterbewegung'' and ``spin Zitterbewegung''. The former emits electromagnetic waves, while the later emits (spin) angular-momentum waves. Because a neutron has no charge but possesses spin, we can change the probe particle $B$ to be a neutron, which cannot be influenced via the electromagnetic field from the source particle $A$. In this way, one can distinguish the contribution from electromagnetic waves and (spin) angular-momentum waves in the thought experiment.
      \end{proof}
      \begin{proof}[Comment (4).]
            Discussion about the motivation, theoretical principle, and experimental relationship.

            \emph{Answer}:
            People might ask: What are the motivation, theoretical principle, and experimental relationship of this work?

            \emph{Motivation.}---Since Maxwell's equations predict the well-known electromagnetic waves, and the Yang-Mills equations are their natural generalization. It gives rise to a scientific question: What types of waves do the Yang-Mills equations predict? Thus the motivation of our work is finding a theoretical prediction of quantum Yang-Mills equations. As we have discussed at the beginning of the last paragraph in the main text (see page 4), ``\emph{for a solid physical theory, its predictions are often reliable. e.g., Maxwell's equations predicted the electromagnetic wave, special relativity predicted the mass-energy relation, general relativity predicted the gravitational waves, and quantum mechanics predicted Heisenberg uncertainty relations, Aharonov-Bohm effect, Berry phase, etc.}''. These sound theories enlighten us to consider the predictions of Yang-Mills equations, which are as the non-Abelian generalization of Maxwell's equations.

            \emph{Theoretical principle.}---Our theory is based on the Yang-Mills equations and quantum mechanics (the 1st quantization only), while not involves quantum field theory, where the Yang-Mills field is quantized again, i.e., the 2nd quantization.

            \emph{Experimental relationship.}---Since electromagnetic waves are the solutions of Maxwell's equations even in vacuum without source, hence we calculate the solutions of Yang-Mills equations. The calculating results about the angular-momentum wave motivate us to devise a possible (thought) experiment to observe such a type of wave based on ``spin Zitterbewegung'' (i.e., a kind of spin oscillation in relativistic quantum mechanics), in analogy with the charge oscillation to produce electromagnetic wave.
      \end{proof}
      \begin{proof}[Comment (5).]
            Discussion about the letter from Sakurai to C. N. Yang.

            \emph{Answer}: As mentioned via C. N. Yang in \cite{2005CNYang},
            ``The gauge theory idea was being applied in the late 1950s to strong interaction theory as well as to weak interactions. In 1960, J. Sakurai published a very enthusiastic paper proposing a non-Abelian gauge theory of strong interactions \cite{1960Sakurai}.

            ...he wrote me a letter expressing his irritation that I `seem to be detached from not too sympathetic toward, the sort of program that J. Sakurai have been pursuing in the past two years.' He asked why I was so detached and went on:

            \vspace{3mm}

             \hspace{2cm}\emph{\textcolor{blue}{You often tell young theoreticians that one of the paramount tasks of theoreticians is}}

             \hspace{1.2cm} \emph{\textcolor{blue}{to suggest a good experiment. Yet when you proposed the Yang-Mills theory in 1954, you}}

              \hspace{1.2cm} \emph{\textcolor{blue}{did not encourage the experimentalists to look for the `Yang-Mills particle'. Why is it?}}

            \vspace{3mm}

            I remember sitting on this letter for a long time, not knowing how to answer him. I don't remember whether I eventually did answer. If I did, it must have been a polite note with no substance.''

            From the commentary of Prof. C. N. Yang we can see that ``Yang-Mills particle'' is an unknown topic then, even to date. Whereas the difficulty of hunting ``Yang-Mills particle'' does not shrink the importance thereof. Just as Feynman commented on the Aharonov-Bohm effect in \cite{2010Feynman}  that: ``\emph{\textcolor{blue}{It is interesting that something like this (AB effect) can be around for thirty years but, because of certain prejudices of what is and is not significant, continues to be ignored.}}'' Remarkably, what we have done in this manuscript is a bold try to seek the ``Yang-Mills'' wave [i.e., the (spin) angular-momentum wave as predicted in our work], or from the perspective of the wave-particle duality, the term ``Yang-Mills particle'' coined by J. Sakurai. We have also suggested a good thought-experiment to detect such a spin angular momentum wave as ``\emph{one of the paramount tasks of theoreticians}'' described by J. Sakurai. The focal conception of ``Yang-Mills particle'' has been disregarded for a long time, so it's exciting to explain why we are motivated with this work.
      \end{proof}
      \begin{proof}[Comment (6).] Discussion about reducing Yang-Mills equations to Maxwell's equations.

      \emph{Answer}: (i) Let us discuss how Yang-Mills equations reduce to the standard Maxwell's equations.

      (i-1) From the viewpoint of field-strength tensors. The field-strength tensors in Yang-Mills theory are defined by
         \begin{eqnarray}\label{eq:E-1}
                        \mathcal{F}_{\mu\nu}=\partial_\mu \mathbb{A}_{\nu}
                              -\partial_\nu \mathbb{A}_{\mu}
                              +{\rm i}\,g[\mathbb{A}_{\mu},\
                                    \mathbb{A}_{\nu}],
         \end{eqnarray}
         where $g$ is the coupling parameter. When the following two conditions
         \begin{subequations}
         \begin{eqnarray}
          && g=0,  \label{eq:E-2a} \\
          &&[\mathbb{A}_{\mu}, \mathbb{A}_{\nu}]=\mathbb{A}_{\mu}\mathbb{A}_{\nu}-\mathbb{A}_{\nu}\mathbb{A}_{\mu}=0,  \label{eq:E-2b}
         \end{eqnarray}
         \end{subequations}
         are satisfied, then the definition of $\mathcal{F}_{\mu\nu}$ in Eq. (\ref{eq:E-1}) reduces to the Abelian case, i.e.,
          \begin{equation}\label{eq:E-3}
                  F_{\mu\nu} =\partial_\mu A_{\nu} -\partial_\nu A_{\mu},
            \end{equation}
          which are the field-strength tensors in Maxwell's theory with $[{A}_{\mu}, {A}_{\nu}]=0$.

          (i-2) From the viewpoint of equations (in the language of field-strength tensors). The Yang-Mills equations are given by Eq. (\ref{eq:E-0a}) and Eq. (\ref{eq:E-0b}). When the two conditions in Eq. (\ref{eq:E-2a}) and Eq. (\ref{eq:E-2b}) are satisfied, then the  Yang-Mills equations reduce to the standard Maxwell's equations, i.e.,
          \begin{subequations}
                  \begin{eqnarray}
                        && \partial_\mu F^{\mu\nu} = 0, \\
                        && \partial_\mu F_{\nu\gamma}+\partial_\nu F_{\gamma\mu}
                              +\partial_\gamma F_{\mu\nu} = 0.
                  \end{eqnarray}
            \end{subequations}

          (i-3) From the viewpoint of equations (in the language of electric field, magnetic field, vector potential and scalar potential). The Yang-Mills equations are given by
          \begin{subequations}
                  \begin{eqnarray}\label{eq:E-4}
                  && \vec{\nabla}\cdot\vec{\mathcal{E}}\textcolor{red}{
                        -{\rm i}\,g\Biggl\{\bigg[\vec{\mathcal{A}}\cdot\Big(
                                          \dfrac{1}{c} \dfrac{\partial\,
                                                \vec{\mathcal{A}}}{\partial\,t}\Big)
                                    -\Big(\dfrac{1}{c} \dfrac{\partial\,
                                          \vec{\mathcal{A}}}{\partial\,t}\Big)\cdot
                                                \vec{\mathcal{A}}\bigg]
                              +\vec{\mathcal{A}}\cdot\bigl(\vec{\nabla}\varphi\bigr)
                              -\bigl(\vec{\nabla}\varphi\bigr)\cdot\vec{\mathcal{A}}
                              \Biggr\}}
                        {\color{blue}+g^2 \biggl\{\vec{\mathcal{A}}\cdot\left[
                                    \varphi,\ \vec{\mathcal{A}}\right]
                              -\left[\varphi,\ \vec{\mathcal{A}}\right]
                                    \cdot\vec{\mathcal{A}}\biggr\}}=0, \label{eq:DivEYM3-n1a} \\
                  && -\dfrac{1}{c} \dfrac{\partial}{\partial\,t} \vec{\mathcal{B}}
                        -\vec{\nabla}\times\vec{\mathcal{E}}
                        \textcolor{red}{+{\rm i}\,g\Biggl\{\bigg[\vec{\mathcal{A}}
                                          \times\Big(\dfrac{1}{c}
                                                \dfrac{\partial\,\vec{\mathcal{A}}}{
                                                      \partial\,t}\Big)
                                    +\Big(\dfrac{1}{c} \dfrac{\partial\,
                                          \vec{\mathcal{A}}}{\partial\,t}\Big)\times
                                                \vec{\mathcal{A}}\bigg]
                              +\Bigl[\varphi,\ \big(\vec{\nabla}\times\vec{\mathcal{A}}
                                    \big)\Bigr]
                              +\bigl(\vec{\nabla}\varphi\bigr)\times\vec{\mathcal{A}}
                              +\vec{\mathcal{A}}\times\bigl(\vec{\nabla}\varphi\bigr)
                              \Biggr\}} \notag \\
                        &&\qquad\qquad\qquad\qquad\ {\color{blue}+g^2 \biggl\{\Bigl[\varphi,\ \bigl(
                                    \vec{\mathcal{A}}\times\vec{\mathcal{A}}\bigr)\Bigr]
                              -\vec{\mathcal{A}}\times\left[\varphi,\
                                    \vec{\mathcal{A}}\right]
                              -\left[\varphi,\ \vec{\mathcal{A}}\right]
                                    \times\vec{\mathcal{A}}\biggr\}}=0, \label{eq:CurlEYM3-n1b} \\
                  && \vec{\nabla}\cdot\vec{\mathcal{B}}
                        \textcolor{red}{-{\rm i}\,g\,\vec{\nabla}\cdot\bigl(\vec{\mathcal{A}}
                              \times\vec{\mathcal{A}}\bigr)}=0, \label{eq:DivBYM3-n1c} \\
                  && -\dfrac{1}{c} \dfrac{\partial}{\partial\,t} \vec{\mathcal{E}}
                        +\vec{\nabla}\times\vec{\mathcal{B}}
                        \textcolor{red}{-{\rm i}\,g\Biggl\{\bigg[\varphi,\
                                    \dfrac{1}{c} \dfrac{\partial\,\vec{\mathcal{A}}}{
                                          \partial\,t}\bigg]
                              +\Big[\varphi,\ \bigl(\vec{\nabla}\varphi\bigr)\Big]
                              -\vec{\mathcal{A}}\times\bigl(\vec{\nabla}\times
                                    \vec{\mathcal{A}}\bigr)
                              -\bigl(\vec{\nabla}\times\vec{\mathcal{A}}\bigr)\times
                                    \vec{\mathcal{A}}\Biggr\}} \notag \\
                        &&\qquad\qquad\qquad\qquad\ {\color{blue}+g^2 \biggl\{\Big[\varphi,\ \big[\varphi,\
                                    \vec{\mathcal{A}}\big]\Big]
                              +\vec{\mathcal{A}}\times\bigl(\vec{\mathcal{A}}
                                    \times\vec{\mathcal{A}}\bigr)
                              +\bigl(\vec{\mathcal{A}}\times\vec{\mathcal{A}}\bigr)
                                    \times\vec{\mathcal{A}}\biggr\}}=0. \label{eq:CurlBYM3-n1d}
            \end{eqnarray}
            \end{subequations}
           Note that in this situation, $\vec{\mathcal{E}}$ and $\vec{\mathcal{B}}$ can be operators, and their corresponding vector potential $\vec{\mathcal{A}}$ and scalar potential $\varphi$ can be non-Abelian, e.g., $\vec{\mathcal{A}}\times \vec{\mathcal{A}}$ can be nonzero.

           When the two conditions Eq. (\ref{eq:E-2a}) and Eq. (\ref{eq:E-2b}) are satisfied, the Yang-Mills equations Eq. (\ref{eq:DivEYM3-n1a})-Eq.(\ref{eq:CurlBYM3-n1d}) reduce to the the standard Maxwell's equations, namely,
          \begin{subequations}
                  \begin{eqnarray}
                  && \vec{\nabla}\cdot\vec{E} = 0, \label{eq:DivE-n1a} \\
                  && \vec{\nabla}\times\vec{E} = -\dfrac{1}{c}
                              \dfrac{\partial\,\vec{B}}{\partial\,t}, \label{eq:CurlE-n1b} \\
                  && \vec{\nabla}\cdot\vec{B} = 0, \label{eq:DivB-n1c} \\
                  && \vec{\nabla}\times\vec{B} = \dfrac{1}{c}
                              \dfrac{\partial\,\vec{E}}{\partial\,t}. \label{eq:CurlB-n1d}
                  \end{eqnarray}
            \end{subequations}
            Note that in this situation, $\vec{E}$ and $\vec{B}$ are some functions of $\vec{r}$ and $t$ (or these functions multiply the identity operator $\openone$ if one views them as operators), and their corresponding vector potential $\vec{A}$ and scalar potential $\varphi$ are Abelian, i.e., $\vec{A}\times \vec{A}=0$ and $[\vec{A}, \varphi]=0$.

            (ii) Let us discuss how the solutions of angular momentum waves reduce to the electromagnetic waves. Here we take the Yang-Mills equations under the weak-coupling approximation as an example.

            The solutions of potentials $\{\vec{\mathcal{A}}, \varphi\}$ and the corresponding angular momentum waves $\{\vec{\mathcal{B}}, \vec{\mathcal{E}}\}$ are given by
            \begin{subequations}
            \begin{eqnarray}
                 && \vec{\mathcal{A}}(\vec{r}, t)=\vec{\tau}\,{\rm e}^{
                        {\rm i}(\vec{k}\cdot\vec{r}-\omega\,t)}, \label{eq:WCA-1a-n1a}\\
                 && \varphi(\vec{r}, t) =\bigl(\vec{\tau}\cdot\hat{k}\bigr){\rm e}^{
                        {\rm i}(\vec{k}\cdot\vec{r}-\omega\,t)}, \label{eq:WCA-1a-n1b}\\
                 && \vec{\mathcal{B}}(\vec{r}, t)=\vec{\nabla}\times\vec{\mathcal{A}}-{\rm i}
                        \,g\bigl(\vec{\mathcal{A}}\times\vec{\mathcal{A}}\bigr)
                  ={\rm i}\bigl(\vec{k}\times\vec{\tau}\bigr){\rm e}^{
                              {\rm i}(\vec{k}\cdot\vec{r}-\omega\,t)}
                        +g\,\hbar\,\vec{\eta}\,{\rm e}^{
                              {\rm i}\,2(\vec{k}\cdot\vec{r}-\omega\,t)}, \label{eq:WCA-1a-n1c}\\
                 && \vec{\mathcal{E}}(\vec{r}, t)=\dfrac{1}{c}
                        \dfrac{\partial\,\vec{\mathcal{A}}}{\partial\,t}
                              -\vec{\nabla}\varphi-{\rm i}\,g\left[
                                    \varphi,\ \vec{\mathcal{A}}\right]
                  =-{\rm i}\,k\Big[\hat{k}\times\bigl(\hat{k}
                                    \times\vec{\tau}\bigr)\Big]{\rm e}^{
                                          {\rm i}(\vec{k}\cdot\vec{r}
                                                -\omega\,t)}
                              +g\,\hbar\,\vec{\xi}\,{\rm e}^{{\rm i}\,2(
                                    \vec{k}\cdot\vec{r}-\omega t)}, \label{eq:WCA-1a-n1d}
            \end{eqnarray}
            \end{subequations}
            with
            \begin{subequations}
            \begin{eqnarray}
                  &&\vec{\tau}=\vec{R}_0 \openone+\vec{R}_1 S_x +\vec{R}_2 S_y
                        +\vec{R}_3 S_z, \label{eq:WCA-1b-n1a}\\
                  && \vec{\eta}=\big(\vec{R}_2 \times\vec{R}_3\big)S_x
                        +\big(\vec{R}_3 \times\vec{R}_1\big)S_y
                        +\big(\vec{R}_1 \times\vec{R}_2\big)S_z,\label{eq:WCA-1b-n1b}\\
                  &&  \vec{\xi}=-\hat{k}\times\vec{\eta}=\vec{\eta}\times \hat{k}, \label{eq:WCA-1b-n1c}\\
                  &&\vec{k}\cdot\bigl(\vec{R}_1 \times\vec{R}_2\bigr)
                  =\vec{k}\cdot\bigl(\vec{R}_2 \times\vec{R}_3\bigr)
                  =\vec{k}\cdot\bigl(\vec{R}_3 \times\vec{R}_1\bigr)
                  =0.\label{eq:WCA-1b-n1d}
            \end{eqnarray}
            \end{subequations}
            Here $\hat{k}=\vec{k}/|\vec{k}|$ is the direction of propagation, $\vec{R}_j$'s ($j=0,1,2,3$) are some constant vectors that do not depend on $\vec{r}$.  Eq. (\ref{eq:WCA-1b-n1d}) indicates that four vectors $\{\vec{R}_1, \vec{R}_2, \vec{R}_3, \vec{k}\}$ locate on the same plane, and $\vec{R}_0$ is arbitrary. The vectorial operator $\vec{\tau}$ has been expanded by the identity operator $\openone$ and three $SU(2)$ generators, which satisfy the following commutative relations
            \begin{eqnarray}
                        [S_x, S_y]={\rm i}\hbar S_z, \;[S_y, S_z]={\rm i}\hbar S_x, \; [S_z, S_y]={\rm i}\hbar S_x,
            \end{eqnarray}
            or in the vector form as
            \begin{eqnarray}
                        \vec{S} \times \vec{S} ={\rm i} \hbar \, \vec{S},
            \end{eqnarray}
            with $\vec{S}=(S_x, S_y, S_z)$, $\hbar=h/2\pi$, and $h$ being Planck's constant. For instance, for the spin-1/2 particle, the angular momentum operator $\vec{S}$ can be realized by
 \begin{eqnarray}
 \label{S-1-n1a}
  && \vec{S}=\frac{\hbar}{2} \vec{\sigma},
  \end{eqnarray}
where $\vec{\sigma}$ is the vector of Pauli matrices, whose three components read
 \begin{eqnarray}
 \label{S-3-n1a}
  && \sigma_x=
  \left(
    \begin{array}{cc}
      0 & 1 \\
      1 & 0 \\
    \end{array}
  \right),\;\;\;\sigma_y=
  \left(
    \begin{array}{cc}
      0 & -{\rm i} \\
      {\rm i} & 0 \\
    \end{array}
  \right),\;\;\;  \sigma_z=
  \left(
    \begin{array}{cc}
      1 & 0 \\
      0 & -1 \\
    \end{array}
  \right).
   \end{eqnarray}

   \begin{remark}Let us consider the first condition shown in Eq. (\ref{eq:E-2a}), then Eq. (\ref{eq:WCA-1a-n1a})-Eq. (\ref{eq:WCA-1a-n1d}) become
   \begin{subequations}
            \begin{eqnarray}
                 && \vec{\mathcal{A}}(\vec{r}, t)=\vec{\tau}\,{\rm e}^{
                        {\rm i}(\vec{k}\cdot\vec{r}-\omega\,t)}, \label{eq:WCA-1a-m1a}\\
                 && \varphi(\vec{r}, t) =\bigl(\vec{\tau}\cdot\hat{k}\bigr){\rm e}^{
                        {\rm i}(\vec{k}\cdot\vec{r}-\omega\,t)}, \label{eq:WCA-1a-m1b}\\
                 && \vec{\mathcal{B}}(\vec{r}, t)=\vec{\nabla}\times\vec{\mathcal{A}}
                  ={\rm i}\bigl(\vec{k}\times\vec{\tau}\bigr){\rm e}^{
                              {\rm i}(\vec{k}\cdot\vec{r}-\omega\,t)}, \label{eq:WCA-1a-m1c}\\
                 && \vec{\mathcal{E}}(\vec{r}, t)=\dfrac{1}{c}
                        \dfrac{\partial\,\vec{\mathcal{A}}}{\partial\,t}
                              -\vec{\nabla}\varphi
                  =-{\rm i}\,k\Big[\hat{k}\times\bigl(\hat{k}
                                    \times\vec{\tau}\bigr)\Big]{\rm e}^{
                                          {\rm i}(\vec{k}\cdot\vec{r}
                                                -\omega\,t)}, \label{eq:WCA-1a-m1d}
            \end{eqnarray}
            \end{subequations}
            with
            \begin{eqnarray}
                  &&\vec{\tau}=\vec{R}_0 \openone+\vec{R}_1 S_x +\vec{R}_2 S_y
                        +\vec{R}_3 S_z, \label{eq:WCA-1b-m1a}
            \end{eqnarray}
            Let us consider the second condition shown in Eq. (\ref{eq:E-2b}), which means
            \begin{subequations}
            \begin{eqnarray}
                 && \vec{\mathcal{A}}(\vec{r}, t) \times \vec{\mathcal{A}}(\vec{r}, t)=0, \nonumber\\
                 && \left[\vec{\mathcal{A}}(\vec{r}, t), \varphi(\vec{r}, t)\right] =0,
            \end{eqnarray}
            \end{subequations}
            i.e.,
            \begin{subequations}
            \begin{eqnarray}\label{eq:tau-1}
                 && \vec{\tau} \times \vec{\tau}=0, \nonumber\\
                 && \left[\vec{\tau}, \vec{\tau}\cdot \hat{k}\right] =0.
            \end{eqnarray}
            \end{subequations}
            A direct solution reads
            \begin{eqnarray}\label{eq:tau-1a}
                 && \vec{\tau} = \vec{R}_0 \openone,
            \end{eqnarray}
            this leads to
            \begin{eqnarray}\label{eq:E-5}
               && \vec{\mathcal{B}}(\vec{r}, t)
                  =\openone \,  {\rm i}\bigl(\vec{k}\times \vec{R}_0\bigr){\rm e}^{
                              {\rm i}(\vec{k}\cdot\vec{r}-\omega\,t)}, \nonumber\\
                 && \vec{\mathcal{E}}(\vec{r}, t)
                  =-\openone \, {\rm i}\,k\Big[\hat{k}\times\bigl(\hat{k}
                                    \times \vec{R}_0\bigr)\Big]{\rm e}^{
                                          {\rm i}(\vec{k}\cdot\vec{r}
                                                -\omega\,t)}.
            \end{eqnarray}
           Apart from the identity operator $\openone$, by setting the amplitude $\vec{A}_{01}=-\Big[\hat{k}\times\bigl(\hat{k}
                                    \times \vec{R}_0\bigr)\Big]$, one can find that Eq. (\ref{eq:E-5}) recovers the solutions of the standard Maxwell's equation, i.e., Eq. (\ref{eq:A-6}), as below
            \begin{eqnarray}\label{eq:E-5a}
                  \vec{B} &=& {\rm i} (\vec{k} \times \vec{A}_{01})\,{\rm e}^{
                        {\rm i}(\vec{k}\cdot\vec{r}-\omega t)}, \nonumber\\
                  \vec{E} &=& {\rm i}\, k\; \vec{A}_{01}\, {\rm e}^{
                        {\rm i} (\vec{k}\cdot\vec{r}-\omega t)}.
            \end{eqnarray}
   $\blacksquare$
   \end{remark}

   \begin{remark} A careful reader may notice that the solution of Eq. (\ref{eq:tau-1}) can have a more general form, i.e.,
    \begin{eqnarray}\label{eq:E-6}
                  &&\vec{\tau}=\vec{R}_0 \openone+\vec{R}_1 S_x,
            \end{eqnarray}
   or $\vec{\tau}=\vec{R}_0 \openone+\vec{R}_2 S_y$, or $\vec{\tau}=\vec{R}_0 \openone+\vec{R}_3 S_z$ (namely, at most one component of $\vec{S}$ is survived). In this situation, we still have $\vec{\tau}\times \vec{\tau}=0$ and $\left[\vec{\tau}, \vec{\tau}\cdot \hat{k}\right] =0$. One can have
     \begin{eqnarray}\label{eq:E-7}
     \vec{k}\times \vec{\tau}&=&\left(\vec{k}\times \vec{R}_0\right)\, \openone+\left(\vec{k}\times \vec{R}_1\right)\, S_x, \nonumber\\
      \vec{k}\times (\vec{k}\times \vec{\tau})&=&\vec{k}\times \left(\vec{k}\times \vec{R}_0\right)\, \openone+\vec{k}\times\left(\vec{k}\times \vec{R}_1\right)\, S_x \nonumber\\
      &=& \left[\left(\vec{k}\cdot{\vec{R}_0}\right)\,\vec{k}-\vec{k}^2\, \vec{R}_0\right]\, \openone+\left[\left(\vec{k}\cdot{\vec{R}_1}\right)\,\vec{k}-\vec{k}^2\, \vec{R}_1\right]\,S_x, \nonumber\\
     \vec{k}\times\left[\vec{k}\times (\vec{k}\times \vec{\tau})\right]&=&\vec{k}\times \left\{\left[\left(\vec{k}\cdot{\vec{R}_0}\right)\,\vec{k}-\vec{k}^2\, \vec{R}_0\right]\, \openone+\left[\left(\vec{k}\cdot{\vec{R}_1}\right)\,\vec{k}-\vec{k}^2\, \vec{R}_1\right]\,S_x\right\}\nonumber\\
     &=&-k^2 \left[\left(\vec{k}\times \vec{R}_0\right)\, \openone+\left(\vec{k}\times \vec{R}_1\right)\, S_x\right]=-k^2(\vec{k}\times\vec{\tau}).
     \end{eqnarray}
     If we let
     \begin{eqnarray}\label{eq:E-8}
                  && \mathbb{M}\,\vec{A}_{01}=-\Big[\hat{k}\times\bigl(\hat{k}\times \vec{\tau}\bigr)\Big],
     \end{eqnarray}
     then for a fixed vector $\vec{A}_{01}$, we have $\mathbb{M}$ as a constant matrix, i.e.,
     \begin{eqnarray}\label{eq:E-9}
                  && \mathbb{M}=-\vec{A}_{01}\cdot\Big[\hat{k}\times\bigl(\hat{k}\times \vec{\tau}\bigr)\Big].
     \end{eqnarray}
     Then from Eq. (\ref{eq:WCA-1a-m1c}) and Eq. (\ref{eq:WCA-1a-m1d}) we have
     \begin{eqnarray} \label{eq:E-10}
                  \vec{B} &=& \mathbb{M} \,{\rm i} (\vec{k} \times \vec{A}_{01})\,{\rm e}^{
                        {\rm i}(\vec{k}\cdot\vec{r}-\omega t)}, \nonumber\\
                  \vec{E} &=&\mathbb{M}\, {\rm i}\, k\; \vec{A}_{01}\, {\rm e}^{
                        {\rm i} (\vec{k}\cdot\vec{r}-\omega t)}.
            \end{eqnarray}
     However, this constant matrix $\mathbb{M}$ does not tell us some more new knowledge, because if Eq. (\ref{eq:E-5a}) satisfy the standard Maxwell's equations, then a constant matrix $\mathbb{M}$ multiply Eq. (\ref{eq:E-5a}) will automatically satisfy the standard Maxwell's equations. Thus, when one considers the problem on reducing the Yang-Mills equations to Maxwell's equations, it is sufficient to set the vector $\vec{\tau}$ as shown in Eq. (\ref{eq:tau-1a}). $\blacksquare$
     \end{remark}

     (iii) One may ask a question: Why the vector $\vec{\tau}$ is written in such a form as shown in Eq. (\ref{eq:WCA-1b-m1a})? The following is our answer. The simple and direct reason is that the solutions of Yang-Mills equations are general non-Abelian, therefore one needs to introduce some non-commutative operators to solve the equations, while the $SU(2)$ angular momentum operators are the simplest non-commutative operators.

    Explicitly, the vector potential is given by $\vec{\mathcal{A}}(\vec{r}, t)=\vec{\tau}\,{\rm e}^{{\rm i}(\vec{k}\cdot\vec{r}-\omega\,t)}$, where the vector $\vec{\tau}$ serves as the amplitude of $\vec{\mathcal{A}}(\vec{r}, t)$. For Maxwell's theory, since the vector potential is Abelian, thus one may set $\vec{\tau}=\vec{R}_0 \openone$. However, for the Yang-Mills theory the vector potential is non-Abelian. Thus in this situation, the vector $\vec{\tau}$ is modified to adding $\vec{R}_0 \openone$ with some operators such that $\vec{\tau}\times\vec{\tau}\neq 0$. The simplest case is setting $\vec{\tau}=\vec{R}_0 \openone+\vec{R}_1 S_x +\vec{R}_2 S_y +\vec{R}_3 S_z$, where $\vec{S}$ is the $SU(2)$  angular momentum operator. A more complicate case is setting $ \vec{\tau} = \vec{R}_0 \openone+\sum_{l=1}^{N^2 -1} \vec{R}_l\,\hat{G}_l$, where $G_j$'s are the generators of $SU(N)$, as we have discussed in Sec. VI. Based on such a unified view, the electromagnetic waves are the $U(1)$ wave (i.e., the Abelian waves), the spin angular momentum waves are the $SU(2)$ waves (i.e., the simplest non-Abelian waves), and the more complicate non-Abelian waves are $SU(N)$ waves.
      \end{proof}
      \begin{proof}[Comment (7).] Discussion about the Poynting vector.

            \emph{Answer}: Poynting developed the Poynting theorem in 1884 \cite{Poyntingthe}, which describes the conservation law in electromagnetic field,
            \begin{eqnarray}
                  -\dfrac{\partial\,u}{\partial t}=\nabla\cdot\vec{\mathbb{S}}
                        +\vec{\mathbb{J}}\cdot\vec{E}
                  =\nabla\cdot\vec{\mathbb{S}}
            \end{eqnarray}
            in vacuum without source; where $u$ is the energy density in the volume one concerns, $\vec{\mathbb{S}}$ is the Poynting vector, $\vec{\mathbb{J}}$ indicates the current density, and the term $\vec{\mathbb{J}}\cdot\vec{E}$ is the rate that the field does work on charges in the volume. In classical electrodynamics, the Poynting vector is defined as (see e.g., \cite{EMWave})
            \begin{equation}
                  \vec{\mathbb{S}}=\dfrac{c}{4\,\pi} (\vec{E}\times\vec{H})
                  =\dfrac{c}{4\,\pi} (\vec{E}\times\vec{B})\propto\vec{E}\times\vec{B}
            \end{equation}
            in vacuum without source (Gaussian unit). The Poynting vector depicts the intensity of energy flow through a unit area perpendicular to the direction of flow per unit time.

            For the classical electromagnetic wave we considered, i.e., \Eq{eq:A-6},
            \begin{eqnarray}
                  && \vec{B}={\rm i}(\vec{k}\times\vec{A}_{01}){\rm e}^{
                        {\rm i}(\vec{k}\cdot\vec{r}-\omega t)},\notag \\
                  && \vec{E}={\rm i}\,k\,\vec{A}_{01}\,{\rm e}^{
                        {\rm i}(\vec{k}\cdot\vec{r}-\omega t)}.
            \end{eqnarray}
            Actually, $\vec{B}$ and $\vec{E}$ alter promptly, thus we care about the time average of them, which can be realized via extracting their real parts, then taking the norms of them. Let
            \begin{eqnarray}
                  && \vec{B}_{\rm R} \equiv{\rm Re}(\vec{B})=-(\vec{k}\times\vec{A}_{01}
                        )\sin(\vec{k}\cdot\vec{r}-\omega\,t),\notag \\
                  && \vec{E}_{\rm R} \equiv{\rm Re}(\vec{E})=-k\,\vec{A}_{01}\,\sin(
                        \vec{k}\cdot\vec{r}-\omega\,t),
            \end{eqnarray}
            which means
            \begin{eqnarray}
                  \vec{E}_{\rm R} \times\vec{B}_{\rm R}
                  &=& k\big[\vec{A}_{01}\times(\vec{k}\times\vec{A}_{01})\big]{\sin^2}(
                        \vec{k}\cdot\vec{r}-\omega\,t)\nonumber\\
                  &=& k\big[\vec{A}_{01}^2 \vec{k}
                        -(\vec{A}_{01} \cdot\vec{k})\vec{A}_{01}\big]{\sin^2}(
                              \vec{k}\cdot\vec{r}-\omega\,t) \notag \\
                  &=& k\,\vec{A}_{01}^2 \vec{k}\,{\sin^2}(\vec{k}\cdot\vec{r}
                        -\omega\,t) \notag \\
                  &=& k^2 \vec{A}_{01}^2 \big[{\sin^2}(\vec{k}\cdot\vec{r}-\omega\,t)
                        \big]\hat{k}.
            \end{eqnarray}
            In this case,
            \begin{eqnarray}\label{eq:PoyntEM}
                   \vec{\mathbb{S}}&=&\dfrac{c}{4\,\pi} (
                        \overline{\vec{E}_{\rm R} \times\vec{B}_{\rm R}})
                  =\dfrac{c}{4\,\pi} k^2 \vec{A}_{01}^2\,\overline{
                        {\sin^2}(\vec{k}\cdot\vec{r}-\omega\,t)}\,\hat{k}\nonumber\\
                  &=&\dfrac{c}{4\,\pi} k^2 \vec{A}_{01}^2\,\dfrac{1}{T} \Big[\int_0^T
                        {\sin^2}(\vec{k}\cdot\vec{r}-\omega\,t){\rm d}t\Big]\hat{k}
                        \notag \\
                  &=& \dfrac{c}{4\,\pi} k^2 \vec{A}_{01}^2\,\dfrac{1}{\omega\,T}
                        \Big[\int_0^{\omega\,T} {\sin^2}(\omega\,t
                              -\vec{k}\cdot\vec{r}){\rm d}(\omega\,t)
                              \Big]\hat{k}\nonumber\\
                  &=&\dfrac{c}{4\,\pi} k^2 \vec{A}_{01}^2\,\dfrac{1}{\omega\,T} \Big[
                        \int_{-\vec{k}\cdot\vec{r}}^{\omega\,T-\vec{k}\cdot\vec{r}}
                              {\sin^2}(\omega\,t-\vec{k}\cdot\vec{r}){\rm d}(
                                    \omega\,t-\vec{k}\cdot\vec{r})\Big]\hat{k} \notag
                        \\
                  &=& \dfrac{c}{4\,\pi} k^2 \vec{A}_{01}^2\,\dfrac{1}{\omega\,T}
                        \Biggl(\int_{-\vec{k}\cdot\vec{r}}^{
                                    \omega\,T-\vec{k}\cdot\vec{r}}
                              \biggl\{\dfrac{1-\cos\big[2(\omega\,t
                                    -\vec{k}\cdot\vec{r})\big]}{2}\biggr\}{\rm d}(
                                          \omega\,t-\vec{k}\cdot\vec{r})
                              \Biggr)\hat{k} \notag \\
                 &=& \dfrac{c}{8\,\pi} k^2 \vec{A}_{01}^2\,\hat{k},
            \end{eqnarray}
            since $T=2\,\pi/\omega$.

            Likewise, for the angular-momentum waves 
            \begin{eqnarray}
                  && \vec{\mathcal{B}}={\rm i}(\vec{k}\times\vec{\tau}){\rm e}^{
                              {\rm i}(\vec{k}\cdot\vec{r}-\omega\,t)}
                        +g\,\hbar\,\vec{\eta}\,{\rm e}^{
                              {\rm i}\,2(\vec{k}\cdot\vec{r}-\omega\,t)}, \notag \\
                  && \vec{\mathcal{E}}=\vec{\mathcal{B}}\times\hat{k}
                        ={\rm i}\big[(\vec{k}\times\vec{\tau})\times\hat{k}
                              \big]{\rm e}^{{\rm i}(\vec{k}\cdot\vec{r}-\omega\,t)}
                        +g\,\hbar(\vec{\eta}\times\hat{k}){\rm e}^{
                              {\rm i}\,2(\vec{k}\cdot\vec{r}-\omega\,t)}.
            \end{eqnarray}
            we have
            \begin{eqnarray}
                  \vec{\mathcal{B}}_{\rm R} &\equiv& (\vec{k}\times\vec{\tau})
                                    {\rm Re}\big[{\rm i}{\rm e}^{
                                          {\rm i}(\vec{k}\cdot\vec{r}-\omega\,t)}\big]
                              +g\,\hbar\,\vec{\eta}\,{\rm Re}\big[{\rm e}^{
                                    {\rm i}\,2(\vec{k}\cdot\vec{r}-\omega\,t)}\big] \nonumber\\
                       & = &-(\vec{k}\times\vec{\tau})\sin(\vec{k}\cdot\vec{r}-\omega\,t)
                              +g\,\hbar\,\vec{\eta}\,\cos\big[2(\vec{k}\cdot\vec{r}
                                    -\omega\,t)\big]\notag \\
                  \vec{\mathcal{E}}_{\rm R} &=& \vec{\mathcal{B}}_{\rm R} \times\hat{k}
                        =-\big[(\vec{k}\times\vec{\tau})\times\hat{k}\big]\sin(
                                    \vec{k}\cdot\vec{r}-\omega\,t)
                              +g\,\hbar(\vec{\eta}\times\hat{k})\cos\big[
                                    2(\vec{k}\cdot\vec{r}-\omega\,t)\big].                       
            \end{eqnarray}
            Notice
            \begin{align}
                  & \big[(\vec{k}\times\vec{\tau})\times\hat{k}\big]\times(
                        \vec{k}\times\vec{\tau})
                  =\big[(\vec{k}\times\vec{\tau})\cdot(\vec{k}\times\vec{\tau})
                        \big]\hat{k}-\big[(\vec{k}\times\vec{\tau})\cdot\hat{k}\big](
                              \vec{k}\times\vec{\tau})
                  =\big[(\vec{k}\times\vec{\tau})\cdot(\vec{k}\times\vec{\tau})
                        \big]\hat{k},
            \end{align}
            \begin{align}
                  & (\vec{\eta}\times\hat{k})\times\vec{\eta}
                  =(\vec{\eta}\cdot\vec{\eta})\hat{k}-(\vec{\eta}\cdot\hat{k})\vec{\eta}
                  =(\vec{\eta}\cdot\vec{\eta})\hat{k},
            \end{align}
            and
            \begin{eqnarray}
                   (\vec{\eta}\times\hat{k})\times(\vec{k}\times\vec{\tau})
                        +\big[(\vec{k}\times\vec{\tau})\times\hat{k}\big]\times\vec{\eta}
                 & =&\big[\vec{\eta}\cdot(\vec{k}\times\vec{\tau})\big]\hat{k}
                        -\vec{\eta}\big[\hat{k}\cdot(\vec{k}\times\vec{\tau})\big]
                        +\big[(\vec{k}\times\vec{\tau})\cdot\vec{\eta}\big]\hat{k}
                        -(\vec{k}\times\vec{\tau})(\hat{k}\cdot\vec{\eta}) \notag \\
                 & =& \big[(\vec{k}\times\vec{\tau})\cdot\vec{\eta}
                        +\vec{\eta}\cdot(\vec{k}\times\vec{\tau})\big]\hat{k},
            \end{eqnarray}
            which means
            \begin{eqnarray}
                   \vec{\mathcal{E}}_{\rm R} \times\vec{\mathcal{B}}_{\rm R} 
                  &=& \Bigl\{-\big[(\vec{k}\times\vec{\tau})\times\hat{k}\big]\sin(
                                    \vec{k}\cdot\vec{r}-\omega\,t)
                              +g\,\hbar(\vec{\eta}\times\hat{k})\cos\big[
                                    2(\vec{k}\cdot\vec{r}-\omega\,t)\big]
                        \Bigr\}\times \nonumber\\
                  &&\Bigl\{-(\vec{k}\times\vec{\tau})\sin(
                                    \vec{k}\cdot\vec{r}-\omega\,t)
                              +g\,\hbar\,\vec{\eta}\,\cos\big[
                                    2(\vec{k}\cdot\vec{r}-\omega\,t)\big]\Bigr\} \notag
                        \\
                  &=& \Bigl\{\big[(\vec{k}\times\vec{\tau})\times\hat{k}\big]\times(
                              \vec{k}\times\vec{\tau})\Bigr\}{\sin^2}(
                                    \vec{k}\cdot\vec{r}-\omega\,t)
                        +g^2 \hbar^2 \big[(\vec{\eta}\times\hat{k})\times\vec{\eta}\big]
                              {\cos^2}\big[2(\vec{k}\cdot\vec{r}-\omega\,t)\big] \notag
                              \\
                       & & -g\,\hbar\Big\{(\vec{\eta}\times\hat{k})\times(
                                    \vec{k}\times\vec{\tau})
                              +\big[(\vec{k}\times\vec{\tau})\times\hat{k}
                                    \big]\times\vec{\eta}\Big\}\sin(\vec{k}\cdot\vec{r}
                                          -\omega\,t)\,\cos\big[
                                                2(\vec{k}\cdot\vec{r}-\omega\,t)\big]
                              \notag \\
                 & =& \big[(\vec{k}\times\vec{\tau})\cdot(\vec{k}\times\vec{\tau})
                              \big]{\sin^2}(\vec{k}\cdot\vec{r}-\omega\,t)\,\hat{k}
                        +g^2 \hbar^2 (\vec{\eta}\cdot\vec{\eta}){\cos^2}\big[
                              2(\vec{k}\cdot\vec{r}-\omega\,t)\big]\,\hat{k} \notag \\
                      &  & -g\,\hbar\big[(\vec{k}\times\vec{\tau})\cdot\vec{\eta}
                              +\vec{\eta}\cdot(\vec{k}\times\vec{\tau})\big]\sin(
                                    \vec{k}\cdot\vec{r}-\omega\,t)\,\cos\big[
                                          2(\vec{k}\cdot\vec{r}-\omega\,t)
                                          \big]\,\hat{k}.
            \end{eqnarray}
           Then we have
            \begin{align}
                  & \vec{\mathbb{S}}_{\rm AMW} =\dfrac{c}{4\,\pi} (\overline{
                        \vec{\mathcal{E}}_{\rm R} \times\vec{\mathcal{B}}_{\rm R}})
                        \notag \\
                  =& \dfrac{c}{4\,\pi} \big[(\vec{k}\times\vec{\tau})\cdot(\vec{k}\times\vec{\tau})
                              \big]\overline{{\sin^2}(\vec{k}\cdot\vec{r}-\omega\,t)}\,
                                    \hat{k}
                        +g^2 \hbar^2 \dfrac{c}{4\,\pi} (\vec{\eta}\cdot\vec{\eta})\overline{{\cos^2}\big[
                              2(\vec{k}\cdot\vec{r}-\omega\,t)\big]}\,\hat{k} \notag \\
                        & -g\,\hbar\dfrac{c}{4\,\pi} \big[(\vec{k}\times\vec{\tau})\cdot\eta
                              +\eta\cdot(\vec{k}\times\vec{\tau})\big]\overline{\sin(
                                    \vec{k}\cdot\vec{r}-\omega\,t)\,\cos\big[
                                          2(\vec{k}\cdot\vec{r}-\omega\,t)\big]
                                          }\,\hat{k} \notag \\
                  =& \dfrac{1}{2} \dfrac{c}{4\,\pi} \big[(\vec{k}\times\vec{\tau})\cdot(
                              \vec{k}\times\vec{\tau})\big]\hat{k}
                        +g^2 \hbar^2 \dfrac{c}{4\,\pi} (\vec{\eta}\cdot\vec{\eta})\dfrac{1}{T} \Bigl\{
                              \int_0^T {\cos^2}\big[2(\vec{k}\cdot\vec{r}-\omega\,t)
                              \big]{\rm d}t\Bigr\}\hat{k} \notag \\
                        & -g\,\hbar\dfrac{c}{4\,\pi} \big[(\vec{k}\times\vec{\tau})\cdot\eta
                              +\eta\cdot(\vec{k}\times\vec{\tau})\big]\dfrac{1}{T}
                                    \Bigl\{\int_0^T \sin(\vec{k}\cdot\vec{r}-\omega\,t)
                                          \,\cos\big[2(\vec{k}\cdot\vec{r}-\omega\,t)
                                          \big]{\rm d}t\Bigr\}\,\hat{k} \notag \\
                  =& \dfrac{1}{2} \dfrac{c}{4\,\pi} \big[(\vec{k}\times\vec{\tau})\cdot(
                              \vec{k}\times\vec{\tau})\big]\hat{k}
                        +g^2 \hbar^2 \dfrac{c}{4\,\pi} (\vec{\eta}\cdot\vec{\eta})\dfrac{1}{4\,\omega\,T}
                              \Bigl\{\int_{-4(\vec{k}\cdot\vec{r})}^{4(\omega\,T
                                    -\vec{k}\cdot\vec{r})
                                    } {\cos^2}\big[2(\omega\,t-\vec{k}\cdot\vec{r})
                                          \big]{\rm d}\big[4(\omega\,t
                                                -\vec{k}\cdot\vec{r})\big]
                                    \Bigr\}\hat{k} \notag \\
                        & +g\,\hbar\dfrac{c}{4\,\pi} \big[(\vec{k}\times\vec{\tau})\cdot\eta
                              +\eta\cdot(\vec{k}\times\vec{\tau})
                              \big]\dfrac{1}{\omega\,T} \Bigl\{
                                    \int_{\cos(\vec{k}\cdot\vec{r})}^{
                                          \cos(\omega\,T-\vec{k}\cdot\vec{r})
                                          } \cos\big[2(\omega\,t-\vec{k}\cdot\vec{r})
                                                \big]{\rm d}\big[\cos(\omega\,t
                                                      -\vec{k}\cdot\vec{r})\big]
                                          \Bigr\}\hat{k} \notag \\
                  =& \dfrac{1}{2} \dfrac{c}{4\,\pi} \big[(\vec{k}\times\vec{\tau})\cdot(
                              \vec{k}\times\vec{\tau})\big]\hat{k}
                        +g^2 \hbar^2 \dfrac{c}{4\,\pi} (\vec{\eta}\cdot\vec{\eta})\dfrac{1}{4\,\omega\,T}
                              \Biggl(\int_{-4(\vec{k}\cdot\vec{r})}^{4(\omega\,T
                                    -\vec{k}\cdot\vec{r})
                                    } \biggl\{\dfrac{1+\cos\big[4(\omega\,t
                                          -\vec{k}\cdot\vec{r})\big]}{2}
                                          \biggr\}{\rm d}\big[4(\omega\,t
                                                -\vec{k}\cdot\vec{r})\big]
                                    \Biggr)\hat{k} \notag \\
                        & +g\,\hbar\dfrac{c}{4\,\pi} \big[(\vec{k}\times\vec{\tau})\cdot\eta
                              +\eta\cdot(\vec{k}\times\vec{\tau})
                              \big]\dfrac{1}{\omega\,T} \Bigl\{
                                    \int_{\cos(\vec{k}\cdot\vec{r})}^{
                                          \cos(\omega\,T-\vec{k}\cdot\vec{r})
                                          } \big[2\,{\cos^2}(\omega\,t
                                                -\vec{k}\cdot\vec{r})-1\big]{\rm d}\big[
                                                      \cos(\omega\,t
                                                            -\vec{k}\cdot\vec{r})\big]
                                          \Bigr\}\hat{k} \notag \\
                  =& \dfrac{c}{8\,\pi} \big[(\vec{k}\times\vec{\tau})\cdot(
                              \vec{k}\times\vec{\tau})\big]\hat{k}
                        +g^2 \hbar^2 \dfrac{c}{8\,\pi} (\vec{\eta}\cdot\vec{\eta}
                              )\hat{k}
                        +g\,\hbar\dfrac{c}{4\,\pi} \big[(\vec{k}\times\vec{\tau})\cdot\eta
                              +\eta\cdot(\vec{k}\times\vec{\tau})
                              \big]\dfrac{1}{\omega\,T} \dfrac{2}{3} {\cos^3}(\omega\,t
                                                -\vec{k}\cdot\vec{r})
                                    \bigg|_{\cos(\vec{k}\cdot\vec{r})}^{
                                          \cos(\omega\,T-\vec{k}\cdot\vec{r})}
                                    \hat{k} \notag \\
                  =& \dfrac{c}{8\,\pi} \Bigl\{\big[(\vec{k}\times\vec{\tau})\cdot(
                              \vec{k}\times\vec{\tau})\big]
                        +g^2 \hbar^2 (\vec{\eta}\cdot\vec{\eta})\Bigr\}\hat{k}.
            \end{align}
            \begin{remark}
                  When $g=0$, choose
                  \begin{align}
                        & \vec{A}_{01} =-\big[\hat{k}\times(\hat{k}\times\vec{R}_0)
                              \big]
                        =-\big[(\hat{k}\cdot\vec{R}_0)\hat{k}-\vec{R}_0\big],
                  \end{align}
                  i.e., $\vec{R}_0 =\vec{A}_{01} +(\hat{k}\cdot\vec{R}_0)\hat{k}$. Further set $\vec{\tau}=\vec{R}_0 \openone$, and it follows that
                  \begin{align}
                        & (\vec{k}\times\vec{\tau})\cdot(\vec{k}\times\vec{\tau})
                        =(\vec{k}\times\vec{R}_0)\cdot(\vec{k}\times\vec{R}_0)\openone
                        =k^2 (\hat{k}\times\vec{R}_0)\cdot(\hat{k}\times\vec{R}_0
                              )\openone \notag \\
                        =& k^2 \Bigl\{\hat{k}\times\big[\vec{A}_{01}
                                    +(\hat{k}\cdot\vec{R}_0)\hat{k}\big]
                              \Bigr\}\cdot\Bigl\{\hat{k}\times\big[\vec{A}_{01}
                              +(\hat{k}\cdot\vec{R}_0)\hat{k}\big]\Bigr\}\openone
                        =k^2 \big[(\hat{k}\times\vec{A}_{01})\cdot(
                              \hat{k}\times\vec{A}_{01})\big]\openone \notag \\
                        =& k^2 \big[(\hat{k}\times\vec{A}_{01})\cdot(
                              \hat{k}\times\vec{A}_{01})\big]\openone
                        =-k^2 \big[(\vec{A}_{01} \times\hat{k})\cdot(
                              \hat{k}\times\vec{A}_{01})\big]\openone
                        =-k^2 \vec{A}_{01} \cdot\big[\hat{k}\times(
                              \hat{k}\times\vec{A}_{01})\big]\openone \notag \\
                        =& -k^2 \vec{A}_{01} \cdot\big[\hat{k}\times(
                              \hat{k}\times\vec{A}_{01})\big]\openone
                        =-k^2 \vec{A}_{01} \cdot\big[(\hat{k}\cdot\vec{A}_{01})\hat{k}
                              -\vec{A}_{01}\big]\openone \notag \\
                        =& k^2 \vec{A}_{01}^2 \openone.
                  \end{align}
                  Then we obtain
                  \begin{align}
                        \vec{\mathbb{S}}_{\rm AMW} =\dfrac{c}{8\,\pi} \Bigl\{
                              \big[(\vec{k}\times\vec{\tau})\cdot(
                                    \vec{k}\times\vec{\tau})\big]
                              +g^2 \hbar^2 (\vec{\eta}\cdot\vec{\eta})\Bigr\}\hat{k}
                        =\dfrac{c}{8\,\pi} \big[(\vec{k}\times\vec{\tau})\cdot(
                              \vec{k}\times\vec{\tau})\big]\hat{k}
                        =\openone\dfrac{c}{8\,\pi} k^2 \vec{A}_{01}^2 \hat{k},
                  \end{align}
                  which coincides with the Poynting vector of planar electromagnetic wave, i.e., \Eq{eq:PoyntEM}.
            \end{remark}
      \end{proof}